\documentclass[a4paper,pdftex,12pt]{scrbook}
\pdfoutput=1
\usepackage{a4wide}
\usepackage{listings}
\usepackage{graphicx}
\usepackage{url}
\usepackage{subfigure}
\usepackage{placeins}
\usepackage{palatino}
\usepackage{fancyhdr}
\usepackage[Conny]{fncychap}

\ChNameUpperCase
\ChTitleUpperCase
\ChNameVar{\Huge\rm\bfseries}
\ChNumVar{\Huge}
\ChTitleVar{\Huge\rm}

\begin{document}

\makeatletter
\gdef\lst@PrintOtherKeyword#1\@empty{%
    \lst@XPrintToken
    \begingroup
      \lst@modetrue \lsthk@TextStyle
      \let\lst@ProcessDigit\lst@ProcessLetter
      \let\lst@ProcessOther\lst@ProcessLetter
      \lst@lettertrue
      #1%
      \lst@SaveToken
    \endgroup
        \lst@RestoreToken
        \global\let\lst@savedcurrstyle\lst@currstyle
        \let\lst@currstyle\lst@gkeywords@sty
    \lst@Output
        \let\lst@currstyle\lst@savedcurrstyle}
\makeatother

\lstdefinelanguage{XOCL}{
        morekeywords={context,self,throw,try,catch,let,then,
                      if,else,elseif,and,or,implies,
                      parserImport,in,end,import,not,true,false,when,do},
        otherkeywords={@,|>,<|,<,>,|,::=,{,},[|,|],[,],;,:,
                       .,*,-,+,->,=,<>,>=,<=},
        sensitive=false,
        comment=[l]//,
        morecomment=[s]{/*}{*/},
        basicstyle=\ttfamily\small,
        stringstyle=\ttfamily\small,
        frame=lines,
        aboveskip=20pt,
        belowskip=20pt,
        breaklines=true,
        showspaces=false,
        showtabs=false
}

\lstset{language=XOCL}

\frontmatter


\begin{titlepage}
 
\begin{center}

 
\textsc{}\\[2cm] 
 
\textsc{\LARGE \textsf{Applied Metamodelling}}\\[0.5cm]
 
\textsc{\Large \textsf{A Foundation for Language Driven Development}}\\[1.0cm]
\textsc{\Large \textsf{Third Edition}}\\[2cm]

\includegraphics[width=0.3\textwidth]{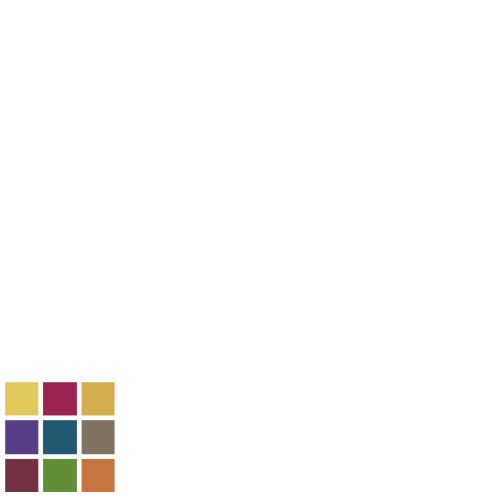}\\[2cm]
 
\large
\textsf{Tony Clark, Paul Sammut, James Willans}\\[0.7cm]
 
\vfill
 
 
\end{center}
 
\end{titlepage}

\tableofcontents
\chapter*{Preface}
\LARGE{\emph{(Third Edition)}}\\[2cm]
\normalsize
\addcontentsline{toc}{chapter}{Preface (Third Edition)}

The Third Edition of Applied Metamodelling represents a small increment since the Second Edition was produced in 2008.
The book continues to be referenced in scholarly articles with 212 citations on Google Scholar including nearly 150 since 2008.
The open-source release of the accompanying technologies XMF and XMF-Mosaic did not generate the same level of
interest, partly due to a lack of exposure and associated tutorial materials. 
Recently, interest in the field of multi-level modelling has increased
and has led to publications \cite{clark2014foundation,clark2013ocl,clark2013domain,henderson2013search,clark2012software} and a Dagstuhl Seminar \cite{clark2013meta} based on the ideas of meta-languages and tool-modelling. An overview of the historical development of XMF, the ideas in this book, and the the birth and death of an associated startup company has been published in \cite{clark2012exploiting} as part of the $10^{th}$ anniversary edition of the the Journal of Software and Systems Modeling.
A project based on the foundations developed in this book and the tooling, now rebranded XModeler, is underway and the aim is to advance the field of multi-level language-based 
system engineering.

\chapter*{Preface}
\LARGE{\emph{(Second Edition)}}\\[2cm]
\normalsize
\addcontentsline{toc}{chapter}{Preface (Second Edition)}

Since publishing the first edition in 2004 we were pleasantly surprised by how much momentum
\emph{Applied Metamodelling} created.  The fundamental tenets of the book have been cited in a large
number of papers and books, and key ideas have been influential in a number of technology
initiatives including the Eclipse Modeling Project \cite{EMP}.

A recurring response that we
commonly receive is a frustration that although the book gives readers the knowledge to apply
metamodelling, they cannot easily do this using the XMF technology which was used to construct
the examples in the book.  This is about to change with the release of XMF as an open source
technology in the first quarter of 2008.

Much has happened since 2004 but the fundamental problem of complexity in software and systems
remains.  This second version of the book makes a number of corrections to the original text and
includes a number of new chapters including two additional case studies which demonstrate the
application and benefits of metamodelling.

\chapter*{Preface}
\LARGE{\emph{(First Edition)}}\\[2cm]
\normalsize
\addcontentsline{toc}{chapter}{Preface (First Edition)}

Modern day system developers have some serious problems to contend
with. The systems they develop are becoming increasingly complex
as customers demand richer functionality delivered in ever shorter
timescales. They have to manage a huge diversity of implementation
technologies, design techniques and development processes:
everything from scripting languages to web-services to the latest
'silver bullet' design abstraction. To add to that, nothing stays
still: today's 'must have' technology rapidly becomes tomorrow's
legacy problem that must be managed along with everything else.

How can these problems be dealt with? In this book we propose that
there is a common foundation to their resolution: {\em languages}.
Languages are the primary way in which system developers
communicate, design and implement systems. Languages provide
abstractions that can encapsulate complexity, embrace the
diversity of technologies and design abstractions, and unite
modern and legacy systems.

\section*{Language-Driven Development}

Understanding how we can manage languages to best fit the needs of
developers is the key to improving system development practises.
We call this {\em Language-Driven Development}. The right
languages enable developers to be significantly more productive
than using traditional development practices. Rather than dealing
with a plethora of low level technologies, developers can use
powerful language abstractions and development environments that
support their development processes. They can create models that
are rich enough to permit analysis and simulation of system
properties before completely generating the code for the system.
They can manipulate their models and programs in significantly
more sophisticated ways than they can code. Moreover, provided the
language definitions are flexible, they can adapt their languages
to meet their development needs with relative ease.

\section*{Metamodelling}

In order to realise Language-Driven Development, we need the
ability to rapidly design and integrate semantically rich
languages in a unified way. {\em Metamodelling} is the way to
achieve this. A metamodel is a model of a language that captures
its essential properties and features. These include the language
concepts it supports, its textual and/or graphical syntax and its
semantics (what the models and programs written in the language
mean and how they behave). Metamodels unify languages because the
same metamodelling language is used in each case. Significant
advantage can be made of this unification to construct powerful
and flexible design environments for languages. These enable the
rapid assembly of Language-Driven Development tools that give
developers the power they need to design their systems faster,
cheaper and more flexibly.

\section*{Purpose of this Book}

The aim of this book is to advance the state of the art in
metamodelling to the point at which it can realise the
Language-Driven Development vision. Traditionally, metamodelling
has focused on the design of data centric models of language
concepts. In this book, we go way beyond that, showing how they
can capture all aspects of languages, including concrete syntax,
abstract syntax and semantics in a sound and pragmatic way.

This book also aims to fill a large gap in the metamodelling
literature, providing a technically rich book on an a subject that
is mentioned often, but for which there is little concrete
material available. Metamodels are increasingly being used across
wider application domains, and it is the intention that this book
will provide good advice to metamodellers irrespective of the
languages and tools they are using.

\section*{Scope of this Book}

The scope of this book is deliberately not restricted to software
systems. Many other types of domains, from systems engineering to
business, manufacturing and even physical engineering, can benefit
from the ideas presented here.

\section*{Intended Audience}

This book should be useful to anyone who has an interest in the
design of languages, language and model-driven development, and
metamodelling. Our intention is that the book has an industrial
focus, and we have tried hard to ensure that it is relevant to
real practitioners. In particular, everything in this book has
been implemented, therefore ensuring that it is has been tried and
tested.

\section*{Relationship to Other Approaches}

We do not claim that the ideas presented in this book are new.
Approaches to rapidly designing languages in flexible ways have
been around since the early days of LISP, ObjVLisp and Smalltalk.
Meta-case tools have been widely touted as a means of designing
tailored tools and languages. Efforts by the Object Management
Group (OMG) to standardise facilities for capturing language
meta-data are already influencing the way vendors build tools.
More recently, work on domain specific languages has highlighted
the benefits of rapidly designing languages targeted at specific
application domains.

In this book we have combined many of these approaches on top of
existing standards to facilitate the definition of languages in a
general and complete way. An important emphasis has been on
raising the level of abstraction at which complex aspects of
language design such as the definition of concrete syntax and
semantics are expressed. Thus we {\em model} languages, but in
sufficient detail that these models can be turned into
semantically rich development environments and tools. This
capability has not been achieved in such a complete way before.

\section*{Organisation of this Book}

This book is organised into three parts. The first and shortest
part gives an overview of challenges facing the system development
industry and proposes Language-Driven Development as a way forward
to addressing those challenges.

The middle part provides a detailed treatment of metamodelling. It
contains the following chapters:

\begin{description}
\item[Metamodelling]: introduces the key features of languages and
describes what metamodels are and how they can capture those
features. \item[A Metamodelling Facility]: presents an overview of
an executable metamodelling facility that provides a number of
powerful languages for capturing the syntax and semantics of
languages. \item[Abstract Syntax]: describes how metamodels can be
used to define the concepts that are provided by the language.
\item[Concrete Syntax]: describes how the textual and
diagrammatical syntaxes of language can be modelled. \item [
Semantics]: introduces semantics and the motivation for having
them in language definitions. The chapter goes on to describe four
different approach to describing semantics. \item[Executable
Metamodelling]: discusses how the addition of a small number of
action primitives to a metamodelling language turn it into a
powerful metaprogramming environment for Language-Driven
Development. \item[Mappings]: motivates and presents two languages
for transforming and relating metamodels. \item[Reuse]: this
chapter describes a number of different approaches to reusing
existing language metamodels.
\end{description}

The final part provides a number of indepth case studies each describing a
specific example of metamodelling.  These examples range from the design of
a small general purpose language to a domain specific language for modelling
interactive television based applications.  The case studies are a significant
resource for demonstrating metamodelling best practice.

\section*{Acknowledgements}

We are grateful to the following people for giving us feedback on
early versions of the book and the technology it is based on:
Manooch Azmoodeh, Steve Crook-Dawkings, Tom Dalton, John Davies,
Kevin Dockerill, Desmond D'Souza, Karl Frank, Nektarios Georgal,
Robert France, Sheila Gupta, Martin Hargreaves, Girish Maskeri,
Stephen Mellor, Joaquin Miller, Alan Moore, Alphonse Orosun, John
Rowlands, Bran Selic, Laurence Tratt, Andrew Watson, Stephen
Whymark.

\mainmatter

\chapter{Language-Driven Development}
\label{lddchapter}

This chapter provides an introduction to Language-Driven
Development. It outlines current problems facing software and
systems developers today, and explains how an integrated
architecture of semantically rich, evolvable languages can provide
huge productivity benefits to industry.

Language-driven development is fundamentally based on the ability
to rapidly design new languages and tools in a unified and
interoperable manner. We argue that existing technologies do not
provide this capability, but a language engineering approach based
on {\em metamodelling} can. The detailed study of metamodelling
and how it can realise the Language-Driven Development vision will
form the focus for the remainder of this book.


\section{Challenges Facing Developers Today} \label{challenges}

When discussing software and systems engineering, it is only a
matter of time before the topic of managing complexity arises. The
desire to manage complexity was the driving force behind the
emergence of the aforementioned disciplines, and despite many
valiant attempts to master it, the problem is still with us today.
However, we believe that the nature of today's systems are quite
different to those developed when those disciplines emerged, and
in turn the developers of today's systems face different
challenges to those in previous decades. In particular, it is no
longer sufficient to manage complexity alone. Instead, we believe
that most of today's development challenges boil down to a
combination of three important factors: complexity, diversity and
change.

\begin{figure}[htb]
\begin{center}
\includegraphics[width=5cm]{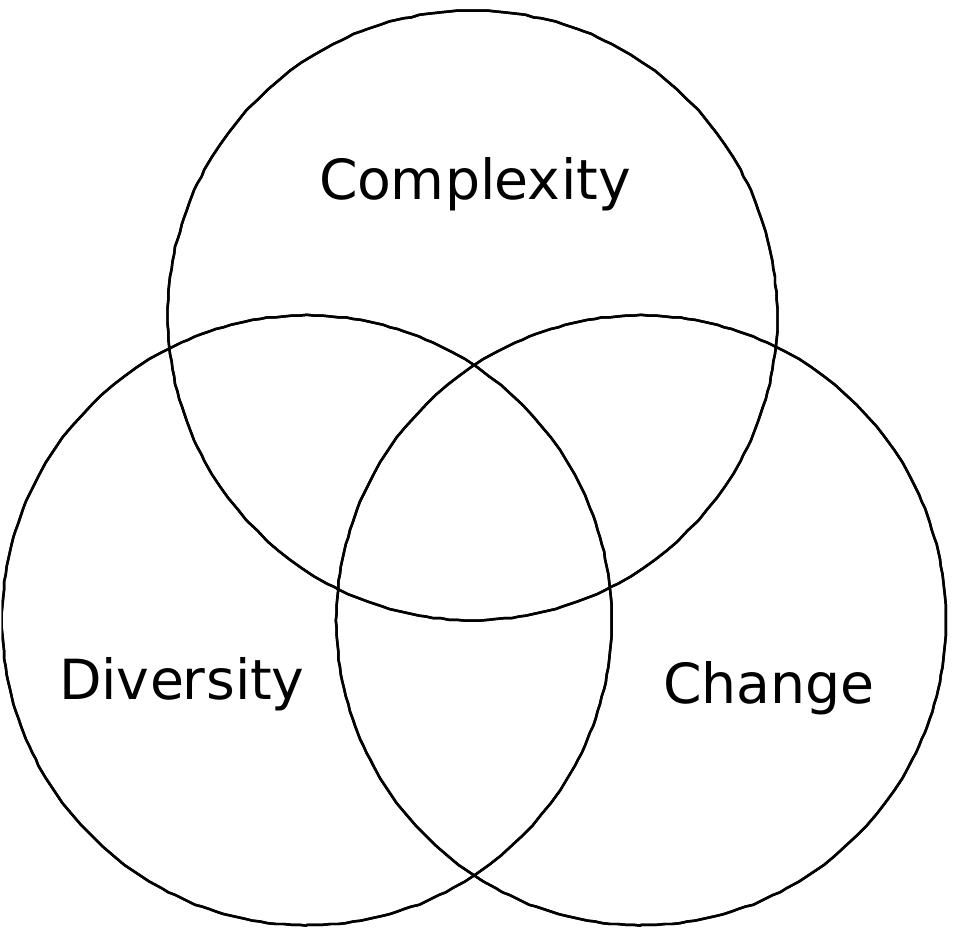}
\caption{Challenges Facing Developers Today}
\label{figchallenges}
\end{center}
\end{figure}

The remainder of this section describes each of these challenges in more detail.

\subsection{Coping with Complexity} \label{complexity}

As hardware costs have plummeted and development and manufacture
techniques have improved, the demands for more sophisticated
systems have been relentless. Of course, the more sophisticated
the requirements of a system are, the larger and more complex the
deployed system is likely to be. Increased system complexity
typically brings with it the following problems:
\begin{itemize}
\item longer development times;
\item more complex assembly due to number of components and number of people involved;
\item increased cost and time for testing;
\item increased maintenance costs.
\end{itemize}

Overall, this results in an increased time to market for any
system, and increased development and maintenance costs in order
for there to be any confidence that the quality of the system is
not compromised.

For software systems, as well as the problems outlined above which
relate to the fundamental increase in lines of code, there is an
additional qualitative difference to the systems being developed
today compared to those of decades past. Modern systems are
increasingly distributed in nature, as demonstrated by the
ubiquity of enterprise applications. This adds another dimension
to software complexity, and brings added challenges of
communication and security to those listed above.

Since the challenge of managing complexity is the main topic of
many software and systems engineering books (such as
\cite{sommerville,boochoo,jacobsonoo}, it will not be discussed in
further detail here. Potential solutions to the complexity
challenge are described in section \ref{abstraction}.

\subsection{The Challenge of Diversity} \label{diversity}

The challenge of diversity reflects how developers have to manage
in a non-homogenous environment. Life would be much easier if
there was only one programming language and one deployment
platform, but of course this is not the case, and for very good
reasons. Diversity is not really a single challenge, but a
category of challenges, outlined below. Section \ref{integration}
describes how diversity as a whole can be managed.

\subsubsection*{Diverse Domains}
The requirements of large systems often relate to a variety of
domains that need to be reconciled by different stakeholders.
These requirements may range far and wide, including functional,
safety, security and performance considerations. Each domain often
has its own specialist approach for dealing with appropriate
requirements, but their specialist nature inevitably precludes
them from being applied in other domains.

\subsubsection*{Diverse Customer Requirements}
The 'one size fits all' approach to software and systems is
increasingly inappropriate in today's market. Vendors who offer
products that can be tailored to the specific needs of a customer
have a strong commercial advantage, but developing products that
are truly customisable such that they can meet the demands of a
broad customer base is a far from trivial matter. Despite the fact
that two products being offered to different customers may share
significant functionality, large rewrites and redesign are often
required because the functional components of a system are too
tightly coupled to allow large scale reuse. In addition, many
systems are designed at too lower a level of abstraction to yield
optimum flexibility.

\subsubsection*{Diverse Implementation Technologies}
Systems are often deployed across a number of different
implementation technologies which need to be integrated, or need
to be deployed for a number of separate implementation
technologies. These implementation technologies each have their
own requirements and languages. However, the separation between
the core functionality and the requirements of the deployment
platform is rarely kept clean during the development of the
system. It has been recognised that in order to support
redeployment and integration (or indeed other customisation such
as that described above), systems need to be designed at a high
level of abstraction; thus software and system modelling has
become popular, but these models are rarely complete (this is
described more fully in section \ref{abstraction}). Software
models in particular seldom get beyond the specification of their
behaviour, such that code cannot be completely generated. Even
when code is generated, full testing and validation is usually
required, which consumes a significant chunk of development
effort.

\subsubsection*{Diverse Tools and Artefact Types}
During the development of a system, a large number of artefacts
are created, including requirements specifications, design
documentation, design and analysis models, analysis and simulation
data and of course the code (for software systems). Unfortunately,
these artefacts are often prevented from being truly valuable
assets because:
\begin{itemize}
\item they are often created by different incompatible tools or
using different languages, some of which may no longer be
supported, such that the artefacts become unmaintainable; \item
the forms of artefacts are incompatible, and many are written in
informal languages, such that there is no clear way to integrate
the information they contain; \item artefacts such as design
models are rarely kept in step with changes to the implementation
artefacts such as code, because there is no automatic way to do
so, vastly reducing their value as assets; \item the artefacts
that are kept up to date, such as code, are often tightly coupled
to the integration technology, reducing their value as reusable
assets; \item many artefacts only exist on paper rather than
electronic form, so any maintenance or integration tasks has to be
manually.
\end{itemize}

This may be fine for one-off systems, but systems are rarely built
from scratch - they are often based on existing systems, and as
such would ideally reuse as much as possible from the baseline
system.

\subsection{The Only Constant is Change} \label{change}

Nearly all systems evolve over time. Typical reasons for this are:
\begin{itemize}
\item change in customers requirements or market trends;
\item change in implementation technologies or deployment platforms;
\item support for additional functionality and features;
\item availability of more effective implementation solutions;
\item bug fixes.
\end{itemize}

One can see a parallel between the challenges of change and
diversity described above. The distinction is that diversity is
meant to reflect the potential variations of a system at one point
in time, whereas change is meant to reflect the variation of a
single system over time.

Once again the problem of managing change is intertwined with the
problem of complexity. Traditionally systems have been developed
at too lower level of abstraction, and code is not always the
ideal starting point for managing change.

Ultimately, managing change is costly and timely - system
maintenance is well known to be an expensive activity, and that is
only part of the bigger challenge of managing change. Problems are
compounded when a tool, platform or other technology involved in
the design, development and deployment of the system becomes
obsolete. It can either become even more expensive or in some
cases impossible to continue to maintain a system. It is clear
then that any technique or technology that can aid in managing
change will have a direct beneficial effect on the bottom line and
shorter lead times for delivery. This is the topic of section
\ref{evolvability}.


\section{Language-Driven Development - Providing the Solution} \label{lddvision}

\subsection{Languages are the Future} \label{languagesarethefuture}

One of the distinguishing features of being human is our use of
language. Languages are fundamental to the way we communicate with
others and understand the meaning of the world around us.

Languages are also an essential part of systems development
(albeit in a more formalised form than natural languages).
Developers use a surprisingly varied collection of languages. This
includes high-level modelling languages that abstract away from
implementation specific details, to languages that are based on
specific implementation technologies. Many of these are
general-purpose languages, which provide abstractions that are
applicable across a wide variety of domains. In other situations,
they will be domain specific languages that provide a highly
specialised set of domain concepts.

In addition to using languages to design and implement systems,
languages typically support many different capabilities that are
an essential part of the development process. These include:
\begin{itemize}
\item \emph{Execution:} allows the model or program to be tested, run and deployed;
\item \emph{Analysis:} provides information of the properties of models and programs;
\item \emph{Testing:} support for both generating test cases and validating them must be provided;
\item \emph{Visualisation:} many languages have a graphical syntax, and support must be provided for this via the user interface to the language;
\item \emph{Parsing:} if a language has a textual syntax, a means must be provided for reading in expressions written in the language;
\item \emph{Translation:} languages don't exist in isolation. They are typically connected together whether it is done informally or automatically through code generation or compilation;
\item \emph{Integration:} it is often useful to be able to integrate features from one model or program into another, e.g. through the use of configuration management.
\end{itemize}

Languages are the true universal abstractions, and hold the key to
managing the challenges described in section \ref{challenges}.
This section describes how particular facets of languages can help
to solve the individual problems described above, and how they can
combine to from the holistic solution of Language-Driven
Development.

\subsection{Rich Organised Abstraction} \label{abstraction}

Abstraction has long been used as a means to allow humans to cope
with complexity. Abstraction concerns distilling the essential
characteristics of something relative to a particular perspective
of the viewer. The two key ideas here are that some non-essential
details are ignored, and that a particular context needs to be
defined in order for the abstraction to make sense. Often
abstraction involves the `chunking' and organisation of
information in a particular problem domain in order to allow the
viewer to better comprehend the problem, by separating concerns.
It is this information chunking that is the fundamental means for
overcoming the limited human capacity for complexity, and
languages are the means for capturing abstractions.

Organised abstraction is the key tool that formed the basis of the
disciplines of software and systems engineering from the outset
right through to recent trends in model-driven development.
However, there has been a backlash against modelling and concerns
that high-level abstraction doesn't work for complex large scale
systems. This has come from a recognition that current
model-driven technologies have failed to deliver the increased
productivity that was promised. However, we argue that it
abstraction is still a crucial tool - it's just that the wrong
abstractions have been used in the past.

This is partly because there has been a tendency for inappropriate
languages to be used for capturing abstractions - this is covered
in section \ref{multipleLanguages}. More significantly, modelling
languages often use 'high-level' as an excuse to suggest that
their abstractions need not be unambiguous, complete, meaningful
or executable. This simply does not work. Abstraction is a means
of \emph{hiding} detail appropriately from various stakeholders,
but that detail must still be there. Also, if such abstractions
have no meaning or its meaning is ambiguous, then the potential
applications on that abstraction are severely limited -
validation, verification, translation, integration, execution and
simulation rely heavily on semantically precise abstractions.

\subsection{Appropriate Abstraction Through Multiple Languages} \label{multipleLanguages}

Section \ref{diversity} highlighted how diversity lies at the
heart of modern system development. Going a step further, we
suggest that the challenge really boils down to a diversity of
languages:

\begin{itemize}
\item specialists require languages to address the particular facets of the problem that lie within their domain - often within each specialist domain there are numerous languages, with new languages being developed all the time;
\item there are countless implementation languages - some differences are due to the continuing trend of increasing abstraction, some are due to the fact that different paradigms or individual languages are better suited to a particular problem-solution pair than others, and some are simply down to competitive commercial interests;
\item the languages and syntax that capture the artefacts created during the development lifecycle.
\end{itemize}

We propose that rather than trying to subdue this diversity by
forcing everyone to talk (or model) using the same language, we
should embrace it and allow everyone to use whatever language best
suits their needs. In many cases, this may be a general-purpose
modelling or programming language, as these will be widely
supported by tools and techniques, but in some cases more
specialised languages may be more appropriate. An example of this
might be an inventory-based system, where developers consistently
have to express their models in terms of inventory type concepts
such as resources, services and products. By allowing engineers
and domain experts to express themselves in the languages that
they are both most comfortable with and that will give them the
most expressive power, productivity can increase with
corresponding gains for industry.

The argument against this is that by having a single standard
language, there is only one language for developers to learn, so
everyone can communicate more easily, and interoperability between
tools will be much simpler. Whilst this is undoubtedly true, in
order to make a modelling language that suits the needs of
everybody (not just software engineers), it will suffer from the
following problems:
\begin{itemize}
\item it will necessarily be a very large, bloated language;
\item there are often contradictory needs of a language from different domains that cannot be reconciled in a single language;
\item any gain made in widening the applicability of a language to different domains will be at the expense of the richness of the language that makes it so suitable for a particular domain.
\end{itemize}

The compromises that can happen due to conflicting requirements of
a language can be seen clearly in programming languages. These
languages sit uncomfortably between the realms of the computer
hardware and the human developer. As humans, we want readable,
maintainable and reusable code, but ideally we also want to
produce a set of efficient machine instructions to keep the
computer happy. The trend of increasing abstraction that has
resulted in Object-Oriented Programming has resulted in more
usable languages, but at the expense of performance. Thus a single
language is not enough.

\subsection{Integration - Weaving the Rich Tapestry of Languages} \label{integration}

In section \ref{multipleLanguages}, we suggested that large
productivity gains could be achieved by opening up the full
spectrum of languages. However, this vision will not work with
isolated language islands - we need to find a way for all the
models and other artefacts for a system written in these disparate
languages to make sense as a meaningful whole. In order for that
to happen, the languages themselves must be integrated.

Language integration between two languages means that some or all
 of the language constructs of each language are in some way mapped
to corresponding constructs of the other language. Some common
applications of language integration are outlined below:

\subsubsection*{Transformation}
The most publicised application of language integration is that of
transformation, where an artefact in one language is transformed
into an artefact of another. This type of activity is of prime
importance in MDA (see section \ref{mda}), as reflected by
dominance of the concept of transforming Platform Independent
Models (PIMs) to Platform Specific Models (PSMs). Language-Driven
Development goes a step further by enabling high-level models to
be transformed directly into fully compiled executable systems, so
long as the appropriate languages that capture such views of a
system are integrated appropriately. Transformation activities
also include reverse engineering, and generation of any secondary
artefacts such as documentation or even full test beds for
systems.

\subsubsection*{Artefact Integration}
If a system is comprised of a number of subsystems from different
domains, and these different system aspects are described in
different languages, then by integrating the languages, those
aspects can themselves be weaved together to form a unified view
of the system. This is typically used to integrate language
artefacts that are at a similar level of abstraction.

\subsubsection*{Equivalence Verification}
Sometimes it is important to check whether an artefact written in
one language is equivalent to one written in another. For example,
it may be important to check whether an implemented system
conforms to a precise system specification written in a high-level
language. Again if the two languages are integrated appropriately,
then this will be possible.

\subsubsection*{Synchronisation}
Language integration can also enable language artefacts to be
synchronised. For example, whilst in some cases it might be
appropriate to generate code for a system from a high-level model
as a one-shot activity, in many cases it is desirable to keep the
model and code in step. Similarly, if you have a diagramming
language that allows graphical representation of a modelling
languages, it is important to keep the graphical entities in step
with any changes to the model.

\subsection{Evolvability - The Key to Managing Change} \label{evolvability}

Languages evolve in the same way as systems, and the best way to
protect systems against change and obsolescence is to protect the
languages that describe them. They should be flexible and
extensible to cope with changing requirements, and when a new
version of a language is developed, mappings (as described in
section \ref{integration} should be provided to provide full
traceability between versions. In this way, an artefact written in
an earlier version should able to be transformed into the new
version. With some legacy systems, the language is no longer
well-supported. In these cases, the language should be described
and integrated within the Language-Driven Development framework,
such that corresponding artefacts can be transformed into
artefacts in a new more current and appropriate language. In
summary, good language design (see Chapter \ref{langchapter})
together with language integration enables both languages and
systems to evolve in a controlled fashion.

\subsection{Language-Driven Development - The Complete Solution} \label{lddComplete}

This section has described how languages can provide the overall
solution to the challenges described in section \ref{challenges} -
more specifically an integrated framework of semantically rich,
flexible and evolvable languages appropriate to their needs. This
Language-Driven Development framework will:
\begin{itemize}
\item allow the contruction of agile abstractions that are resistant to change;
\item enable those abstractions to be transformed into, integrated with, validated against or synchronised with abstractions written in other languages;
\item support powerful applications (editors, analysis and simulation tools, the aforementioned transformers and integrators) to be written and applied on those abstractions.
\end{itemize}

The right languages enable developers to be significantly more
productive  than using traditional development technologies
because engineers and domain experts can speak in the languages
they understand. Rather than dealing with low level coding issues,
developers can use powerful language abstractions and development
environments that support their development processes. They can
create models that are rich enough to permit analysis and
simulation of system properties before completely generating the
code for the system, and are more reusable and agile. They can
manipulate their models and programs in significantly more
sophisticated ways than they can code. Moreover, provided the
language definitions are flexible, they can adapt their languages
to meet their development needs with relative ease.

Language-Driven Development is the next generation development
paradigm which can provide a step gain in productivity through the
recognition that languages, rather than objects or models, are the
abstractions needed for today's development environment.


\section{From Model-Driven to Language-Driven Development} \label{mdd2ldd}

Much of what has been described in this chapter has a lot in
common with  model-driven development approaches such as the OMG's
Model Driven Architecture (MDA). However, there are two prime
motivations for distinguishing Language-Driven Development from
model-driven approaches:
\begin{itemize}
\item the term \emph{model} suggests a focus on high-level abstractions and \emph{modelling} languages, with other artefacts seen as of lesser value. We feel that languages itself are the truly central abstractions, and that modelling languages form an undoubtedly useful yet partial subset of the spectrum of useful languages in system development. Consequently, all language artefacts, not just models, have a crucial role to play in the process;
\item the prominent model-driven approach, MDA, is limited in its scope of application, compared to the full potential of Language-Driven Development (see section \ref{mda}.
\end{itemize}

The remainder of this section examines two key model-driven
technologies from the Object Management Group, UML and MDA, and
assesses their suitability for the basis of Language-Driven
Development.\footnote{Two other key technologies underpinning MDA
are the Meta-Object Facility (MOF) and the
Query/View/Transformations language (QVT). MOF is the
metamodelling language for MDA, and QVT is the mappings language
for MOF - both are described in section \ref{differences}.}

\subsection{The Unified Modelling Language} \label{uml}

The Unified Modelling Language (UML) came out of a desire to
consolidate  all the notations in the various object-oriented
methodologies that had arisen in the eighties and nineties, such
as Schlaer-Mellor \cite{schlaermellor} and OMT \cite{omt}. UML
consists of a number of different notations that allow different
views of a software system to be modelled at different stages of
the development lifecycle. Both static and dynamic aspects of a
system can be captured, and facilities are also provided to enable
model management and limited extensibility. A textual constraint
language (OCL) is included to allow the state space represented by
a model to be further constrained in ways that are too complex to
be captured by the graphical notations alone.

As highlighted earlier, there are certainly advantages of having a
common language such as UML, particularly with regard to
communication. In line with this, UML has been well-received and
is now the \emph{de facto} software modelling language. However,
it has some major shortcomings:

\subsubsection*{Imprecise semantics}

The UML 1.x specification \cite{umlspec} falls some way short of
providing a precise semantics. Whilst its syntax is mostly well
specified, the semantics of those syntactic elements is either
missing or provided informally using English. This has led to a
situation where, as of version 1.3, no tool could claim to be UML
compliant \cite{pumlrfi}. This in turn has inhibited model
interchange between tools, leading back to the situation of vendor
lock-in. In addition, as explained earlier, models written in such
an informally specified language are open to misinterpretation, a
potentially dangerous or expensive problem. Whilst a major
revision of UML will be released soon, draft versions of the UML
2.0 standard do not indicate major improvements with regard to
semantics.

\subsubsection*{Limited scope and flexibility}

UML has been successfully applied across the software development
community, and it is increasingly being applied to non-software
domains such as systems engineering\cite{umlse}, and specialised
software domains such as real time and high integrity systems. The
diverse modelling requirements that this widespread use brings
makes defining what a unified modelling language should be a
considerable problem. Early attempts to enhance UML to support new
requirements adopted a 'mud-packing' approach\cite{kobryn}, which
involved making direct amendments to the monolithic definition of
UML itself. This resulted in a language that became increasingly
large, unwieldy to use, incomprehensible, and difficult to
maintain and test for consistency.

In order to overcome these problems, UML was refactored from a one-size-fits-all modelling language into a family of languages. The foundation of the UML family is a stable core UML metamodel, consisting of minimal modelling concepts that are supported by all family members. Each dialect of UML consists of the UML core
metamodel and one or more extensions to the core known as `profiles'. The profile mechanism is quite straightforward to apply, but is limited as it is based upon constraining existing language constructs rather then modifying or adding new language constructs.

\subsubsection*{Non-executability}

UML is not in itself executable - it was designed to be a
declarative  language. In other words you cannot run a UML model
as defined in the specification, merely define a specification to
which any executable program must conform. This is certainly
useful, but does not (in the general case) allow executable code
to be generated automatically from the model. This was deemed to
be a desirable application of models, so an Action Semantics
extension was provided. Whilst this was a step in the right
direction, like much of UML, the semantics of this extension is
weakly defined.

\vspace{1cm} \noindent These shortcomings are constantly being
addressed by revisions of the language. At the time of writing,
UML 2.0 is due to be released in the near future. This addresses
some of the problems of imprecise semantics, and improves the
profile mechanism of UML 1.4, but it is still limited by the
fundamental flaw of trying to have a one-size-fits-all language.
UML started out as general purpose object-oriented modelling
language, and was good at describing high level object-oriented
software models. But as a consequence of its popularity, attempts
were made to tailor for more and more highly specialised uses for
which it was not originally intended. Developing it as an
extensible language was a major step forward, but the core that
the profiles are built upon is still an object-oriented core,
which does not suit the needs of all languages. We are not
proposing that UML should be scrapped, simply that it used where
it makes sense to use it - and use other languages where the
abstractions provided by UML do not fit.

\subsection{MDA} \label{mda}

The Model Driven Architecture is framework for unifying a number
of technologies based around OMG standards such as UML, MOF, CWM
and CORBA. It is founded on the metamodelling language MOF, which
is used to define other languages such as UML and CWM.

Primarily MDA concerns models and mappings between those models.
The most widely recognised application of MDA is the mapping or
transformation between Platform Independent Models (PIMs) and
Platform Specific Models (PSMs). A key idea is that system models
are constructed that realise all the functional requirements, but
are completely independent of platform, programming language and
other implementation issues (PIMs). Instead of producing code for
a system manually, a model that contains all the constructs and
details needed for the system to operate on the intended
implementation technology (the PSM) is generated from the
appropriate PIM using a mapping. Because the core functionality of
a system is captured in the PIM, if that system needs to be
deployed on to a new platform, a new PSM can be generated simply
by changing the PIM to PSM mapping. Thus faster platform migration
and platform independence are achieved through the large scale
reuse that PIMs provide \cite{mdaexec}.

MDA is an ambitious vision that could change the way software is
developed in the future. However, as with UML, it has some
problems:
\begin{itemize}
\item whilst the MDA vision is grand, the technology for implementing it is very vaguely specified. So weak in fact that any modelling tool which has some simple code generation facility can (and in most cases does) claim to implement MDA. MDA is more useful as a marketing tool than anything else;
\item MDA is too fixed on the notion of \emph{platform}. What constitutes a \emph{platform} is unclear at best - the transition from the most abstract model of a system to the most refined model may include several stages of models, each which could considered Platform Specific when compared to the previous stage, or Platform Independent when compared to the following stage. In any case, PIM to PSM mappings are just one of a whole spectrum of potential applications of Language-Driven Development;
\item MDA is built on a weak inflexible architecture. This will be discussed in the context of metamodelling in section \ref{metaarch}.
\end{itemize}

Language-Driven Development is not just about PIM to PSM mappings
- it is about being able to capture all aspects of the software
and systems development process in a unified way, through the rich
tapestry of languages described in section \ref{lddvision}.





\section{Language Engineering and Metamodelling}

In order for a development process to be truly adaptable, it is
not simply  a case of enabling it to support a number of
pre-defined languages. If a development process limits itself to
the application of a fixed set of languages, it will still
necessarily limit the range of problems that it can address as
well as the potential solutions it can provide. Instead, a
development process should incorporate the ability to adopt and
construct whatever languages provide the best fit. In other words,
on top of the disciplines of Software and System Engineering,
there needs to be a new discipline for Language Engineering.

Language engineering is required whenever the integrated language
framework does not support the problem-solution pair. For example,
if Language-Driven Development is required on a problem domain
that has its own specialist language or if a new programming
language is developed, then that language must be captured in an
appropriate form to support Language-Driven Development
technologies. However language engineering involves not just the
construction of semantically rich languages for capturing
appropriate abstractions (section \ref{abstraction}). It also
involves the integration of such languages within the language
framework (section \ref{integration}) and the evolution of such
languages (section \ref{evolvability}). Thus language engineering
provides the foundation for all we have described in this chapter.

Language engineering is a more complex activity than software and
system engineering needing specialised skills, however only a
fraction of Language-Driven Development practitioners will be
involved in this activity. For most system developers, it will be
sufficient to know that languages need not be static entities, and
that languages can be customised, extended and created as needed.
Some of these language engineering tasks they may be able to carry
out themselves, and some (particularly the creating of new
languages entirely) will have to be carried out by language
specialists.

In order to be able to engineer languages, we need a language for
capturing, describing and manipulating all aspects of languages in
a unified and semantically rich way. This language is called a
metamodelling language. Metamodels (models of languages) are the
primary means by which language engineering artefacts are
expressed, and are therefore the foundation for Language-Driven
Development. While we have motivated Language-Driven Development
in this chapter, the rest of the book will explore how
metamodelling (the process of creating metamodels) can realise the
Language-Driven Development vision.


\section{Conclusion}

This chapter has outlined some of that the key challenges facing
developers today are complexity, diversity and change. It has
proposed that Language-Driven Development can help developers to
manage these challenges by utilising the following tools:

\begin{itemize}
\item abstraction through rich languages helps to manage complexity;
\item integration of multiple appropriate languages help to manage diversity;
\item flexible, evolvable languages help manage change.
\end{itemize}

An outline as to how Language-Driven Development differs from
model-driven development was then given, along with an overview of
existing model-driven technologies and their limitations. The
chapter closed with an introduction to the discipline of Language
Engineering, which this book is fundamentally about, and is
described in more detail in the following chapter.

Language-Driven Development provides practitioners with an
integrated framework of rich evolvable languages appropriate to
their needs. Productivity can be increased because engineers and
domain experts can speak in the languages they understand, and
both the problem space and solution space are opened up to their
full extent, and artefacts developed in this way will be more
agile, powerful, reusable and integrated. This approach offers a
paradigm shift beyond object-oriented programming and modelling
that has major implications for industry in terms of cost
reduction and productivity.

\chapter{Metamodelling}
\label{metamodellingChapter}

\section{Introduction}

The previous chapter described the benefits of using semantically
rich languages to precisely capture, relate and manipulate
different aspects of a problem domain. These languages may be
general purpose languages, domain specific languages, modelling
languages or programming languages. In order to realise these
benefits, a way must be found of defining languages in a unified
and semantically rich way. In this chapter we begin exploring a
means of achieving this using {\em metamodels}.

This chapter sets out to explain a number of key aspects of
metamodelling that lay the foundation for the rest of this book.
An important starting point is to understand the features of
languages that a metamodel must be capable of describing. A
definition of a metamodel is then given, and the type of language
necessary to construct metamodels with is explored. This language,
a metamodelling language, is just another example of a language.
As we will shall see later in the book, all language metamodels
can be described in this language: thus facilitating the unified
definition of the languages that underpins Language-Driven
Development.

\section{Features of Languages} \label{syntaxsemantics}

Whilst the nature, scope and application of the languages used in
systems development is naturally diverse, there are a number of
key features they all share. Understanding these features is a
first step towards developing a generic approach to modelling
languages.

\subsection{Concrete Syntax}

All languages provide a notation that facilitates the presentation
and construction of models or programs in the language. This
notation is known as its \emph{concrete syntax}. There are two
main types of concrete syntax typically used by languages: textual
syntax and visual syntax.

A textual syntax enables models or programs to be described in a
structured textual form. A textual syntax can take many forms, but
typically consists of a mixture of declarations, which declare
specific objects and variables to be available, and expressions,
which state properties relating to the declared objects and
variables. The following Java code illustrates a textual syntax
that includes a class with a local attribute declaration and a
method with a return expression:

\lstset{language=Java}
\begin{lstlisting}
public abstract class Thing
  {
  private String nameOfThing;
  public String getName()
    {return nameOfThing;}
  }
\end{lstlisting}
\lstset{language=XOCL}

An important advantage of textual syntaxes is their ability to
capture complex expressions. However, beyond a certain number of
lines, they become difficult to comprehend and manage.

A visual syntax presents a model or program in a diagrammatical
form. A visual syntax consists  of a number of graphical icons
that represent views on an underlying model. A good example of a
visual syntax is a class diagram, which provides graphical icons
for class models. As shown in Figure \ref{classDiagramView} it is
particularly good at presenting an overview of the relationships
and concepts in a model:

\begin{figure}[htb]
\begin{center}
\includegraphics[width=6cm]{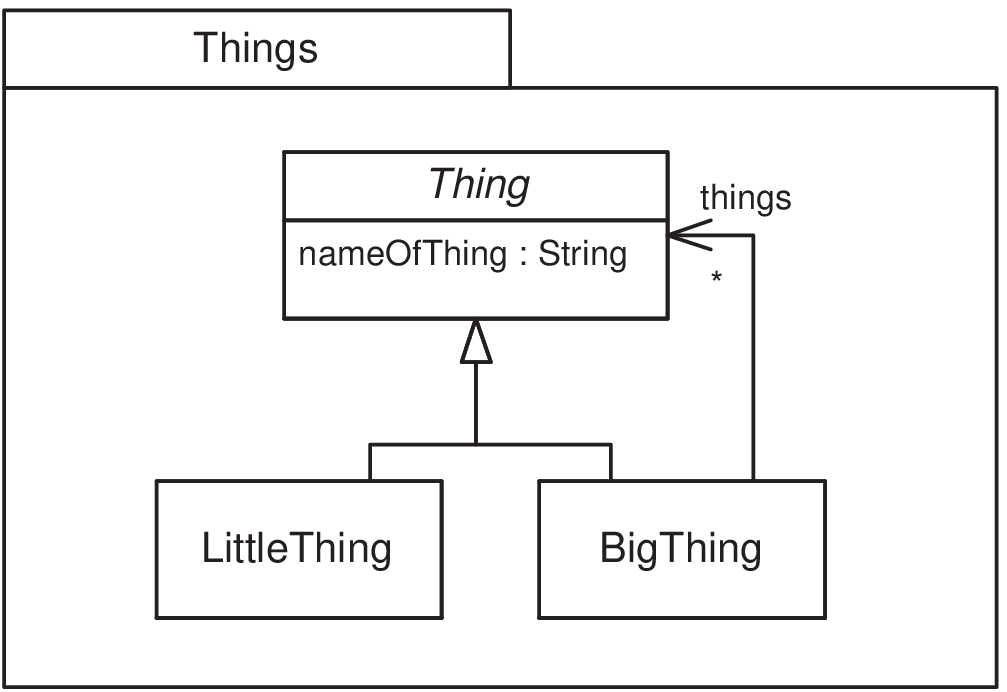}
\caption{Visualising concepts and relationships}
\label{classDiagramView}
\end{center}
\end{figure}

The main benefit of a visual syntax is its ability to express
large amounts of detail in an intuitive and understandable form.
Its obvious weakness is that only certain levels of detail can be
expressed beyond which it becomes overly complex and
incomprehensible.

In practice, utilising a mixture of diagrammatical and textual
syntaxes gains the benefits of both forms of representation. Thus,
a language will often use visual notations to present a higher
level view of the model, whilst textual syntax will be used to
capture detailed properties.

\subsection{Abstract Syntax}

The {\em abstract syntax} of a language describes the vocabulary
of concepts provided by the language and how they may be combined
to create models. It consists of a definition of the concepts, the
relationships that exist between concepts and well-formedness
rules that state how the concepts may be legally combined.

Consider a simple state machine language. An abstract syntax model
of this language may include concepts such as State, Transition
and Event. In addition, there will be relationships between
concepts, such as a Transition being related to a source and
target State. Finally, well-formedness rules will be defined that
ensure, for example, that no two transitions may be triggered by
the same event.

It is important to emphasise that a language's abstract syntax is
independent of its concrete syntax and semantics. Abstract syntax
deals solely with the form and structure of concepts in a language
without any consideration given to their presentation or meaning.

\subsection{Semantics}

An abstract syntax conveys little information about what the
concepts in a language actually mean. Therefore, additional
information is needed in order to capture the semantics of a
language. Defining a semantics for a language is important in
order to be clear about what the language represents and means.
Otherwise, assumptions may be made about the language that lead to
its incorrect use. For instance, although we may have an intuitive
understanding of what is meant by a state machine, it is likely
that the detailed semantics of the language will be open to
misinterpretation if they are not defined precisely. What exactly
is a state? What does it mean for transition to occur? What
happens if two transitions leave the same state. Which will be
chosen? All these questions should be captured by the semantics of
the language.

It is critical that semantics should be captured in a way that is
precise and useful to the user of the language. An abstract
mathematical description has little benefit if it cannot be
understood or used. Instead, a semantic definition that provides
rich ways of interacting with the language should be the goal of
the language designer: An executable language should have an
operational semantics that allows it be run; A language which
contains type concepts, such as classes, should permit the
creation of objects according to the rules of instantiation, and
so on.

\subsection{Mappings} In the real world, languages
do not exist in isolation. They will have a relationships to other
languages. This may be via translation (concepts in one language
are translated into concepts in another language); semantic
equivalence (a language may have concepts whose meaning overlaps
with concepts in another language) or abstraction (a language may
be related to another language that is at a different level of
abstraction). Capturing these relationships is an important part
of a language's definition as it serves to place the language in
the context of the world around it. Furthermore, mappings exist
between the internal components of languages, such as between a
concrete and abstract syntax, and are an important part of a
language's architecture (see section \ref{families}).

\subsection{Extensibility} Languages are not static entities: they
change and evolve over time. For instance, new concepts may be
added that enable common patterns of model or code to be expressed
more succinctly, whilst unused elements of the language will
eventually die out. The ability to extend a language in a precise
and well-managed way is vital in order to be able to support
adaptability. It allows the language to adapt to new application
domains and to evolve to meet new requirements. Furthermore,
extensibility enables the commonality and differences between
languages to be precisely captured.

\section{Modelling Languages vs. Programming Languages}
\label{progvsmodel} A strong distinction has traditionally been
made between modelling languages and programming languages (a fact
reflected by the two distinct modelling and programming
communities!). One reason for this is that modelling languages
have been traditionally viewed as having an informal and abstract
semantics whereas programming languages are significantly more
concrete due to their need to be executable.

This is not the case in this book. Here, we view modelling
languages and programming languages as being one and the same.
Both have a concrete syntax, abstract syntax and semantics. If
there is a difference, it is the level of abstraction that the
languages are targeted at. For instance, UML tends to focus on
specification whilst Java emphasises implementation. However, even
this distinction is blurred: Java has been widely extended with
declarative features, such as assertions, whilst significant
inroads have been made towards developing executable versions of
UML.

Another common distinction made between modelling and programming
languages is their concrete syntax. Modelling languages tend to
provide diagrammatical syntaxes, whilst programming languages are
textual. However, the representational choice of a language should
not enforce this distinction. There is nothing to say that a
modelling language cannot have a textual syntax or that
programming language cannot have a visual syntax: it is purely a
matter of representational choice. Indeed there is already a human
readable textual form of UML and tools that provide visual front
ends to programming languages like Java are commonplace.

If modelling languages and programming languages are essentially
the same, why can't the mature techniques used to define
programming languages be used to design modelling languages? The
answer is that they can - indeed many of the techniques presented
here have their foundation in programming language design.
However, there is one important element that is missing from many
approaches to defining programming languages, and that is
unification. It is the ability to define multiple languages that
co-exist in a unified meta-architecture that make metamodelling
such a powerful technology.

Thus, the techniques that are developed in this book are equally
as applicable to programming languages as they are to modelling
languages. A critical failing of modelling languages is that they
have not, until now, been given the precise, executable
definitions that programming languages enjoy.

\section{What is a Metamodel?}

In its broadest sense, a metamodel is a model of a modelling
language. The term "meta" means transcending or above, emphasising
the fact that a metamodel describes a modelling language at a
higher level of abstraction than the modelling language itself.

In order to understand what a metamodel is, it is useful to
understand the difference between a metamodel and a model. Whilst
a metamodel is also a model (as defined in chapter
\ref{lddchapter}), a metamodel has two main distinguishing
characteristics. Firstly, it must capture the essential features
and properties of the language that is being modelled. Thus, a
metamodel should be capable of describing a language's concrete
syntax, abstract syntax and semantics. Note, how we do this is the
major topic of the rest of this book!

Secondly, a metamodel must be part of a {\em metamodel
architecture}. Just as we can use metamodels to describe the valid
models or programs permitted by a language, a metamodel
architecture enables a metamodel to be viewed as a model, which
itself is described by another metamodel. This allows all
metamodels to be described by a single metamodel. This single
metamodel, sometimes known as a meta-metamodel, is the key to
metamodelling as it enables all modelling languages to be
described in a unified way. How metamodels can be {\em described}
by a meta-metamodel is discussed in more detail in section
\ref{metaarch}.

It is important to be aware that there is a good deal of confusion
about what is meant by a metamodel in the literature. Many
standards such as UML \cite{umlspec}, CWM \cite{cwmspec} and MOF
\cite{mofspec} provide `metamodels' that claim to define the
standard, yet they only focus on the abstract syntax of the
languages. They should really be viewed partial metamodels (or
even just models) as they do not provide a complete language
definition.

\section{Why Metamodel?}

As discussed in chapter \ref{lddchapter}, system development is
fundamentally based on the use of languages to capture and relate
different aspects of the problem domain.

The benefit of metamodelling is its ability to describe these
languages in a unified way. This means that the languages can be
uniformly managed and manipulated thus tackling the problem of
language diversity. For instance, mappings can be constructed
between any number of languages provided that they are described
in the same metamodelling language.

Another benefit is the ability to define semantically rich
languages that abstract from implementation specific technologies
and focus on the problem domain at hand. Using metamodels, many
different abstractions can be defined and combined to create new
languages that are specifically tailored for a particular
application domain. Productivity is greatly improved as a result.

\subsection{Metamodels and Tools}

The ability to describe all aspects of a language in a metamodel
is particularly important to tool developers.

Imagine the benefits of loading a metamodel of a language into a
tool that defined all aspects of the language. The tool would
immediately understand everything relating to the presentation and
storage of models or programs in the language, the users'
interaction with and creation of models or programs, and how to
perform semantically rich activities, such as execution, analysis
and testing. Furthermore, any number of other languages could also
be loaded in the same way, enabling the construction of
semantically rich development environments. Because all the
languages are defined in the same way, interoperability between
the tools would be straightforward. This flexibility would not
just be restricted to user level languages. Another example might
be loading an extension to the meta-metamodel, such as a new kind
of mapping language. This language would then be immediately
available to capture mappings between different languages.

Allowing all aspects of tools to be modelled in a single, platform
independent metamodelling language will have big implications for
the software engineering industry. Firstly, the interoperability
and flexibility of tools will be drastically increased. This will
lead to a marketplace for tool metamodels. Metamodels that provide
partial definitions of languages could be easily extended to
provide many other capabilities by vendors with expertise in a
specific modelling domain.

Secondly, rich metamodels will have a significant benefit to the
standards community. As we have argued, there is currently no
means of capturing complete language definitions in existing
metamodelling languages. As a result, standards that use
metamodels to describe the languages they define suffer because
their informal definitions can be interpreted in many ways.
Complete metamodels of standards such as UML would greatly enhance
the rigour by which the standard is implemented and understood -
something that all stakeholders will benefit from.

\section{Where do you find Metamodels?}

Metamodels have been around for many years in a wide variety of
different application domains and under various pseudonyms: "data
model", "language schema", "data schema" are all terms we have
seen. Wherever there is a need to define a language, it is common
to find a metamodel. This is particularly the case for standards,
which by virtue of being a standard must have a precise
definition. Examples include AP233 and SysML (systems
engineering), SPEM (process modelling), OSS (telecoms) and CWM
(data warehousing). The Object Management Group (OMG) has been
particularly involved in their use in the standards arena. One of
the largest metamodels (about 200 pages long) is contained in the
UML specification \cite{umlspec}. With the advent of MDA
\cite{mda} and the increasing need for standardisation across the
systems development community, the number of applications of
metamodels is set to grow significantly.

Finally, although many developers may view metamodels as being
un-connected with their daily work, it is interesting to note that
many are already using metamodels without knowing it! Many
developers have already experienced the benefits of designing
frameworks containing a vocabulary of language concepts. For
example, developers of financial systems will use concepts such as
financial transactions, accounts, and so on. In reality, they are
defining a language for their domain.

\section{Metamodelling Languages}

A metamodel is written in a metamodelling language, which is
described by a meta-metamodel. As described above, the aim is that
the same metamodelling language (and meta-metamodel) is used to
describe any number of different languages. Thus, provided that
the modelling languages have been defined in the same
metamodelling language, it is possible to treat their definitions
in a unified manner. For example, they can be stored and
manipulated in the same way, or related by mappings.

What distinguishes a metamodelling language from a general purpose
programming language like Java or a modelling language like UML?
The answer is that a metamodelling language is a language
specifically designed to support the design of languages. An
essential requirements of a metamodelling language therefore is
its ability to concisely capture all aspects of a modelling
language, including its syntax and semantics.

\noindent The next chapter will examine in detail the required
components of a metamodelling language.

\section{Metamodel Architectures}
\label{metaarch}

A metamodelling language places requirements on there being a
specific metamodelling archictecture. This architecture provides a
framework within which some key features of a metamodel can be
realised. An important property of a metamodel architecture is
that it describes a classification hierarchy. Models written in
the language are {\em instances} of the concepts that are defined
in the metamodel - the structure of the instances is classified by
the metamodel. Furthermore, many  languages have their own notion
of classification (although they need not), and the pattern is
repeated until a point is reached at which further classification
does not occur. This repeating pattern of
classification/instantiation contributes to what is commonly known
as a meta-level architecture - a concept that will be described in
more detail in the next sections.

\subsection{Traditional Metamodel Architecture}

The traditional metamodel architecture, proposed by the original
OMG MOF 1.X standards is based on 4 distinct meta-levels. These
are as follows:

\begin{description}
\item[M0] contains the data of the application (for example, the
instances populating an object-oriented system at run time, or
rows in relational database tables). \item[M1] contains the
application: the classes of an object-oriented system, or the
table definitions of a relational database. This is the level at
which application modeling takes place (the type or model level).
\item[M2] contains the metamodel that captures the language: for
example, UML elements such as Class, Attribute, and Operation.
This is the level at which tools operate (the metamodel or
architectural level). \item[M3] The meta-metamodel that describes
the properties of all metamodels can exhibit. This is the level at
which modeling languages and operate, providing for interchange
between tools.
\end{description}

\begin{figure}[htb]
\begin{center}
\includegraphics[width=10cm]{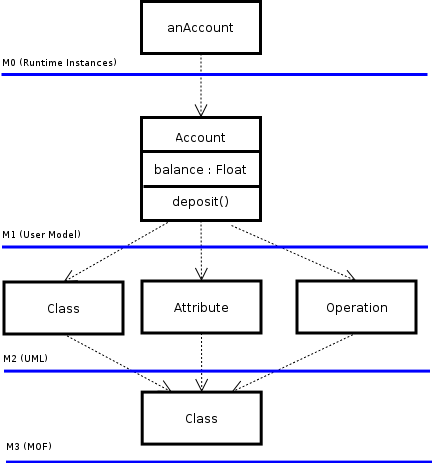}
\caption{Example of the 4 Layer Meta Architecture} \label{4layer}
\end{center}
\end{figure}

Each level in this hierarchy represents an instance of a
classifier relationship. As shown in figure \ref{4layer}, elements
at M0 are instances of classes at M1, which themselves can be
viewed as instances of metamodel classes, which can be viewed as
instances of meta-metamodel classes.

The unifying factor in this architecture is the meta-metamodel. It
defines the simplest set of concepts required to define any
metamodel of a language.

\subsection{Golden Braid Metamodel Architecture}
\label{goldenbraid}

Although the 4-layer metamodel is widely cited, its use of
numbering can be confusing. An alterative architecture is the
golden braid architecture \cite{escher}. This architecture
emphasises the fact that metamodels, models and instances are all
relative concepts based on the fundamental property of
instantiation.

The idea was first developed in LOOPS (the early Lisp Object
Oriented Programming System, and then became a feature of both
ObjVLisp \cite{objVlisp} and also CLOS (the Common Lisp Object
System).

Underpinning the golden braid architecture is the relationship
between a Class and an Object. A Class can be instantiated to
create an Object. An Object is said to be an instance of a Class.
This fact can be determined through a distinct operation, of(),
that returns the Class that the Object was created from.

In addition, a Class is also a subclass of Object. This means that
a Class can also be instantiated by another Class: its meta Class.
This relationship is key to the meta-architecture, as it enables
an arbitrary number of meta-levels to be described through the
instantiation relationship.

In practice, there will be a distinct Class that all elements in
the meta-architecture are instances of. This is the
meta-metaclass, which is effectively used to bootstrap the entire
metamodel architecture. This class will be defined as part of the
meta-metamodel (the model of the metamodelling language used to
model all languages).

In terms of the 4-layer metamodel, it is clear that it can be
viewed as the result of stamping out the golden braid architecture
over a number of different levels. Thus, there is no notion of a
meta-metamodel: it is just a metamodel that describes models,
which themselves may be metamodels.

The golden braid architecture offers a great deal of flexibility.
Thus it forms the foundation of the XMF metamodelling language,
which will be presented in chapter \ref{xmfchapter}.

\subsubsection{Meta Object Protocol}

A related aspect of the golden braid architecture is its use of a
meta-object protocol (MOP). A meta-object protocol is a set of
classes and methods that allow a program to inspect the state of,
and alter the behaviour of its meta-metamodel at run-time. These
make it possible to easily adapt the metamodelling language to
support different types of behaviours. For instance, changing the
way that inheritance works, or modifying how the compiler works
without having to change the code for the compiler. This adds
further flexibility to the metamodelling process.

\section{The Metamodelling Process}

The task of creating a metamodel for a language is not a trivial
one. It will closely match the complexity of the language being
defined, so for example, a language containing rich executable
capabilities will be much more complex to define than a simple
static language.

However, there {\em is} a clearly defined process to constructing
metamodels, which does at least make the task a well-defined, if
iterative, process. The process has the following basic steps:

\begin{itemize}
\item defining abstract syntax \item defining well-formedness
rules and meta-operations \item defining concrete syntax \item
defining semantics \item constructing mappings to other languages
\end{itemize}

Much of the remainder of the book will focus on the detail
involved in this process. Initially, we will present the tools
necessary to create metamodels in the first place; the armoury of
metamodelling facilities that is the metamodelling language.

\section{Five levels of Metamodelling}

We are often asked by clients how they can assess the quality of a
metamodel. To help them, we have found the following five levels
to be useful:

\begin{description}
\item [Level 1] This is the lowest level. A simple abstract syntax
model must be defined, which has not been checked in a tool. The
semantics of the language it defines will be informal and
incomplete and there will be few, if any, well-formed rules.
\item [Level 2] At this level, the abstract syntax model will
be relatively complete. A significant number of well-formedness
rules will have been defined, and some or all of the model will
have been checked in a tool. Snapshots of the abstract syntax
model will have been constructed and used to validate its
correctness. The semantics will still be informally defined.
However, there may be more in the way of analysis of the language
semantics.
\item [Level 3] The abstract syntax model will be completely tried
and tested. Concrete syntax will have been defined for the
language, but will only have been partially formalised. Typically,
the concrete syntax will be described in terms of informal
examples of the concrete syntax, as opposed to a precise concrete
syntax model. Some consideration will have been given to the
extensibility of the language architecture, but it will not be
formalised or tested.
\item [Level 4] At level 4, the concrete syntax of the language
will have been formalised and tested. Users will be able to create
models either visually and textually and check that they result in
a valid instance of the abstract syntax model. The language
architecture will have been refactored to facilitate reuse and
extensibility. Models of semantics will have begun to appear.
\item [Level 5] This is the topmost level. All aspects of the
language will have been modelled, including its semantics. The
semantic model will be executable, enabling users of the language
to perform semantically rich operations on models written in the
language, such as simulation, evaluation and execution. The
language architecture will support good levels of reuse, it will
have been proven to do so through real examples. Critically, the
completed metamodel will not be reliant on any external technology
- it will be a fully platform independent and self contained
definition of the language that can be used `as is' to generate or
instantiate tools.
\end{description}

Most of the metamodels we have seen do not achieve a level greater
than 2. Even international standards such as UML do not exceed
level 3. Yet, reaching level 5 must be an aspiration for all
language developers.

\section{Conclusion}

This chapter has outlined some of the key features of system
development languages. All languages have a concrete syntax, which
defines how they are presented, an abstract syntax that describes
the concepts of the language, and a semantics that describes what
the concepts mean. A metamodel is a model of all these different
aspects of a language. Crucially, a metamodel can also be thought
of as a model, written in a metamodelling language, which is
itself is also described by a metamodel. This enables all
metamodels to be described in the same way. This facilitates a
truly unified approach to language definition.

\chapter{A Metamodelling Facility}
\label{xmfchapter}

\section{Introduction}

In order to be able to construct semantically rich models of
languages, a facility that fully supports language definition is
required. This is known as a {\em metamodelling facility}. A
metamodelling facility should provide the ability to capture the
key features of a language, including its syntax and semantics in
a unified and platform independent way, along with support for
other important language design requirements such as extensibility
and executability.

This chapter gives a brief introduction to a metamodelling
facility called XMF (eXecutable Metamodelling Facility). XMF
extends existing standards such as MOF, OCL and QVT with rich
executable metamodelling capabilities. It provides a number of
languages for metamodelling, all based around a core executable
meta-architecture and metamodelling framework.

XMF will form the foundation for exploring metamodelling
throughout the rest of this book.

\section{Requirements of a Metamodelling Facility}

Before introducing XMF, it is important to understand some of the
key requirements of a metamodelling facility.

Firstly, as discussed in the previous chapter, a metamodelling
facility should provide metamodelling languages that can capture
all the essential features of a language. They should include
languages that can capture the abstract syntax, concrete syntax
and semantics of a language. In addition, it should provide
facilities for manipulating metamodels, including the ability to
map them to other metamodels and extend metamodels to support new
language definitions.

Secondly, it should provide a meta-architecture in which the
metamodelling languages (including their semantics) are themselves
described in terms of a core metamodelling language. This ensures
that the languages are self defined and complete, and enables
their definitions to be readily reused to create new language
definitions.

Finally, to gain maximum flexibility, a metamodelling facility
must be platform independent. In other words, its metamodels
should be self sufficient and independent of implementation
specific descriptions of the language being modelled. Thus,
reliance on the implementation of the language's behaviour in a
programming language or the assumption that there will be an
external database that manages object creation and persistence is
completely avoided.

\subsection{XMF}

XMF aims to provide a rich environment for language design that
supports the key requirements of a metamodelling facility. It
combines and extends a number of standard object-oriented
modelling facilities to provide a minimal, but expressive,
platform independent language for metamodelling. It includes the
following features:

\begin{itemize}

\item Support for core OO modelling concepts such as packages, classes, and
associations to describe language concepts and their relationship to one another.

\item A constraint language, which can used to describe well-formedness rules.

\item A set of action primitives, which can be used to capture the behavioural
semantics of a language and for manipulating metamodels. This
turns it into a meta-programming language.

\item A concrete syntax language which can be used to model the concrete
syntax of any modelling language.

\item A generic metamodel framework, which supports standard plug-points
and machinery for expressing model element instantiation,
execution, expression evaluation and reflection.

\item Conformance to the golden-braid metamodel architecture
described in section \ref{goldenbraid}, ensuring that the
language, including its semantics is completely self described.

\item A collection of richer metamodelling facilities. In particular,
languages for expressing mappings (both uni-directional and
bi-directional) between metamodels.

\end{itemize}

The following sections present an overview of the key features of
XMF. This is not a full definition, but will reference fuller
descriptions of the components of the language where appropriate.

\section{XMF Features}

\subsection{Core OO Modelling Concepts}
\label{coreconcepts}

\begin{figure}[htb]
\begin{center}
\includegraphics[width=10cm]{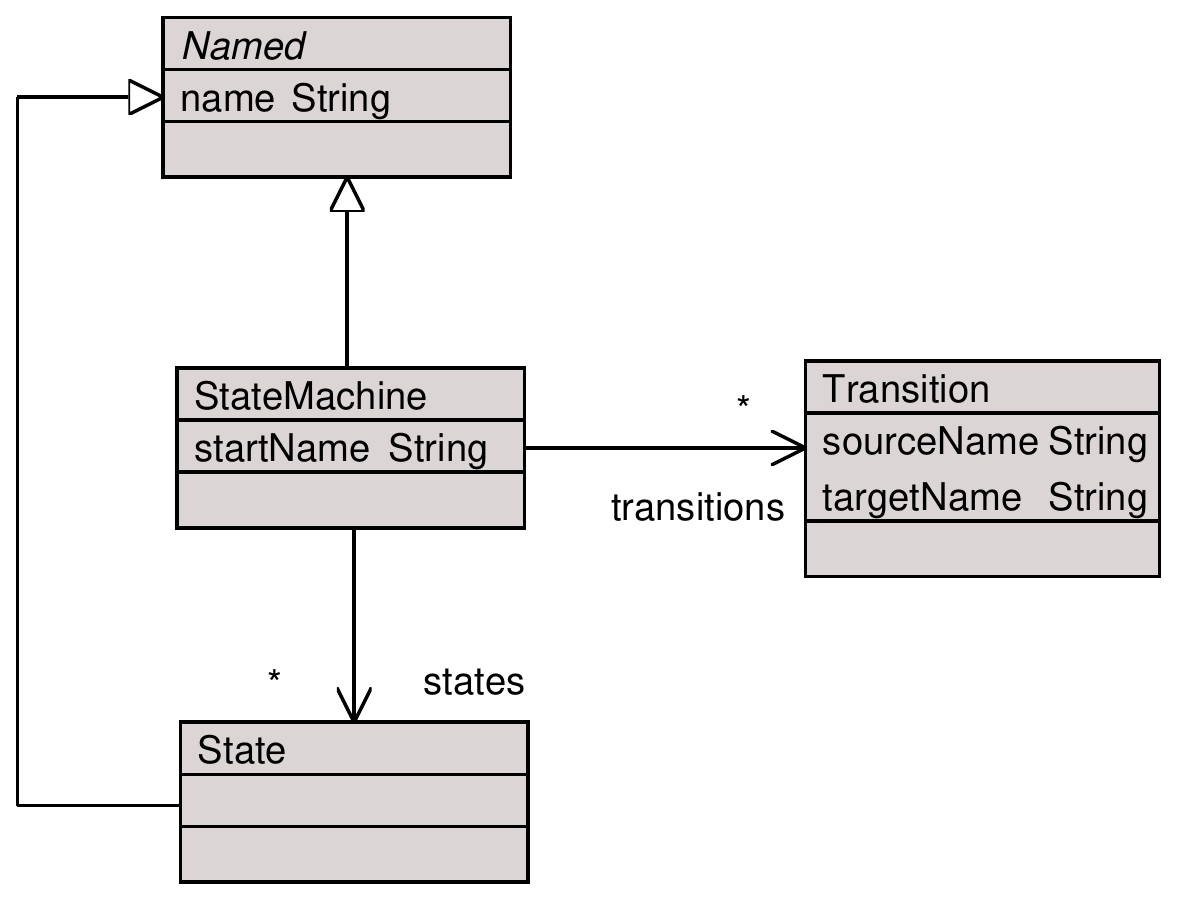}
\caption{An example of a class diagram}
\label{classDiagram}
\end{center}
\end{figure}

XMF provides the standard OO modelling concepts that are supported
by MOF and UML, including packages, classes and associations.
These are visualised using class diagrams. Figure
\ref{classDiagram} shows an example of a class diagram of a simple
model consisting of a package with a number of classes and
associations. This model describes a simple StateMachine, where a
StateMachine contains States and Transitions, and Transitions have
source and target states.

Note that associations are defined in terms of attributes. Thus, a
uni-directional association (with a single arrow head) is visual
syntax for an attribute of the target type belonging to the source
class. A bi-directional association (with no arrow heads) is
visual syntax for a pair of attributes plus a constraint that
ensures they are inverses of one another.

XMF also provides a concrete syntax for packages and classes. The
equivalent concrete representation of the StateMachine class
diagram shown in figure \ref{classDiagram} is shown below.

\begin{lstlisting}
1  @Package StateMachines
2    @Class isAbstract Named
3      @Attribute name : String end
4    end
5    @Class StateMachine extends Named
6      @Attribute startName : String end
7      @Attribute states : Set(State) end
8      @Attribute transitions : Set(Transition) end
9    end
10   @Class State extends Named
11   end
12   @Class Transition
13     @Attribute sourceName : String end
14     @Attribute targetName : String end
15   end
16 end
\end{lstlisting}Line 1 shows the start of a package named StateMachines, which
contains all the sub-definitions relating to StateMachines.

Line 2 contains the start of a class definition for the class
Named,  which is an abstract class for a named element. Line 5 is
the start of the definition of the StateMachine class. It defines
three attributes. The first attribute is the name of the starting
state of the StateMachine. The other two attributes are states and
transitions, and their types are Set(State) and Set(Transition).
These types correspond to the "*" multiplicity of the equivalent
association ends in the class diagram.

Lines 10 and 12 are the start of the State and Transition class
definitions. The State class specialises the class Named, and
therefore inherits a name attribute. A Transition has two
attributes sourceName and targetName, which are the names of its
source and target states.

As chapter \ref{concretechapter} will show, the concrete
representation of a modelling language should be clearly
distinguished from its abstract syntax representation. Here, two
different concrete syntaxes are being used to capture the same
information.

\subsection{Imports}

Just as in UML and MOF, a package can be imported into another
package. The result is that all referenceable elements in the
imported package can be referenced by elements in the importing
package. Consider the following package:

\begin{lstlisting}
@Package X
  @Class Y end
end

@Package A imports X
  @Class B
    @Attribute b : Y end
  end
end
\end{lstlisting}
Because the package A imports the package X, elements in the
package X can be referenced without the need to provide a full
path name.

\subsection{Constraints}

It is often necessary to state well-formedness rules about
concepts in a model. These rules are often made informally, for
example in the context of figure \ref{classDiagram} it might be
useful to specify that \emph{all transitions have unique names}. A
constraint language provides a means of succinctly and
unambiguously expressing complex well-formedness rules. The
well-formedness rule mentioned above can be added to the class
StateMachine as follows:

\begin{lstlisting}
context StateMachine
  @Constraint NoTwoTransitionsWithTheSameName
    transitions->forAll(t1 |
      transitions->forAll(t2 |
        t1.name = t2.name implies t1 = t2))
  end
\end{lstlisting}Another well-formedness rule requires that the starting state must
be one of the states of the state machine:

\begin{lstlisting}
context StateMachine
  @Constraint ValidStartingState
      states.name->includes(startName)
  end
\end{lstlisting}The constraint language used in XMF is OCL \cite{oclBook}. The
primary difference between the OCL used here and standard OCL is
the use of a different syntax for declarations. In this case
''@Constraint'' is used as opposed to the ''inv:'' declaration
used in standard OCL. The reason for this choice will become
apparent later when we consider the need for a flexible parsing
language.

\subsection{Queries}

OCL can be used to write queries. A query is used to produce a
value in the context of a current object state; it does not cause
any side effects. The following is an example of a query:

\begin{lstlisting}
context StateMachine
  @Operation getState(n:String):State
    self.states->select(s | s.name = n)->sel
  end
\end{lstlisting}
This will filter the states of a state machine, selecting those
states whose name matches the string n. Here, sel, is an in-built
operation that selects a single element from a collection. Again,
note that the declaration of a query differs from that in the OCL
standard.

\noindent Another example of query returns true if there exists a
state with the name n:

\begin{lstlisting}
context StateMachine
  @Operation isState(n:String):Boolean
    self.states->exists(s | s.name = n)
  end
end
\end{lstlisting}\subsection{Actions and XOCL}

OCL is by design a static language and does not change the state
of objects it refers to.  In some situations this is a good thing
because it guarantees side-effect free evaluation.  However this
limitation makes it very difficult to describe operational
behaviour in a way which can be readily executed.  Standard OCL
provides pre-/post- conditions as a way of specifying the effect
of an operation, however in general these cannot be executed.

An alternative approach is to augment OCL with action primitives.
The XOCL (eXecutable OCL) language extends OCL with a number of
key behaviour primitives. This is a essential step towards making
XMF a true meta-programming environment (see chapter
\ref{execChapter}).

An example of the use of actions in XOCL can be seen in the state
machine of figure \ref{classDiagram}. If it was required to add
new states to the state machine dynamically then the following
XOCL statement can be written:

\begin{lstlisting}
context StateMachine
  @Operation addState(name:String)
    self.states := self.states->including(StateMachines::State(name))
  end
end
\end{lstlisting}New instances of classes can be created by calling its {\em
constructor}. A constructor is an operation that has the same name
as the class. It takes a sequence of values as an argument, which
are then used to intialise the object's slots. In the case of the
state machine, three constructors are required, one for each
concrete class. The first intialises the name of a state, the
second assigns a source and target state name to a transition, and
the third initialises a new state machine with a name, a starting
state, a set of transitions and set of states. Constructors also
support getters and setters via the $?$ and $!$ notations - a $?$
results in the creation of a getX() operation, while a $!$ results
in an addX() operation where X is a non-singleton attribute. Note,
that if the body of the constructor is empty, the default action
is to set the attribute values (slots) of the created object with
the values of the parameters.

\begin{lstlisting}
context State
  @Constructor(name)
  end

context Transition
  @Constructor(sourceName,targetName)
  end

context StateMachine
  @Constructor(name,startName,states,transitions) ?
  end
\end{lstlisting}
The body of the next example illustrates how OCL conditional
expressions and logical operators can be combined with XOCL
actions to define the operation of adding a transition. This
operation takes the name of the new transition and the name of a
source and target state. The isState() query is then used to check
that the states belong to the StateMachine before creating a
transition between them. The last line of the if statement shows
how XOCL deals with the printing of strings to the console.

\begin{lstlisting}
context StateMachine
@Operation
addTransition(source:String,target:String)
  if self.isState(source) and self.isState(target) then
    self.transitions := self.transitions->
      including(StateMachines::Transition(source,target))
  else
    "Invalid State in addTransition()".println()
  end
end
\end{lstlisting}As this example shows, augmenting OCL with a small number of
action primitives results in a powerful and expressive programming
language. Furthermore, as we will see in later chapters of this
book, because the language works at the XMF level, it also
provides a rich meta-programming environment that can be used to
construct sophisticated facilities such as parsers, interpreters
and compilers. Indeed, it is so expressive that it has been used
to implement XMF itself.

\subsection{Instantiation}

\begin{figure}[htb]
\begin{center}
\includegraphics[width=11cm]{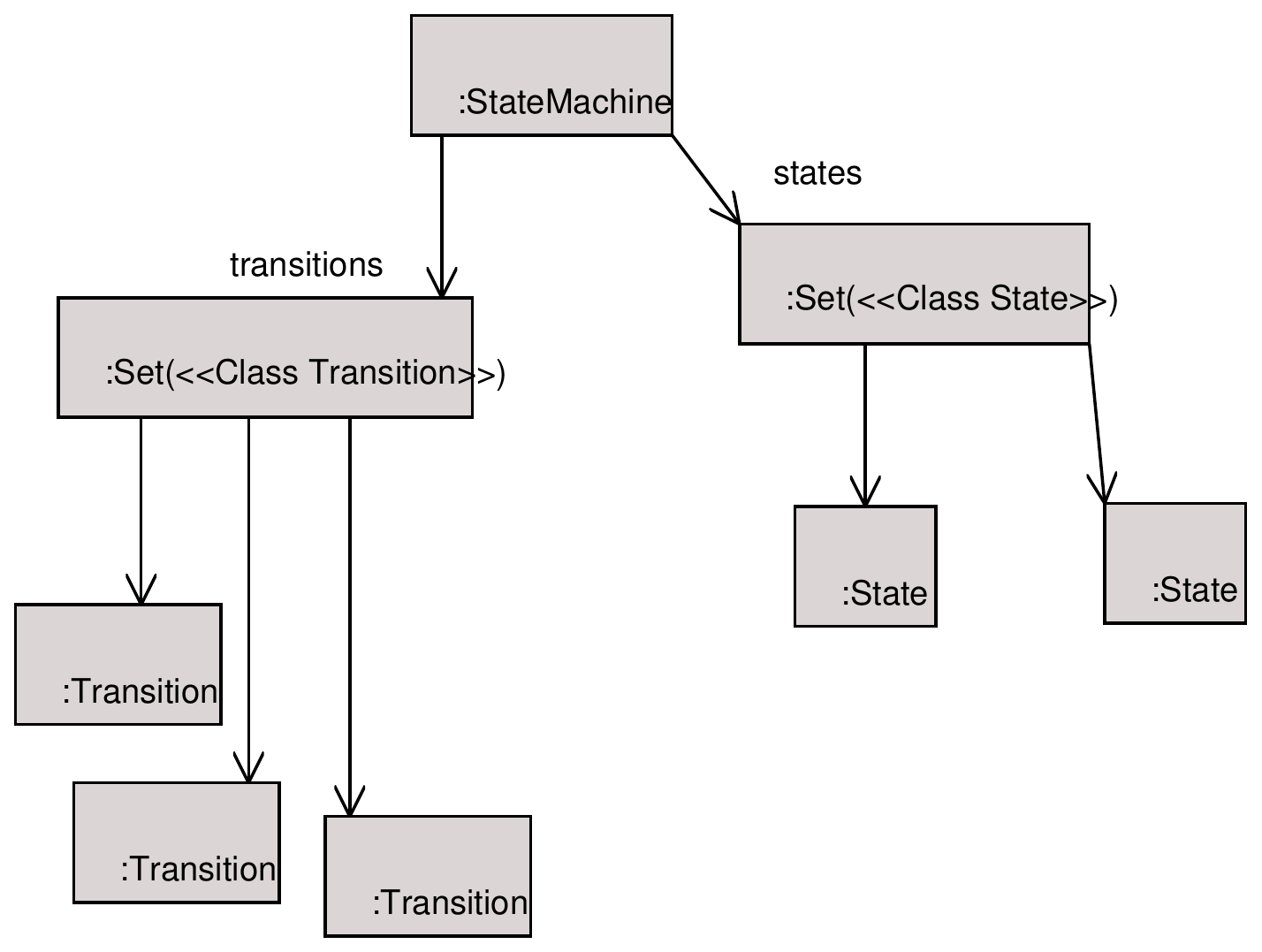}
\caption{A snapshot showing an instance of the statemachine
metamodel} \label{snapshotExample}
\end{center}
\end{figure}

A useful notation for visually representing the instances of a
metamodel is a snapshot - a notation popularised by the Catalysis
method \cite{Catalysis}. A snapshot shows objects, the values of
their slots (instances of attributes) and links (instances of
associations). A snapshot of a metamodel will thus show objects
and links that are instances of elements in the metamodel. The
example shown in figure \ref{snapshotExample} is a snapshot of the
StateMachine metamodel. It shows an instance of a StateMachine
containing two states and three transitions.

Of course, because XMF provides an interpreter, instances of
models can be easily created and tested via its interpreter
console.

\subsection{Concrete Syntax}

The concrete syntax of a modelling language defines the notation
that is used to present models in the language. A notation may be
textual, or diagrammatical, or a mixture of both. A key part of
any metamodelling language is the ability to model both these
aspects in a platform independent way, and as a result facilitate
the rapid creation of parsers and diagram editors for a new
language.

\subsubsection{Textual Syntax}

XMF provides a generic parser language that can be used to model
the textual syntax of any language. This language, called XBNF
(the reader will be starting to see a pattern to our naming
conventions by now!) allows new textual constructs to be defined
that can be used as input to a model parser. These new constructs
to be defined as:

\begin{lstlisting}
@<NAME>
  <BODY>
end
\end{lstlisting}
\noindent where {\tt <NAME>} is the name of the construct and{\tt
<BODY>} is an XBNF expression.

An XBNF expression consists of a number of EBNF definitions within
which XOCL variables are embedded, followed by an XOCL expression.
When a construct is parsed, the XBNF expression is used to match
each parsed element with the variables. These variables can then
be used within the XOCL action to create instances of classes that
populate a model of the abstract syntax of a language.

Imagine that we wish to create a concrete syntax for the simple
StateMachine example. One part of this task would be to create a
concrete syntax for states. This might take the form:

\begin{lstlisting}
@State On
end

@State Off
end
\end{lstlisting}The following XBNF defines the syntax for parsing states and
populating instances of the class State.

\begin{lstlisting}
State ::= name = Name {[| State(name) |]}
\end{lstlisting}\noindent Now we have a new construct. When we type:

\begin{lstlisting}
@State X end
\end{lstlisting}\noindent we will get an instance of the class State named X.

\noindent Chapter \ref{concretechapter} will provide more detail
on the use of XBNF.

\subsubsection{Diagrammatical Syntax}

In order to model the diagrammatical syntax of a modelling
language, XMF provides a generic model of diagram elements such as
boxes, lines, etc, that can be tailored to support specific types
of diagrams. Bi-directional mappings are used to model the
relationship between a specific diagram model and the model of the
language's abstract syntax. This ensures that whenever changes are
made to a diagram, they are reflected in the model and vice versa.
By providing an interpreter (written in XMF) for displaying
instances of a diagrammatical syntax model, it is possible to
construct new diagram editors for a specific modelling language in
very short timescales. Chapter \ref{concretechapter} provides more
detail on this aspect.

\subsection{Mappings}

Chapter \ref{lddchapter} highlighted the fact that one important
use case of languages is to transform or relate them to other
languages. Mappings describe how models or programs in one
language are transformed or related to models or programs in
another. In order to describe these mappings, mapping languages
are required.

Two types of mapping languages are included in XMF: a
uni-directional pattern oriented mapping language called XMap, and
a bi-directional synchronisation language called XSync.

\subsubsection{Uni-directional Mappings}

XMap, is a declarative, executable language for expressing
uni-directional mappings that is based on pattern matching.

\begin{figure}[htb]
\begin{center}
\includegraphics[width=11cm]{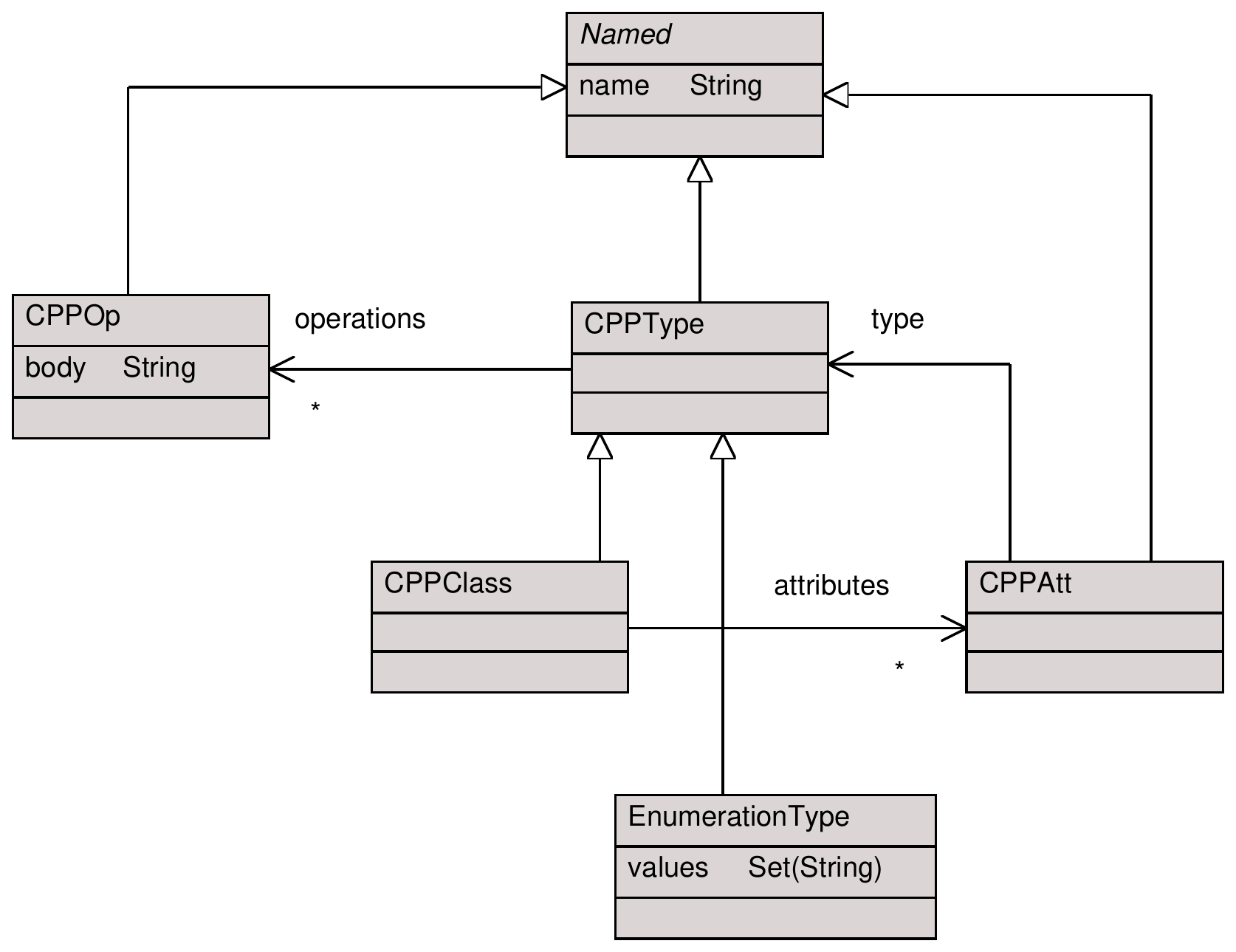}
\caption{A simple model of C++ classes}
\label{cppExample}
\end{center}
\end{figure}

To illustrate the use of XMap, figure \ref{cppExample} shows a
simple model of {C++} classes which will be used as the target of a
mapping from a state machine.

A {C++} class is a namespace for its attributes and operations
(methods). An attribute has a type, and for the purposes of this
example, its type may either be another class or an enumeration
type. An enumeration type has a value, which is the sequence of
strings in the enumeration. An operation has a name and a body,
which contains a simple string representation of the body of the
operation.

A mapping from from the StateMachine model in figure
\ref{classDiagram} to the {C++} in \ref{cppExample} can be
defined. This maps a StateMachine to a {C++} class, where each
state in the state machine is mapped to a value in an enumerated
type called STATE. Each transition in the state machine is mapped
to a {C++} operation with the same name and a body, which changes
the state attribute to the target of the transition.

\begin{figure}[htb]
\begin{center}
\includegraphics[width=14cm]{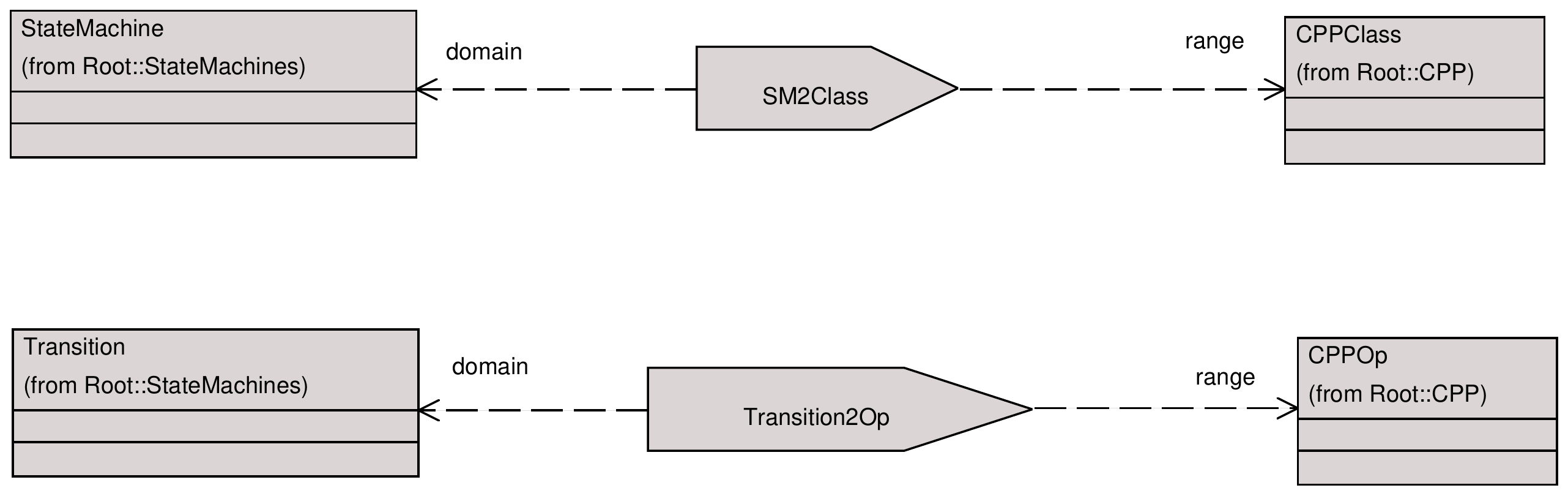}
\caption{A mapping between state machines and C++ classes}
\label{cppmapping}
\end{center}
\end{figure}

The mapping can be modelled in XMap as shown in figure
\ref{cppmapping}.  The arrows represent mappings between elements
of the two languages. A mapping has a domain (or domains), which
is the input to the mapping, and a range, which is the output. The
first mapping, SM2Class, maps a state machine to a C++ class. The
second mapping, Transition2Op, maps a transition to an operation.

In order to describe the details of the mapping, XMap uses a
textual  mapping language based on pattern matching. Working
backwards, the definition of the mapping between a transition and
an operation is as follows:

\begin{lstlisting}
context Transition2Op
  @Clause Transition2Op
    StateMachines::Transition
      [sourceName = S,
       targetName = T]
    do
    CPP::Operation
      [name = S+T,
       body = B]
    where
      B = "state = " + T
  end
end
\end{lstlisting}A mapping consists of a collection of clauses, which are pattern
matches between source and target objects. Whenever a source object
is successfully matched to the input of the mapping, the resulting
object in the do expression is generated. Variables can be used
within clauses, and matched against values of slots in objects.
Because XMap builds on XOCL, XOCL expressions can also be used to
capture complex relationships between variables.

In this example, whenever the mapping is given a Transition with a
sourceName equal to the variable S and a targetName equal to T, it
will generate an instance of the class Operation, whose name is
equal to the concantenation of S and T, and whose body is equal to
the variable B. The where clause is used to define values of
variables, and it is used here to define the variable B to be
concatenation of the text "state = " with the target state name.
For instance, given a transition between the states "On" and
"Off", the resulting operation will have the name "OnOff" and the
body "state = Off". Note that it would be quite possible to model
a part of the syntax of {C++} expressions, and equate B with an
instance of an expression class.

The mapping between state machines and {C++} classes is shown below:

\begin{lstlisting}
 context SM2Class
   @Clause SM2Class
     StateMachines::StateMachine
       [states = S,
        transitions = TS]
     do
     CPP::CPPClass
       [attributes =
         Set{CPP::CPPAtt
           [name = "state",
            type = T]},
        operations = O]
     where
       T = CPP::EnumerationType
         [name = "STATE",
          values = S->collect(s | s.name)];
       O = TS->collect(t | Transition2Op(t))
  end
\end{lstlisting}
Here, a state machine with a set of states, S, and a set of
transitions, TS, is mapped to a {C++} class with a distinguised
attribute state of type T, and a set of operations, O. The value
of T is an EnumerationType whose name is ``STATE'' and whose
values are the names of the states in S. Finally, the transitions
are mapped to operations by iterating through the transitions in
TS and applying the Transition2Op mapping.

This mapping illustrates the power of patterns in being able to
match arbitrarily complex structures of objects. As shown by the
class CPPClass, objects may be matched with nested objects to any
depth. It also shows the necessity of a rich expression language
like OCL for capturing the complex navigation expressions that are
often encountered when constructing mappings.

\subsubsection{Synchronised Mappings}

While it is common to want to translate from one language to
another, there is also a requirement to keep different models in
sync. For example, an abstract syntax model will need to be kept
in sync with a model of its concrete syntax, or a model may need
to be synchronised with code. To achieve this, XMF provides a
bi-directional mapping language called XSync. This enables rules
to be defined that state how models at either end of a mapping
must change in response to changes at the other end, thus
providing a declarative language for synchronisation. This
language will be explored in greater detail in chapter
\ref{mappingchapter}.

\section{XMF Architecture}

As outlined in the previous chapter, XMF follows the golden braid
metamodel architecture, and therefore is defined in its own
language. Understanding the architecture of XMF is important for
many reasons. Firstly, it provides a good example of a language
definition in its own right - the fact that XMF is self describing
makes a strong statement about the expressibility of the language.
Secondly, XMF acts as a foundation for many other metamodel
definitions. For instance, it can be viewed as a subset of the UML
metamodel, and many other types of modelling languages can also be
viewed as extensions of the core architecture.

\begin{figure}[htb]
\begin{center}
\includegraphics[height=8cm]{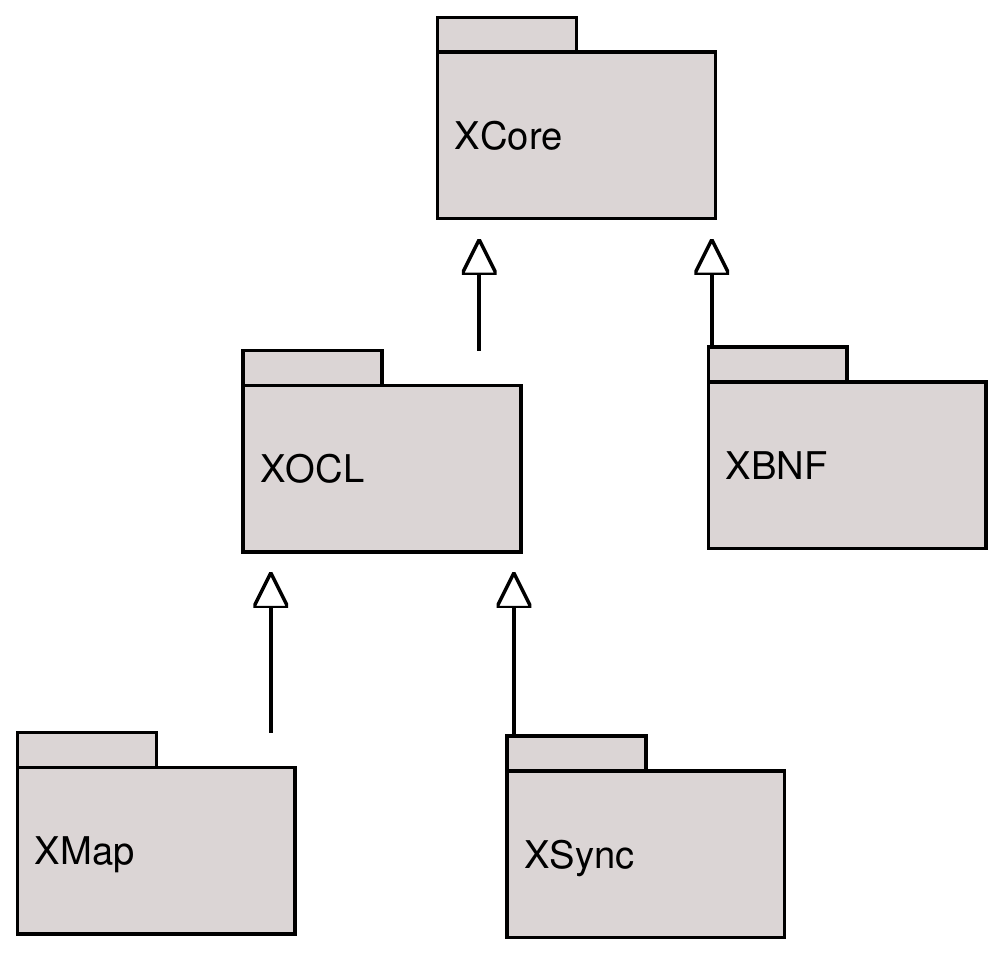}
\caption{Overview of XMF Architecture} \label{XMFOverview}
\end{center}
\end{figure}

Figure \ref{XMFOverview} shows the key components of the XMF
architecture. At the heart of XMF is the XCore metamodel. This
provides the core modelling concepts of the metamodelling language
(as described in section \ref{coreconcepts}). This metamodel also
provides a framework for language extension (a topic discussed in
more detail in chapter \ref{langchapter}). Around it, sit the
metamodels for the OCL, XOCL, XBNF, XMap and XSync languages.

The classes and relationships in these metamodels correspond
precisely to the modelling features that have been used in this
chapter. For example, the package and classes shown in the
StateMachine abstract syntax model in figure \ref{classDiagram}
are concrete syntax representations of instances of the classes
XCore::Package and XCore::Class. Expressions in a concrete syntax
model are themselves instance of XBNF::Grammar, and so on.

\subsection{XCore Metamodel}

\begin{figure}[htb]
\begin{center}
\includegraphics[height=21cm]{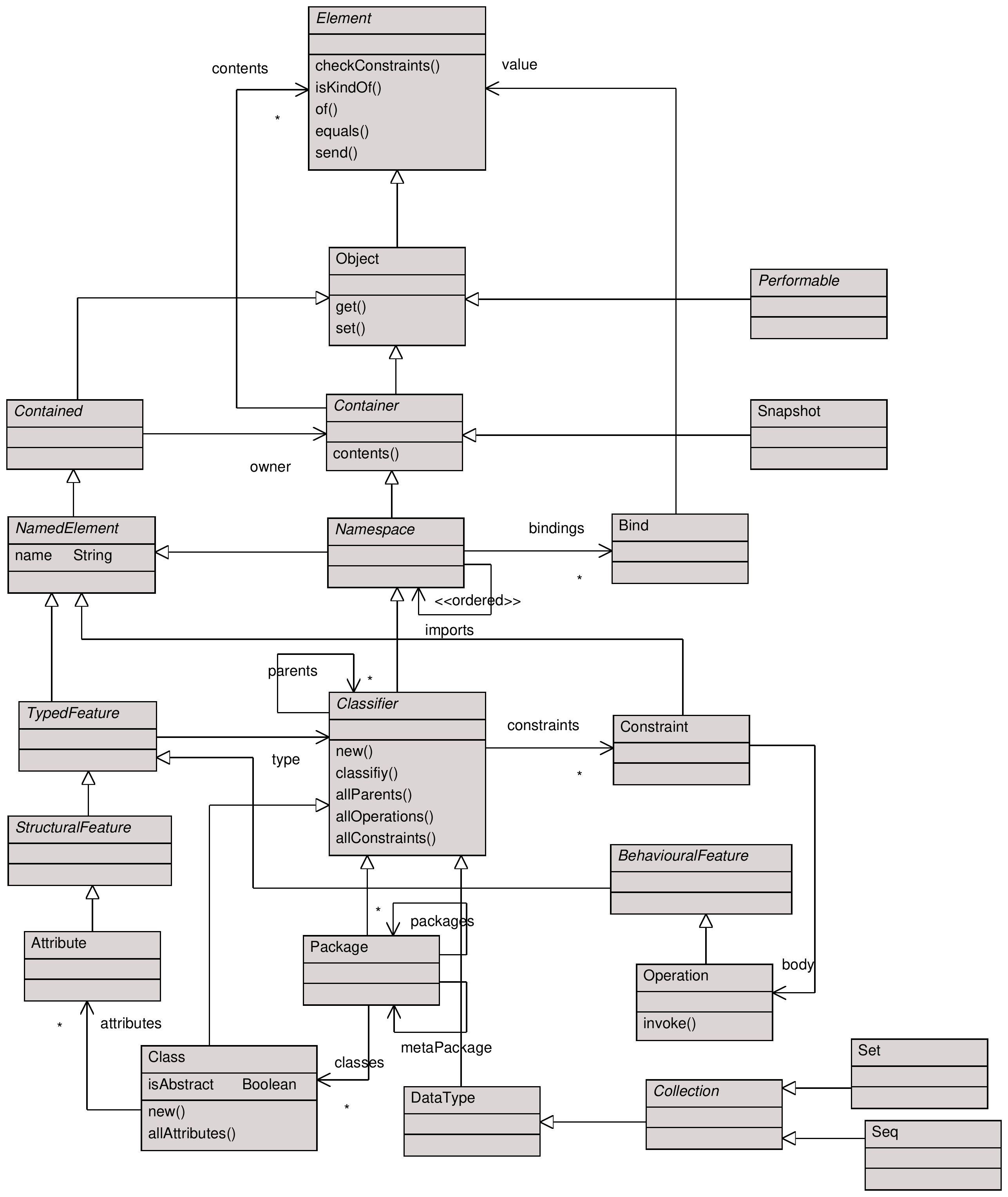}
\caption{The XCore Metamodel} \label{coreMetamodel}
\end{center}
\end{figure}

As shown in figure \ref{coreMetamodel}, the classes that are
defined in the XCore metamodel provide the core modelling concepts
used in XMF such as Class and Package. As we discuss metamodelling
in more detail in later chapters, many of the features of this
metamodel will discussed in detail.

\noindent There are however, some key parts of this model that are
worth pointing out here:

\begin{description}
\item[Elements and Objects] The most fundamental modelling
concepts in XMF are elements and objects. Elements are the root of
all modelling concepts, in other words, every type of modelling
concept in XMF is an element. All elements are instances of a
classifier. Elements do not have structure or state. Objects on
the other hand are elements that encapsulate data values - called
slots in XMF. An object's slots must conform to the name and type
of the attributes of the class it is an instance of.
\item[Performable] The performable class represents the root of
all modellng concepts that can be evaluated in the context of an
environment (a collection of variable bindings) to return a
result. An OCL expression is a good example of a performable
concept. \item[Classification] In XMF some types of modelling
elements are modelled as Classifiers. A Classifier is an element
that can be instantiated via the new() operation to create new
instances. Good examples of these elements are classes and
datatypes, which can be instantiated to create objects and
datavalues. All elements have an of() operation, which returns the
classifier that the element was instantiated from. For example the
objects statemachine1, statemachine2 will return the class
StateMachine while the object fido might return the class Dog.
\item[Reflection] An important property of all XMF models is that
classes are also objects. This apparently strange assumption means
that classes can also be viewed as instances of classes. This
capability is an important one, as it enables any operation that
can be carried out on an object such as executing its operations,
or mapping it to another object can also be applied to classes.
\item[XOCL] The XOCL class is an extension of OCL, and is the root
of all imperative expressions that can change the state of a model
(not shown here). \item[Snapshot] A snapshot is a collection of
elements, and may therefore include any type of element that is
available in XMF. \item[Grammar] A grammar can be evaluated in the
context of sequence of tokens to generate an instance of a model.
\end{description}

In addition, the XCore metamodel defines a number of abstract
classes that provide a framework for language design. This
framework will be discussed in greater detail in chapter
\ref{langchapter}.

\section{Relationship to Existing Metamodelling Languages}
\label{differences}

The requirement to be able to construct metamodels is not a new
one, and not surprisingly a variety of languages have been
proposed as metamodelling languages. The most notable of these are
UML and the MOF (the Meta Object Facility). The MOF is the
standard language for capturing meta-data, and goes further than
the UML in meeting the requirements of a metamodelling language.
Nevertheless, whilst the MOF provides many of the features
necessary to define metamodels, there are a number of crucial
areas in which it needs to be extended.

\begin{itemize}
\item The MOF does not explicitly support executable metamodelling
in a platform independent way. Although it does provide OCL, this
does not support the construction of operations that change the
state of a model. An alternative might be to use the Action
semantics. This is a platform independent language for executing
models that is currently a part of the the UML 1.5 metamodel.
There are a number of problems with this language however.
Firstly, it does not have a concrete syntax, which has slowed its
adoption. Secondly, it is a complex language that includes many
constructs that are not relevant to metamodelling, such as
concurrent actions. Whilst work is ongoing to resolve the first
issue, XOCL resolves both by the minimal extension of OCL,
resulting in a powerful executable metaprogramming language. \item
The MOF does not support an extensible grammar language. This is
important in being able to model new concrete syntax grammars in
MOF. \item The MOF currently is not defined fully in terms of
itself. Its semantics and syntax are stated in a form (a mixture
of OCL, and informal English) that prevents it from being
completely self describing and self supporting. \item The MOF
currently does not support mapping languages such as the XMap and
XSync languages described above. Work is currently proceeding on
an extension to MOF called QVT (Queries, Views, Transformations),
which aims to define a language for doing uni-directional
mappings, but it is unclear at this stage whether it will support
all the capabilities of XMap in the short to medium term. There is
unlikely to be support for synchronised mappings in the near
future.
\end{itemize}

XMF aims to provide these extensions in the most concise and
minimal fashion necessary to support precise, semantically rich
metamodelling.

\section{Conclusion}

This chapter has described some of the essential features of XMF,
an extended MOF like facility that provides the ability to define
platform independent metamodels of semantically rich languages.
These facilities will be used throughout the rest of this book to
illustrate the metamodelling process in greater detail.

\chapter{Abstract Syntax}
\label{abschapter}

\section{Introduction}

Just as the quintessential step in object oriented design is
constructing a model of system concepts, constructing an abstract
syntax model is an essential first step in the design of a
modelling language. An abstract syntax model describes the
concepts in the language and their relationships to each other. In
addition, it also defines the rules that determine whether a model
written in the language is valid or not. These are the
well-formedness rules of the language.

Imagine a business modelling language suitable for modelling high
level business rules about business data. An appropriate language
for this domain might provide modelling concepts such as "data
model", "data entity", and "business rule". In addition, there
will be relationships between these concepts: a "data model" may
be composed of a number of "data entities". There will also be
rules describing the valid models that may be constructed in the
language, for instance, "a datamodel cannot contain data entities
with the same name" might be one such rule.

The concepts, relationships and rules identified during this step
will  provide a vocabulary and grammar for constructing models in
the language. This will act as a foundation upon which all other
artefacts of the language design process will be based.

This chapter describes the steps required to construct abstract
syntax models, along with examples of their application to the
definition of the abstract syntax model of a simple modelling
language.

\section{Modelling Abstract Syntax}

As stated, the purpose of an abstract syntax model is to describe
the concepts in a language and the relationships that exist
between those concepts. In the context of a language definition, a
concept is anything that represents a part of the vocabulary of
the language. The term abstract syntax emphasises a focus on the
the abstract representation of the concepts, as opposed to their
concrete representation. As a result, abstract syntax models focus
on the structural relationship that exist between language
concepts. Note that it is not the purpose of the abstract syntax
model to describe the semantics of the language. These aspects
will be described later.

An abstract syntax model should also describe  the rules by which
a model written in the language is deemed to be well-formed, i.e.
is syntactically valid. These provide a more detailed description
of the syntactical rules of the language than is possible by
describing the concepts and relationships alone. Well-formedness
rules are particularly useful when it comes to implementing a tool
to support the language as they can be used to validate the
correctness of models as they are created.

Constructing an abstract syntax model has much in common with
developing an abstract grammar for a programming language, with
the exception that the language is more expressive than that used
in programming language design.

Abstract syntax models are written in a metamodelling language. As
described in chapter \ref{xmfchapter}, the metamodelling language
we will use, called XMF, provides a number of modelling
abstractions suitable for modelling languages. For the purposes of
modelling abstract syntax only a subset of XMF will be required.
This is the subset suitable for capturing the static properties of
language concepts and well-formedness rules, and includes:

\begin{itemize}
\item Classes to describe the concepts in the language.
\item Packages to partition the model into manageable chunks where
necessary.
\item Attributes and associations to describe the relationships between
concepts.
\item Constraints, written in OCL, to express the well-formedness rules.
\item Operations, written in XOCL, to describe operations on the
state of a model.
\end{itemize}

\section{The Process}

There are a number of stages involved in the development of an
abstract syntax model: concept identification; concept modelling;
model architecting; model validation and model testing. These
stages are described below.

\subsection{Concept Identification}
\label{conceptidentification}

The first stage in modelling the abstract syntax of a language is
to utilise any information available to help in identifying the
concepts that the language uses, and any obvious rules regarding
valid and invalid models.

There are a number of useful techniques that can be used to help in
this process:

\begin{itemize}

\item Construct a list of candidate concepts in the language.
Focus on determining whether the concepts make sense as part of
the language's vocabulary. In particular, identify concepts
that match the following criteria:

\begin{itemize}
  \item Concepts that have names.
   \item Concepts that contain other concepts, e.g. a class containing attributes.
  \item Concepts that record information about relationships with other concepts, e.g. named associations between classes.
  \item Concepts that play the role of namespaces for named concepts.
  \item Concepts that exhibit a type/instance relationship.
  \item Concepts that are recursively decomposed.
  \item Concepts that are a part of an expression or are associated with expressions.
\end{itemize}

\item Build examples of models using the language.

\begin{itemize}
\item Utilise any notation that you think appropriate to represent
each type of concept. Use this notation to build models of meaningful/real world examples.
\item In the case of a pre-existing language there should be
resources already available to help in the identification of
concepts. These may include BNF definitions of the language
syntax, which can be translated into a metamodel and examples of
usage. Other sources of inspiration include existing tools, which
will provide automated support for building example models. Such
tools may also provide additional facilities for checking valid
and invalid models, which will be useful when identifying
well-formedness rules or may provide more detailed examples of the
language syntax.
\end{itemize}

\end{itemize}

Once some examples have been constructed, abstract away from them
to identify the generic language concepts and relationships between
concepts. It is often useful to annotate the examples with the
modelling concepts as a precursor to this step. Examples of invalid
models can also be used to help in the identification of
well-formedness rules.

It is important during this process to distinguish between the
concrete syntax of a language and its abstract syntax. Whilst it is
common for the structure of the abstract syntax to reflect its
concrete syntax, this need not be the case. For example, consider a
language which provides multiple, overlapping views of a small
number of core modelling concepts. In this case, the abstract
syntax should reflect the core modelling concepts, and not the
concrete syntax.

'Modelling the diagrams' is a common mistake made by many novice
modellers. If in doubt, ask the following questions when
identifying concepts:

\begin{itemize}
\item Does the concept have meaning, or is it there purely for
presentation? If the latter, then it should be viewed as concrete
syntax. An example might be a "Note", which clearly does not
deserve to be modelled as an abstract syntax concept.
\item Is the concept a derived concept or is it just a view
on a collection of more primitive concepts? If the latter, a
relationship should be defined between the richer concept and the
more primitive concept.
\end{itemize}

In general, the best abstract syntax models are the simplest ones.
Any complexity due to diagrammatical representation should be
deferred to the concrete syntax models.

\subsection{Use Cases}

A useful technique to aid in the identification of modelling
language concepts is to consider the different use cases
associated with using the language. This is almost akin to writing
an interface for a metamodel, i.e. a collection of operations that
would be used when interacting with the metamodel. Some typical
use cases might include creating and deleting model elements, but
may also include semantically rich operations such as transforming
or executing a model. The detailed descriptions that result from
the use case analysis can then be mined for modelling language
concepts.

\subsection{Concept Modelling}

Once concepts have been identified, standard object-oriented
modelling features, classes, packages, and associations are
applied to model the concepts in the language. There are many
examples of developing conceptual models available in the
literature, for instance \cite{larman} devotes a significant
amount of time to this subject.

In general, concepts will be described using classes, with
appropriate attributes used to capture the properties of the
concepts. Some concepts will have relationships to other concepts,
and these will be modelled as associations. Where it makes sense
to define categories of concepts, generalization can be used to
separate them into more general and more specific types of
concepts.

A useful strategy to apply at this stage is to reuse existing
language definitions where possible. There are a number of
techniques that can be used to achieve this, including:

\begin{itemize}
\item Extending or tailoring an existing metamodel to fit the
new language. This can be achieved by specializing classes from the
existing metamodel or by using package extension to import and
extend whole packages of concepts.
\item Utilising language patterns. A language pattern may be realised
as a framework of abstract classes that capture a repeatedly used
language structure, or by a package template (a parameterized package).
\end{itemize}

\noindent A full discussion of reuse techniques will be presented
in chapter \ref{langchapter}.

\subsection{Well-formedness Rules}

Once a basic model has been constructed, start identifying both
legal and illegal examples of models that may be written in the
language. Use this information to define a set of well-formedness
rules in OCL that rule out the illegal models. When defining
rules, look to identify the most general rules possible, and
investigate the implications of what happens when rules are
combined - sometimes rules can conflict in unexpected ways. There
are many books available that can help guide this process (see
\cite{oclBook} for example). Reusing existing language components
can also minimise the effort involved in writing well-formedness
rules, as constraints can be reused as well via inheritance.

\subsection{Operations and Queries}

Operations and queries should also be defined where appropriate.
Examples of operations include general utility operations such as
creating new model elements and setting attribute values, or
operations that act as test harnesses. Examples of queries might
include querying properties of models for use in constraints and
operations, or for validation purposes. Operations that change the
state of models should take any constraints into account, ensuring
that if a constraint holds before the operation is invoked, it
will continue to hold afterwards. Note that in general it is best
to focus on using operations to test out the model - avoid at this
point writing operations that implement other features of the
language such as semantics.

\subsection{Validation and Testing}

It is important to validate the correctness of the abstract syntax
model. Investing in this early on will pay dividends over full
language design lifecycle. A useful (static) technique for doing
this is to construct instances of the abstract syntax model that
match those of example models. An object diagram is a useful way
of capturing this information as it shows instances of classes
(objects) and associations (links). There are tools available that
will both help create object diagrams, and also check they are
valid with respect to a model and any OCL constraints that have
been defined.

The best means of testing the correctness of a language definition
is to build a tool that implements it. Only then can the language
be tested by end users in its entirety. The architecture of a tool
that facilitates the rapid generation of tools from metamodels
will be discussed in detail in later versions of this book.

\section{Case Study}

The best way to learn about metamodelling is to tackle a real
example. An example of a simple, but widely known modelling
language, is a StateMachine. StateMachines are widely used as a
notation for modelling the effect of state changes on an object.
StateMachines are a good starting point for illustrating
metamodelling as they are simple, but have enough features to
exercise many facets of the metamodelling process.

Many different types of State Machine languages are described in
the literature. This chapter will concentrate on a simplified form
of StateMachines. Later chapters will show this language can be
extended with richer modelling capabilities.

\begin{figure}[htb]
\begin{center}
\includegraphics[width=11cm]{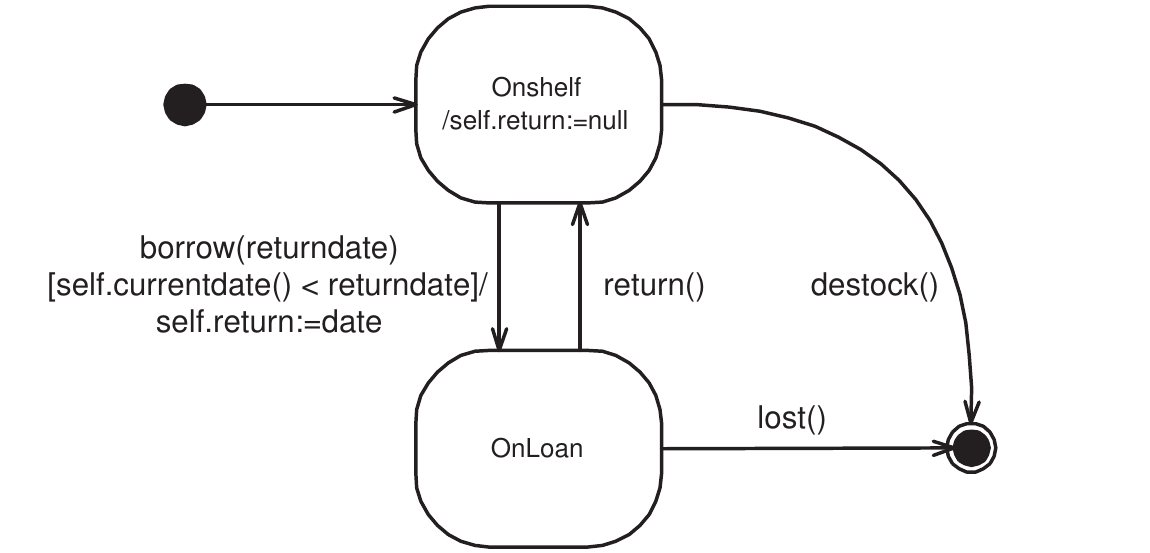}
\caption{An example of a StateMachine} \label{smexample}
\end{center}
\end{figure}

As shown in figure \ref{smexample}, a StateMachine is essentially a
visual representation of states and transitions. The StateMachines
described here are a form of object-oriented (OO) StateMachines. As
such, they describe the state changes that can occur to an instance
of a class, in this case a Library Copy.

StateMachines have an initial state and an optional final state,
which are represented by a filled circle and double filled circle
respectively. In addition, StateMachines provide guards, actions
and events. Guards are owned by transitions, and are boolean
expressions that must evaluate to true before a transition can be
invoked. Guards are represented by square brackets after an event
name. An example of a guard is shown in figure \ref{smexample}
attached to the borrow event.

In order to describe the effect of a transition on the state on an
object, actions can also be attached to transitions. Actions are
written in an action language, which in this case is XOCL. Actions
can also be attached to states, and are invoked when the state is
first entered. An example of an action that sets the returndate of
a Copy is shown in the figure \ref{smexample}.

Finally, events are associated with transitions via an event.
Events have a name and some optional parameters. It is the receipt
of an event that causes a transition to be triggered. Events can be
generated by StateMachine actions.

\subsection{Identification of Concepts}

Based on the example shown in figure \ref{smexample}, the
following candidate concepts can be immediately identified:

\begin{description}
\item [State]  A named representation of the state of an object at
a particular point in time.

\item [Initial State] The initial state that the object is
in when it is created.

\item [Final State] The final state of the object - typically when
the object has been destroyed.

\item [Transition] A state change. A transition has a source and target
state.

\item [Event] A named event that causes a transition to occur.

\item [Action] An executable expression that may be attached to
a transition or a state. Actions are invoked whenever the
transition occurs, or a state is entered.

\item [Guard] A boolean expression which must be evaluated to
be true before a transition can occur.

\end{description}

\begin{figure}[htb]
\begin{center}
\includegraphics[width=13cm]{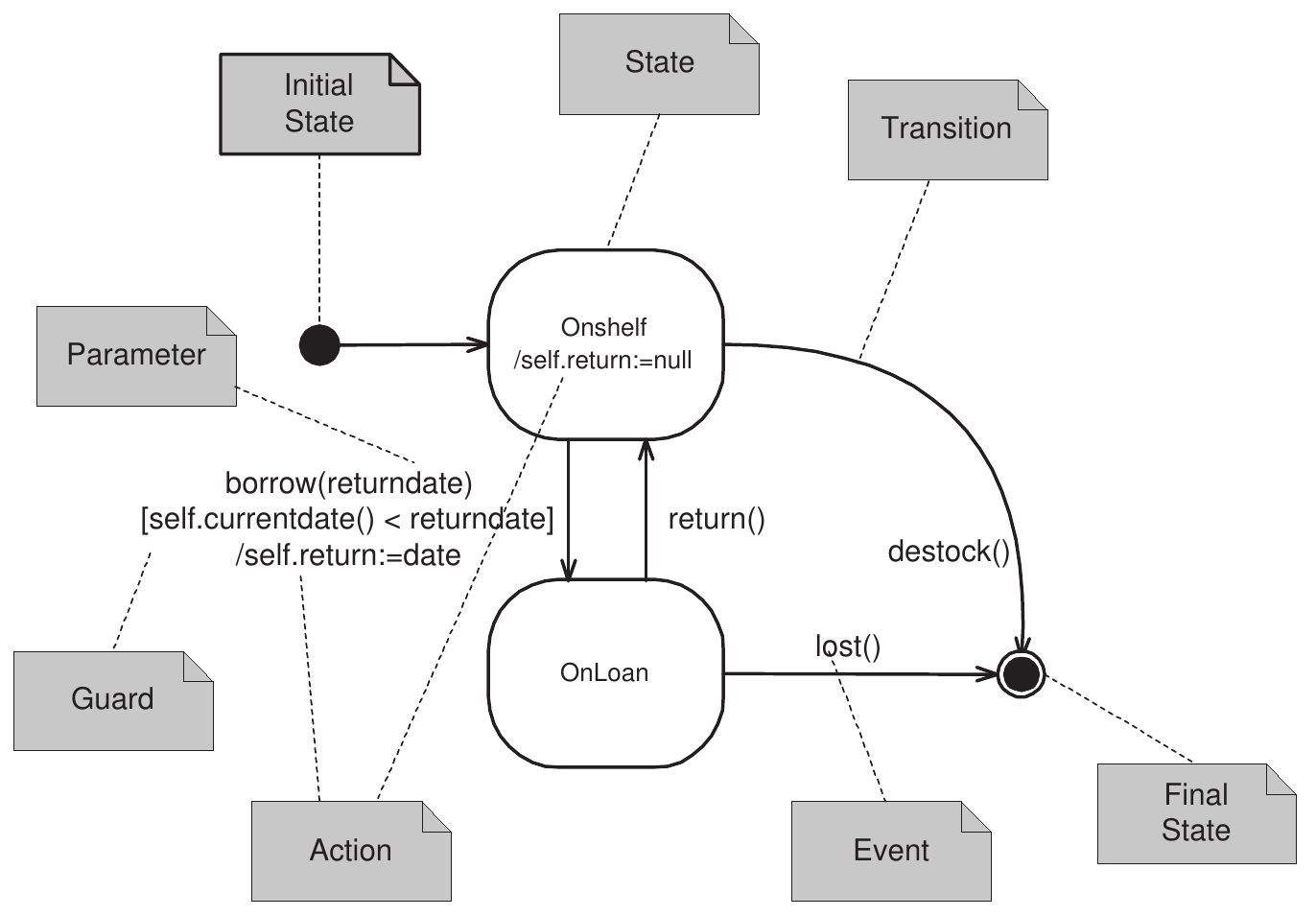}
\caption{An annotated example of a StateMachine}
\label{stmannotated}
\end{center}
\end{figure}

Figure \ref{stmannotated} shows the same StateMachine model 'marked
up' with some of the identified concepts.

At this stage it would also be possible to start listing the
concepts that are used in the body of guards and actions, such as
equality ("="), assignment (":="). This would be a precursor to
constructing models of the expression language they belong to.
However, this is not necessary because we assume that the
expression languages are already defined (in this case OCL and
XOCL).

\subsection{The Model}

Once some candidate concepts have been identified, the next step
is to construct the model.

An important question to ask at this point is whether each concept
is a distinct abstract syntax concept. States and Transitions are
clearly core to StateMachines, as they capture the central concepts
of state and state change. However, initial states and final states
could be argued not to be concepts in their own right. An initial
state can be modelled as an attribute of the StateMachine, whilst a
final state can be viewed as a property of the instance (the
instance has been deleted).

\begin{figure}[htb]
\begin{center}
\includegraphics[width=16cm]{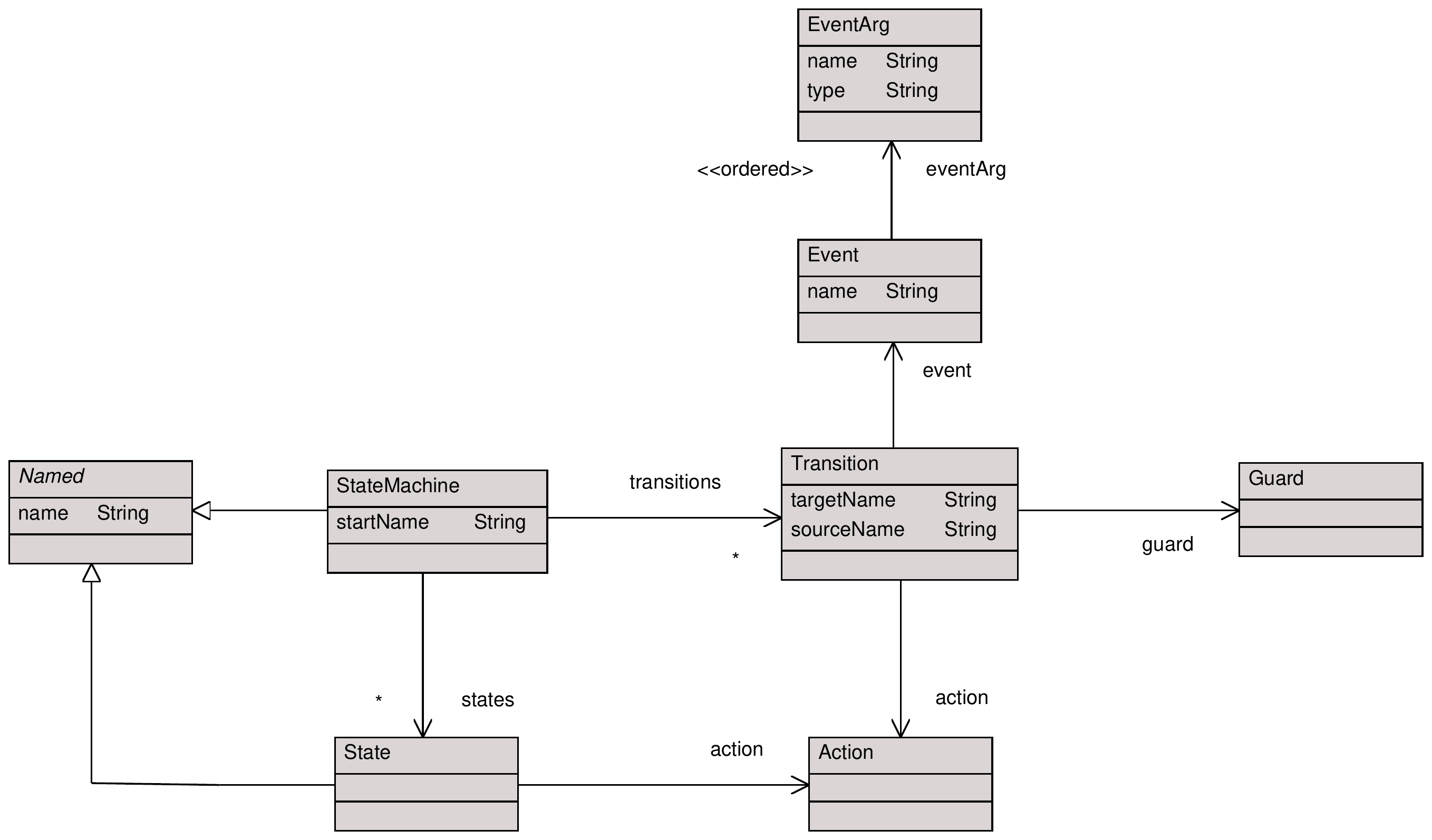}
\caption{An abstract syntax metamodel for StateMachines}
\label{stmabs1}
\end{center}
\end{figure}

The result of constructing the abstract syntax model is shown in
figure \ref{stmabs1}.

There are a number of points to note about the model:

\begin{itemize}
\item Transitions reference their source and target states by
name. \item The actions and guards of states and transitions are
associated with the classes Action and Guard respectively. These
will be extended in the next section to deal with the language
needed to construct action and guard expressions. \item The
initial state of a StateMachine is represented by the attribute
{\em startName}. This is a good example of where the abstract
syntax does not match the concrete syntax one to one (in other
words, the visual representation of an initial state does not have
a counterpart class concept in the abstract syntax model). \item
Transitions are optionally associated with an event, which has a
name, and a sequence of arguments, which have a name and a type.
\end{itemize}

\subsection{Metamodel Reuse}

Once the basic concepts have been identified, the next step is to
identify opportunities for reusing existing metamodelling
concepts. There are two main places where this occurs.

Firstly, elements may specialise existing metamodel concepts. As
described in chapter \ref{langextension}, XMF provides a framework
of classes that are specifically designed for reuse when
metamodelling. In this example, one type of element that can be
reused is a NamedElement. Rather than inventing a new concept, we
can specialise this class wherever we want a concept to have a
name.

Before proceeding we must be clear about the exact properties of
the concept we are specialising. A brief look at the XCore
framework in section \ref{framework} tells us that named elements
also have owners. Moreover, named elements also inherit the
ability to be navigated to via pathnames (this is defined in their
concrete syntax definition). While there are a number of classes
that have a name, it does not make sense for them all to have
owners and to be accessible via pathnames. In particular, concepts
like Event, Message and their arguments types do not have this
property. The classes State and StateMachine on the other hand,
do. A State has a name and it is likely that it will need to know
about the state machine it belongs to. A StateMachine will be
owned by a class.

Another concept that we may wish to specialise is a Contained
element. A contained element has an owner, but does not have a
name. In this example, Transition fits that criteria.

Reuse can also be achieved by referencing existing metamodel
concepts. For example, in order to precisely write guards and
actions, a metamodel of an appropriate expression language will be
required. It is much easier to reuse an existing metamodel that
provides the necessary expression types than to build this model
from scratch. In this case the XOCL metamodel provides a good
match: it includes both constraint style (OCL) expressions and
action style expressions.

\begin{figure}[htb]
\begin{center}
\includegraphics[width=16cm]{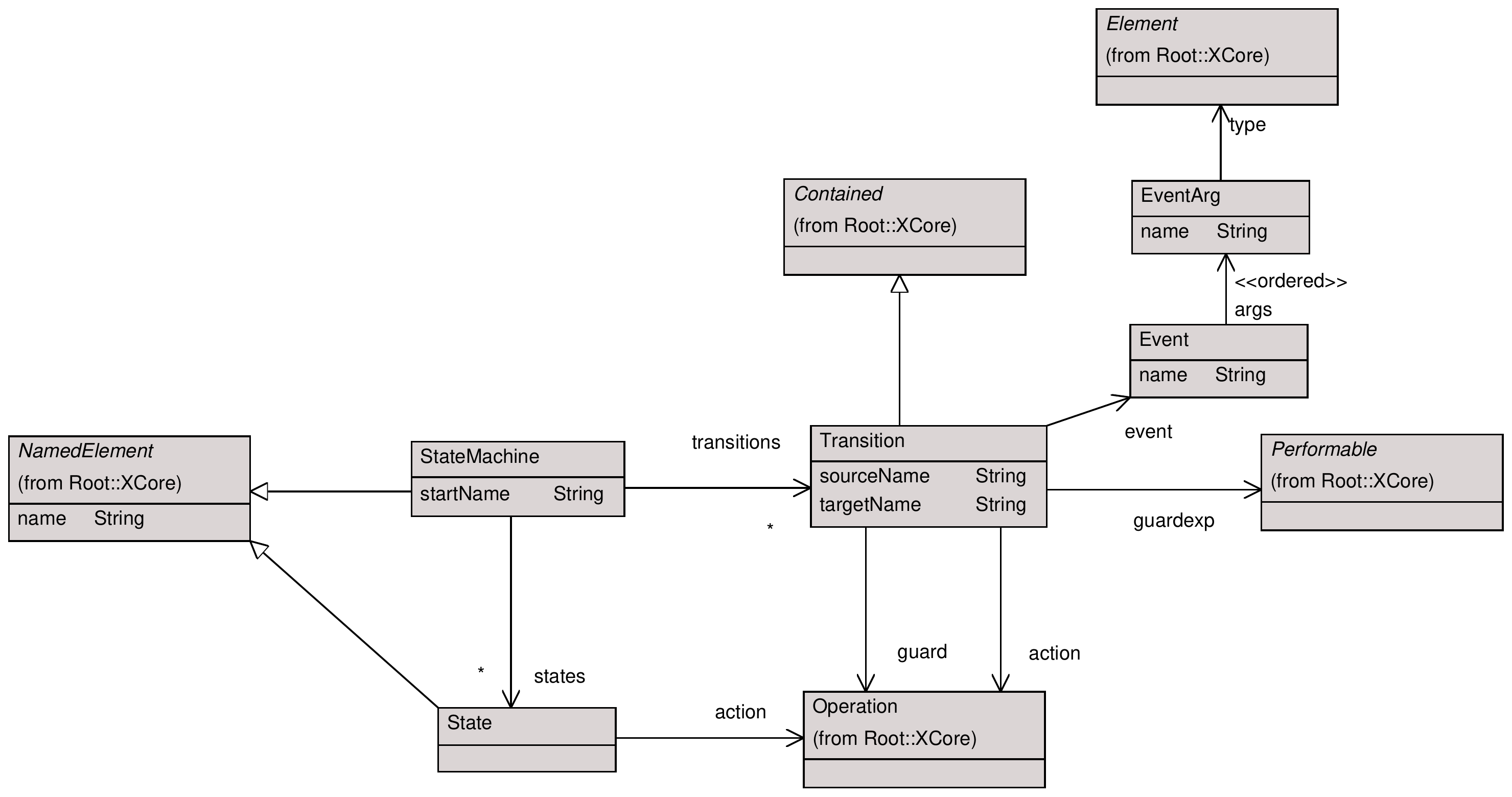}
\caption{An extended abstract syntax metamodel for StateMachines}
\label{stmabs2}
\end{center}
\end{figure}

Thus, the model needs to be changed so that actions and guards are
associated with XOCL expressions. A convenient way of achieving
this is to use the class XMF::Operation to encapsulate the
expression bodies. The resulting metamodel is shown in figure
\ref{stmabs2}

Finally, the type of a parameter is replaced with the class
XMF::Element. This enables parameter types to be of any XMF type,
thereby reusing XMF's type machinery.

\subsection{Well-formedness Rules}

Once the concepts and relationship in the StateMachine language
have been identified,  well-formedness rules can be defined. Here
are some examples:

\noindent Firstly, it must be the case that all states have a
unique names:

\begin{lstlisting}
context StateMachine
  @Constraint StatesHaveUniqueNames
    states->forAll(s1 |
      states->forAll(s2 |
        s1.name = s2.name implies s1 = s2))
  end
\end{lstlisting}\noindent Secondly, the initial state of the StateMachine must be
one the StateMachine's states:

\begin{lstlisting}
context StateMachine
  @Constraint StatesIncludeInitialState
    states->exists(s | s.name = startName)
  end
\end{lstlisting}\subsection{Operations}

Next, operations are added that will be used to create elements of
a StateMachine. The first adds a state to a StateMachine and sets
the owner of State to be the StateMachine. Note the operation
cannot be invoked if the name of the new State conflicts with an
existing state in the StateMachine, thus ensuring that
StatesHaveUniqueNames constraint is not broken.

\begin{lstlisting}
context StateMachine
  @Operation addState(state:StateMachines::State)
    if not self.states->exists(s | s.name = state.name) then
      self.states := states->including(state);
      self.state.owner := self
    else
      self.error("Cannot add a state that already exists")
  end
\end{lstlisting}\noindent The second adds a transition (again setting the owner to
be the StateMachine):

\begin{lstlisting}
context StateMachine
  @Operation addTransition(transition:StateMachines::Transition)
    self.transitions := transitions->including(transition);
    self.transition.owner := self
  end
\end{lstlisting}\noindent Similar operations will need to be defined for deleting
states and transitions.

The following query operations are defined. The first returns the
initial state of the StateMachine provided that it exists, or
returns an error. An error is a pre-defined operation on XMF
elements that is used to report errors.

\begin{lstlisting}
context StateMachine
  @Operation startingState()
    if states->exists(s | s.name = startName) then
       states->select(s | s.name = startName)->sel
    else
      self.error("Cannot find starting state: " + startName)
    end
  end
\end{lstlisting}\noindent The second returns the set of transitions starting from a
particular state:

\begin{lstlisting}
context StateMachine
  @Operation transitionsFrom(state:String)
     transitions->select(t | t.sourceName = state)
  end
\end{lstlisting}\subsection{Validating the StateMachine Metamodel}

\begin{figure}[htb]
\begin{center}
\includegraphics[width=11cm]{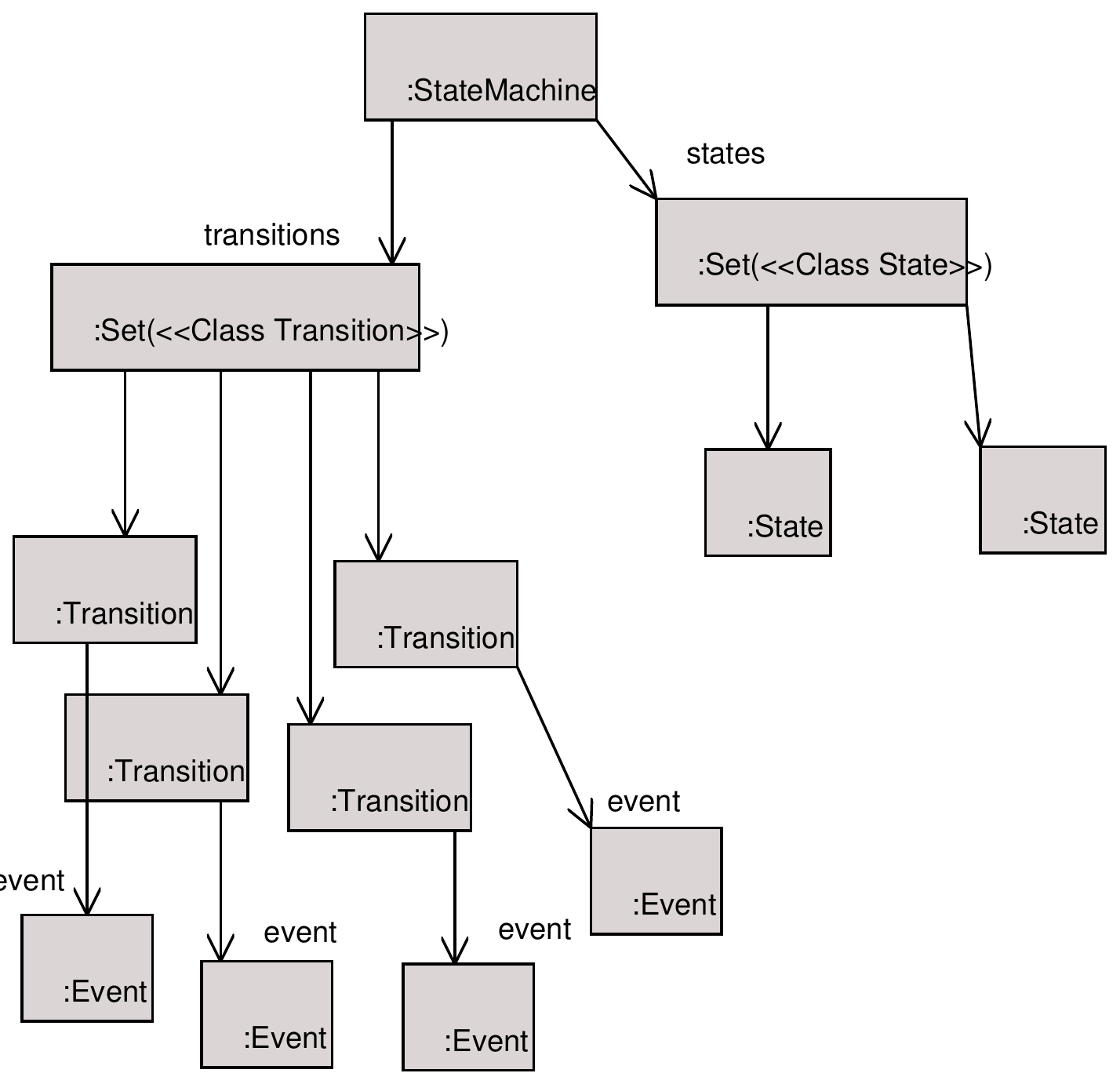}
\caption{A snapshot (object diagram) corresponding to figure
\ref{smexample}} \label{smsnapshot}
\end{center}
\end{figure}

In figure \ref{smsnapshot} a partial snapshot corresponding to the
StateMachine in figure \ref{smexample} is shown.

A number of different properties of the model can be usefully
tested by snapshots. These include checking the well-formedness
rules and query operations by running them against the snapshot.

However, this is about as far as we can go. At this point, the
model is just a model of the abstract syntax and nothing more.
Other aspects of the StateMachine language, including its concrete
syntax and semantics, must be modelled before a precise, self
contained language definition is obtained. These aspects will be
explored in the following chapters.

\section{Conclusion}

This chapter has examined the process and modelling language
required to model the abstract syntax of languages. The result is
a definition of the concepts in a language and the relationship
and rules that govern them. However, this is just a first step
towards a complete definition of the language. Only once the
concrete syntax and semantics of the language are modelled does it
become a complete definition of the StateMachine language.

\chapter{Concrete syntax}
\label{concretechapter}

\section{Introduction}

Although an abstract syntax describes the underlying vocabulary and grammar of a language, it does not define how the abstract syntax is presented to the end user, that detail is described by a concrete syntax which can either be in diagrammatic or textual form.  Many contemporary modelling languages use diagrammatic concrete syntax such as state machines and class diagrams, but diagrammatic syntaxes are often limited by the real estate of the user's display and some languages such as OCL have only a textual syntax.

Dealing with concrete syntax is a two stage process.  The first stage involves interpreting the syntax and ensuring that it is valid.  In the second stage, the concrete syntax is used to build the abstract syntax.  These stages are equally applicable to both text and diagrams although there is an important difference in the way the concrete syntaxes are constructed by the end user. Diagrams are commonly constructed interactively and therefore incrementally, consequently the syntax must be interpreted in parallel to the user's interaction.  On the other hand, textual syntax is usually interpreted in batch, the user constructs the complete model using the syntax and only then is it passed to the interpreter.  This style of processing syntax is comparable to programming language compilers and interpreters.

The first part of this chapter describes the XBNF textual parser component of XMF and demonstrates how it can be used to realise the interpreting of text based concrete syntax and the construction of the abstract syntax.  The second part of the chapter discusses how diagrammatic syntax is defined and how it is linked and synchronised with abstract syntax using XSync.

\section{Textual Syntax}

In this section we describe the parsing of text using XBNF.  The XBNF language is based on EBNF which is a popular approach to describing the grammars of languages, because of this the grammars of most textual languages are easily available (Ada, Java, SQL, for example).  Once a grammar for a language has been defined, the next step is to define how abstract syntax is constructed (or synthesised) as a result of the parsing input using the grammar.  Although the approach described in this section is oriented towards producing the type of abstract syntax structures described in chapter \ref{abschapter}, the mechanisms used are generic one and can be generally applied to a wide range of text parsing problems.

\subsection{Parsing text using EBNF}

The interpretation of textual syntax involves defining the rules by which streams of ASCII characters are deemed to be valid concrete syntax.  These rules are commonly referred to as the grammar rules of a language.  A grammar consists of a collection of clauses of the form:

\begin{verbatim}
NAME ::= RULE
\end{verbatim}

\noindent where a RULE defines how to recognise a sequence of input characters.  An example of a rule is the following:

\begin{verbatim}
Calculator ::= Mult '='
\end{verbatim}

\noindent which defines that to satisfy the rule Calculator the input characters must first satisfy Mult, which is a non-terminal because it refers to another rule, followed by a terminal \emph{'='}.  A terminal is understood to be a sequence of characters in quotes.  A Mult is a multiplicative expression possibly involving the multiplicity (*) and division (/) operators, the grammar for Mult is defined as:

\begin{verbatim}
Mult ::= Add ('*' Mult | '/' Mult)
\end{verbatim}

The rule for Mult shows a number of typical grammar features.  A Mult is successfully recognised when an Add is recognised followed by an optional \emph{'*'} or \emph{'/'} operator.  The choice is described by separating the two or more options (terminals or non-terminals) using the vertical bar.  Consequently this rule defines three possibilities:

\begin{itemize}

\item The input stream satisfies Add followed by a '*' followed by Mult.
\item The input stream satisfies Add followed by a '/' followed by Mult.
\item The input stream satisfies Add.

\end{itemize}

The grammar for Add is the same as Mult except Add recognises addition expressions:

\begin{verbatim}
Add ::=  Number ('+' Add | '-' Add)
\end{verbatim}

\subsection{Parsing text using XBNF}

XBNF augments EBNF with details concerned with managing multiple grammars.  Firstly it is necessary to put grammars somewhere so that they can be used when required, secondly it is necessary to be able to specify given an ASCII stream which grammar should be used to understand the stream.  In order to illustrate XBNF we will define the grammar for a state machine.  The following is an example of a state machine's textual concrete syntax:

\begin{lstlisting}
@StateMachine(Off)
  @State Off
  end

  @State On
  end

  @Transition(On,Off)
  end

  @Transition(Off,On)
  end
end
\end{lstlisting}\noindent As in this example, all textual syntax dealt with by XMF is littered with @ symbols.  These symbols are
special because the word immediately after the symbol indicates where the top level grammar rules can be found which validate the syntax immediately after the declaration.  In the above example, this specifies that the syntax \emph{(Off)} should be understood in the context of the grammar defined in class \emph{StateMachine}, that \emph{Off} and \emph{On} should be understood in terms of the grammar defined in the class \emph{State} and that \emph{(On,Off)} and \emph{(Off,On)} should be interpreted in the context of the grammar defined in the class \emph{Transition}.  The three classes to interpret this collective syntax is defined below\footnote{Note that \emph{Name} is a built in rule which matches any combination of alphanumeric characters}:

\begin{lstlisting}
@Class StateMachine
  @Grammar
    StateMachine ::= '(' startName = Name ')' elements = Exp*
  end
end

@Class State
  @Grammar
    State ::= name = Name
  end
end

@Class Transition
  @Grammar
    Transition ::= '(' sourceName = Name ',' targetName = Name ')'
  end
end
\end{lstlisting}\noindent The grammar rules are embedded in the body of the \emph{@Grammar}.  These say that the textual syntax for a
\emph{StateMachine} consists of a \emph{Name} (the name of the starting state) followed by zero or more \emph{Exp} referring to the whichever grammar rules parses the body of the \emph{StateMachine} (which is likely to be \emph{State} or \emph{Transition}).  A \emph{State} consists only of a \emph{Name} (its name).  A \emph{Transition} consists of two names parsing the source and target state of the transition.

The @ grammar reference symbol as a means of referring to different grammars offers massive flexibility since language grammars can be mixed.  It is possible to use the state machine language in the context of regular XOCL expressions for example (since XOCL and all XMF languages are defined using precisely the method outlined in this book).  Sometimes however the @ symbol can be inconvenient when an existing language must be parsed since the text of that language must be augmented with the @ grammar reference symbol in order to parse it using XBNF.  In this case it is possible to collectively define the grammar rules such that only a single top-level @ is required.  For instance, consider the case of the following state machine grammar definition:

\begin{lstlisting}
@Class StateMachine
  @Grammar
    StateMachine ::= '(' startName = Name ')' elements = (State | Transition)*.

    State ::= name = Name.

    Transition ::= '(' sourceName = Name ',' targetName = Name ')'.
  end
end
\end{lstlisting}\noindent This grammar will support the parsing of the following syntax which has a single @ denoting the single required rule:

\begin{lstlisting}
@StateMachine(Off)
  State Off
  end

  State On
  end

  Transition(On,Off)
  end

  Transition(Off,On)
  end
end
\end{lstlisting}\subsection{Building Abstract Syntax}

\begin{figure}[htb]
\begin{center}
\includegraphics[width=12cm]{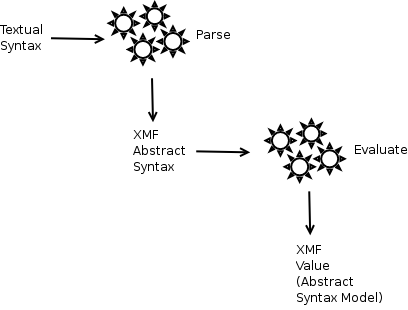}
\caption{The process of translating textual syntax to model-based abstract syntax}
\label{process}
\end{center}
\end{figure}

Having established whether a given textual syntax is valid, the next step is to model the construction of the abstract syntax.  The steps taken by XBNF are illustrated in figure \ref{process}.  As can be seen from this, XBNF produces XMF abstract syntax as a result of the parse and evaluates this to produce an abstract syntax model.  At first glance this may seem confusing since there are two different types of abstract syntax and an evaluation step from one to the other.  The need for this is based on the generality of the XBNF mechanisms, XBNF can be used in the context of a more conventional interpreter where textual input such as 5 + 5 evaluates to the result 10.  The grammar produces the abstract syntax of 5 + 5 and the evaluator evaluates this abstract syntax to produce the value 10.  In the definition of a model based language however, the value required from the evaluator is the model based abstract syntax of the type defined in chapter \ref{abschapter}.  In which case the grammar rules must produce XMF abstract syntax that evaluates to the value of model based abstract syntax.  To avoid confusion, for the remainder of this chapter XMF abstract syntax will be refereed to as \emph{abstract syntax} and abstract syntax model will be refereed to as \emph{abstract syntax value}.

Raw abstract syntax can be defined as a result of grammar rules however defining this can be verbose and time consuming, an alternative is to use a powerful XMF tool which eases the process of this definition.  This tool is designed around the observation that it is often the case that the required abstract syntax can be expressed in terms of an existing XMF based concrete syntax.  For example the state machine abstract syntax value described in chapter \ref{abschapter} is expressed in terms of the syntax for classes and operations.  The tool, called quasi-quoting, allows the description of new concrete syntax in terms of old concrete syntax, because most languages tend to be built from existing languages (and rarely from scratch) we deal with this method in detail in the next sections.

\subsubsection{Quasi-quoting}

Rather then describing the abstract syntax that gives rise to the abstract syntax value, quasi-quotes enables the user to instead say "give me the abstract abstract syntax that you get if you parsed this existing concrete syntax".  Quasi-quotes can be used to find the abstract syntax of any concrete textual syntax by enclosing the textual syntax as exemplified in the following way:

\begin{lstlisting}
[| ... |]
\end{lstlisting}\noindent For example, the following quasi-quotes will return the
abstract syntax for \emph{StateMachine}:

\begin{lstlisting}
[| StateMachine() |]
\end{lstlisting}Often the abstract syntax returned by quasi-quotes is not the required abstract syntax, instead it is desirable to be able to \emph{drop} values into this which have been parsed and stored by the grammar rules.  Syntactically a drop is enclosed by \emph{$<$} and \emph{$>$}.  For example, the grammar for \emph{StateMachine} parses the starting state \emph{startName} which, when evaluated to produce the abstract syntax, should be passed to the constructor of \emph{StateMachine}.  This is done in the following way (ignoring \emph{elements} for the time being):

\begin{lstlisting}
context StateMachine

@Grammar
  StateMachine ::= '(' startName = Name ')' elements = Exp {
    [| StateMachine(<startName>) end |]
  }.
end
\end{lstlisting}The same approach can be used to construct the abstract syntax for a state:

%
\begin{lstlisting}
context State

@Grammar
  State ::= name = Name {
    [| State(<name>) end |]
  }.
end
\end{lstlisting}\noindent Here the abstract syntax has the \emph{name} dropped into the quasi-quotes, and the body of an operation.  When this is evaluated to produce the abstract syntax value, these will be passed as parameters to the constructor of \emph{State}.  \emph{Transition} and \emph{StateMachine} can be defined in a similar way:

\begin{lstlisting}
context Transition

@Grammar
  Transition ::= '(' sourceName = Name ',' targetName = Name ')' {
    [| Transition(<sourceName>,<targetName>) |]
  }.
end

context StateMachine

@Grammar
  StateMachine ::= '(' startName = Name ')' elements = Exp* {
    [| StateMachine(<startName>,<elements>) |]
  }.
end
\end{lstlisting}

\subsubsection{Sugar}

One of the side effects of using the quasi-quoting mechanism, as illustrated in the previous section, or creating abstract syntax directly, is that the original concrete syntax is lost since the translation does not record details of the original concrete syntax.   Often it is desirable to be able to make the transition back from the abstract syntax to the concrete syntax, a good example of where this is useful is during the process of debugging a model.  In order to support this XMF provides a further tool called \emph{Sugar}\footnote{The process of \emph{Sugaring} and \emph{Desugaring} (often used in programming language theory) is used to describe the process of translating to and from new syntax using existing syntax.  It captures the idea that the new syntax is not adding any fundamentally new semantics to the language, merely making existing syntax more palatable (or sweeter!) to the end user of the syntax.}.  Instead of the grammar rules constructing the abstract syntax directly, they instantiate classes of type \emph{Sugar} which then create the abstract syntax and provide methods for doing other things such as translating the consequent abstract syntax back to the concrete syntax.

An example of \emph{Sugar} class is illustrated below:

\begin{lstlisting}
@Class StateSugar extends Sugar
  @Attribute name : String end

  @Constructor(name) end

  @Operation desugar()
    [| State(<name>) |]
  end

  @Operation pprint(out,indent)
    format(out,"@State ~S",Seq{name})
  end
end
\end{lstlisting}Any class extending \emph{Sugar} must override the abstract operation \emph{desugar()} which should return the required abstract syntax.  The additional (and optional) method \emph{pprint} is used to \emph{pretty print} the original concrete syntax from the abstract syntax.  This method takes an output stream (to print the text to) and an indent variable which denotes the current indentation of the textual syntax.  In the above example, the text \emph{@State} and the state name is sent to the output stream.

The grammar rule constructs \emph{Sugar} classes as appropriate:

\begin{lstlisting}
context State
  @Grammar
    State ::= name = Name action = Exp* {
      StateSugar(name,action)
    }.
  end
\end{lstlisting}
\section{Diagrammatic Syntax}

As indicated in the introduction to this chapter, the major difference between diagrammatic syntax and textual syntax is that diagrammatic syntax is interpreted incrementally whereas textual syntax is interpreted in batch.  This presents a different set of challenges for diagrammatic syntax to that of its textual counterpart, the incremental nature of its construction must be taken into account when devising an architecture in order to support it.  A further challenge for diagrammatic languages is being able to specify the valid concrete syntax of the diagram in a
textual language such as XMF.  As we have discussed in the previous section, this is well understood for textual syntax using notations such as EBNF, but for diagrams it is less clear how the equivalent can be achieved.  In the following sections we discuss the XMF approach to doing this.

\subsection{Parsing Diagrams}

\begin{figure}[h]
\begin{center}
\includegraphics[width=15cm]{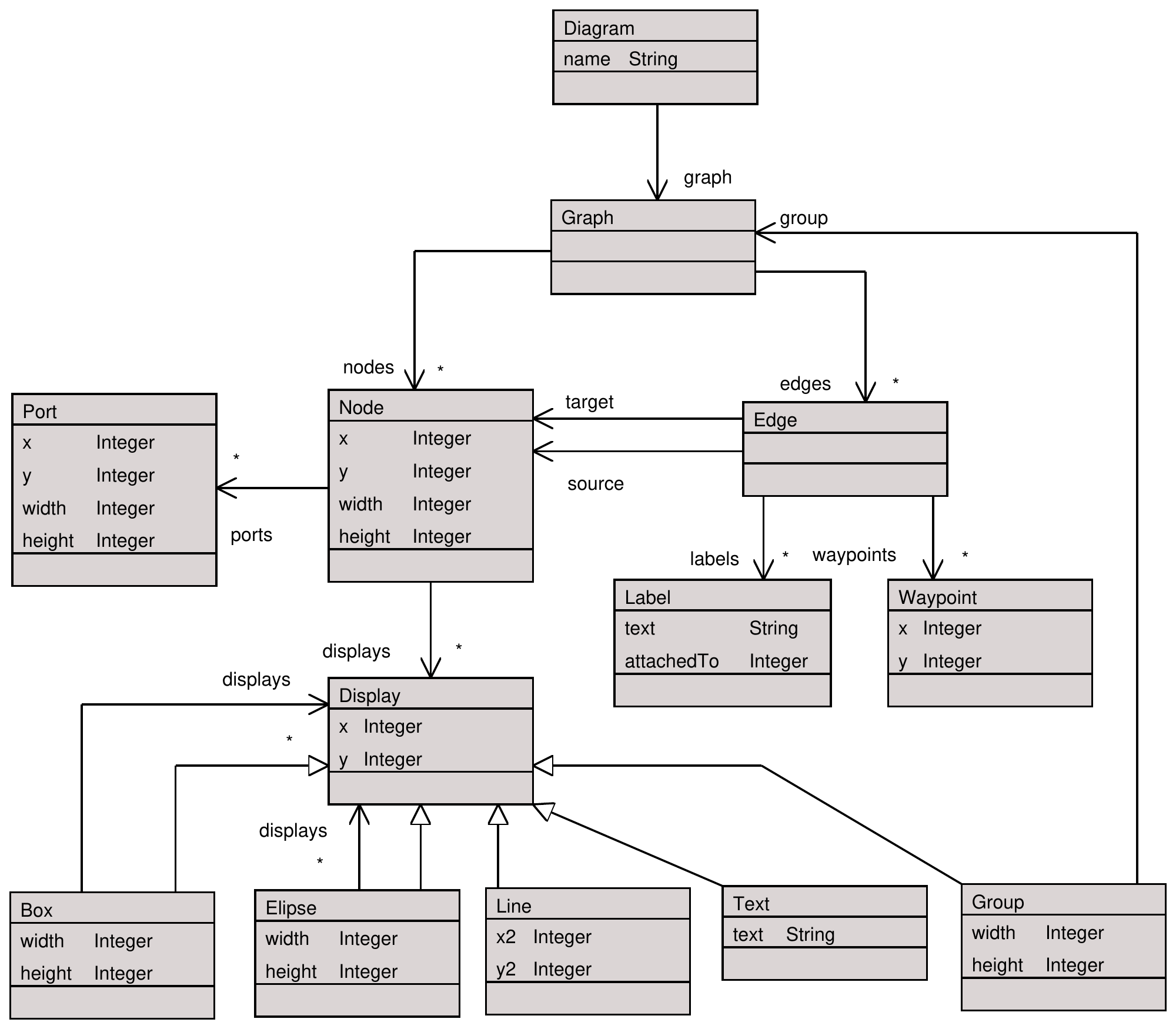}
\caption{A model of diagrams}
\label{modelOfDiagrams}
\end{center}
\end{figure}

In order to describe how to interpret a diagram, we must first define what it means to be a diagram at some (useful) level of abstraction.  This is achieved by forming a model of diagrams in XMF as illustrated in figure \ref{modelOfDiagrams}, this model shares many parallels with OMG's diagram interchange model \cite{}. A key characteristics which it shares with the diagram interchange model is that its generality enables it to capture the concrete syntax concepts, and the relationship between these, for a broad spectrum of diagram types ranging from  sequence diagrams to state machines to class diagrams.

Clearly the model shown in figure \ref{modelOfDiagrams} does abstract from details that may be important such as their rendering and the paradigm of interaction that gives rise to their state being changed.  This detail is outside the scope of the model, but is an important component of realising this approach in the context of a real tooling environment.  One approach to dealing with this detail is to interface to some external tool that understands how
to render the concepts and respond to user interaction.  Since this is an implementation detail we do not dwell on it here, instead we make the assumption that the instantiating of concepts in figure \ref{modelOfDiagrams} results in the concrete display of the denoted element.  Moreover we assume that interaction with the concrete display, moving a node for example, results in the state of its model-based counterpart being updated.  A suitable implementation of this scenario can be black boxed such that new diagram types are constructed only as a result of dealing with the model.

\begin{figure}[h]
\begin{center}
\includegraphics[width=16cm]{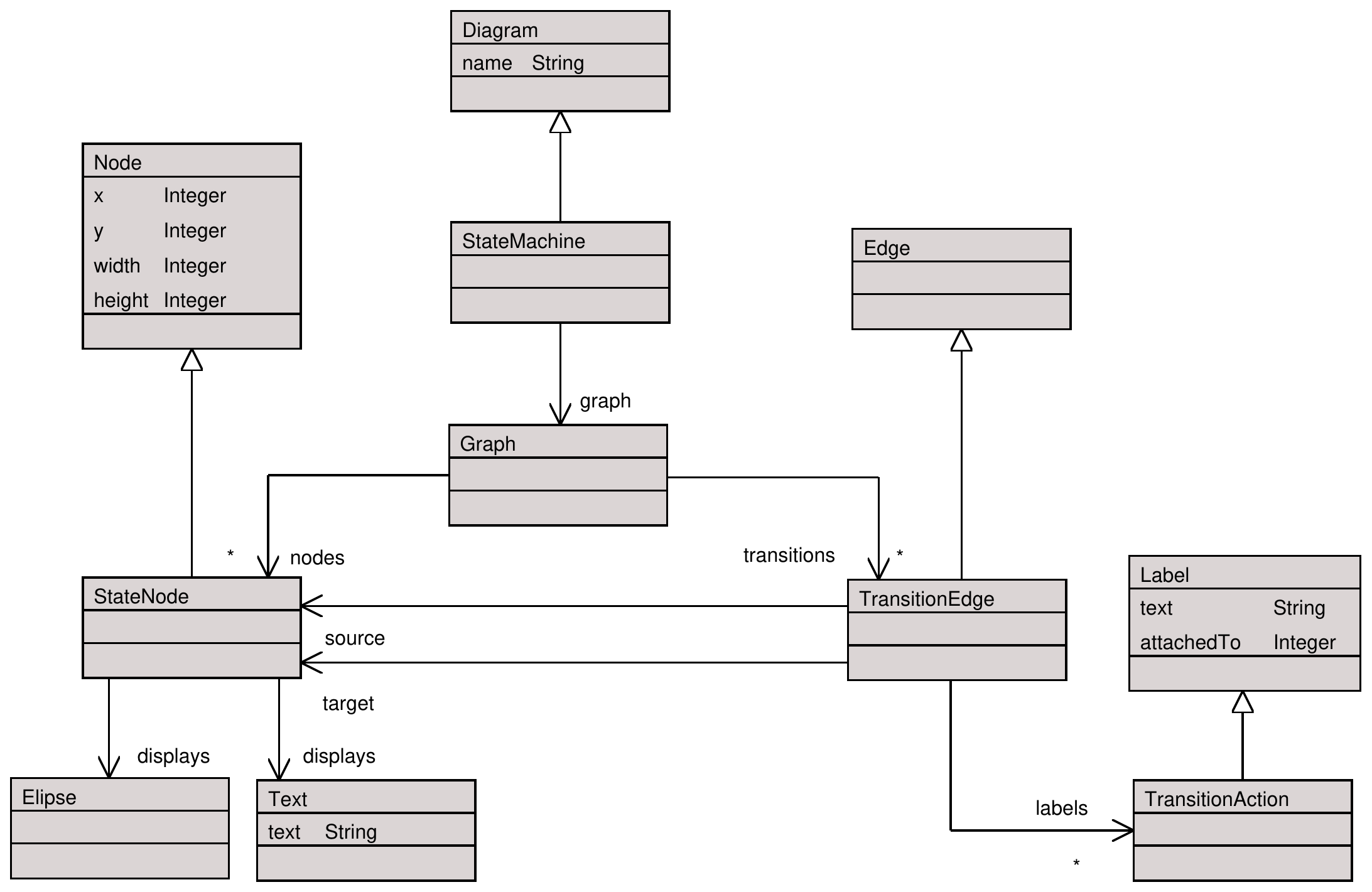}
\caption{Specialising the model of diagrams for state machines}
\label{stateMachineDiagram}
\end{center}
\end{figure}

Specific diagramming types are described by specialising the model of diagrams, this is shown in figure \ref{stateMachineDiagram} for state machines.  In this a \emph{StateDiagram} is a type of \emph{Diagram}, the \emph{Graph} of a \emph{StateMachine} is constrained to contain \emph{TransitionEdge}s and \emph{StateNode}s
(which are specialised \emph{Edge}s and \emph{Node}s respectively). A \emph{TransitionEdge} is constrained to be the \emph{source} and \emph{target} of \emph{StateNode}s, and a \emph{StateNode} is displayed using an \emph{Elipse}.

\subsection{Building Abstract Syntax}

\begin{figure}[htb]
\begin{center}
\includegraphics[width=10cm]{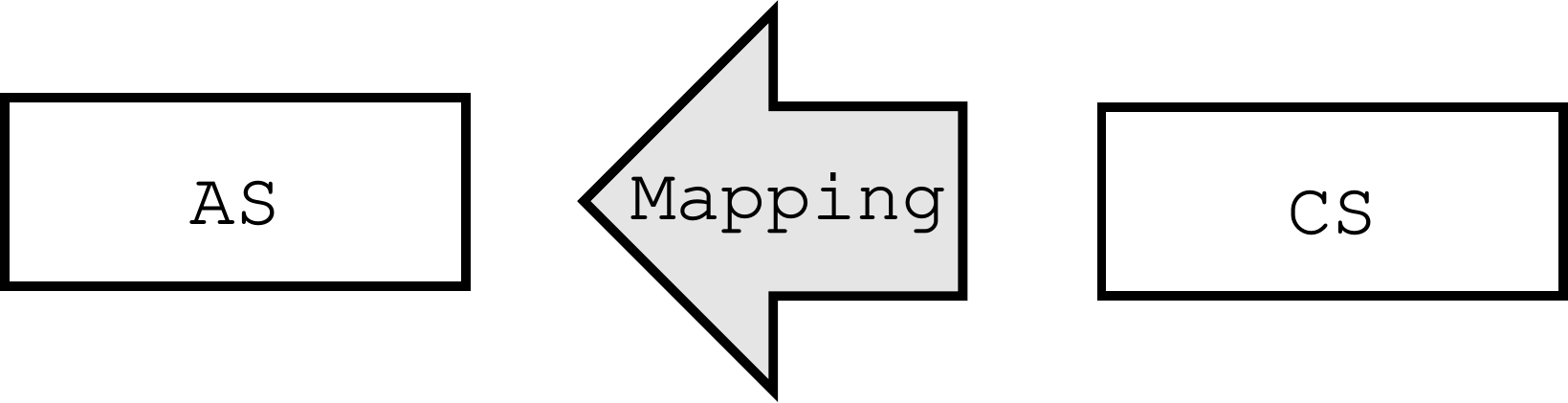}
\caption{Mapping between diagram syntax and abstract syntax}
\label{uniDirMapping}
\end{center}
\end{figure}

As the user interactively constructs instances of the model of diagrams the abstract syntax should be concurrently constructed as appropriate.  One approach to achieving this is to embed operations in the concrete syntax which generate abstract syntax on the fly.  Although this approach would work well in practice, a deficiency is that it couples the abstract syntax to the concrete syntax.  In practice it may be desirable to have a single diagram
type generating different abstract syntaxes under different circumstances.  A much more flexible approach is to form a mapping between the concrete syntax and abstract syntax which neither side know about.  This scenario is illustrated in figure
\ref{uniDirMapping}.

In this the mapping listens to changes in the concrete syntax for changes that impact the abstract syntax and updates the abstract syntax appropriately.  Given a \emph{StateDiagram} and an
\emph{StateMachine}:

\begin{lstlisting}
@Class StateDiagram extends Diagram
  ...
end

@Class StateMachine

  @Attribute startName : String end
  @Attribute states : Set(State) end
  @Attribute transitions : Set(Transition) end

  ...
end
\end{lstlisting}\noindent A mapping is defined:

\begin{lstlisting}
@Class StateMachineXStateDiagram

  @Attribute statemachine      : StateMachine end
  @Attribute diagram           : StateDiagram end
  @Attribute stateMaps         : Set(StateXNode) end
  @Attribute transitionMaps    : Set(TransitionXTransitionEdge) end
  ...
\end{lstlisting}
\noindent The mapping knows about the top level constructs of the statemachine's concrete and abstract syntax.  When the mapping is constructed, a listener (or daemon) is placed on the \emph{diagram}'s graph, such that when a state added or removed, a method is invoked adding or removing a \emph{State} from the abstract syntax \emph{statemachine}:

\begin{lstlisting}
  @Constructor(statemachine,diagram)
    self.addDaemons()
  end

  @Operation checkDiagramDaemons()
    if not diagram.hasDaemonNamed("stateAdded") then
      @SlotValueChanged + stateAdded(
        diagram.graph,"nodes",newStateNode)
          self.stateNodeAdded(newStateNode)
      end
    end;
    if not diagram.hasDaemonNamed("stateRemoved") then
      @SlotValueChanged - stateRemoved(
        diagram.graph,"nodes",removedStateNode)
          self.stateNodeAdded(removedStateNode)
      end
    end
  end
\end{lstlisting}
\noindent Listeners to detect the addition and removal of transitions are implemented in a similar way.  The methods for adding and removing the \emph{State} from the \emph{statemachine} are specified as follows:


\begin{lstlisting}
  @Operation stateNodeAdded(newStateNode)
    if not stateMaps->exists(map | map.stateNode = newStateNode)
    then
      let name = self.newStateName() then
        state = State(name)
        in newStateNode.setName(name);
          self.add(StateXNode(state,newStateNode,self));
          statemachine.add(state)
      end
    end
  end

  @Operation classRemoved(class)
    @Find(map,classMaps)
      when map.class = class
      do self.remove(map);
        map.node.delete()
    end
  end
end
\end{lstlisting}
\FloatBarrier

\begin{figure}[htb]
\begin{center}
\includegraphics[width=14cm]{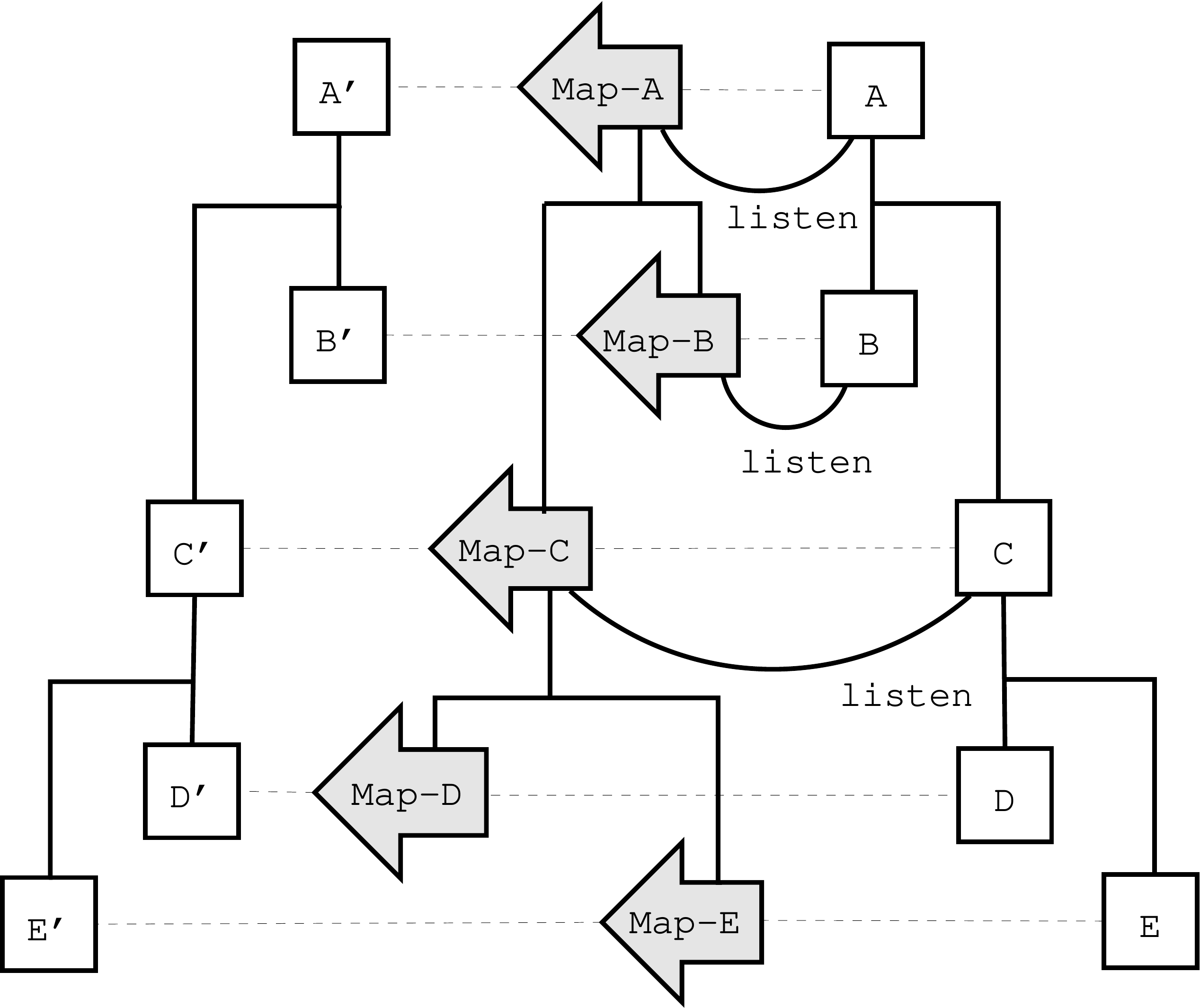}
\caption{An example of how listeners can generate mappings which can themselves listen and generate mappings}
\label{hierarchy}
\end{center}
\end{figure}

The style of mapping concrete syntax to abstract syntax can be nested  within containership relationships such that a new mapping is generated from a listener, which in turn listens to the newly mapped elements and generates new mappings as concrete syntax elements are created ... This gives rise to the hierarchy exemplified in figure \ref{hierarchy} where only the top level mapping between concrete syntax element \emph{A} and abstract syntax element \emph{A'} is generated statically (i.e. not via listeners).

\subsection{Building Diagrams from Abstract Syntax}

\begin{figure}[htb]
\begin{center}
\includegraphics[width=16cm]{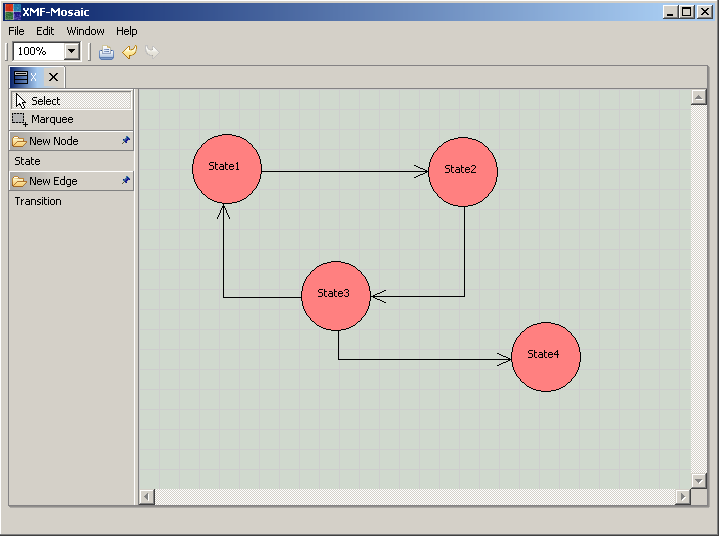}
\caption{}
\label{stateMachineTool}
\end{center}
\end{figure}

The approach to mapping concrete syntax to abstract syntax described in the  previous section can equally be applied to the inverse mapping of abstract syntax to concrete syntax, potentially this can be done simultaneously so that changes to either side of the mapping result in an update to the other.  Such bi-directional mappings are particularly desirable in a scenario where an abstract syntax is the source of multiple concrete syntaxes (which are effectively views on the abstract syntax) a change to one concrete syntax will propagate the change to another concrete syntax via the shared abstract syntax.  A screenshot of the XMF-Mosaic using this style of bi-directional concrete syntax definition is shown in \ref{stateMachineTool}.

\section{Conclusion}

In order to support the rapid creation of concrete syntaxes for a modelling language, higher level modelling languages are required. This chapter has described XBNF, a language for modelling textual syntaxes. In order to model diagrammatical syntaxes, a framework for modelling common diagrammatic primitives is required, along with a language for describing how the diagrammatic syntax is synchronised with the underlying abstract syntax model of the language.  In chapter \ref{mappingchapter} XSync is introduced which is a declarative language for defining the synchronisation of data and languages.  XSync can be used to succinctly define the types of synchronisations between abstract and concrete syntax discussed in this chapter.

\chapter{Semantics}
\label{semanticschapter}

\section{Introduction}

The purpose of this chapter is to describe how the semantics of
modelling and programming languages can be described using
metamodels. The semantics of a modelling language describes what
the language means in terms of its behaviour, static properties or
translation to another a language. The chapter begins by
motivating the need for semantics and then describes a number of
different approaches to defining semantics using metamodels. These
approaches aim to provide a way of constructing platform
independent models of semantics - thus enabling the semantics of a
language to be interchanged between metamodelling tools.  An
example of how the approaches can be used to define the semantics
of the StateMachine language is then presented.

\section{What is Semantics?}

In general terms, a language semantics describes the {\em meaning}
of concepts in a language. When using a language, we need to
assign a meaning to the concepts in the language if we are to
understand how to use it. For example, in the context of a
modelling language, our understanding of the meaning of a
StateMachine or the meaning of a Class will form a key part of how
we choose to use them to model a specific aspect of a problem
domain.

There are many ways of describing meaning of a language concept.
Consider the ways in which the meaning of concepts in natural
languages can be described:

\begin{itemize}
\item In terms of concepts which already have a well defined
meaning. For instance "a car consists of a chassis, four wheels,
an engine, body, and so on". This is only meaningful if the
concepts themselves are well defined. \item By describing the
properties and behaviour of the concept: "a car can be stationary,
or can be moving, and pressing the accelerator increases its
speed". \item As a specialisation of another concept. For
instance, "a truck is a vehicle with a trailer". \item By
describing the commonly shared properies of all possible instances
of a concept. For example, the concept of a car could be described
in terms of the valid properties that every instance of a car
should have.
\end{itemize}

In a natural language, semantics is a correlation or mapping
between concepts in a language with thoughts and experiences of
concepts in world around us. Although a more formal approach to
semantics is required for modelling and programming languages,
there is a close parallel to the approaches used to express
natural language semantics described above. In both cases, a key
requirement of a semantics is that it should be of practical use
in understanding the meaning of a language.

\section{Why Semantics?}

A semantics is essential to communicate the meaning of models or
programs to stakeholders in the development process. Semantics
have a central role to play in defining the semantically rich
language capabilities such as execution, analysis and
transformation that are required to support Language-Driven
Development. For example, a language that supports behaviour, such
as a StateMachine, requires a semantics in order to describe how
models or programs written in the language execute.

Traditionally, the semantics of many modelling languages are
described in an informal manner, either through natural language
descriptions or examples. For instance, much of the UML 1.X
specification \cite{umlspec} makes use of natural language
descriptions of semantics.

However, an informal semantics brings with it some significant problems:

\begin{itemize}
\item Because users have to assign an informal or intuitive
meaning to models, there is significant risk of misinterpretation
and therefore misuse by the users of the modelling language.

\item An informal semantics cannot be interpreted or understood by tools. Tool
builders are thus required to implement their own interpretation
of the semantics. Unfortunately, this means that the same language
is likely to be implemented in different ways. Thus, two different
tools may offer contradictory implementations of the same
semantics, e.g. the same StateMachine may execute differently
depending on the tool being used!

\item An informal semantics makes the task of defining new
languages difficult. It  makes it hard to identify areas where
concepts in the languages are semantically equivalent or where
there are contradictions. It is harder still to extend an existing
language if the semantics of the language are not defined.

\item Standards require precise semantics. Without them, a
standard is open to misunderstanding and misuse by practitioners
and tool developers while significantly limiting interoperablity.
\end{itemize}

\section{Semantics and Metamodels}

While it is clear that semantics is crucial part of a language
definition, the question is how should semantics be described? One
approach is to express semantics in terms of a formal,
mathematical language. Many academic papers have been written
describing the semantics of modelling languages in this way. Yet,
the complex mathematical descriptions that result are hard to
understand and are of limited practical use. Another approach is
to express semantics in terms of an external programming language.
This is a far more practical solution. Nevertheless, it results in
a definition which is tied to a specific programming language thus
compromising its platform independence. Furthermore, being forced
to step out of the metamodelling environment into a programming
language makes for a very non-intuitive language definition
process.

An alternative strategy is to describe the semantics of languages
using metamodels. There are some significant benefits to this
approach. Firstly, the semantics of a language is fully integrated
in the language's definition, which means the task of relating it
to other language design artifacts (concrete syntax, mappings,
abstract syntax, etc) is immediately simplified. Secondly, because
the same metamodelling language is used across all languages,
semantic definitions become reusable assets that can be integrated
and extended with relative ease. Finally, and most crucially,
semantics definitions are platform independent - they can be
interchanged in the same way, and if they are understood by the
tool that imports them, they can be used to drive the way that the
tool interacts with the language. For instance, a semantics that
describes how a StateMachine executes can be used to drive
simulators across a suite of compliant tools. The ability to
define semantics in a platform independent way is crucial to the
success of Language-Driven Development. It enables tools to be
constructed for languages that are both portable and interoperable
irrespective of the environment in which they are defined.

It is important to note that a metamodel semantics is quite
different  from the abstract syntax model of a language, which
defines the structure of the language. However, an abstract syntax
model is a pre-requisite for defining a semantics, as a semantics
adds a layer of meaning to the concepts defined in the abstract
syntax. Semantics in this sense should also be distinguished from
static semantics, which are the rules which dictate whether or not
an expression of the language is well-formed. Static semantics
rules are those employed by tools such as type checkers.

\section{Approaches}

There are many different approaches to describing the semantics of
languages in a metamodel. This section examines some key
approaches and gives examples of their application. All the
approaches are motivated by approaches to defining semantics that
have widely been applied in programming language domains. The main
difference is that metamodels are used to express the semantic
definitions.

The approaches include:

\begin{description}
\item [Translational] Translating from concepts in one language
into concepts in another language that have a precise semantics.
\item [Operational] Modelling the operational behaviour of
language concepts. \item [Extensional] Extending the semantics of
existing language concepts.\item [Denotational] Modelling the
mapping to semantic domain concepts.

\end{description}

Each approach has its own advantages and disadvantages. In
practice, a combination of approaches is typically used, based on
the nature of individual concepts in the language. In each case,
it is important to note that the semantics are described in the
metamodelling language - no external formal representation is
used. The following sections describe each of the approaches,
along with a small example of their application.

\section{Translational Semantics}

Translational semantics is based on two notions:

\begin{itemize}
\item The semantics of a language is defined when the language is
translated into another form, called the target language. \item
The target language can be defined by a small number of primitive
constructs that have a well defined semantics.
\end{itemize}

The intention of the translational approach is to define the
meaning  of a language in terms of primitive concepts that have
their own well defined semantics. Typically, these primitive
concepts will have an operational semantics (see later).

The advantage of the translational approach is that provided there
there is a machine that can execute the target language, it is
possible to directly obtain an executable semantics for the
language via the translation. This approach is closely related to
the role played by a compiler in implementing a richer programming
language in terms of more primitive executable primitives.

The main disadvantage of the approach is that information is lost
during the transformation process. While the end result is a
collection of primitives, it will not be obvious how they are
related to the original modelling language concepts. There are
ways to avoid this, for instance information about the original
language concepts can be "tagged" onto the target language, thus
enabling it to retain information about the structure of the
original language. Alternatively, one may consider maintaining
information about the mapping between the two models.

The translational approach can be incorporated in a language
definition in many different ways:

\begin{itemize}
\item Within a language metamodel by translating one concept into
a concept that has a semantics, e.g. translating a case statement
into a sequence of if statements, where the if statements have an
operational semantics. Another example might be translating a rich
mapping language into a collection of primitive mapping functions
with additional information tagged onto the primitives to indicate
the structure of the source language. \item Between language
metamodels, by translating one metamodel into another. For
example, an abstract syntax metamodel for UML can be mapped to a
metamodel of a small well-defined language such as XCore, or to a
programming language such as Java.
\end{itemize}

\subsection{Approaches}

There are a number of different approaches to implementing a
transformational semantics in XMF:

\begin{itemize}
\item A mapping can be defined between the abstract syntax of the
source and target languages. This could be written in a mapping
language (such as XMap), or implemented by appropriate operations
on the source model. \item The result of parsing the concrete
syntax of the source language can be used to directly create an
instance of the target language. \item The result of parsing the
concrete syntax can be translated directly into machine code that
can be run on the XMF virtual machine.
\end{itemize}

In the second approach, the intermediate step of generating an
abstract syntax model for the source language can be omitted. The
concrete syntax of the source language thus acts as {\em sugar}
for the target language. This approach is particularly useful when
constructing new types of expressions that can be desugared into a
number of more primitive concepts, e.g. translating a case
statement into a collection of if expressions.

The third approach requires the construction of machine code
statements but has the advantage of resulting in a highly
efficient implementation.

\subsection{Example}

\begin{figure}[htb]
\begin{center}
\includegraphics[width=14cm]{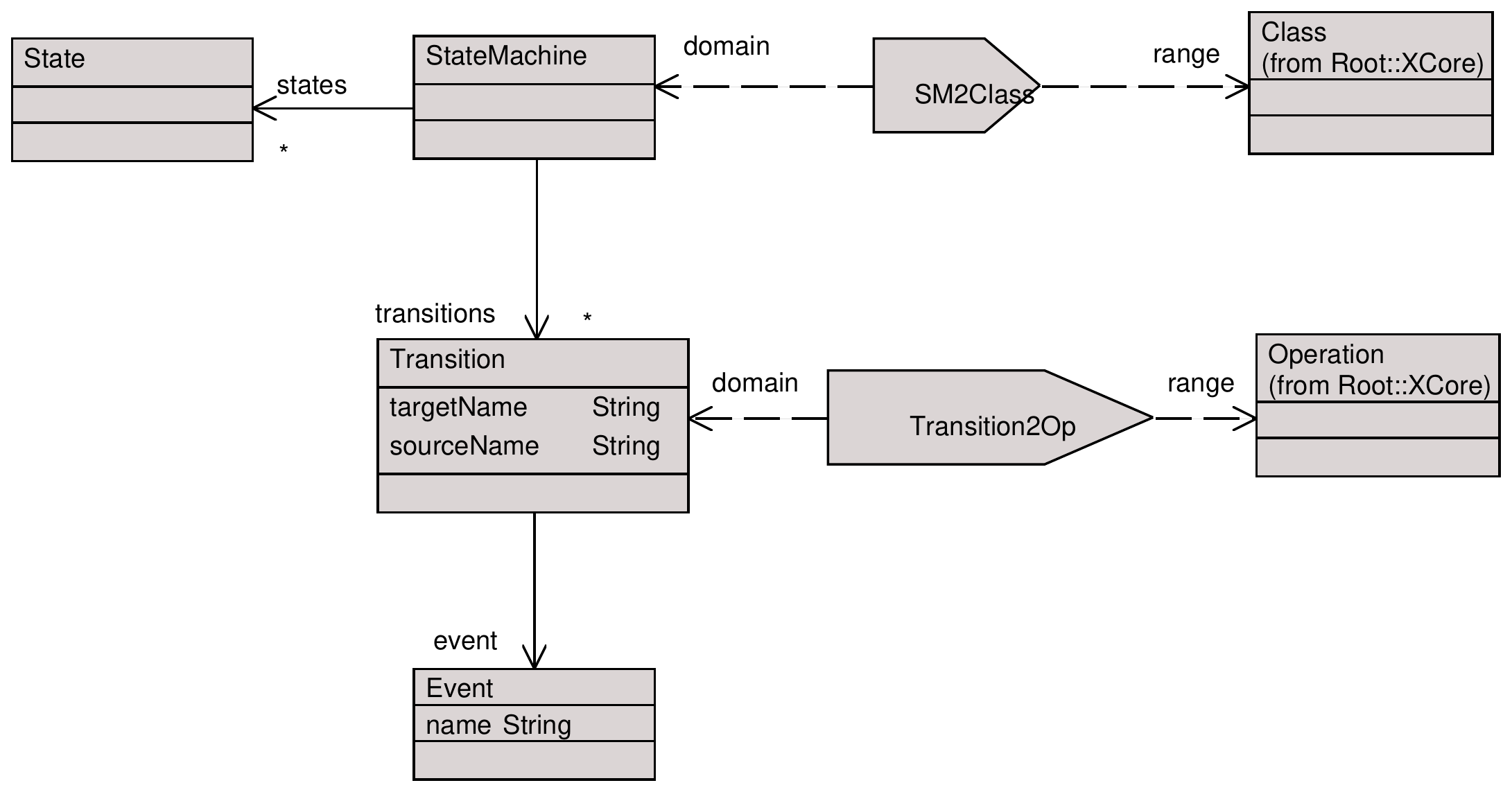}
\caption{Overview of the translational mapping example}
\label{translationExample}
\end{center}
\end{figure}

Because XMF provides a number of well defined primitives, it is a
good target for translation. As an example, consider defining a
semantics for StateMachines via a translation from the
StateMachine abstract syntax model (described in chapter
\ref{abschapter}). The mapping language used in this example is
described in detail in chapter \ref{mappingchapter}.  A summary of
the approach taken is shown in figure \ref{translationExample}.

The aim is to translate a StateMachine into an instance of a XCore
Class such that semantics of the StateMachine is preserved. This
is achieved by ensuring that the Class simulates the behaviour of
the StateMachine:

\begin{itemize}
\item The StateMachine is translated into a Class containing an
attribute called {\em state}, whose permitted values are the
enumerated state names of the StateMachine.
\item The Class will inherit all the attributes and operations
of the StateMachine's context class.
\item Each transition of the StateMachine is translated
into an operation of the Class with the same name as the
transition's event.
\item The transition will contain code that will simulate the
invocation of the transition's guard and actions, causing the value
of {\em state} to change.
\end{itemize}

\noindent The following pattern of code will thus be required in
the body of each operation:

\begin{lstlisting}
  if <guard> then
    self.state := <target-state-name>;
    <action>
  end
\end{lstlisting}

Where $<$guard$>$ and $<$action$>$ are the bodies of the
corresponding transition's guard and action, and
$<$target-state-name$>$ is the name of the state that is the target
of the transition.

The most challenging part of this translation is creating the
appropriate code bodies in each operation. The following code
shows how XOCL can be used to define an operation, transitionOp(),
which given the guard, action, event name and target name of a
transition, returns an XCore operation that simulates the
transition. This operation is called by the transition2Op mapping.
Note that in the operation transitionOp() an operation cannot be
assigned a name until it is declared, thus setName() is used to
assign it a name after it has been declared.

\begin{lstlisting}
@Map Transition2Op
  @Operation
  transitionOp(g:Operation,a:Operation,eventName,targetName)
    let op = @Operation()
      if g() then
        a();
        self.state := targetName
      end
    end
    in
      op.setName(eventName);
      op
    end
  end

  @Clause Transition2Op
    Transition
      [event = Set{
        Event
          [name = N]
        },
       name = T,
       guard = G,
       action = A]
    do
      self.transitionOp(G,A,N,T.name)
  end
end
\end{lstlisting}

This definition of transitionOp() makes use of the fact that in
XMF an Operation is also an Object and can therefore be assigned
to a variable.

As an example, consider the simple traffic light model shown in
figure \ref{trafficlight}. The result of applying the
Transition2Op mapping to the GreenRed() transition will be the
following XOCL operation:

\begin{figure}[htb]
\begin{center}
\includegraphics[width=11cm]{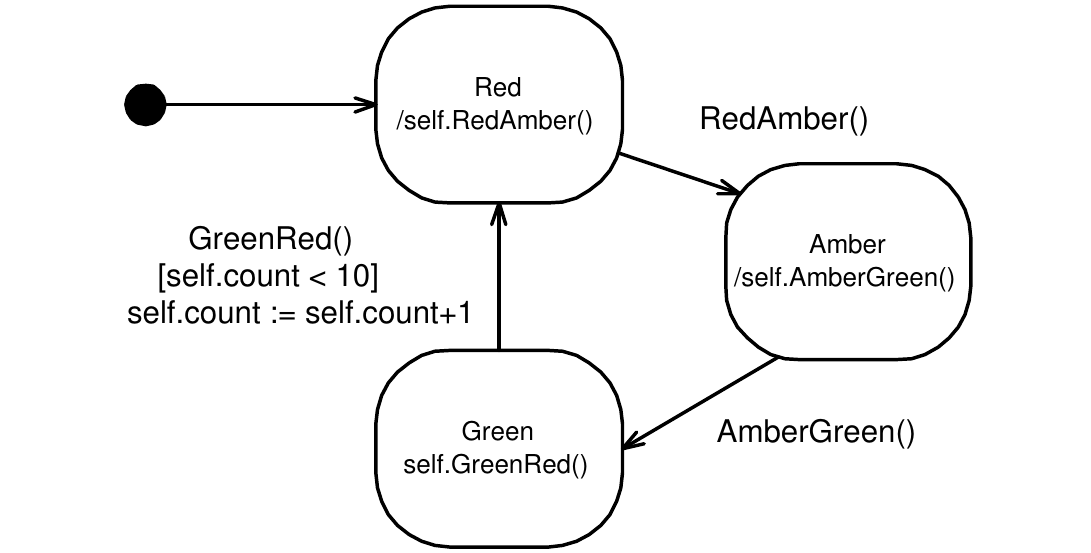}
\caption{The traffic light example} \label{trafficlight}
\end{center}
\end{figure}

\begin{lstlisting}
@Operation GreenRed()
  if g() then
    a();
    self.state := "Red"
  end
end
\end{lstlisting}

\noindent where g and a are the following anonymous operation
definitions:

\begin{lstlisting}
@Operation anonymous()
 self.count < 10
end

@Operation anonymous()
  self.count := self.count + 1
end
\end{lstlisting}

\section{Operational Semantics}

An operational semantics describes how models or programs written
in a language can be directly executed. This involves constructing
an interpreter. For example, an assignment statement "V := E" can
be described by an interpreter that executes the steps that it
performs: Evaluate the expression E and then change the value
bound to the variable V to be the result.

The advantage of an operational semantics is that it is expressed
in terms of operations on the language itself. In contrast, a
translational semantics is defined in terms of another, possibly
very different, language. As a result, an operational semantics
can be easier to understand and write.

Writing an interpreter as part of a metamodel relies on the
metamodelling language itself being executable. Provided this is
the case, concepts can define operations that capture their
operational behaviour.

Typically, the definition of an interpreter for a language follows
a pattern in which concepts are associated with an operational
description as follows:

\begin{itemize}
\item Operations will be defined on concepts that implement their
operational semantics e.g. an action may have an run() operation
that causes a state change, while an expression may have an eval()
operation that will evaluate the expression. \item The operations
typically take an environment as a parameter: a collection of
variable bindings which will be used in the evaluation of the
concepts behaviour, and a target object, which will be the object
that is changed as a result of the action or which represents the
context of the evaluation. \item The operations will return the
result of the evaluation (a boolean in the case of a static
expression) or change the value of the target object (in the case
of an action).
\end{itemize}

\subsection{Example}

A StateMachine can be given an operational semantics by defining
an interpreter that executes a StateMachine. It is implemented by
constructing a run() operation for the StateMachine class. We also
add an attribute messages, which records the messages that are
pending on the StateMachine as a result of a send action on a
transition:

\begin{figure}[htb]
\begin{center}
\includegraphics[width=15cm]{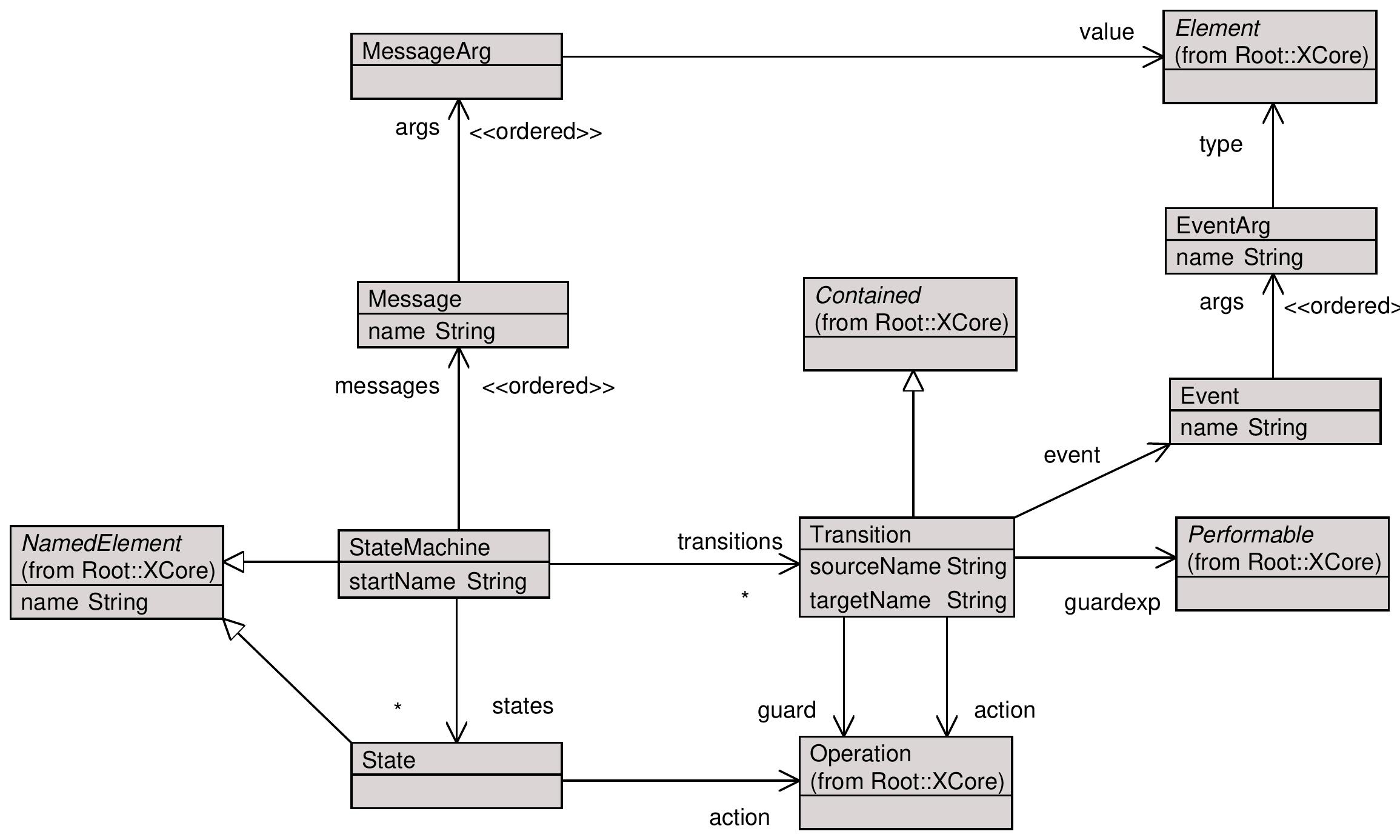}
\caption{StateMachine model extended with message queues}
\label{messages}
\end{center}
\end{figure}

\noindent The code that executes the run() operation is below.
More details about the executable language used here, and the use
of executability as a means of defining semantics, will be
explained in chapter \ref{execChapter}.

\begin{lstlisting}
@Operation run(element)
      let state = self.startingState() in
        @While state <> null do
          let result = state.activate(element) then
            transitions = self.transitionsFrom(state) then
            enabledTransitions = transitions->select(t |
            t.isEnabled(element,result) and
            if t.event <> null then
              messages->head().name = t.event.name
            else
              true
            end) in
              if enabledTransitions->isEmpty then
                state := null
              else
              let transition = enabledTransitions->sel in
                transition.activate(element,result + self.messages->head().args.value);
                state := transition.target();
                if transition.event <> null then
                  self.messages := self.messages->tail()
                end
              end
            end
          end
        end
      end
    end
\end{lstlisting}

The operation takes the element that the StateMachine is being
applied to. It first sets the state to be the starting state,
then enters a while loop. Provided that the StateMachine has not
terminated (state $<>$ null) the following is peformed:

\begin{enumerate}
\item The entry action on the current state is invoked by calling
its activate operation. \item The collection of enabled
transitions is determined by selecting all the transitions that
leave the current state such that the evaluation of their guard is
true, and that an event is waiting on the machine that corresponds
to the event on the transition. \item If there are no enabled
transitions, the state is set to null, and the run() operation
terminates. \item If there are enabled transitions, one of them is
chosen and its action is invoked, before assigning the state to be
the target of the transition.
\end{enumerate}

Consider the traffic light example shown in figure
\ref{trafficlight}. If the current state of the StateMachine is
Green, then the above semantics ensures that the guard on the
GreenRed() transition will be evaluated and if it returns true the
action on transition will be executed.

As this example shows, complex behaviour can be captured in terms
of operational definitions. Moreover, the definition can
immediately be tested and used within a tool to provide
semantically rich means of validating and exercising models and
programs written in the language.

\section{Extensional Semantics}

In the extensional approach, the semantics of a language is
defined as an extension to another language. Modelling concepts in
the new language inherit their semantics from concepts in the
other language. In addition, they may also extend the semantics,
adding new capabilities for example.

The benefit of the approach is that complex semantic concepts can
be reused with minimum effort. For example, a business entity need
not define what it means to create new instances, but can inherit
the capability from Class.

The extensional approach has some commonality with the notion of a
profile (\cite{umlspec}). A profile provides a collection of
stereotypes, which can be viewed as sub-classes of UML or MOF
model elements. However, by performing the extension at the
metamodel level, greater expressibility is provided to the user,
who can add arbitrarily rich semantic extensions to the new
concept. A suitable tool can make use of this information to
permit the rapid implementation of new modelling languages. It may
recognise that an extension has occurred, and use stereotype
symbols to tailor the symbols of the original modelling language
to support the new language (see section \ref{stereotypes}).

\subsection{Example}

\begin{figure}[htb]
\begin{center}
\includegraphics[width=11cm]{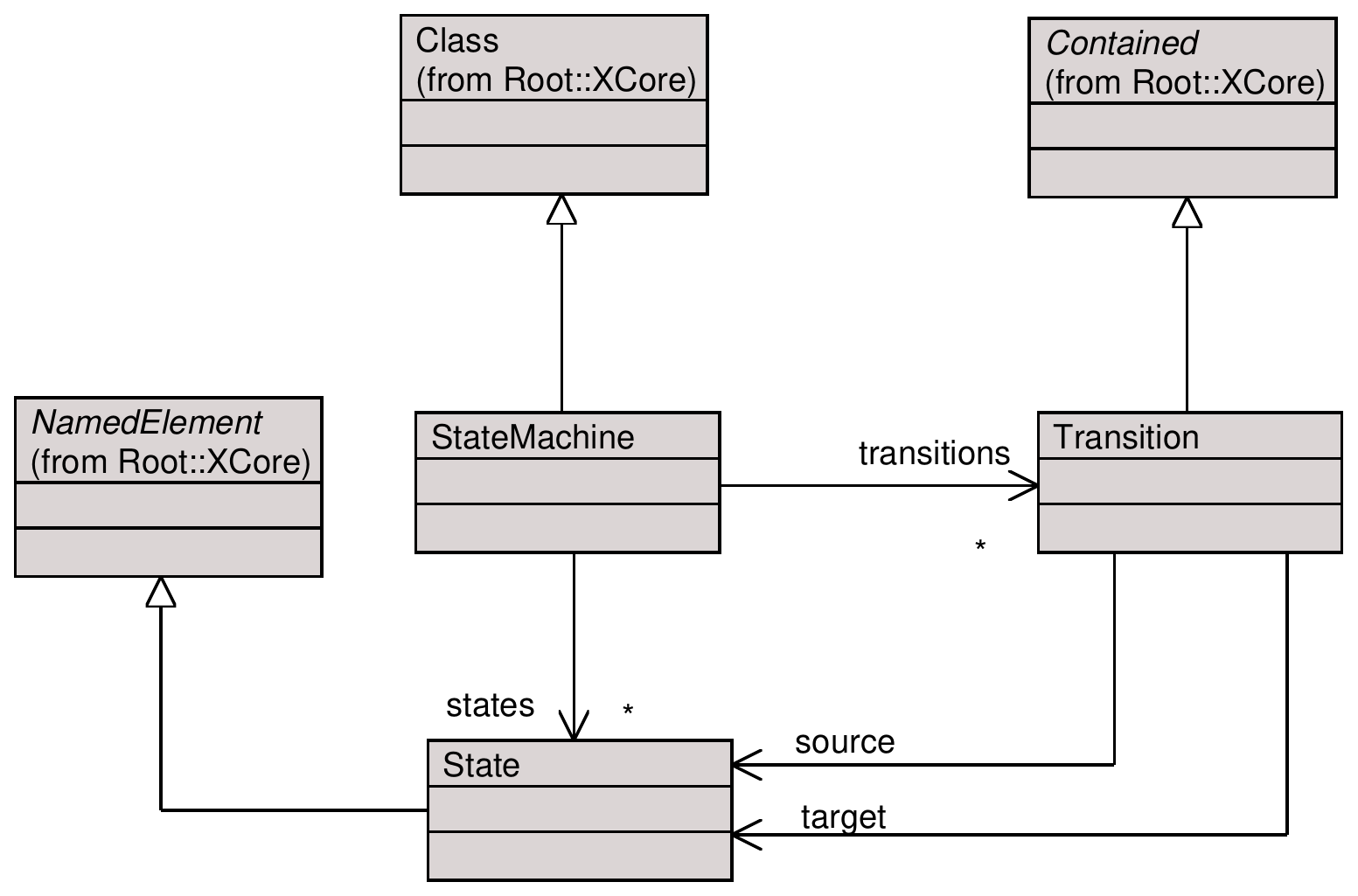}
\caption{Example of extending the class Class}
\label{extensionExample}
\end{center}
\end{figure}

As an example, consider the requirement to be able to create
multiple instances of the same StateMachine. This can be achieved
by specialising the class Class from the XCore metamodel (see
figure \ref{extensionExample}. Because Class is instantiable, the
StateMachine will also inherit its semantics.

By specialising the class NamedElement a State can be owned by a
StateMachine and can be navigated to via a pathname. Similarly, a
Transition can owned by a StateMachine, but in this case it cannot
have a name as it specialises the class Contained.

While the extensional approach is appealing in terms of its
simplicity, it does have its disadvantages. Firstly, it is
tempting to try and 'shoe-horn' the complete semantics of a
language into this form. In practice this is usually not possible
as there will be concepts that simply do not fit. In these cases,
other approaches to defining semantics must be used.

\section{Denotational Semantics}

The purpose of denotational semantics is to associate mathematical
objects, such as numbers, tuples, or functions, with each concept
of the language. The concept is/are said to {\em denote} the
mathematical object(s), and the object is called the {\em
denotation} of the concept. The objects associated with the
concept are said to be the {\em semantic domain} of the concept. A
widely used example of this in programming language semantics is
the denotation of the operation + by a number. For instance, the
denotation of 4+5 is 9.

A denotational semantics can be thought of as semantics by
example. By providing all possible examples of a concept's
meaning, it is possible to define precisely what it means. In the
above example, there is only one denotation. However, many
concepts are denoted by a collection of examples. To describe the
semantics of an Integer, the set of all positive numbers would be
required, i.e. the denotation of Integer is 0..infinity.
Denotational descriptions of semantics tend to be static, i.e.
they enumerate valid instances of a concepts, and a non-executable
fashion.

Here are some common examples of denotational relationships found
in metamodels:

\begin{itemize}
\item The denotation of a Class is the collection of all Objects
that may be an instance of it. \item The denotation of an Action
is a collection of all possible state changes that can result from
its invocation. \item The denotation of an Expression is the
collection of all possible results that can be obtained from
evaluating the expression.
\end{itemize}

A denotational semantics can be defined in a metamodel by
constructing a model of the language's abstract syntax and
semantic domain and of the semantic mapping that relates them.
Constraints are then written that describe when instances of
semantic domain concepts are valid with respect to their abstract
syntax. For example, constraints can be written that state when an
Object is a valid instance of a Class.

The advantage of the denotational approach is its declarative
nature. In particular, it captures semantics in a way that does
not commit to a specific choice of operational semantics. For
instance, while an operational semantics would have to describe
how an expression is evaluated, a denotational semantics simply
describes what the valid evaluation/s of the expression would be.

In practice a purely denotational approach is best used when a
high-level specification of semantics is required. It is
particularly useful in a standard where commitment to a specific
implementation is to be avoided. Because they provide a
specification of semantics they can be used to test
implementations of the standard: candidate instances generated by
an implementation can be checked against the denotational
constraints. A good example of the denotational semantics approach
is the OCL 2.0 specification \cite{ocl2}, where they are used to
describe the semantics of OCL expressions.

\subsection{Example}

We can describe the execution semantics of a StateMachine by
modelling the semantic domain concepts that give it a meaning. A
useful way of identifying these concepts is to consider the
vocabulary of concepts we would use to describe the behaviour of a
StateMachine. This might include concepts such as a state change
(caused by a transition), examples of statemachines in specific
states, and message queues.

\begin{figure}[htb]
\begin{center}
\includegraphics[width=14cm]{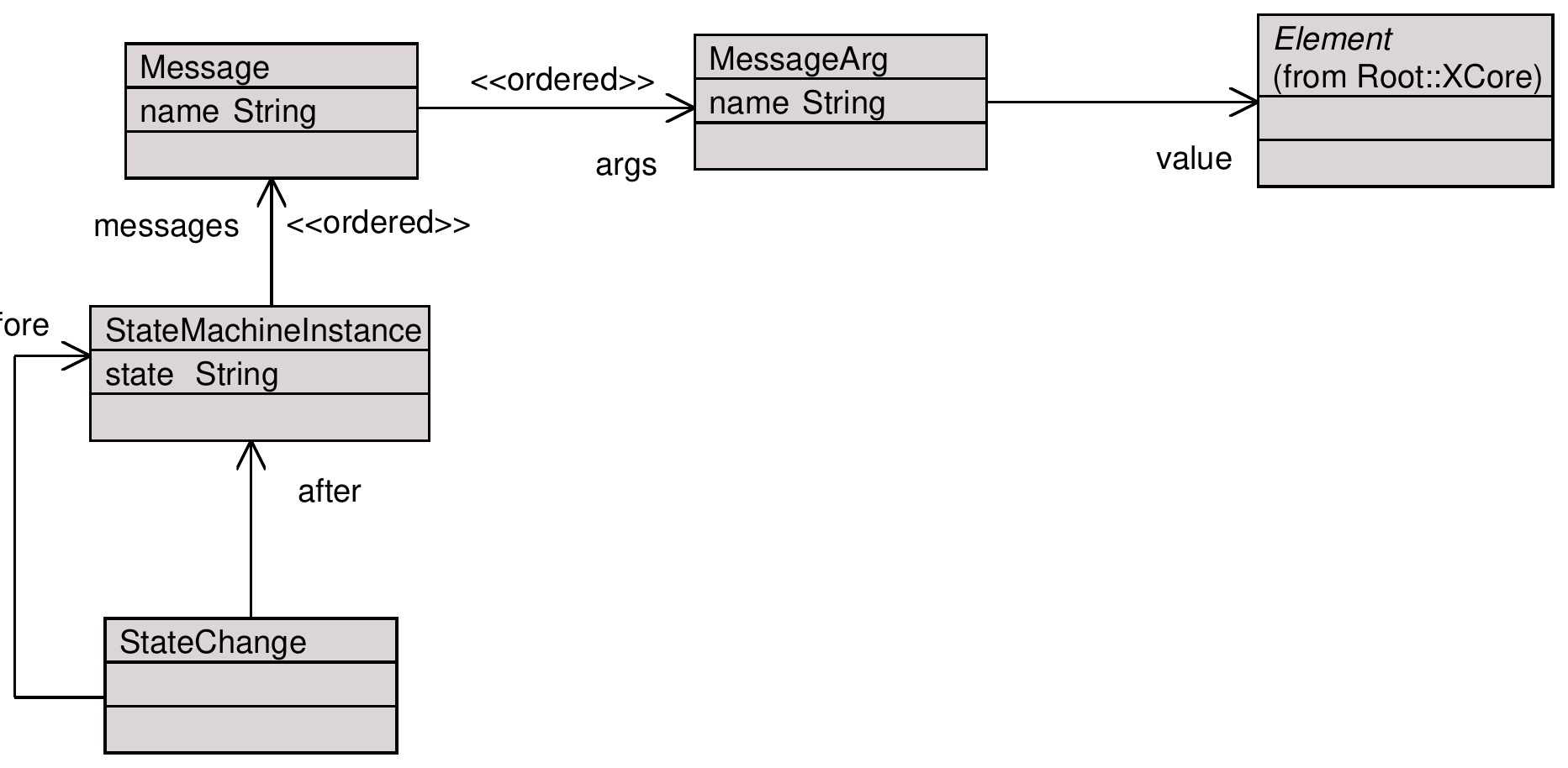}
\caption{A semantic domain for StateMachines} \label{sdExample}
\end{center}
\end{figure}

Figure \ref{sdExample} shows an example of a semantic domain that
might result from generalising these concepts into a model.

The denotation of a StateMachine is essentially a model of the
valid state changes that may occur to the StateMachine at any
point in time. Here a StateMachineInstance is used to model a
StateMachine at a specific point in time. It has a state, and a
sequence of messages which are waiting to be consumed by the
machine. State changes represent the invocation of a transition,
causing the StateMachine to move from one state (the before state)
to another (the after state).

\begin{figure}[htb]
\begin{center}
\includegraphics[width=13cm]{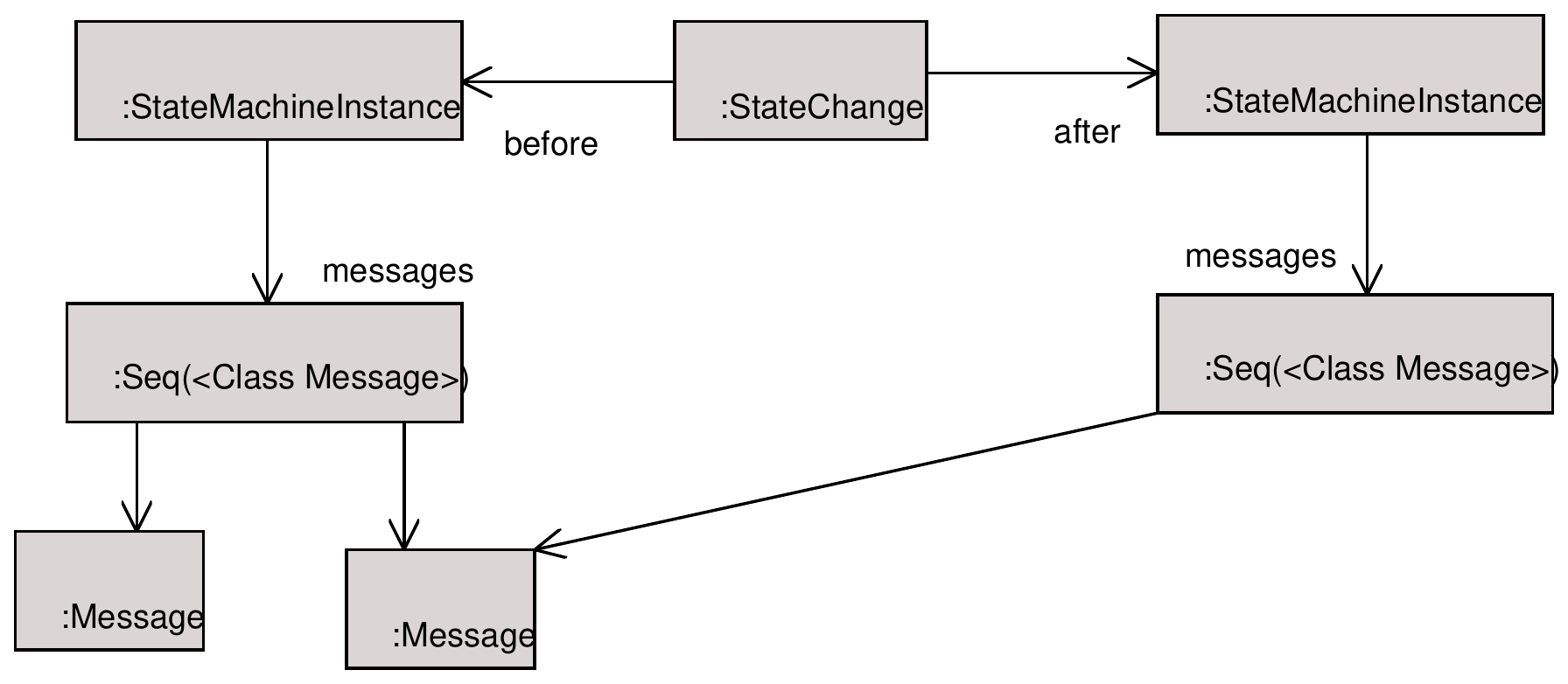}
\caption{An example snapshot of a StateMachine's behaviour}
\label{sdsnapshot}
\end{center}
\end{figure}

Using this model, many different examples of valid StateMachine
behaviours can be tested. The snapshot in figure \ref{sdsnapshot}
shows a fragment of the behaviour of a machine that has two
transitions from the state A to the state B and back again.

In order to complete the model, the relationship between the
semantic domain and abstract syntax model must also be modelled.
This relationship, commonly called a {\em semantic mapping}, is
crucial to defining the semantics of the language. It makes
precise the rules that say when instances of semantic domain
concepts are valid with respect to a abstract syntax model. In
this case, when a particular sequence of state changes is valid
with respect to a specific StateMachine.

\begin{figure}[htb]
\begin{center}
\includegraphics[width=8cm]{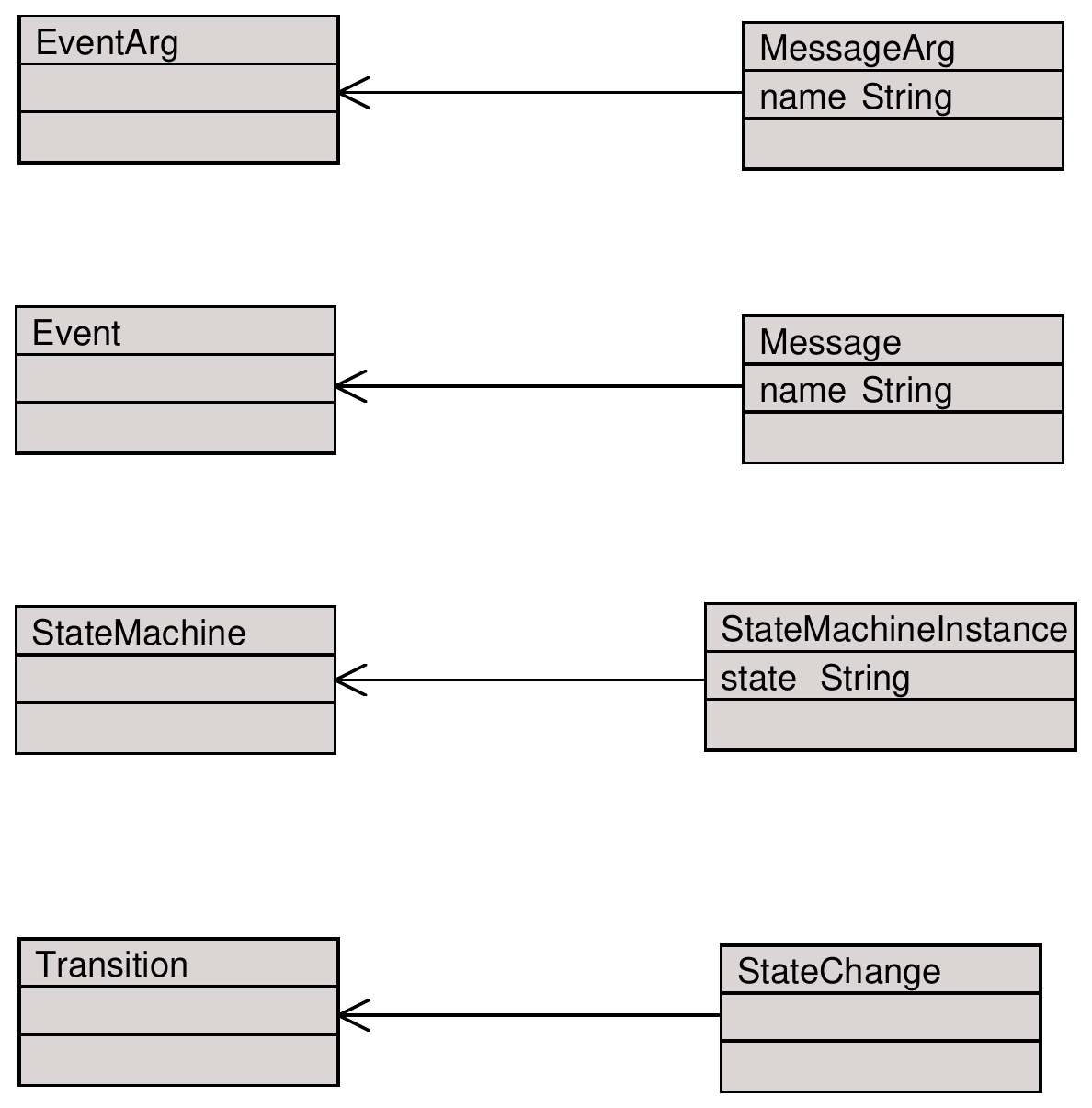}
\caption{Semantic mapping model for the StateMachine language}
\label{semanticmapping}
\end{center}
\end{figure}

Figure \ref{semanticmapping} shows the semantic mapping model. A
StateMachineInstance is associated with its StateMachine, while a
StateChange and Message are associated with the Transition and
Event that they are instances of.

Finally, well-formedness rules will be required. An example is the
rule which guarantees that the before and after states of a
StateChange commutes with the source and target states of the
transition it is an instance of:

\begin{lstlisting}
context StateChange
  self.transition.sourceName = self.before.state.name and
  self.transition.targetName = self.after.state.name
\end{lstlisting}

Other rules will be required to ensure that the state changes
associated with a state machine are valid with respect to its
execution semantics. These will describe such aspects as the
conditions under which transitions are enabled, and the order in
which they may fire.

\section{Process}

The choice of semantics depends on the type of language being
defined. The following pointers provide a guide to choosing the
most appropriate approach.

\begin{itemize}
\item If a declarative, non executable semantics is required, use
the denotational approach. \item If a language has concepts that
need to be evaluated, executed or instantiated then they should be
modelled using an operational, translational or extensional
approach. The choice of approach will be based on the following:
  \begin{itemize}
  \item If a concept is clearly a sugared form of more primitive
  concepts, adopt a translation approach. This avoids having to
  construct a semantics for the concept from scratch - reuse the
  semantic machinery that has already been defined for the primitive
  concepts. This approach should only be used when it is
  acceptable to loose information about the original concept.
  \item If information must be kept about the concept and it is
  possible to reuse an existing concept use the extensional approach.
  \item If there is no convenient means of reusing an existing
  concept use the operational approach to construct an interpreter.
  \end{itemize}
\end{itemize}

Note, there are no hard and fast rules to choosing an approach.
The primary aim should be to develop a semantic model that meets
the needs of the stakeholders of the language.

\section{Conclusion}

Semantics is crucial in being able to understand the meaning of a
modelling language, to be able to interact with it a meaningful
way, and to be able to support truly interoperable language
definitions. This chapter has shown that a language semantics can
be successfully captured with a metamodel. The key advantages are
that the semantics becomes an integrated part of the language
definition while remaining understandable to users and modelling
tools.

\chapter{Executable Metamodelling}
\label{execChapter}

\section{Introduction}

The purpose of this chapter is to describe how the addition of
executable  primitives to a metamodelling language can result in a
powerful meta-programming environment in which many different
operational aspects of languages can be described. This includes
the ability to model the operational behaviour of a modelling or
programming language and the ability to quickly create many types
of applications to manipulate models.

\section{Why Executable Metamodelling?}

Executable metamodelling is a natural evolution of a growing trend
towards executable modelling \cite{Mellor}. Executable modelling
enables the operational behaviour of system models to be captured
in a form that is independent of how the model is implemented.
This is achieved by augmenting the modelling language with an
action language or other executable formalism. Executable models
enable a system to be tested before implementation begins,
resulting in improved validation of the system design.
Furthermore, it is also possible to generate executable code from
an executable model, as there is sufficient information to
generate method bodies, etc, in the target programming language.

The ability to execute {\em metamodels} has additional benefits
over and above those associated with executable modelling. At the
simplest level, many developers need the facility to access and
manipulate models at the metamodel level. For instance, a
developer may need to analyse general properties of their models,
or to globally modify certain features of a model, or write useful
functions that automate a time consuming modelling task. Being
able to quickly add some functionality that will perform the
required task will thus save time and improve the modelling
process.

More generally, the addition of executability to metamodels
facilitates what is almost metamodelling nirvana: the ability to
model all aspects of a modelling language as a unified, executable
entity. A static metamodel cannot model the operational behaviour
of a language. Yet, in the semantic rich tools that are required
by Language-Driven Development, capturing this aspect of a
language is essential. With an executable metamodelling language,
all of this can be described in the metamodel itself.

The advantage is that the modelling language definition becomes
completely self contained, and can be interchanged between any
modelling tool that supports the necessary metamodelling
machinery. Such definitions are not reliant on platform specific
technology but the metamodel architecture that they are defined
within (which as we shall see can itself be modelled independently
of other technologies).

\subsection{Executability and XMF}

Because XMF is also defined in terms of an executable metamodel,
XMF can effectively implement everything associated with
supporting itself as a language definition language, including the
construction of parsers, compilers, interpreters and so on. The
key advantage is that there is unification in the way that all
modelling languages are constructed. Thus, a language for
modelling StateMachines will use entirely the same meta-machinery
as a language for modelling user interactions, business use cases,
and so on. Moreover, because the definition is in effect a
program, it will be as precise and unambiguous as any program
written in a programming language. Of course, it is still
necessary that the machinery developed to support XMF is as
generic as possible to facilitate the rapid construction of new
modelling languages - being executable does not necessarily mean
that it is generic.

A unified executable metamodelling environment offers important
advantages to the language developer. We have seen many
metamodelling tools where the developer has to contend with a
multitude of ill-fitting languages for dealing with each of the
different aspects required for tool design. For instance, a
repository for dealing with meta-data, a scripting language for
model manipulation, a GUI framework for user interface design, and
lexx and yacc for parser construction. An approach in which all
these languages are unified under a single executable
metamodelling umbrella offers a far more effective and productive
development environment.

Executable metamodelling is at the heart of the tool development
vision described in chapter \ref{metamodellingChapter}. By
supporting executability, metamodels are enriched to the point
where they can support a multitude of semantically rich tool
capabilities.

\subsection{Executable Metamodelling and Programming}

How does executable metamodelling relate to programming? Not at
all? In fact it is a natural extension (see section
\ref{progvsmodel}). Rather than thinking about modelling languages
as being different to programming languages, executable
metamodelling really views modelling languages as part of a
spectrum of programming languages, where each language is only
different in the abstractions and behaviour it encapsulates.
Indeed, executable metamodelling can be thought of as next
generation programming: whilst many programming languages are
fixed in what they can represent, executable metamodelling offers
infinite extensibility and flexibility.

\section{Adding Executability}

How can executability be added to a metamodelling language? The
answer (briefly discussed in chapter \ref{xmfchapter}) is to
provide the action primitives necessary to support execution.
These in combination with a suitable model querying and navigation
language, result in a powerful metaprogramming language.

The language proposed here is XOCL (eXecutable OCL). XOCL is a
combination of executable primitives and OCL (the Object
Constraint Language). The motivations for this combination are
discussed in section \ref{differences}.

The action primitives that are provided by XOCL are as follows:

\begin{description}
\item[Slot Update] Assigns a value to a slot of an object via the
assignment expression ":=". An example of slot update might be
{\tt self.x := self.x + 1}. \item[Object Creation] Creates a new
instance of a class via a new operation on a class from which the
object is created. An example might be: {\tt fido := Dog()}.
\item[Sequential Operator] Enables two expressions to be executed
in sequence via the operator ";". For instance, {\tt
self.x:=self.x+1; self.x:=self.x+2} will result in {\tt x} being
incremented by {\tt 3}.
\end{description}

XOCL expressions can be used in the bodies of expressions
belonging to behavioural modelling elements such as operations and
mappings. The following operation is described in the context of
the class X:

\begin{lstlisting}
context X
  @Operation doIt():Integer
    self.x := self.x + 1;
    self.x
  end
\end{lstlisting}

The above primitives provide the minimal actions necessary to add
sequential executability to OCL. Concurrent execution can be
supported by adding a suitable concurrency primitive such as
fork(), which allows multiple threads of execution to be
propagated. While the majority of metamodels only require
sequential execution, there are some cases where concurrency needs
to be supported. Examples include providing models of
communication between clients and a server or in describing the
operational semantics of concurrent modelling languages.

Finally, it is important that XMF provides the necessary
architecture to support execution in an object-oriented
environment. This aspect, called a Meta-Object Protocol or MOP
will be discussed in later versions of this book.

\subsection{XOCL Extensions}

While OCL provides a number of useful imperative style constructs,
such as {\em if} and {\em for} loop expressions, there are a small
number of extensions that we have found to be very useful when
using it as a programming language. These include:
\ \\

\noindent While expressions: standard while loops as provided
by many programming languages:

\begin{lstlisting}
@While x < 10
  do x := x + 1
end
\end{lstlisting}

\noindent Find expressions: a simplified way of traversing
collections of models and finding an element that matches specific
criteria. If xset contains an x whose value is greater than zero,
y will be incremented by 1.

\begin{lstlisting}
@Find (x,set)
  when x > 10
  do
  y := y + 1
end
\end{lstlisting}

\noindent Tables provide efficient lookup over large data
structures:

\begin{lstlisting}
let table = Table() in
    table.put(key,value);
    table.get(key)
end
\end{lstlisting}

\noindent The case statement as provided by many programming
languages:

\begin{lstlisting}
@Case(x)
  x > 1 do x := x + 1;
  x = 0 do x := 2
end
\end{lstlisting}

\noindent TypeCase expressions: selects a case
statement depending on the type of object that is passed to it.

\begin{lstlisting}
@Case(x)
  Boolean do x := false;
  Integer do x := 0
end
\end{lstlisting}

\noindent Because objects are fundamental to XMF, a number of
operations are provided for accessing and manipulating their
properties:

\begin{itemize}
\item The operation of() returns the class that the object is an
instance of, e.g. StateMachines::State.of() returns XCore::Class.
\item The operation getStructuralFeatureNames() returns the names
of the slots (attribute values) belonging to an object, e.g. if x
is an instance of the class StateMachines::State, then
x.getStructuralFeatureNames() will return the set containing
"name". \item The operation get() takes a name and return the
value of the slot of that name, e.g. x.get("y") will return the
value of the slot called "y". \item The operation set() take a
name and a value, and sets the value of the slot called name to be
the value, e.g. x.set("y",10), will set the value of the slot
called "y" to 10.
\end{itemize}

\section{Examples}

The best way to understand the benefits of executable
metamodelling is to look at some real examples. This section
provides a number of examples that provide general utility
operations for manipulating metamodels. Other parts of the book
(notably chapter \ref{semanticschapter}) describe how
executability can be used to define semantics. The case study
chapters at the end the book also provide examples of executable
metamodels.

\subsection{Example 1: Model Merge}
\label{merge}

Merging the contents of two packages is a very useful capability
that has many applications, including versioning (merging
different versions of a model together) and model management
(breaking up large models into composable pieces that can be
merged into a single artefact). The following operation is defined
on a package.

\begin{lstlisting}
context Package
  @Operation merge(p:Package)
    self.contents()->collect(c1 |
      if p.contents()->exists(c2 | c1.isMergable(c2)) then
        c1.merge(c2)
      else
        c1
      end)->union(
        p.contents()->select(c2 | self.contents()->exists(c1 |
          not c1.isMergable(c2))))
  end
\end{lstlisting}

The operation merges an element belonging to the package p with an
element belonging to the contents of the package provided they are
mergeable. Note contents() is an meta-operation belonging to all
containers (see section \ref{framework}. The conditions under
which two elements are mergeable will be defined on a case by case
basis. For example, two elements may be mergeable if they have the
same name and are of the same type. If two elements are mergeable,
then the result of the merge will be defined via a merge()
operation on the elements' types.

\subsection{Example 2: Find and Replace}

This example defines a simple algorithm for finding and replacing
named elements in the contents of a Namespace. It works by
determining whether an object is a specialisation of the class
NamedElement, and then performs the appropriate substitution if
the element name matches a name in the to be replaced in the subs
list:

\begin{lstlisting}
context Namespace
  @Operation replace(subs:Set(Sub))
    @For i in self.contents
    if i.isKindOf(NamedElement) then
      if subs->exists(r | r.from = i.name) then
        i.name := subs->select(s | r.from = i.name)->sel.to
      end
    end
  end
\end{lstlisting}

\noindent Where the class Sub is defined as follows:

\begin{lstlisting}
@Class Sub
  @Attribute from : String end
  @Attribute to : String end
end
\end{lstlisting}

Applying this operation to any Namespace is now straightforward.
For instance, the operation replaceStateName() can be written like
so:

\begin{lstlisting}
context StateMachine
  @Operation replaceStateName(subs:Set(Sub))
    self.replace(subs);
    self
  end
\end{lstlisting}

This operation is a useful mechanism for performing global search
and replace on models and programs.

\subsection{Example 3: Walker}
\label{walker}

The previous example is restricted as it only applies to
Namespaces. In many situations, one wants to be able to walk over
any structure of objects in a metamodel performing arbitrary
tasks. These tasks may include find and replace, but could be
anything from constraint checking to the global invocation of
operations.

This can be achieved using a walker. A walker recursively descends
into an elements structure and dispatches to appropriate operations
depending on the values of component elements. A critical
requirements is that the walker can handle cycles in a structure.
It does this by recording the elements that have been walked, and
then using this information to ignore those elements if they are
met again.

\begin{lstlisting}
@Class Walker
  @Attribute table : Table end
  @Attribute refCount : Integer end
  @Attribute visited : Integer (?) end

  @Constructor()
    self.initWalker()
  end

  @Operation initWalker()
    self.table := Table(1000)
  end
end
\end{lstlisting}

The class Walker contains a hashkey table, in which a list of all
the walked elements is kept along with an integer reference to the
element. A count is kept of the number of elements that have been
walked along with the number of references created.

\begin{lstlisting}
context Walker
  @Operation encounter(e:Element)
    self.encounter(e,self.newRef())
  end

  @Operation encounter(e:Element,v:Element)
    // Save a reference to v against the walked value e.
    table.put(e,v)
  end

  @Operation newRef():Integer
    self.refCount := refCount + 1;
    refCount
  end
\end{lstlisting}

The encounter() operation is called when a new element is
encountered. It creates a new reference for the element, and adds
it to the table.

The following operations deal with walking the tree. The operation
encountered() returns true if the element has already been
encountered.

\begin{lstlisting}
context Walker
  @Operation encountered(e:Element):Boolean
    // Returns true when we have already walked e.
    table.hasKey(e)
  end

  @Operation getRef(e:Element):Element
    table.get(e)
  end

  @AbstractOp reWalk(e:Element,arg:Element):Element end
\end{lstlisting}

The operation walk() performs the task of walking an element. If
the element has already been encountered then the operation
reWalk() is run (this will be specialised for specific
applications, but in most cases it will do nothing). Otherwise,
depending on the type of element that is being walked, appropriate
element walkers will be called.

\begin{lstlisting}
context Walker
  @Operation walk(e:Element,arg:Element):Element
    // Walk the element e with respect to the argument.
    self.visited := visited + 1;
    if self.encountered(e)
      then self.reWalk(e,arg)
    else
      @TypeCase(e)
        Boolean      do self.walkBoolean(e,arg) end
        Integer      do self.walkInteger(e,arg) end
        Null         do self.walkNull(arg) end
        Operation    do self.walkOperation(e,arg) end
        SeqOfElement do self.walkSeq(e,arg) end
        SetOfElement do self.walkSet(e,arg) end
        String       do self.walkString(e,arg) end
        Table        do self.walkTable(e,arg) end
        Object       do self.walkObject(e,arg) end
        else self.defaultWalk(e,arg)
      end
    end
  end
\end{lstlisting}

The most complex of these walkers is the object walker. This gets
all the structural feature names of the object, i.e. the names of
the attributes of the class the object is an instance of. To do
this, it uses the {\tt getStructuralFeatureNames} operation
defined on the class Object to return the names of the structural
features. It then walks over each of the slots that correspond to
each structural feature:

\begin{lstlisting}
context Walker
  @Operation walkObject(o:Object,arg:Element):Element
    self.encounter(o);
    @For name in o.getStructuralFeatureNames() do
      self.walkSlot(o,name,arg)
    end
  end
\end{lstlisting}

Again, walkSlot() will be defined on an application basis, but in
general will simply get the element value of each slot and call its
walker.

This example illustrates the benefits of being able to program at
the meta-level. It allows the designer to produce code that is
reusable across multiple metamodels irrespective of what they
define. It does not matter whether the object that is being walked
is a StateMachine or a BusinessEntity. At the meta-level they are
all viewed as objects.

\subsection{Example 4: Meta-Patterns}

This example shows how meta-operations can stamp out a structure
over existing model elements. This can be used as the basis for
capturing libraries of reusable patterns at the meta-level.

The following operation captures a simple containership pattern,
which can be used to stamp out a containership association between
two classes and adds an add() operation to the owning class.

Given a pair of classes, c1 and c2, the operation first creates an
instance of an attribute, a, whose name is the name of c2, and
whose type is the class c2. Next, an anonymous operation is
created called c2. It takes an object x and sets the value of the
attribute name c2 to the value of x including its existing
contents. It's name is then set to "add" plus the name of c2.
Finally, both the attribute and the operation are added to the
class c1.

\begin{lstlisting}
context Root
@Operation contains(c1 : Class,c2 : Class):Element
      let a = Attribute(Symbol(c2.name),Set(c2));
          o = @Operation anonymous(x : Element):Element
                self.set(c2.name,self.get(c2.name)->including(x))
              end
      in o.setName("add" + c2.name);
         c1.add(a);
         c1.add(o)
      end
    end
\end{lstlisting}

\begin{figure}[htb]
\begin{center}
\includegraphics[width=8cm]{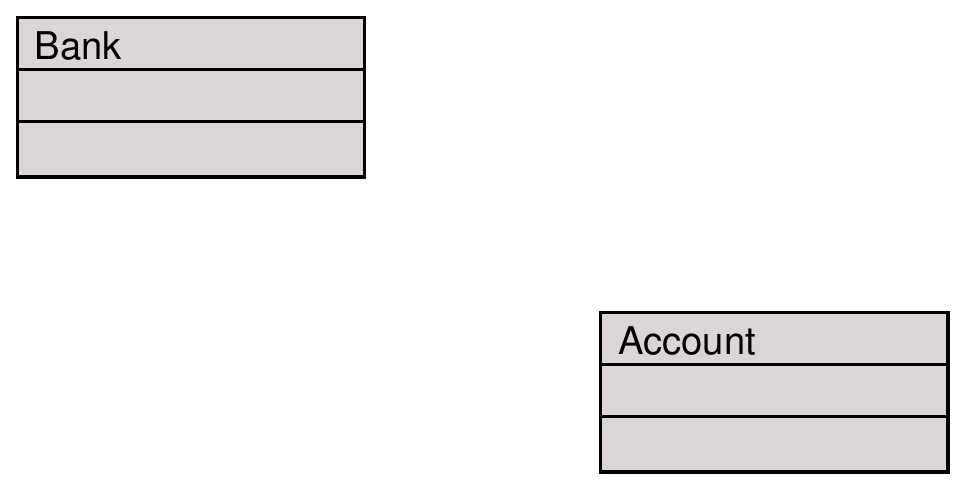}
\caption{An example model before applying the contains operation}
\label{patterns1}
\end{center}
\end{figure}

\begin{figure}[htb]
\begin{center}
\includegraphics[width=8cm]{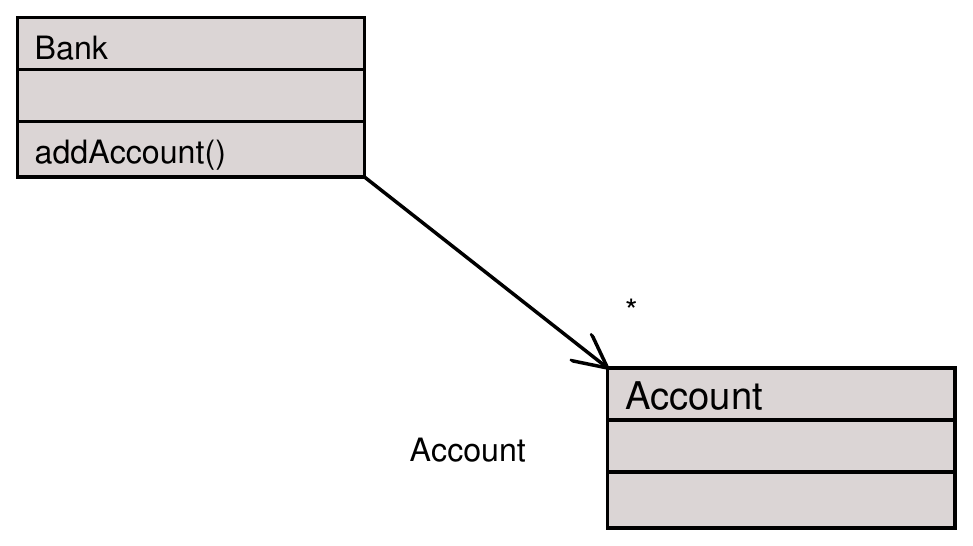}
\caption{An example model after applying the contains operation}
\label{patterns2}
\end{center}
\end{figure}

Figures \ref{patterns1} and \ref{patterns2} show the result of
applying the operation to two classes.

\section{Conclusion}

This chapter has aimed to show the benefits that can be achieved
through the extension of traditionally static metamodelling
languages to fully executable metamodelling. The resulting language
can be used in the definition of useful meta-level utility
operations right through to the construction of complete modelling
language definitions. This has wide-reaching applications for how
we treat metamodels. Rather than viewing them as static entities,
they can now be viewed as highly abstract programming devices for
language design.

\chapter{Mappings} \label{mappingchapter}

\section{Introduction}

A mapping is a relationship or transformation between models or
programs written in the same or different languages. Mappings are
a key part of the metamodelling process because of the important
role they play in describing how instances of one metamodel are to
be mapped to instances of other metamodels.

In the world of system development, mappings are everywhere:
between abstract (platform independent) models and platform
specific implementations, between legacy code and re-engineered
systems, across and within application and organisational domains.
Each of type of mappings places different requirements on the type
of mapping language that is required. Sometimes, a one shot,
unidirectional transformation is required. Other times, a mapping
must maintain consistency between models, perhaps by continuously
reconciling information about the state of both models. A mapping
language has to be able to deal with all these requirements.

At the time of writing, work is already underway, initiated by the
Object Management Group in the form of a request for proposals
(RFP) \cite{qvtrfp} to design a standard mapping language,
provisionally entitled QVT (Queries, Views, Transformations). Even
though there are many proposed meta-models for a mapping language
there are some basic foundations which are fairly independent of
the particular meta-model itself. This chapter begins with a
discussion on applications and existing technologies and
techniques for mappings and moves on to identify the requirements
for an ideal mapping language. Two languages, XMap and XSync are
then described that together address the complete requirements.

\section{Applications Of Mappings}

The application of mappings can be broadly divided into vertical,
horizontal and variant dimensions of a problem domain
\cite{mdatech}.

\subsection{Vertical Mappings}

Vertical mappings relate models and programs at different levels
of abstraction. Examples include mapping between a specification
and a design, and between a design and an implementation. In MDA
the mapping from more abstract models to less abstract, more
detailed models and programs is typically known as a PIM (platform
independent model) to PSM (platform specific model) mapping. Note,
these terms are somewhat relative (machine code can be viewed as
platform specific in relation to Java, yet Java is often viewed as
a platform specific in relation to UML).

The precise nature of a vertical PIM to PSM mapping will be
dependent upon the nature of the target platform. For example, the
following platform properties may be taken into account:

\begin{itemize}
\item Optimisation: improving one or more aspects of the efficiency
of the resultant platform specific code, e.g. efficiency of memory
usage, speed of execution, and usage of dynamic memory allocation.
\item Extensibility: generating platform specific code that
is more open to adaptation, e.g. through the use of polymorphic
interfaces.
\item Language paradigm: Removing (or adding) language features,
for instance substituting single inheritance for multi-inheritance
or removing inheritance altogether.
\item Architecture: if the target platform is a messaging broker,
such as CORBA, rules will be defined for realising the platform
independent mapping in terms of appropriate interface calls to the
message broker.
\item Trustworthiness: visible and clear mappings may allow some
level of reasoning to be applied to the target notation, which may
enable certain properties (e.g. pertaining to safety) to be
automatically established.
\end{itemize}

It is important to note that vertical mappings may also go in the
reverse direction, e.g. from implementation to design. This is
particularly appropriate for reverse engineering. Here the essence
of the functionality of the platform specific language is reverse
engineered into another language. Typically, this process will not
be an automatic one, but must be supported by tools.

\subsection{Horizontal Mappings}

Whilst vertical mappings have received much attention in the MDA
literature, horizontal mappings are just as important. Horizontal
mappings describe relationships between different views of a
problem domain. Examples of these include:

\subsubsection{System Views}

In large systems, many different aspects of the problem domain
will need to be integrated. This may include different aspects of
the business domain such as marketing or sales, or the technology
domain such as safety or security. Critically, many of these
different aspects will overlap. Horizontal mappings can be defined
between languages that capture these different views, allowing
them to be consistently managed.

\subsubsection{Language Views}

Complex systems can rarely be modelled using a single notation. As
the popularity of UML has demonstrated, different notations are
required to precisely and succinctly model different view points
of the system. For instance, in UML, class diagrams are used to
model the static structure of a system and a completely different
language, state machines for example, to model the dynamic view
point. Even though these notations are different they describe the
same system with overlapping views and hence there exists a
relationship between them.

\begin{figure}[htb]
\begin{center}
\includegraphics[width=10cm]{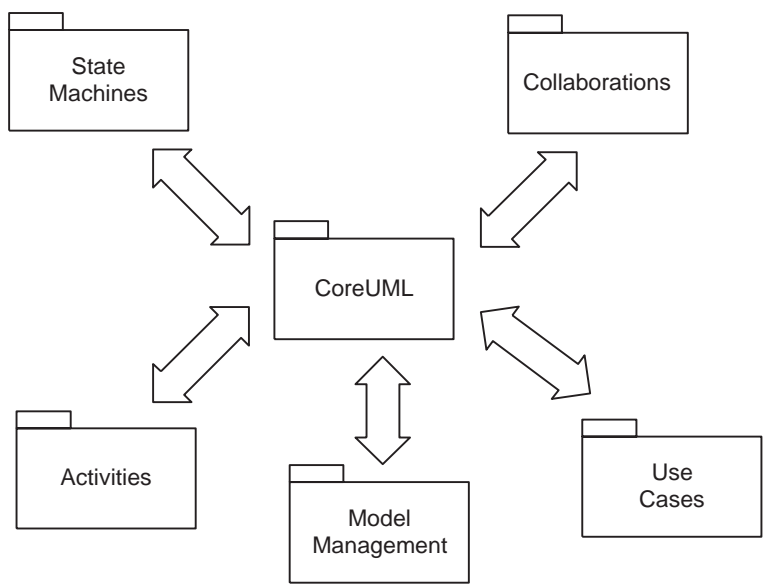}
\caption{An example of a horizontal mapping between core UML and
its diagrams} \label{horizontalmap}
\end{center}
\end{figure}

As shown in chapter \ref{concretechapter}, horizontal mappings
provide a means of integrating language notations in a way that
ensures changes in one view automatically updates other views.
Figure \ref{horizontalmap} shows how this might work for a small
part of UML. Here, it is assumed that there is a core OO modelling
language, with its own, precisely defined semantics. Many of the
different modelling notations provided by UML are modelled as a
view on the underlying OO modelling language. Mappings reconcile
changes to the diagrams by making changes to the underlying OO
model, which in turn may impact models in other diagrams.

Of course, previous chapters have shown that horizontal mappings
are also necessary for integrating different aspects of a
modelling language. Mappings between concrete syntax and abstract
syntax, and between abstract syntax and a semantic domain are all
critical parts of a language definition.

\subsection{Variant Dimensions}

Variant dimensions include product families and product
configurations. Mappings can model the relationship between
variant dimensions enabling each dimension to be precisely related
to one another.

\section{Types of Mappings}

\subsection{Unidirectional Mappings}

Unidirectional mappings take an input model or collection of input
models and generate an output model in one go. A unidirectional
mapping may record information about the relationship between the
input and output model, but there is no dependency of the input
model on the output model. If the input model/s change, then the
entire mapping must be rerun again in order to re-generate the
output model.

An example of a one shot mapping is a code generator, that takes a
platform independent model as its input, and generates a platform
specific model or program.

\subsection{Synchronised Mappings}

Synchronised mappings are a special class of mapping where it is
important to continuously manage consistent relationships between
models. This requirement can occur in many different situations.
One common example is maintaining consistency between a model and
the code that it is being transformed to. This would be a
requirement if a programmer wants to add/change code whilst
maintaining consistency with the model it is generated from. In
this scenario a change to the code would be immediately reflected
back in the model, and vice versa: if the model was changed, it
would be reflected in changes in the code. In the context of
horizontal mappings, synchronised mappings have an important part
to play in maintaining consistency between different viewpoints on
a model.

\section{Requirements of a Mapping Language}

When applying mappings on real projects it becomes apparent that
there are some key requirements that a mapping language must
support if it is to be of practical use. These include the
following:

\begin{itemize}
\item Support for mappings of relatively high complexity. For
example, the mapping language should be able to model complex
mappings, such as transforming an object oriented model into a
relational model including the mapping of the same attribute to
different columns for foreign key membership \cite{qvtrfp}. In
practice, this means that the mapping language must provide good
support for arbitrarily complex queries of models (to be able to
access the information necessary to drive a complex mapping), and
support the ability to modify models  using relatively low level
operations (such operations can, if used sensibly, significantly
reduce the size and complexity of a mapping).

\item Support for reuse. It should be possible to extend and adapt
mappings with ease. This meets the need to be able to reuse
existing mappings rather than having to create them from scratch
each time.

\item Facilitate the merging of models. If the source metamodel of
a mapping represents a graph then any duplicate elements that are
generated by the mapping must be merged.

\item Provide mechanisms that support the structuring of mappings, e.g.
being able to model the fact that a mapping owns or is dependent
on sub-mappings.

\item Be able to record information about a mapping to provide
traceability during the mapping process.

\item Integration within the metamodel architecture, so that
mappings may access models at all levels.

\item Support for execution. It may seem obvious, but a mapping
should be executable in order to support the physical generation
of new models from a mapping. This is contrasted with a
non-executable mapping (see below).

\item Provide diagrammatic notations that can be used to visualize
mappings. Visual models have an important role to play in
communicating the overall purpose of a mapping.

\item Support for bi-directional and persistent mappings. As described
above, this is essential in being able to support mappings where
models must be synchronised with other models.

\item Support for mapping specifications. A mapping specification
is a non-executable description of `what' a mapping does, which
does not commit to`how' the mapping will be implemented. Mapping
specifications are a valuable means of validating the correctness
of an executable mapping, or as a contract between a designer and
an implementor.
\end{itemize}

An important question to ask at this point is whether all these
requirements can be addressed by a single mapping language. In our
experience, it does not make sense to have a 'one size fits all'
mapping language because not all the requirements are
complimentary. In particular, there is a strong distinction to be
made between a bi-directional and unidirectional mapping
languages. Each are likely to be targeted at different types of
problems and thus have different requirements in terms of
expressibility and efficiency.

Instead, it is better to use a small number of mapping languages,
each targeted at a specific mapping capability, yet still capable
of being combined within a single model. This is strategy taken in
this book. In the following sections, two mapping languages, XMap
and XSync are described each aimed at addressing complimentary
mapping capabilities.

\section{XMap}

XMap is a language designed to support unidirectional mappings. It
includes the following features:

\begin{description}
\item [Mappings] Mapping are the used to model unidirectional
transformations from source to target values. Mappings have state
and can be associated with other mappings. \item[Syntax] Mappings
have a visual syntax, enabling them to be drawn between model
elements in a class diagram, and a concrete syntax for describing
the detailed aspects of the mapping. \item[Executability] Mappings
have an operational semantics enabling them to be used to
transform large models. \item[Patterns] Mappings are described in
terms of patterns. A pattern describes what a mapping does in
terms of how a value in the source model is related to a value in
the target model - this provides maximum declarative
expressibility, whilst also remaining executable. Patterns are
described in a pattern language that runs over OCL expressions.
\item[OCL] Mappings can make use of OCL to express complex model
navigations.
\end{description}

In addition, XMap provides {\em Mapping specifications}. These are
multi-directional, non-executable, transformation specifications.
In the general case they are non-executable, but useful restricted
types of mapping specification can be automatically refined into
mappings. Mapping specifications are written in a constraint
language, in this case OCL. Typically they are used in the
specification stages of system development.

The syntax and semantics of XMap are described in exactly the same
way that all languages are defined in XMF: as an XMF metamodel,
with an operational semantics expressed in terms of XOCL.

\section{XMap Syntax}

\begin{figure}[htb]
\begin{center}
\includegraphics[width=11cm]{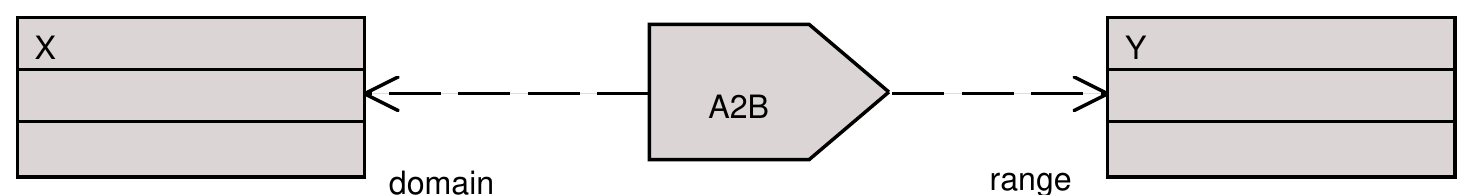}
\caption{An example mapping} \label{examplemapping}
\end{center}
\end{figure}

XMap has a visual syntax and a textual syntax. As shown in figure
\ref{examplemapping} the visual syntax consists of a mapping arrow
that can be associated with other model elements such as classes
in a class diagram. A mapping has a collection of domain or input
elements that are associated with the tail of the arrow, and a
single range or output element that is associated with the end of
the arrow.

The textual syntax of a mapping consists of a mapping declaration,
which is equivalent to a mapping arrow on a diagram:

\begin{lstlisting}
@Map <name>(<domain_1>,<domain_2>,...<domain_n>)-><range>

end
\end{lstlisting}\noindent A mapping contains a collection of clauses of the form:

\begin{lstlisting}
@Clause <name>
  <pattern>
end
\end{lstlisting}\noindent Note, each clause must have a different name.

A pattern is the mechanism that is used to match values in the
domain of the mapping to values in the range. It has the general
form:

\begin{lstlisting}
<exp> do
  <exp>
where
  <exp>
\end{lstlisting}Expressions are written in a mixture of XOCL expressions and
patterns expressions, where a pattern expression is a syntactical
relationship between expressions containing variables that are
bound by pattern matching. A common pattern used in mappings is to
relate object constructors. These describe a pattern match between
a domain object constructor and a range object constructor subject
to there being a pattern match between slot value expressions. An
example of this is:

\begin{lstlisting}
X[a = A] do
  Y [b = A]
\end{lstlisting}
Here the variable, A, is bound to the value of the slot a of any
instance of the class X using the expression a = A. This value is
then matched with the value of the slot b. The result of this
expression will be to match any instance of the class X with the
class Y, subject to the value of the slot b being equal to the
value of a.

Patterns can contain arbitrarily complex expressions. For
instance, this expression matches a with another constructor,
which contains a variable c, which is then matched with b:

\begin{lstlisting}
X[a =
  Z[c = A]] do
  Y [b = A]
\end{lstlisting}\noindent Patterns may also be embedded in sets and sequences:

\begin{lstlisting}
X[a =
  Set{Z[c = A]]} do
  Y [b = A]
\end{lstlisting}In this case, the slot a must be an instance of an attribute of
type Set(Z) and provided it contains an single instances of Z will
be matched.

Patterns can also be used in a very implicit fashion, to state
properties of values that must match. For instance, consider the
requirement to match a with an object belonging to a set, subject
to a slot being of a specific value. This could be expressed as
follows:

\begin{lstlisting}
X[a = S->including(Z[c = 1,d = A]]) do
  Y [b = A]
\end{lstlisting}This will match an object belonging to the set a subject to its
slot c being equal to 1. The value of S will be the remaining
elements of the set.

The 'where' component of a clause can be used to write additional
conditions on the pattern. Consider the following pattern, in
which the relationship between the variables A and B are related
by a {\tt where} expression:

\begin{lstlisting}
X[a = A]] do
  Y [b = B]
  where B = A + 1
\end{lstlisting}Finally, a mapping clause may call other mappings. This is
achieved by creating an instances of the mapping, and passing it
the appropriate domain values:

\begin{lstlisting}
X[a = A]] do
  Y [b = B]
  where B = mapIt(A)
\end{lstlisting}Here, the value of B is assigned the result of passing the value
of A to the mapping called mapIt. Note that a mapping can also be
instantiated and then invoked. This enables values to be passed to
the mapping via its constructor, e.g. mapIt(5)(A).

Finally, it is common to iterate over a collection of values,
mapping each one in turn. This type of mapping would look like so:

\begin{lstlisting}
X[a = A]] do
  Y [b = B]
  where B = A->collect(a | mapIt()(a))
\end{lstlisting}\noindent This would pass each element of A into the mapIt()
mapping collecting together all their values and assigning them to
B.

\section{XMap Examples}

This section demonstrates two mappings written using XMap: a
mapping from StateMachines to Java and a mapping from Java to XML.
Together they aim to demonstrate the main features of XMap and to
give an understanding of the essential requirements of a mapping
language.

\subsection{StateMachines to Java}

The purpose of this mapping is to translate a StateMachine into a
Java class. The source of the mapping is the StateMachines
metamodel described in chapter \ref{abschapter} and the class
diagram for this metamodel is repeated below.

\begin{figure}[htb]
\begin{center}
\includegraphics[width=17cm]{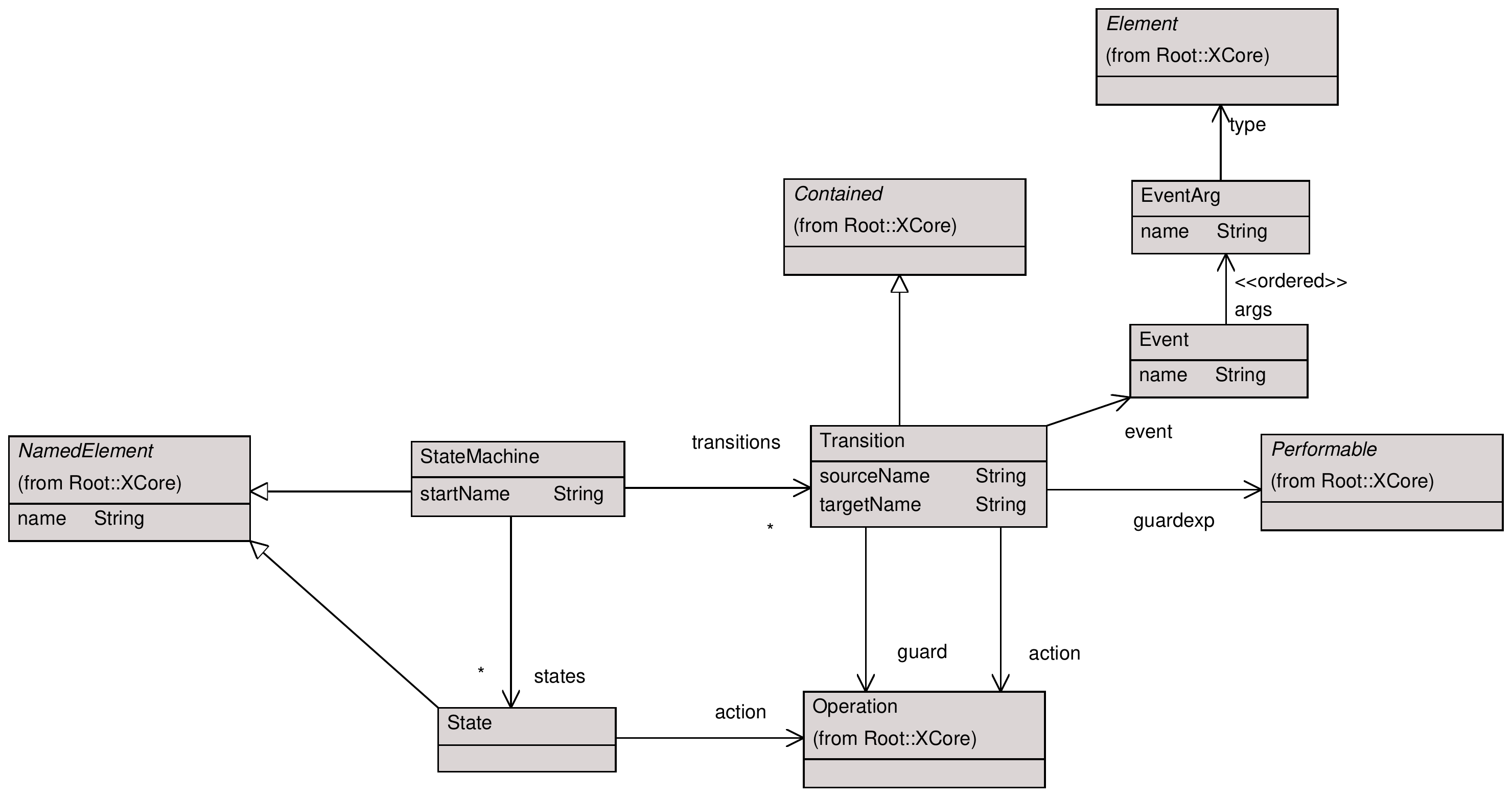}
\caption{The abstract syntax metamodel for StateMachines}
\end{center}
\end{figure}

\begin{figure}[htb]
\begin{center}
\includegraphics[width=15cm]{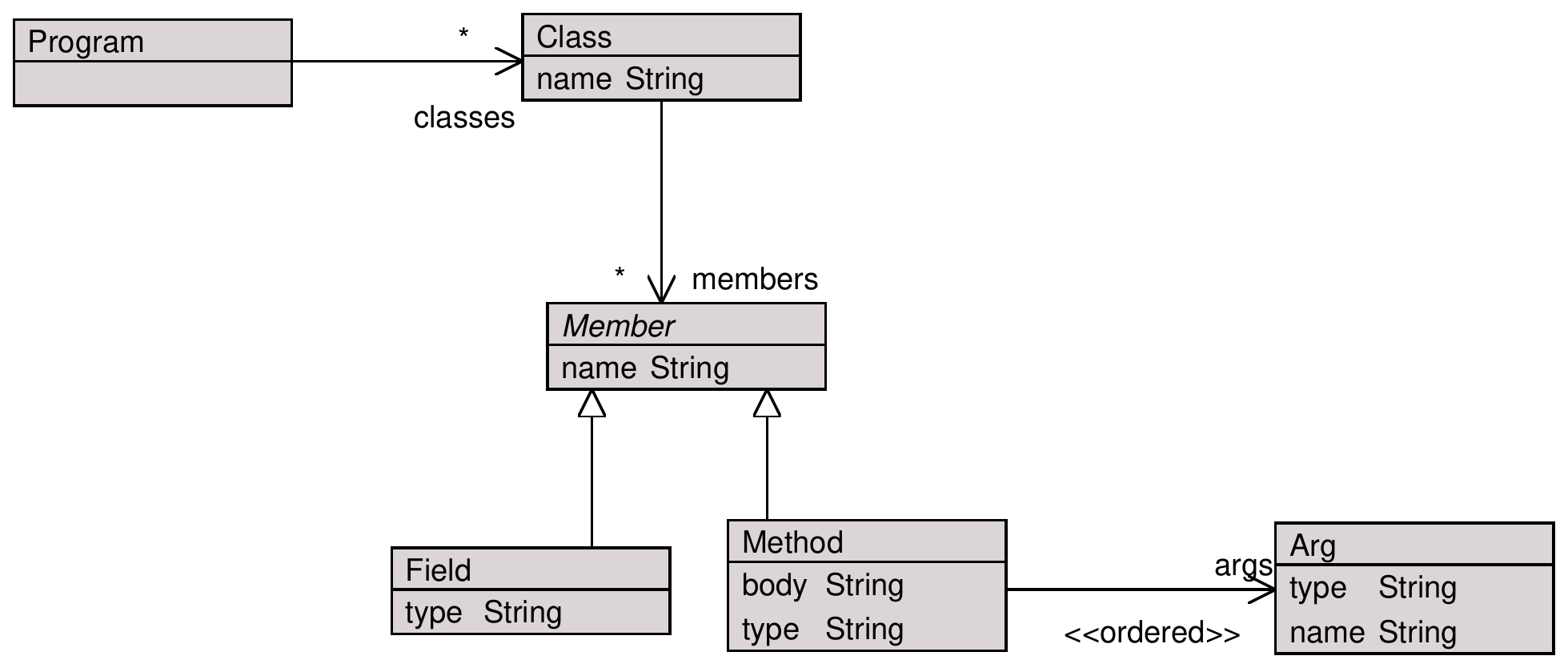}
\caption{The abstract syntax metamodel for Java} \label{javamodel}
\end{center}
\end{figure}

The target of the mapping is a simple model of Java as shown in
figure \ref{javamodel}. A Java program is a collection of named
classes. A Java class has a collection of member properties, which
may be named fields (attributes) or methods. A method has a name,
a body and  a return type and a sequence of arguments that have a
name and a type.

\begin{figure}[htb]
\begin{center}
\includegraphics[width=15cm]{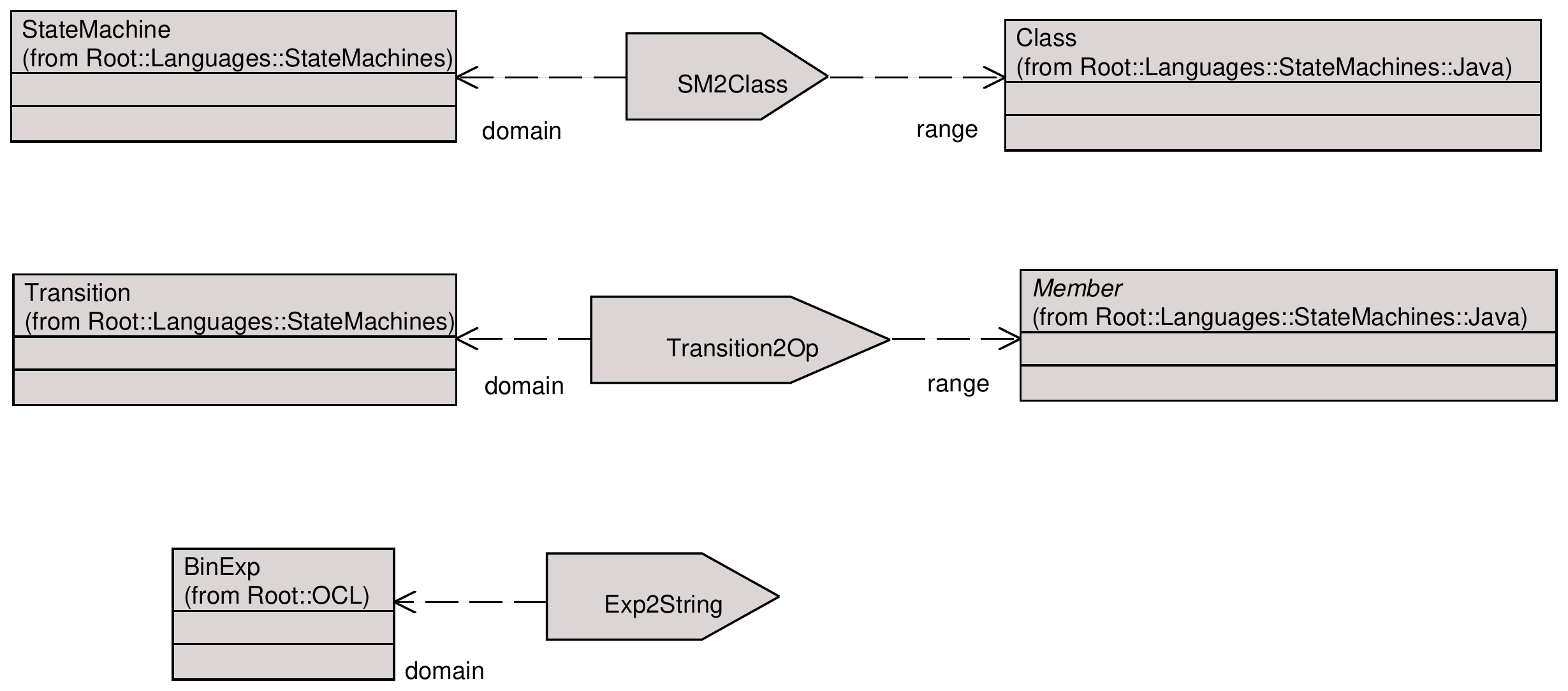}
\caption{The StateMachine to Java mapping} \label{javamapping}
\end{center}
\end{figure}

The mapping is shown in figure \ref{javamapping}. It maps a
StateMachine into a Java class with an attribute called state and
maps each transition to a method that changes the value of state
from the source state to the target state.

The detail of the StateMachine mapping are described by the
following code. A mapping consists of a collection of clauses,
which are pattern matches between patterns of source and target
objects. Whenever a collection of source values is successfully
matched to the input of the mapping, the resulting collection of
values after the do expression is generated. Variables can be used
within clauses, and matched against values of slots in objects.
Because XMap builds on XOCL, XOCL expressions can also be used to
capture complex relationships between variables.

\begin{lstlisting}
@Map SM2Class(StateMachines::StateMachine)->Java::Class
  @Clause ToClass
    s = StateMachine
      [name = N,
       transitions = T,
       states = S] do
    Class
      [name = N,
       members = M->including(
         Field
           [name = "state",
            type = "String" ])]
    where M = T->collect(t | Transition2Method(t))
  end
\end{lstlisting}
In this example, whenever the mapping is given a StateMachine
object with a name equal to the variable N, a set of transitions T
and a set of states S, it will generate an instance of the class
Java::Class. This will have a name equal to N and members that
includes a single state attribute named "state" and a set of
methods, M. The where clause is used to calculate the value M. It
matches M with the results of iterating over the transitions, T,
and applying the Transition2Method mapping.

\noindent The mapping from transitions to Java methods is shown
below:

\begin{lstlisting}
  @Map Transition2Method(Transition)->Java::Method
    @Clause Transition2Method
      t = Transition
        [event = Event[name = N] ] do
      Method
        [name = N,
         body = B]
      where
        B = "if (" + Exp2String()(t.guardexp.performable) + ")\n" +
        "  this.state := " + "\"" + t.targetName + "\"" + "; \n"

    end
  end
\end{lstlisting}
This mapping matches a Transition owning an Event named N to a
Java Method with a name N and a body B. This mapping illustrates
how patterns can capture arbitrarily nested structures: any depth
of object structure could have been captured by the expression.

The definition of the body, B, requires some explanation. Because
the mapping generates a textual body, this expression constructs a
string. The string contains an "if" expression whose guard is the
result of mapping the transition's guard expression (an instances
of the OCL metamodel) into a string (see below). The result of the
"if" expression is an assignment statement that assigns the state
variable to be the target of the transition.

\subsubsection{Mapping Guard Expressions}

Whilst the above mapping deals with the mapping to the
superstructure of Java, it does not deal with mapping the guards
and actions of the StateMachine. However, in order to be able to
{\em run} the Java, these will need mapping across too. As an
example, the following mapping describes how some sub-expressions
of OCL can be mapped into Java code.

\begin{lstlisting}
  @Map Exp2String(OCL::BinExp)->String
    @Clause BinExp2String
      BinExp
      [binOp = N,
       left = L,
       right = R] do
       self(L) + " " + N + " " + self(R)
    end
    @Clause IntExp2String
      IntExp
      [value = V] do
       V
    end
    @Clause Dot2String
      Dot
      [name = N,
       target = S] do
       if S->isKindOf(OCL::Self) then
         "self"
       else
         self(S)
       end + "." + N
    end
    @Clause BoolExp2String
      BoolExp
      [value = V] do
       V
    end
  end
\end{lstlisting}
This mapping contains a number of clauses, which are matched
against an sub-expression of OCL. In order to understand this
mapping, it is necessary to understand the OCL metamodel. The
relevant fragment of this is shown below in figure
\ref{oclfragment}.

\begin{figure}[htb]
\begin{center}
\includegraphics[width=7cm]{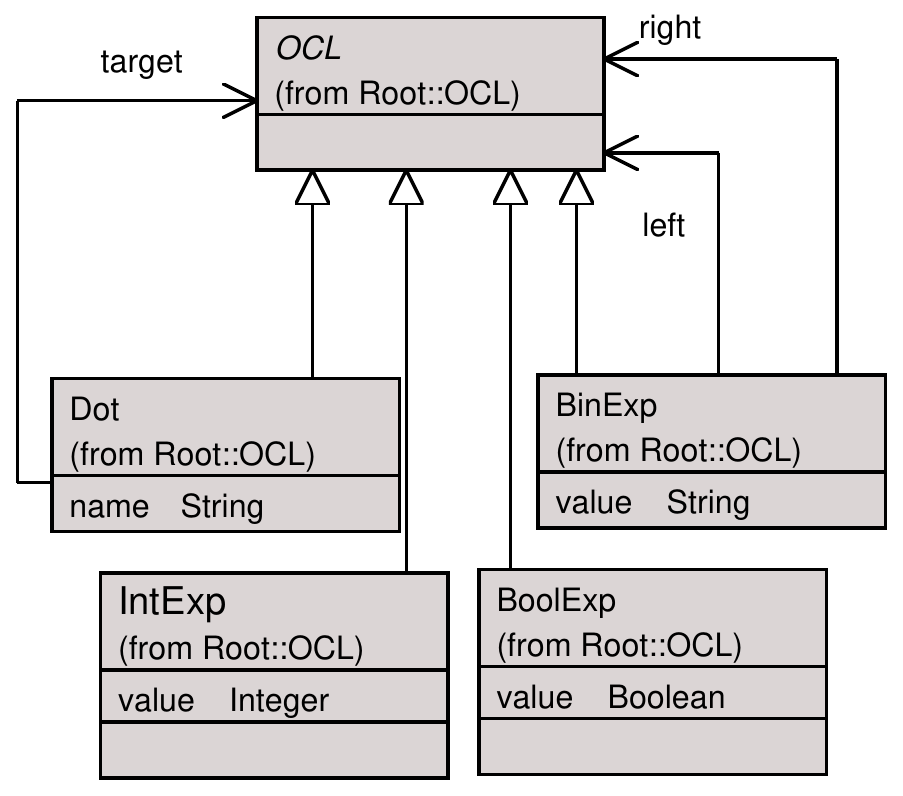}
\caption{Fragment of OCL expression metamodel} \label{oclfragment}
\end{center}
\end{figure}

The first clause of this expression matches against any binary
expression, and generates a string containing the results of
mapping the left hand expression followed by the binary operator
and the result of mapping the right hand expression. The following
clauses describe what should happen for integer expressions, dot
expressions and boolean expressions.

As an example, consider the StateMachine in figure
\ref{trafficlight}. The body of the guard on the GreenRed
transition will be parsed into an instance of an OCL expression.
The mapping described above will translate the transition and its
guard expression into the following Java code:

\begin{lstlisting}
public GreenRed() if (self.count < 10)
  this.state := "Red";
\end{lstlisting}\subsection{Mapping to XML}

This section gives an example of using XMap to map between Java
and XML. The aim of this example is to illustrate the ability of
mappings to record information about the execution of a mapping.

\begin{figure}[htb]
\begin{center}
\includegraphics[width=11cm]{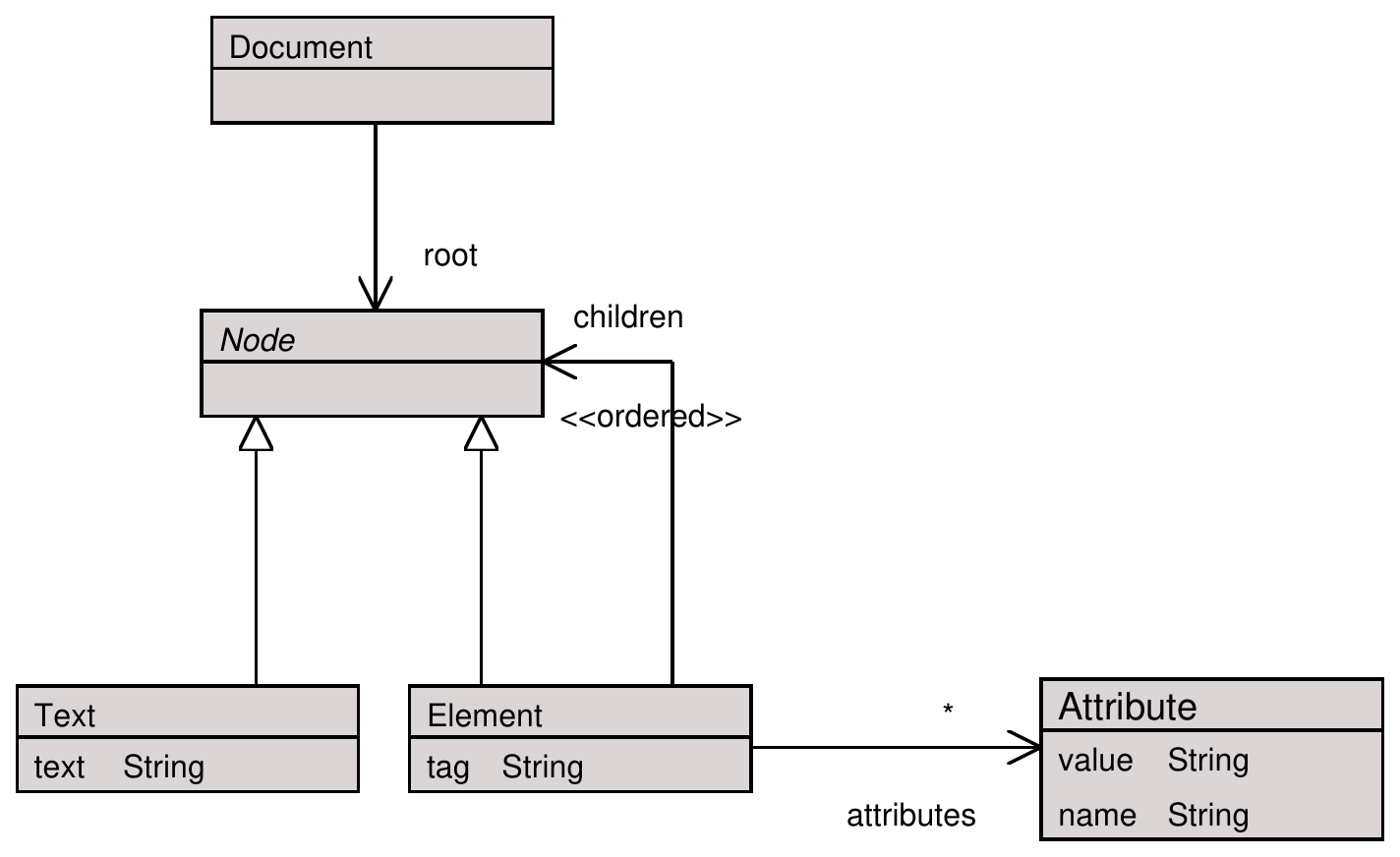}
\caption{Abstract syntax model for XML} \label{xmlmodel}
\end{center}
\end{figure}

A model of XML is shown in figure \ref{xmlmodel}. An XML document
consists of a root node, which may be an element, or a text
string. Elements can have attributes, which have a name and a
value. Elements may also have child nodes.

\begin{figure}[htb]
\begin{center}
\includegraphics[width=14cm]{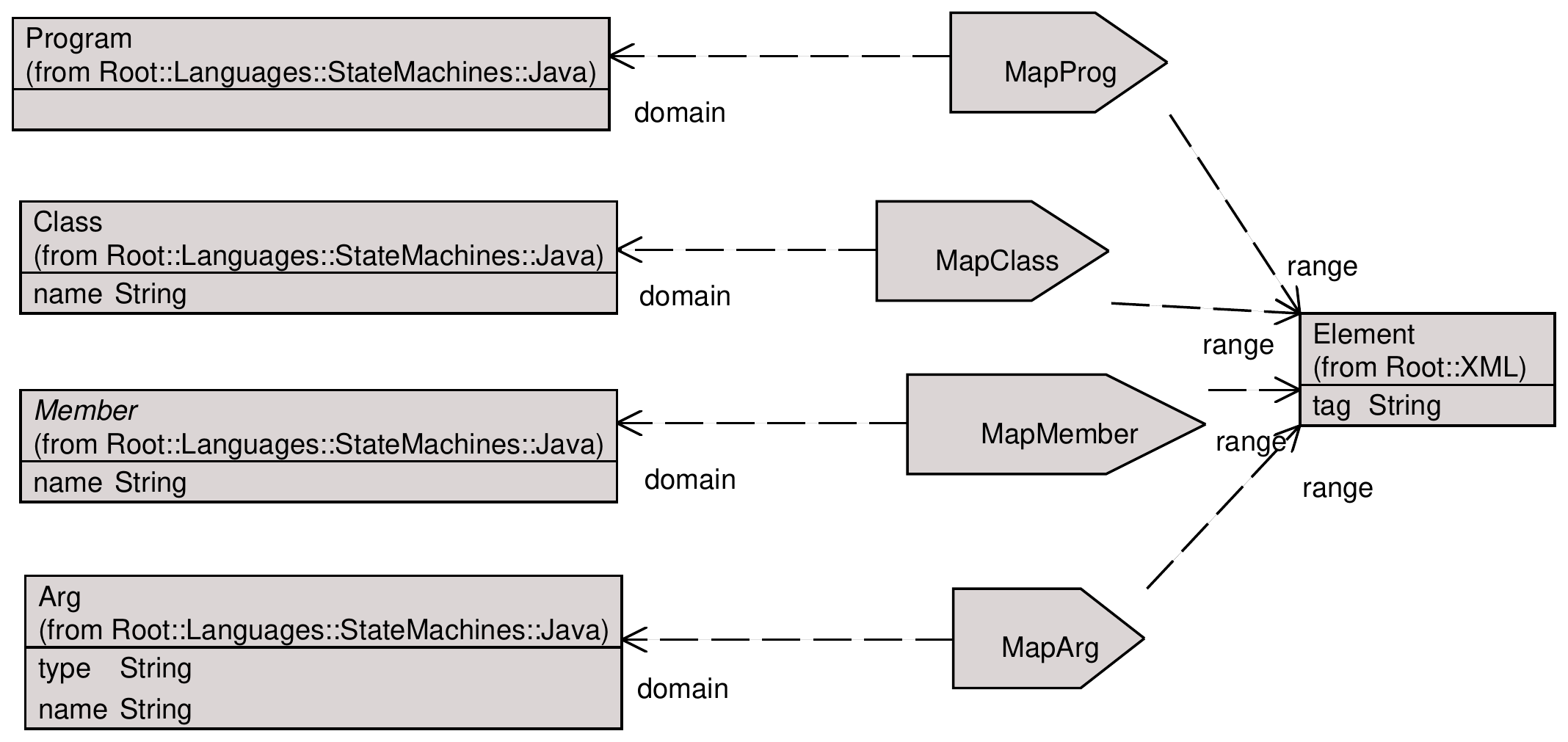}
\caption{Mapping from Java to XML} \label{xmlmapping}
\end{center}
\end{figure}

A mapping between Java and XML maps each element of the Java model
(classes, fields, methods and arguments into XML elements. The
relationship between XML elements and their children matches the
hierarchical relationship between the elements in the Java model.
The mapping diagram is shown in figure \ref{xmlmapping}.

The mapping from a Java program to an element is shown below. The
operation getId() will return an id ref for an element if that
element has already been generated, alternatively it will create a
new one and place it in a table. All subordinate mappings can
reach this mapping via their 'owner' slot.

\begin{lstlisting}
  @Map MapProg(Java::Program)->XML::Element

    @Attribute mapClass : MapClass = MapClass(self) end
    @Attribute idTable : Table = Table(100) end

    @Operation getId(name)

      if idTable.hasKey(name) then
        idTable.get(name)
      else
        idTable.put(name,"id" + idTable.keys()->size.toString());
        idTable.get(name)
      end
    end

    @Clause Program2Element
      Program
        [classes = C]
      do
      Element
        [tag = "Program",
         attributes = Set{},
         children = E]
      where
      E = C->collect(c | mapClass(c))
    end
  end
\end{lstlisting}
The mapping from classes to XML elements is more complicated. It
maps a class with name N and members M to an element tagged as a
"Class" containing two attributes. The first attribute corresponds
to the name attribute of the class, and thus has the name "name"
and value N. The second attribute provide an id for the element,
which is the result of running its getId() operation.

\begin{lstlisting}
@Map MapClass(Java::Class)->XML::Element

  @Attribute owner : MapProg end
  @Attribute mapMember : MapMember = MapMember(self) end

  @Constructor(owner) end

  @Operation getId(name)
    owner.getId(name)
  end

  @Clause Class2Element
    Class
      [name = N,
       members = M] do
    Element
      [tag = "Class",
        attributes =
          Set{}->including(
             Attribute
               [name = "name",
                value = N])->including(
                  Attribute
                    [name = "id",
                     value = self.getId(N)]),
         children = E]
      where
      E = M->collect(m | mapMember(m))
  end
end
\end{lstlisting}\noindent Exactly the same strategy is used to map Field, Methods
and Arguments to XML Elements. The following shows the mapping for
Fields:

\begin{lstlisting}
@Map MapMember(Java::Member)->XML::Element

  @Attribute owner : MapClass end
  @Attribute mapArg : MapArg = MapArg(self) end

  @Constructor(owner) end

  @Operation getId(name)
    owner.getId(name)
  end

  @Clause Field2Element
    Java::Field
      [name = N,
      type = T] do
    Element
      [tag = "Field",
         attributes =
           Set{}->including(
             Attribute
               [name = "name",
                value = N])->including(
                  Attribute
                    [name = "type",
                     value = self.getId(T)]),
      children = Set{}]
  end
end
\end{lstlisting}
\section{Mapping Specifications}

\begin{figure}[htb]
\begin{center}
\includegraphics[width=12cm]{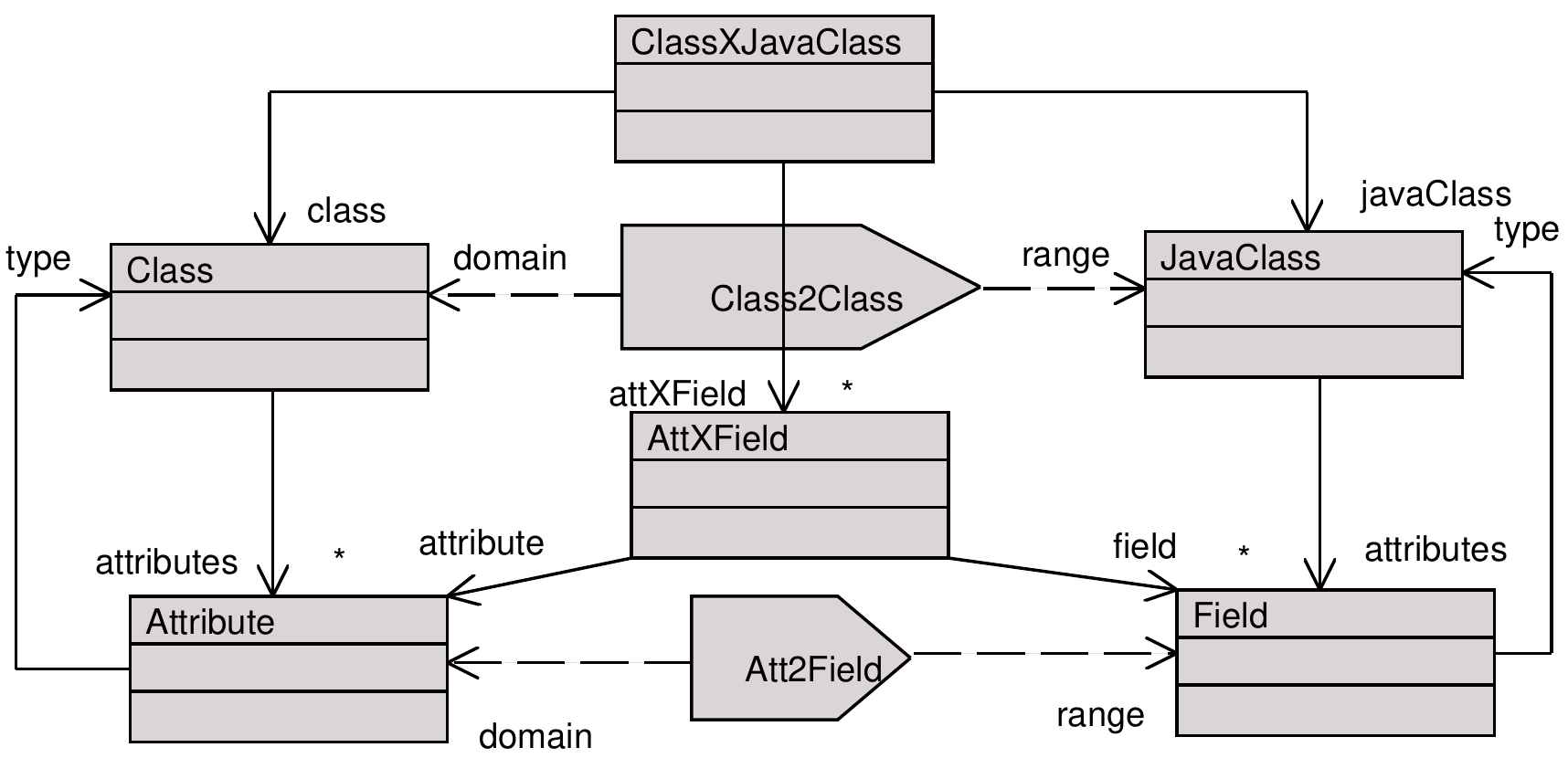}
\caption{An example mapping specification} \label{mappingspec}
\end{center}
\end{figure}

A mapping specification is useful in describing what a mapping
does as opposed to how it is to be implemented. Mapping
specifications do not require any additional modelling facilities
beyond OCL. As shown in figure \ref{mappingspec}, a mapping
specification consists of a number of mapping classes, which sit
between the elements of the models that are to be mapped.

OCL is then used to define the constraints on the relationship
between the models. The following constraint requires that the
names of the two classes must be the same.

\begin{lstlisting}
context ClassXJavaClass
  @Constraint SameNames
    class.name = javaClass.name
  end
\end{lstlisting}\noindent This constraint ensures that there is an AttXField
mapping for each attribute owned by a class:

\begin{lstlisting}
context ClassXJavaClass
  @Constraint AttXFieldForClassAtt
    class.attributes = attXField.attribute
  end
\end{lstlisting}\noindent Similar constraints will be required for Java classes.

Mapping specifications are a useful means of validating a mapping.
This can be achieved by firstly specifying the mapping. Next, an
implementation of the mapping is designed so that whenever it is
run it will generate instances of the appropriate mapping
specification classes. The constraints on the mapping
specification can then checked to test that the mapping
implementation has satisfied the mapping specification.

\section{Mapping Issues}

\subsection{Merging}

The previous example highlights an important issue that often
occurs when writing mappings: how to merge duplicate elements.
Duplicate elements typically occur when mapping graph like
structures. Elements in the graph may have references to other
elements in the graph that have already been mapped. In this
situation, naively following the link and mapping the element will
result in duplicate elements being generated.

\begin{figure}[htb]
\begin{center}
\includegraphics[width=12cm]{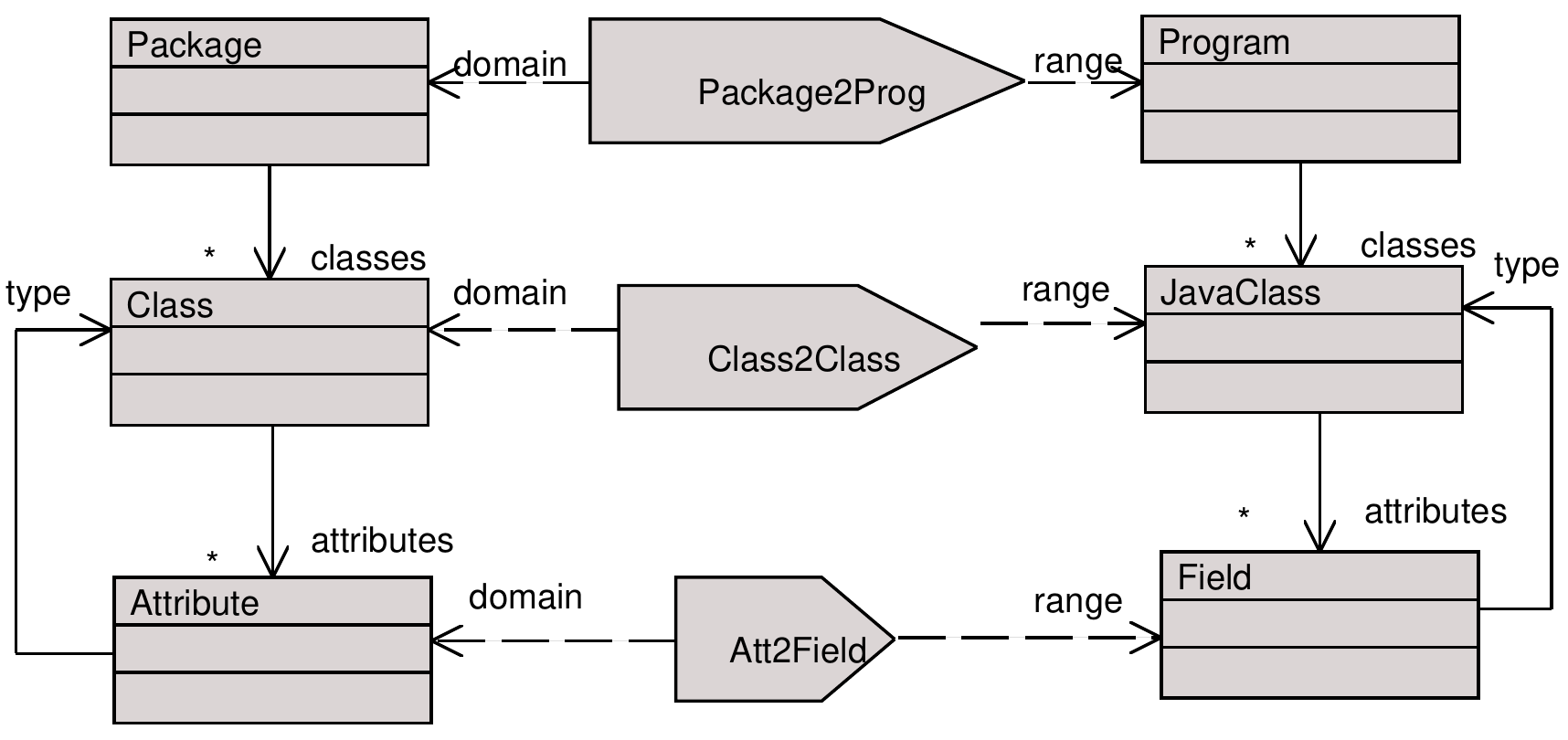}
\caption{A simplified UML to Java mapping} \label{umlmapping}
\end{center}
\end{figure}

A good example is a mapping between UML and Java (a simplified
model is shown in figure \ref{umlmapping}). One way to implement
this mapping is to traverse each of the classes in the package,
and then map each class and its associated attributes to Java
classes and fields. A problem arises because Java classes
reference their types (thus introducing a circular dependency). At
the point at which a Att2Field mapping is executed, the generated
Field's type may already have been generated. If the Att2Field
mapping then generates a new type, it will be duplicated.

\noindent There are a number of solutions to this problem:

\begin{itemize}
\item Maintain a table of generated elements and check it before
generating a new element. The Java to XML mapping described above
is an example of this. \item Run the mapping and then merge
duplicate elements on a case by case basis. An example of how this
is achieved in XOCL can be found in section \ref{merge}. \item Use
a generic mechanism that can run over object graphs merging
duplicate elements. The walker algorithm defined in section
\ref{walker} can form the basis for such an algorithm.
\end{itemize}

In all cases, criteria must be identified for mergeable elements.
In above mapping, the criteria for merging two types is that they
have the same name. In general, the criteria must be defined on a
case by case basis, and accommodated in the mappings.

\subsection{Traceability}

A common requirement when constructing mappings is to keep
information about the mapping. There are two strategies that can
be used to achieve this:

\begin{itemize}
\item Create instances of reverse mappings or mapping
specifications as the mapping is performed. The result will be a
graph of reverse mappings or mapping specifications connecting the
domain and range elements. \item Extend the mapping language so
that it records a trace of the executed mappings in a generic.
\end{itemize}

The former approach is most appropriate when wishing to reverse
the mapping or check that it satisfies a specification of the
mapping. The latter approach is most useful when analysing or
debugging a mapping.

\section{XSync}

Very often mappings are required that are not uni-directional, but
which synchronise elements, ensuring that if changes occur in one
element they are reflected in another. There are many applications
of synchronised mappings, including:

\begin{itemize}
\item Maintaining consistency between views on a common collection
of elements: for instance, keeping diagrams consistent with models
and vice versa. \item Managing multiple models of a system: for
example, a large systems development project might use multiple
tools to design different aspects of a system but be required to
maintain consistency where design data overlaps. \item Supporting
round trip engineering where models are maintained in sync with
code and vice versa.
\end{itemize}

XSync is a mapping language that permits the modelling of
synchronised mappings. It enables rules about the relationship
between concepts in two or more models to be captured at a high
level of abstraction. Synchronised mappings can be run
concurrently, continuously monitoring and managing relationship
between elements.

\subsection{Examples}

The following code describes a simple XSync model in which we want
to maintain consistency between two instances of a class. The
class is defined as follows:

\begin{lstlisting}
context Root
  @Class Point
    @Attribute x : Integer end
    @Attribute y : Integer end
    @Constructor(x,y) ! end
  end
\end{lstlisting}\noindent We create two instances of the class, p1 and p2:

\begin{lstlisting}
Root::p1 := Point(100,100); Root::p2 := Point(1,2);
\end{lstlisting}\noindent Now we create a synchronised mapping:

\begin{lstlisting}
Root::n1 :=

@XSync
  @Scope
    Set{p1,p2}
  end
  @Rule r1 1
      p1 = Point[x=x;y=y];
      p2 = Point[x=x;y=y] when p1 <> p2
    do
      "The points match".println()
  end
  @Rule r2 1
      p1 = Point[x=x1];
      p2 = Point[x=x2] when p1 <> p2 and x1 < x2
    do
      "Incrementing x".println();
      p1.x := x1 + 1
  end
  @Rule r3 1
      p1 = Point[y=y1];
      p2 = Point[y=y2] when p1 <> p2 and y1 < y2
    do
      "Incrementing y".println();
      p1.y := y1 + 1
  end
end;
\end{lstlisting}A synchronised mapping consists of a scope and a collection of
synchronisation rules. The scope of the mapping is the collection
of elements over which the mapping applies. In this case, it is
the two instances of Point, p1 and p2.

A rule describes the condition under which an action should occur.
Actions are used to describe synchronisations but they may also
cause other side effects. This is important because in practice it
is often necessary to perform other tasks such as generating
reports, modifying values and so on.

A rule takes the form of a pattern, a boolean 'when' clause and a
'do' action. Provided that the 'when' clause is satisfied by
variables introduced by a matched pattern, the 'do' action will be
invoked.

In this example, rule r1 checks to see whether p1 and p2 match
against instances of Point that have the same x and y values. If
they do, and  p1 and p2 are not the same element, the 'do' action
is invoked to display a message. Rule r2 matches the x values of
p1 and p2 against the variables x1 and x2. If x1 is less than x2,
the value of p1's x is incremented. Rule r3 does the same for the
y values. The result of running this mapping with two points with
different x or y values will be to increment the values until they
are synchronised.

The previous example is limited as it is fixed to two specific
instances. The following shows how a synchronisation mapping can
be parameterised by an operation:

\begin{lstlisting}
context Root
  @Operation sameName(c1,c2)
    @XSync
      @Scope
        Set{c1,c2}
      end
      @Rule r1 1
        x1 = Class[name = n1] when x1 = c1;
        x2 = Class[name = n2] when x2 = c2 and n1 <> n2
        do x1.name := x2.name
      end
    end
  end
\end{lstlisting}This mapping applies to any pair of classes and synchronises the
name of the first with the second. Such a mapping would be useful
for synchronising concrete syntax classes and abstract syntax
classes, for example see chapter \ref{concretechapter}.

This section has provided a short introduction to the XSync
synchronisation language. Future versions of this book will
explore a number of deeper issues, including its application to
the synchronisation of complex structures, and the different
execution models it supports.

\section{Conclusion}

Mappings are a central part of Language-Driven Development as they
enable key relationships between many different types of problem
domain to be captured. Mappings exist between vertical, horizontal
and variant dimensions of a problem domain. In the context of
languages, mappings are needed to synchronise abstract and
concrete syntaxes and to relate models and programs written in
different languages.

\chapter{Reuse}
\label{langchapter}

\section{Introduction}

Languages are hard to design. The effort that goes into producing
a language definition can be overwhelming, particularly if the
language is large or semantically rich. One way to address this
problem is to find ways of designing languages so that they are
more reusable and adaptable. By reusing, rather than re-inventing,
it is possible to significantly reduce the time spent on
development, allowing language designers to concentrate on the
novel features of the language.

This chapter looks at techniques that can be used to design
reusable metamodels. These include approaches based on the use of
specific extension mechanisms such as stereotyping and class
specialisation, and richer mechanisms that support the large
grained extension of modelling languages, such as meta-packages
and package specialisation.  An alternative approach based on the
translation of new concepts into pre-defined is also discussed.

The following sections examine each of these approaches in turn,
identifying their advantages and disadvantages and offering
practical advice on their application.

\section{Extension Based Approaches}
\label{langextension}

This approach involves extending and tailoring concepts from an
existing metamodel. The most common mechanisms for implementing
extension are class specialisation and stereotyping.

\subsection{Specialisation}

Specialisation has been widely used as a reuse mechanism by OO
programming languages for many years. Since XMF is an OO
metamodelling language, it is not surprising that this mechanism
can also be used to reuse language concepts.

One of the best ways to support this approach is to provide a
collection of classes (a framework) that supports a collection of
reusable language concepts. These concepts can then be specialised
to support the new language concepts.

The XCore framework (see section \ref{framework}) aims to support
the abstractions found in popular metamodelling frameworks such as
MOF \cite{mofspec} and EMF \cite{emf} at a usable level of
granularity. An important difference between XCore and other
frameworks is that XCore is a platform independent, executable
language with a precisely defined syntax and semantics. This means
that these semantics can be extended as well to rapidly define
semantically rich language concepts. For example, if the class
XCore::Class is specialised by the class X, all instances of X
will inherit the properties of a class, including the following:

\begin{itemize}
\item They can be instantiated. \item They can have attributes,
operations and constraints. \item They can be serialized. \item
Their instances can be checked against any constraints. \item
Their operations can be invoked on their instances. \item They
have access to the grammar and diagram syntax defined for classes.

\end{itemize}

Furthermore, because the class Class also specialises Object, all
operations that apply to an Object can also be applied to
instances of X, such as executing its meta-operations, mapping it
to another concept and so on.

Having an executable semantics for a metamodelling framework adds
significant value because semantics can be reused along with the
structural properties of the language concepts.

\subsubsection{Example}

Consider the requirement to model the concept of a mapping. A
mapping has the following properties:

\begin{itemize}
\item It has a domain, which is a collection of input types to the
mapping. \item It has a range, which is the result type of the
mapping. \item A mapping can be instantiated and the state of a
mapping instance can be recorded. \item Operations can be defined
on an mapping and can be executed by its instances. \item A
mapping owns a collection of clauses that define patterns for
matching input values to output values. \item A mapping instance
can be executed, matching inputs to clauses and resulting in an
output value.
\end{itemize}

\begin{figure}[htb]
\begin{center}
\includegraphics[width=10cm]{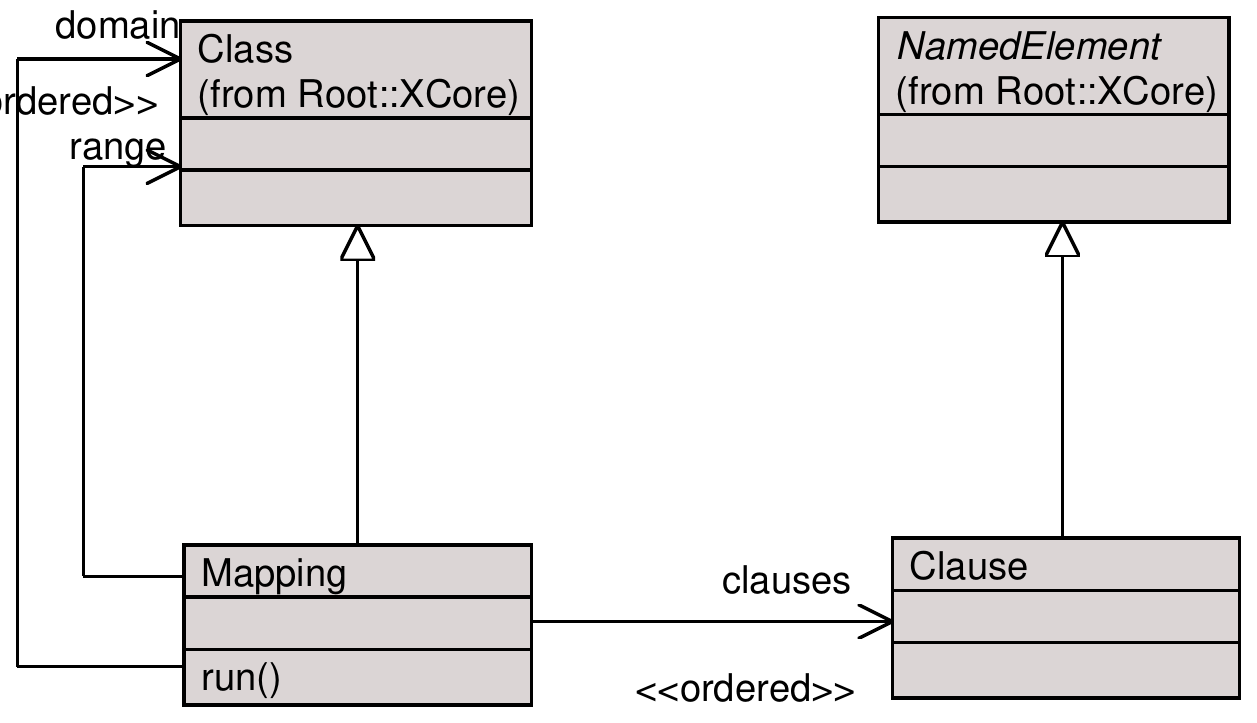}
\caption{Reuse of the class Class to define a new Mapping concept}
\label{mappingspecialisation}
\end{center}
\end{figure}

Many of these properties can be reused from existing XCore
concepts, thus avoiding the need to model them from scratch. The
model in figure \ref{mappingspecialisation} shows how this might
be done by specialising the class Class.

The mapping class reuses all the properties of Class, including
attributes, operations and the machinery to support instantiation
and operation invocation. Other properties, such as having domain
and range types, and executing mappings are layered on top.

\subsection{Stereotyping, Tags and Profiles}
\label{stereotypes}

Stereotypes are a widely used device that enable existing
modelling concepts to be treated as virtual subclasses of an
existing metamodelling class. This is achieved by associating
metamodel classes with information about what stereotypes they may
support. An example of this might be to stereotype the class Class
as a Component and the class Association as a Component Connector.

Tags and tagged values are closely related to stereotypes. These
have the same effect as extending metamodel classes with
additional attributes and information about the values these
attributes may be assigned. An example of tag and tagged value
might be the tag "visibility" attached to an Operation, which can
have the values, "public", "private" or "protected".

Together, stereotypes and tags can be used to create what is
called a {\em profile}: a collection of stereotyped concepts that
form the vocabulary of a new or hybrid modelling language.

\begin{figure}[htb]
\begin{center}
\includegraphics[width=10cm]{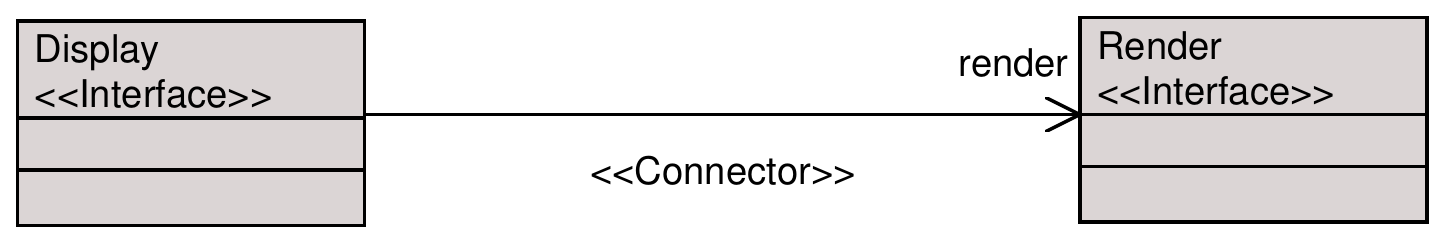}
\caption{Using stereotypes to model interfaces and connectors}
\label{stereotypeExample}
\end{center}
\end{figure}

Many UML tools provide support for stereotypes, tags and tagged
values, allowing them to be visually displayed in an editor with
an appropriate identifier. Figure \ref{stereotypeExample} shows a
class diagram with two stereotyped classes and a stereotyped
association.

The advantage of stereotypes and tags is that they add a
convenient level of tailorability to modelling tools that would
otherwise be unavailable.

\begin{figure}[htb]
\begin{center}
\includegraphics[width=9cm]{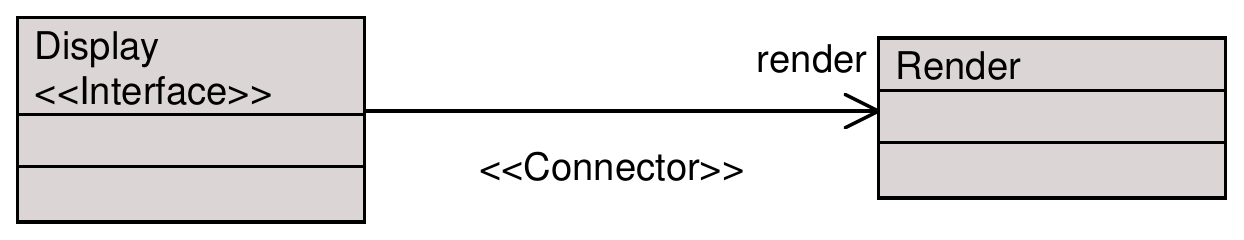}
\caption{An example of inconsistencies that can arise through the
use of stereotypes} \label{wrongstereotypes}
\end{center}
\end{figure}

Nevertheless, stereotypes and tags suffer from being semantically
weak and are not a replacement for a well-defined metamodel of the
language. As the example in figure \ref{wrongstereotypes} shows,
they offer little control over the correct usage of the language:
for instance, a model written in a component modelling language
might not permit interfaces to be connected to classes, yet this
will not be ruled out by stereotypes. Furthermore, stereotypes are
not able to capture the semantics of a language as they cannot
change the properties of meta-classes.

\noindent While UML profiles \cite{umlspec} allow additional
constraints to be defined on stereotypes, there are few if any
tools available that support this capability.

\subsection{Package specialisation and Meta-Packages}
\label{metapackages}

Although specialisation can be used to directly extend metamodel
concepts, this is a significantly more involved process than
defining stereotypes because a concrete syntax for the concepts
will also have to be modelled if they are to be used in a tool. A
better solution would be one that combines the simplicity of
stereotypes with the power of specialisation.

One way of achieving this is through the use of {\em package
specialisation} and {\em meta-packages}. Package specialisation is
a relationship between two packages where the child package
specialises the parent package. The consequence of this is that it
enables a package of language concepts to be clearly identified as
extensions of a package of existing language concepts (the parent
package).

\begin{figure}[htb]
\begin{center}
\includegraphics[width=2.5cm]{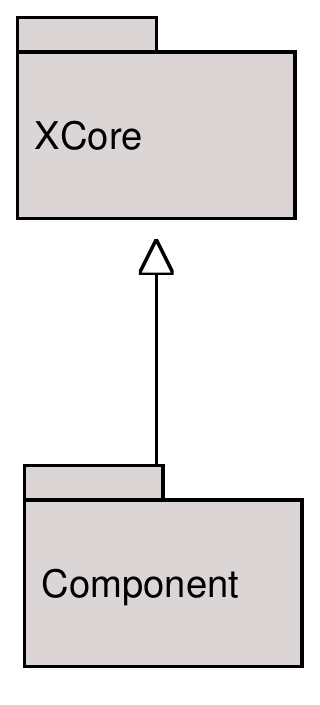}
\caption{Specialising the XCore package}
\label{componentspecialisation}
\end{center}
\end{figure}

As an example, figure \ref{componentspecialisation} shows a
package of component concepts specialising the XCore package.

\begin{figure}[htb]
\begin{center}
\includegraphics[width=10cm]{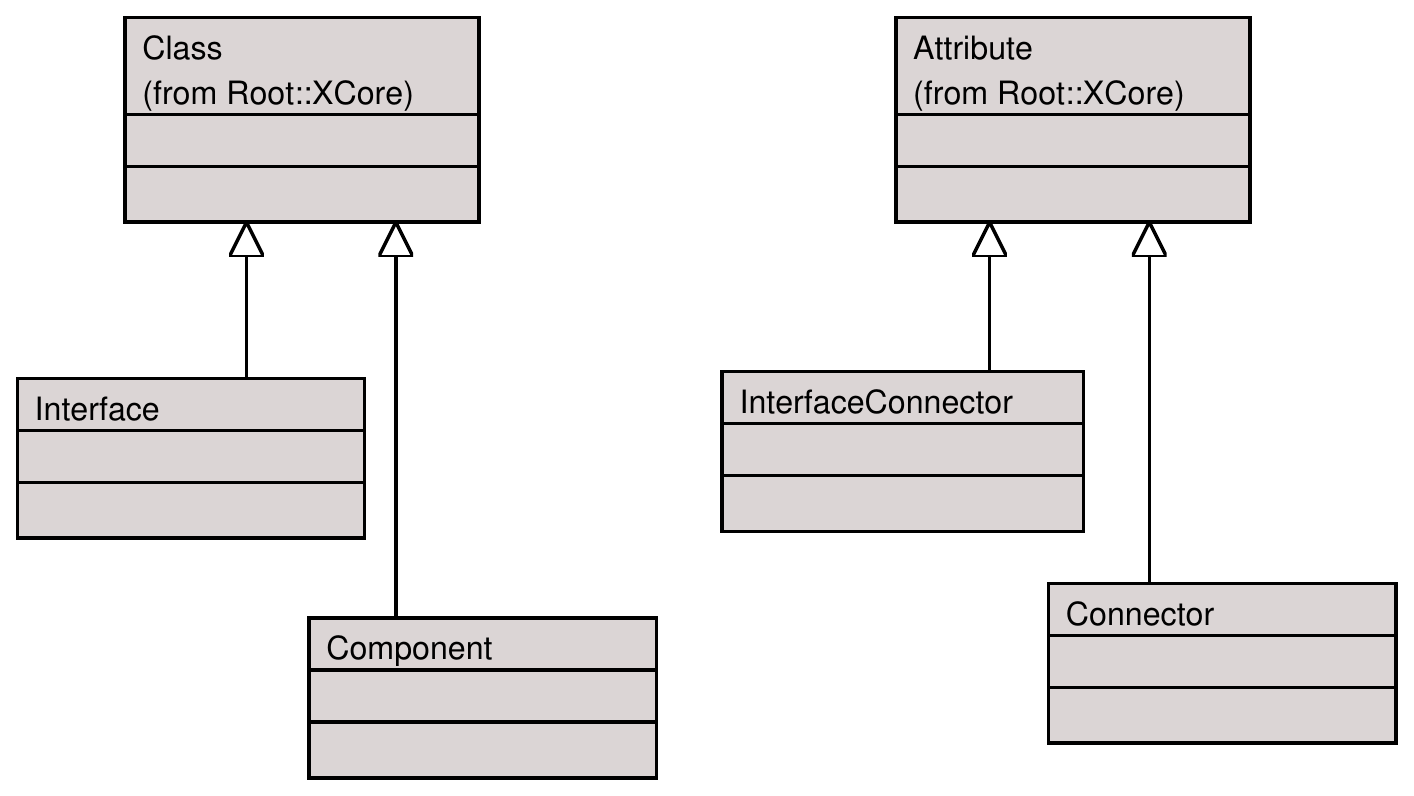}
\caption{specialising XCore concepts}
\label{componentspecialisation2}
\end{center}
\end{figure}

Inside the components package some choices are made about the
XCore classes that are to be specialised. These are shown in
figure \ref{componentspecialisation2}. Additional constraints can
be added to rule out certain combinations of components, for
example, a connector must always connect two interfaces.

\begin{figure}[htb]
\begin{center}
\includegraphics[width=2.5cm]{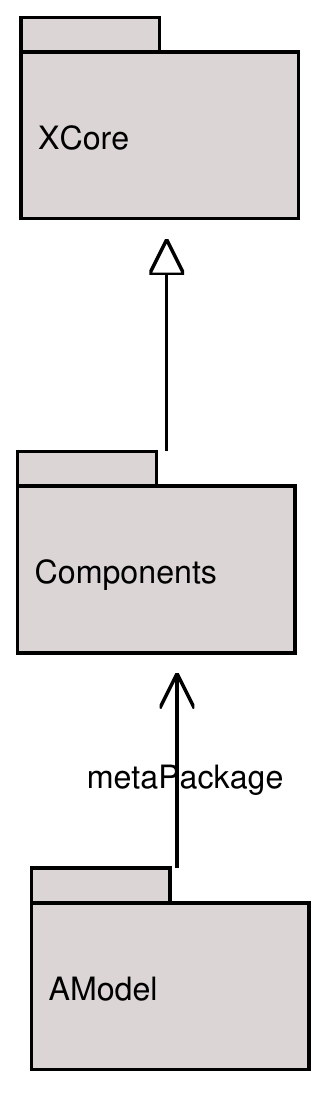}
\caption{A metapackage relationship between a model and the
components package} \label{metapackage}
\end{center}
\end{figure}

This provides sufficient information to determine whether a
concept should be represented as a native concrete syntax element
or as a stereotyped element. However, a way must now be found of
{\em using} the specialised package. This is where meta-packages
come in useful. A meta-package is a package of elements that are
instances of elements in another package (the meta-package). In
the components example, a package containing a model written in
the components language is related to the components package by a
meta-package relationship (see figure \ref{metapackage}).

\begin{figure}[htb]
\begin{center}
\includegraphics[width=10cm]{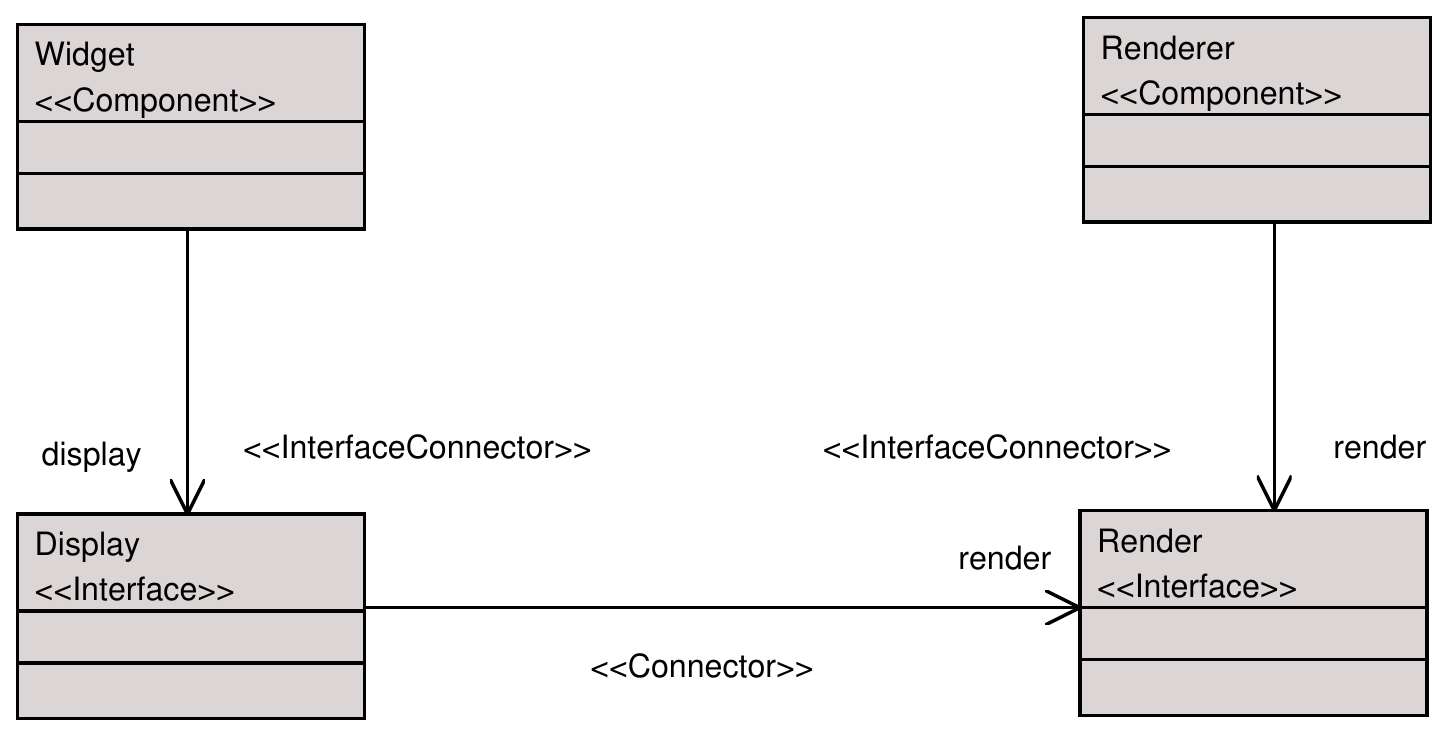}
\caption{A model written in the components language}
\label{amodel}
\end{center}
\end{figure}

If a tool understands that the model is an instance of a package
that specialises the XCore package it can provide appropriate
stereotypes for each of the specialised elements (see figure
\ref{amodel} for an example). These stereotyped elements can be
used to construct models, which can then be checked against any
constraints in the components package. Because the stereotyped
elements are real instances of meta-model elements all the
semantics inherited from XCore will be available as well.

Meta-packages are a more generic concept than just a mechanism for
dealing with concrete syntax. Consider a language expressed as a
meta-package that already has tools developed for the language. If
a new language differs from the existing one in a few minor but
important ways we would like to make use of the existing tools but
clearly work with the new language. The new language can be
defined as an extension of the meta-package. A package whose
meta-package is the new languages can thus be supplied to the
development tools using the standard principle of substitution.

If tools are generic with respect to the meta-package, then it can
tailor itself by providing specific instances of functionality for
each new language feature that is a sub-class of a corresponding
feature in the meta-package. This may cover a wide range of
different aspects of tool functionality.

As an example, imagine that a configuration management tool
expects instances of any sub-package of XCore and provides
facilities such as rollback, model-merge etc, based on all the
sub-classes of Class, Association, Attribute etc then a wide
variety of development tools can be constructed each of which
works with a different language but which all use the same
development code.

\section{Translation Based Approaches}

The aim here is to define new modelling concepts by a translation
to existing concepts that implement their required properties. As
described in chapter \ref{mappingchapter}, there are a variety of
ways of translating between metamodel concepts. These include:

\begin{itemize}
\item Defining a translation from the concrete syntax of a
concept to an appropriate abstract syntax model written in another
language. The new language is thus a sugar on top of the existing
language.
\item Translating from the abstract syntax of the new concept into
the appropriate abstract syntax model.
\item Maintaining a synchronised mapping between concepts.
\end{itemize}

By translating into an existing primitive, the effort required to
model the semantics of the new concept is significantly reduced.
The translation approach is akin to a compilation step, in which
new concepts are compiled into more primitive, well-defined
concepts. The advantage is that layers of abstraction can be built
up, each ultimately based on a core set of primitives support by a
single virtual machine. This greatly facilitates tool
interoperability as the virtual machine becomes a single point of
tool conformance.

\subsection{Example}

A common approach to implementing associations is to view them as
a pair of attributes and a constraint, where associations between
classes are implemented  as attributes on each class plus a
round-trip constraint that must hold between instances of the
classes. A translation can be defined between an association model
(either expressed as concrete or abstract syntax) and the more
primitive concept of class and attribute.

\section{Family of Languages}
\label{families}

As the number of languages built around a metamodelling language
architecture grows, it is very useful to be able to manage them in
a controlled way. One approach is to organise them into families
of languages \cite{prefaces}. In a family of languages, each
member is related to another language through its relationship to
a common parent. The result is a framework of languages that can
be readily adapted and combined to produce new language members:
in other words a language factory or product line architecture.

\begin{figure}[htb]
\begin{center}
\includegraphics[width=8cm]{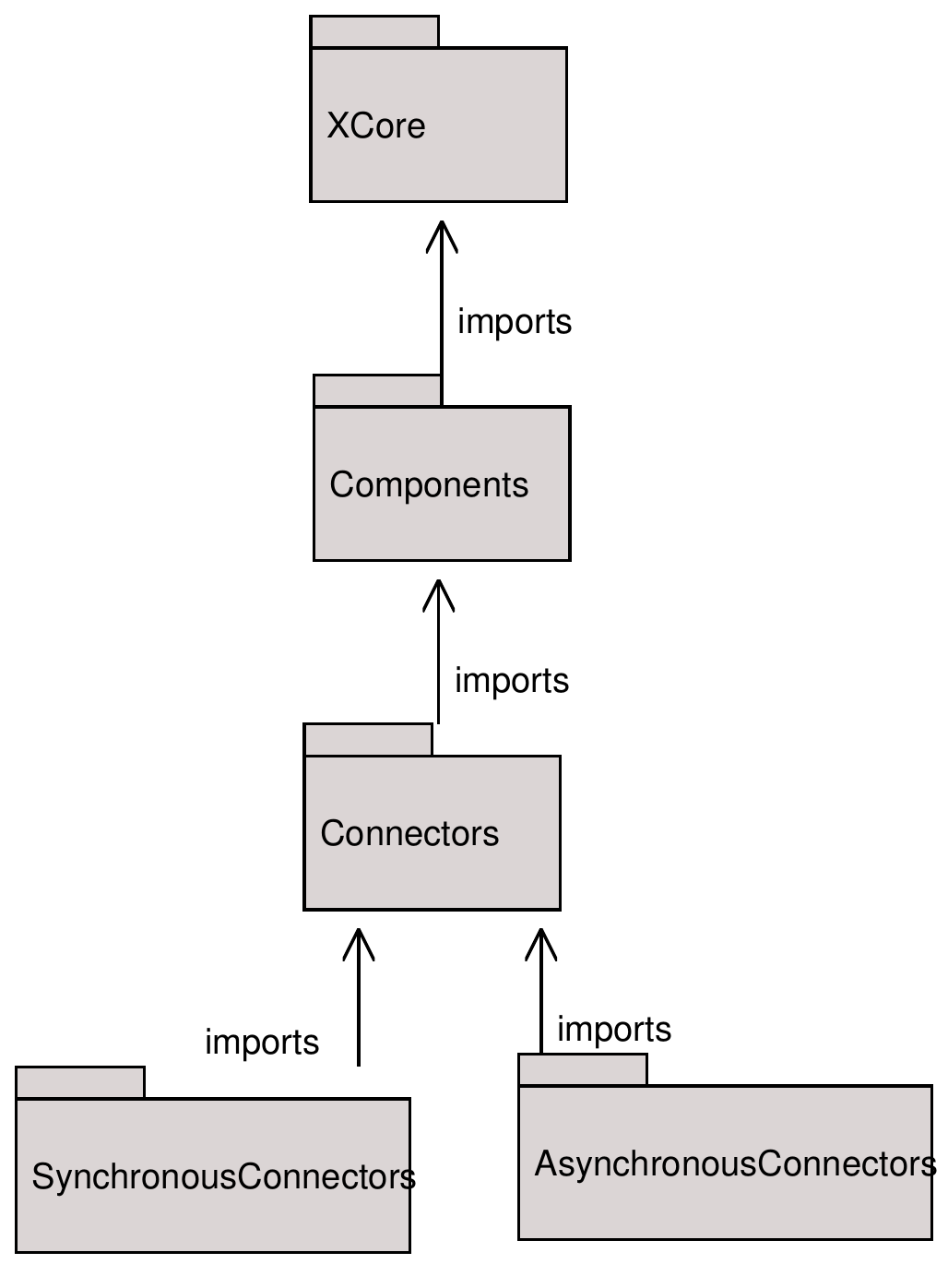}
\caption{A language architecture} \label{langarch}
\end{center}
\end{figure}

There are a number of mechanisms that are useful realising such a
framework. Packages are a good mechanism for dividing languages
into language components, which can be combined using package
import. An example of such an arhictecture is shown in figure
\ref{langarch}.

Packages can be broken down further, both vertically and
horizontally. Vertical decomposition decomposes a language into a
sub-language components that incrementally build on each other. An
example might be an expression language that is arranged into
packages containing different types of expressions: binary, unary
and so on.

\begin{figure}[htb]
\begin{center}
\includegraphics[width=9cm]{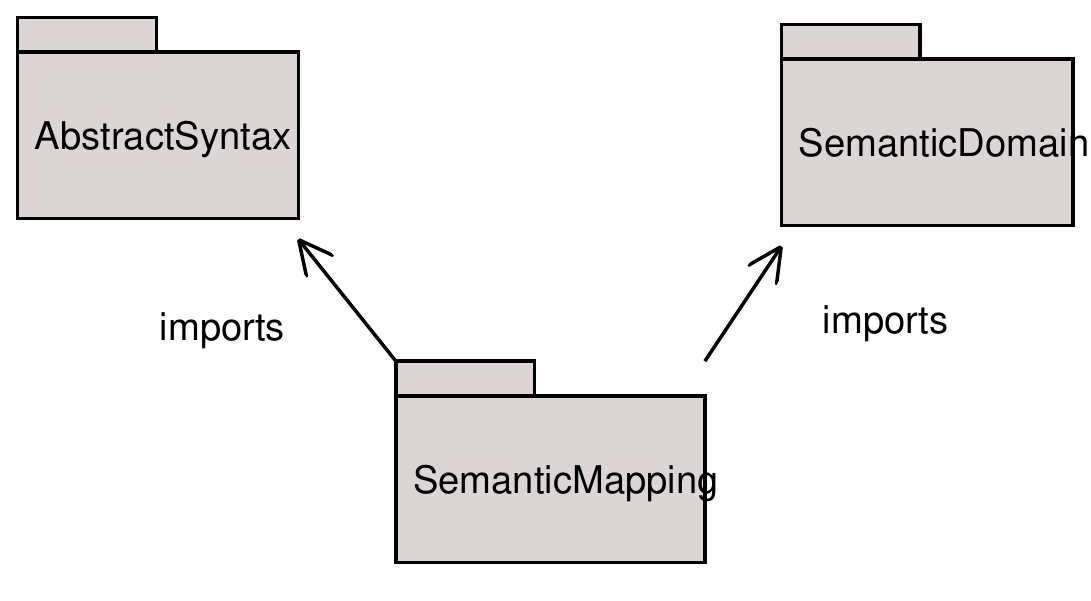}
\caption{Horizontal language components} \label{horizontal}
\end{center}
\end{figure}

A horizontal decomposition breaks a language into different
aspects. It makes sense to make a distinction between the concrete
syntax, abstract syntax and semantic domain of the language
component (see figure \ref{horizontal}).

\section{The XCore Framework}
\label{framework}

\begin{figure}[htb]
\begin{center}
\includegraphics[width=15cm]{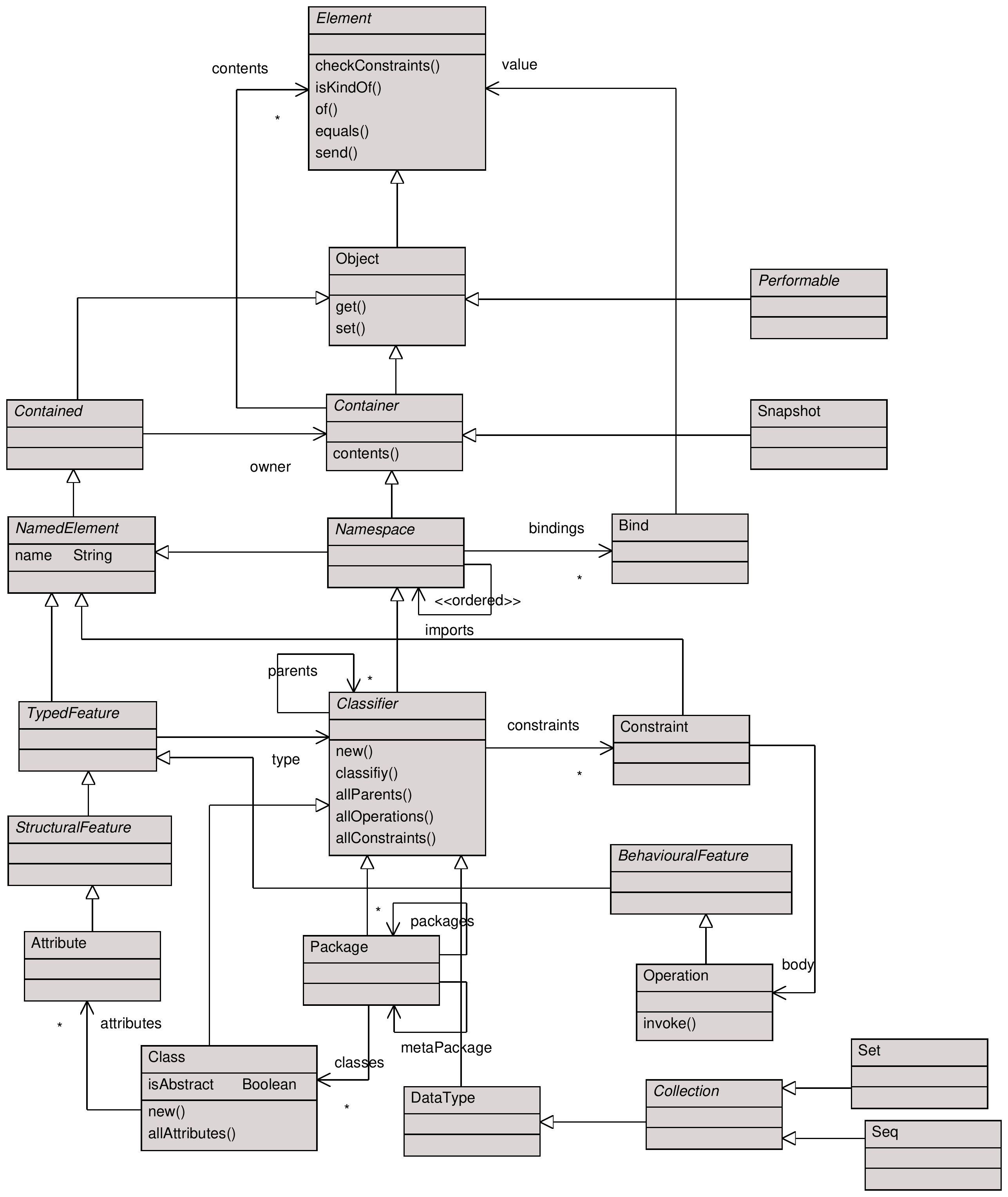}
\caption{Overview of the XCore class framework}
\label{xmfframework}
\end{center}
\end{figure}

Figure \ref{xmfframework} shows the class framework for XCore.
This framework is based on a combination of MOF and other
metamodelling frameworks such as Ecore with necessary extensions
to support executability. As such it should be viewed as an
example of a typical metamodelling language framework.

Within this framework there are a number of abstract classes that
encapsulate the generic properties of common types of language
concepts. The following table identifies some of the key ones:

\begin{description}
\item [Classifier] A classifier is a name space for operations and
constraints. A classifier is generalizable and has parents from
which it inherits operations and constraints. A classifier can be
instantiated via new(). In both cases the default behaviour is to
return a default value as an instance. If the classifier is a
datatype then the basic value for the datatype is returned
otherwise 'null' is returned as the default value. Typically you
will not create a Classifier directly, but create a class or an
instance of a sub-class of Class. \item [Container] A container
has a slot 'contents' that is a table. The table maintains the
contained elements indexed by keys. By default the keys for the
elements in the table are the elements themselves, but sub-classes
of container will modify this feature accordingly. Container
provides operations for accessing and managing its contents. \item
[DataType] DataType is a sub-class of Classifier that designates
the non-object classifiers that are basic to the XMF system. An
instance of DataType is a classifier for values (the instances of
the data type). For example Boolean is an instance of DataType -
it classifies the values 'true' and 'false'. For example Integer
is an instance of DataType - it classifies the values 1, 2, etc.
\item [NamedElement] A named element is an owned element with a
name. The name may be a string or a symbol. typically we use
symbols where the lookup of the name needs to be efficient. \item
[Namespace] A name space is a container of named elements. A name
space defines two operations getElement() and hasElement() that
are used to get an element by name and check for an element by
name. Typically a name space will contain different categories of
elements in which case the name space will place the contained
elements in its contents table and in a type specific collection.
For example, a class is a container for operations, attributes and
constraints. Each of these elements are placed in the contents
table for the class and in a slot containing a collection with the
names 'operations', 'attributes'; and 'constraints' respectively.
The special syntax '::' is used to invoke the getElement()
operation on a name space. \item [StructuralFeature] This is an
abstract class that is the super-class of all classes that
describe structural features. For example, Attribute is a
sub-class of StructuralFeature. \item [TypedElement] A typed
element is a named element with an associated type. The type is a
classifier. This is an abstract class and is used (for example) to
define Attribute.

\end{description}

\section{Conclusion}

This chapter has shown the importance of being able to reuse
language definitions rather than having to design new languages
from scratch. Two approaches were presented: specialisation and
translation. specialisation involves reusing pre-existing language
concepts via class specialisation. A framework of language
concepts is important for this purpose. The advantage of the
approach is that it enables tools to rapidly adapt their
functionality to support new concepts. Translation involves
mapping new concepts to existing concepts. It is particularly
useful when constructing layered definitions in which richer
abstractions are translated down onto more primitive abstractions.

\chapter{Case Study 1: An Aspect-Oriented Language}

\section{Introduction}

This chapter presents a metamodel for a simple aspect-oriented
language (AOL). We construct an abstract syntax model for the
language, along with an operational semantics and a concrete
syntax definition.

\section{AOL}

The AOL enables different aspects of components to be modelled
separately from the component itself, thus facilitating the
separation of concerns. The AOL is a general purpose language
because it can be applied across multiple languages provided that
they can access its metamodel. An example of its syntax is shown
below:

\begin{lstlisting}
@Aspect <name>
  ...
  @Class <path>
     <namedelement>
     ...
  end
end
\end{lstlisting}The aspect named $<$name$>$ adds one or more named elements to the
class referenced by the path $<$path$>$. In the version of AOL
presented here, classes are the components that aspects are added
to, but in practice any element can be viewed as a component.

As an example, consider the requirement to extend the class Box
with the ability to generate SVG (Scalable Vector Graphics).
Rather than merging this information into the class definition, an
aspect can be used to separate out this additional capability:

\begin{lstlisting}
@Aspect ToSVG
  @Class Box
    @Operation toSVG(parentx,parenty,out)
      if self.shown() then
        format(out,"<rect x=\"~S\" y=\"~S\" width=\"~S\" height=\"~S\"
        fill =\"#DBD5D5\" stroke=\"black\" stroke-width=\"1\"/>~\%",
        Seq{parentx+x,parenty+y,width,height});
        @For display in displays do
          display.toSVG(parentx+x,parenty+y,out)
        end
      end
    end
  end
\end{lstlisting}It is important to note that this is a much simpler mechanism than
that used by aspect-oriented programming languages such as
AspectJ. Nevertheless, it is still a useful and practical
mechanism for separating out different aspects of a model.

\section{Language Definition Strategy}

The approach taken to defining the syntax and semantics of this
language is to clearly separate syntax concepts from semantic
concepts. Parsing the concrete syntax for AOL results in an
intermediate abstract syntax definition, which is then desugared
into a model of the AOL's operational semantics.

\section{Abstract Syntax}

\subsection{Identification of Concepts}

Based on the example shown above the following candidates for
concepts can be immediately identified:

\begin{description}
\item [Aspect]  An aspect has a name and contains a collection of
components.

\item [Class] A class is a syntax concept that has a path to the
class that the named element is to be added to. A class is also a
component.

\item [NamedElement] The element that is added to the class
referenced by its path.
\end{description}

\subsection{Abstract Syntax Model}

\begin{figure}[htb]
\begin{center}
\includegraphics[width=14cm]{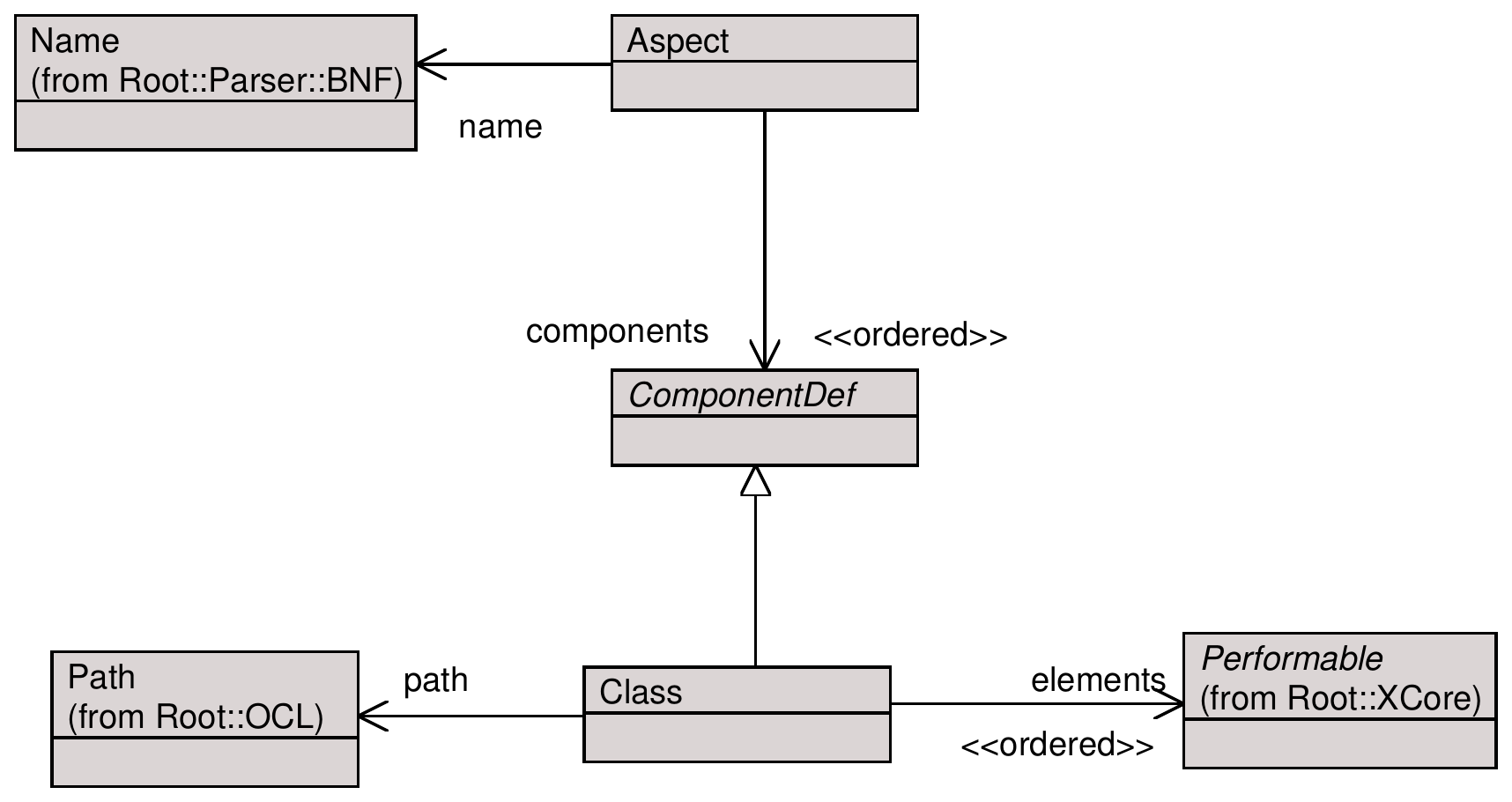}
\caption{An abstract syntax model for AOL} \label{aspectabs2}
\end{center}
\end{figure}

An extended abstract syntax model is shown in figure
\ref{aspectabs2}.

There are a number of points to note about the model:

\begin{itemize}
\item Aspects have a name and are containers of component
definitions. A component definition is the abstract superclass of
all aspect component definitions  \item Class specialises
component and has a path to the class that its elements are to be
added to. The elements of Class are of type Performable (the Root
class for all parsable elements). \item Class is not the class
XCore::Class, but is purely a syntactical concept.
\end{itemize}

\section{Semantics}

The abstract syntax model focuses purely on syntax and does not
define a semantics. This is important for the AOL because there is
a clear separation between the language's syntax and the result of
parsing the syntax, which is to add a new named element to an
existing class.

\begin{figure}[htb]
\begin{center}
\includegraphics[width=14cm]{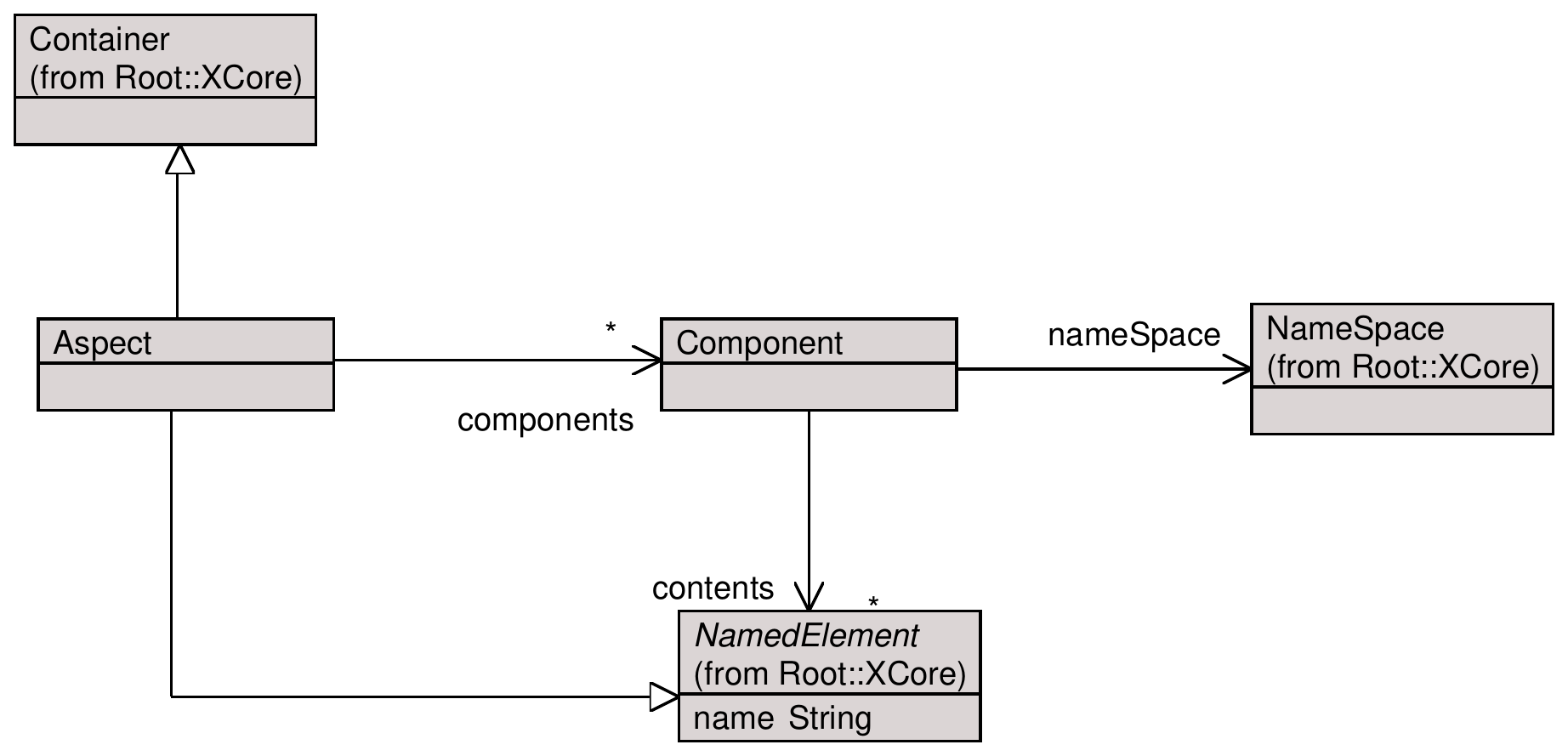}
\caption{A semantic model for the AOL} \label{aspectsem}
\end{center}
\end{figure}

To specify the semantics of the language we must therefore build a
semantic model. This is presented in figure \ref{aspectsem}.

A component has a namespace and a collection of named elements
that are to be added to the namespace.

The following operation is required in order to add elements to a
component:

\begin{lstlisting}
context Component
  @Operation add(e:NamedElement)
    self.contents := contents->including(e)
  end
\end{lstlisting}The semantics of adding a named element to the namespace
referenced by the component is defined by the following operation:

\begin{lstlisting}
context Component
  @Operation perform()
    @For e in self.contents do
      nameSpace.add(e)
    end
  end
\end{lstlisting}\section{Concrete Syntax}

A concrete syntax for the AOL is defined in XBNF as follows. An
aspect is a name followed by a sequence of component definitions.
The result of parsing an aspect is to create an instance of the
abstract syntax class Aspect.

\begin{lstlisting}
  Aspect ::= name = Name components = (ComponentDef)* { Aspect(name,components) }.
\end{lstlisting}An aspect is then desugered into an instance of the semantics
class Aspect:

\begin{lstlisting}
  context Aspect
    @Operation desugar()
      components->iterate(c e = [| Aspects::Semantics::Aspect(<StrExp(name)>) |] | [| <e>.add(<c>) |])
    end
\end{lstlisting}An abstract syntax class definition is a path followed by a
sequence of expressions, which are the result of parsing the
elements that are to be added to the class:

\begin{lstlisting}
  Class ::= path = Path elements = Exp* { Class(path,elements) }.
  Path ::= root = VarExp names = ('::' Name)* { Path(root,names) }.
\end{lstlisting}Desugaring an abstract syntax class creates an instance of the
class Semantics::Component whose namespace is the class given by
the path of the syntax class, and whose elements are the named
elements of the abstract syntax class:

\begin{lstlisting}
  context Class
    @Operation desugar()
      elements->iterate(c e = [| Aspects::Semantics::Component(<path>) |] | [| <e>.add(<c>) |])
    end
\end{lstlisting}Once the desugaring process has occurred, the perform() operation
will be run on each component to add the named elements to the
class referenced by its path.

\section{Conclusion}

This chapter has shown that the key features of a simple
aspect-oriented language can be captured within metamodels. The
strategy taken was to keep the abstract syntax and semantic models
separate and use the concrete syntax definition to firstly create
an instance of the abstract syntax model, and then desugar it into
an instance of the semantic model.

\chapter{Case Study 2: A Telecoms Language}

\section{Introduction}

This chapter presents an example of the definition of a domain
specific modelling language: a telecoms modelling language. The
approach taken is to extend the XCore metamodel and to use package
extension and metapackages to construct stereotyped concrete
syntax elements for the language. Mappings are then constructed
from the language metamodel to a metamodel of Java and a metamodel
of a user interface language. The work presented in this chapter
is based on an earlier case study described in
\cite{ossjcasestudy}.

\section{The Case Study: OSS/J Inventory}

Developing and operating Operational Support Systems (OSS) for
telecommunications companies (telcos) is a very expensive process
whose cost continuously grows year on year. With the introduction
of new products and services, telcos are constantly challenged to
reduce the overall costs and improve business agility in terms of
faster time-to-market for new services and products. It is
recognised that the major proportion of overall costs is in
integration and maintenance of OSS solutions. Currently, the OSS
infrastructure of a typical telco comprises an order of O(1000)
systems all with point-to-point interconnections and using diverse
platforms and implementation technologies. The telcoms OSS
industry has already established the basic principles for building
and operating OSS through the TMF NGOSS programme [NGOSS] and the
OSS through Java initiative \cite{ossj}. In summary, the NGOSS
applies a top-level approach through the specification of an OSS
architecture where:

\begin{itemize}
\item Technology Neutral and Technology Specific Architectures are
separated. \item The more dynamic "business process" logic is
separated from the more stable "component" logic. \item Components
present their services through well defined "contracts" \item
"Policies" are used to provide a flexible control of behaviour in
an overall NGOSS system. \item The infrastructure services such as
naming, invocation, directories, transactions, security,
persistence, etc are provided as a common deployment and runtime
framework for common use by all OSS components and business
processes over a service bus.
\end{itemize}

The case-study was based upon OSS component APIs specified in Java
and J2EE by OSS/J. The case-study was specifically driven by the
OSS/J Inventory component API \cite{Gauthier} and set as its goal
to automatically conduct compliance tests between the API
specification and the results of the case study. This end acquired
more value by the fact that this particular API specification
lacks, as of yet, a reference implementation and compatibility kit
that would permit its practical validation.

The exercise had two main objectives:

\begin{itemize}
\item Construction of a domain specific language metamodel for the
OSS/J Inventory: The OSS/J Inventory specification document
includes a UML class diagram of an inventory meta-model and some
textual, i.e. informal, description of its semantics. The
meta-model defines the types of information/content the inventory
will manage, such as products, services and resources. \item
Automatic generation of a system implementation conforming to
standard OSS/J architectural patterns and design guidelines: In
order to comply with the OSS/J guideline, the case-study aims at
implementing an application tool that allows users to manage the
inventory content through a simple GUI. Example users of such a
tool may be front-desk operators who respond to customer calls and
access the inventory to setup a new or change the state of an
existing product/service instance.
\end{itemize}

\begin{figure}[htb]
\begin{center}
\includegraphics[width=12cm]{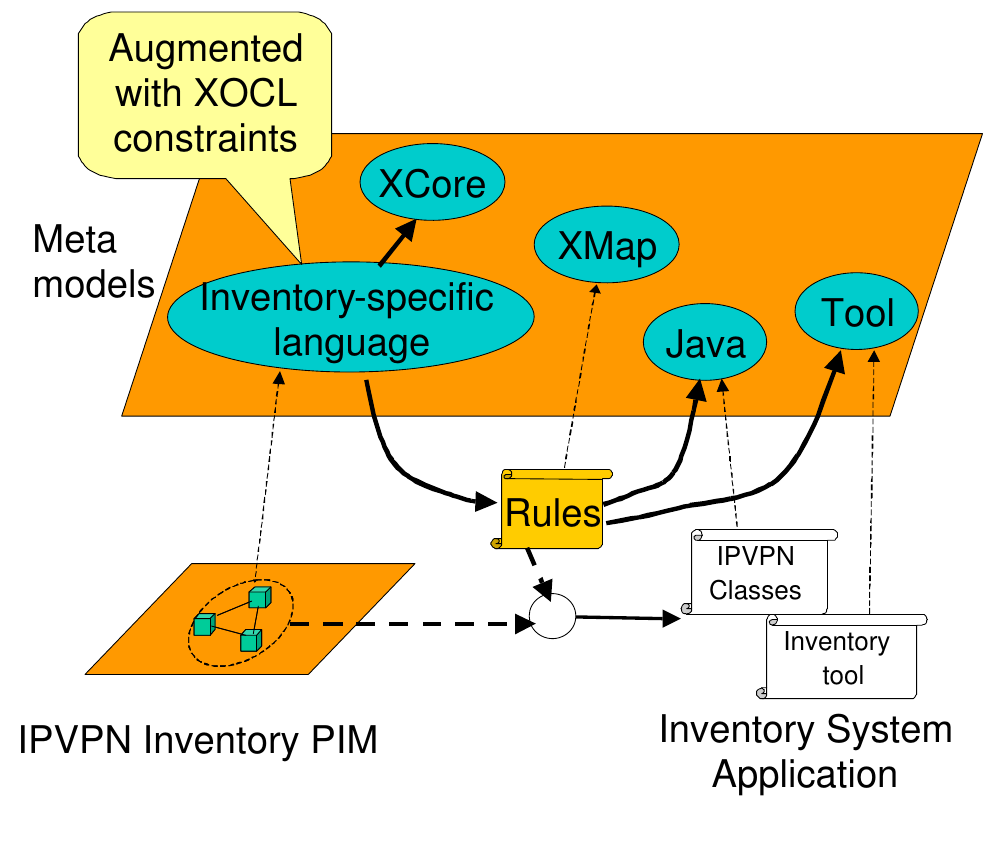}
\caption{Overview of the inventory language definition}
\label{inventoryoverview}
\end{center}
\end{figure}

Figure \ref{inventoryoverview} shows how a language definition for
the inventory modelling language was constructed. Firstly, a
metamodel for the inventory DSL was defined by extending the XCore
meta-model. XOCL was used to specify meta-model constraints so
that models written in the inventory language can be checked for
correctness. That is, by means of XOCL, the meta-model semantics
can be formally captured and automatically enforced, in contrast
to the informal, textual description of the semantics presented in
the OSS/J Inventory API specification document. Next, mapping
rules written in XMap were constructed to transform the inventory
meta-model into meta-models of two target platform specific
languages: EJB and Java.

\section{Abstract Syntax Model}

\begin{figure}[htb]
\begin{center}
\includegraphics[width=14cm]{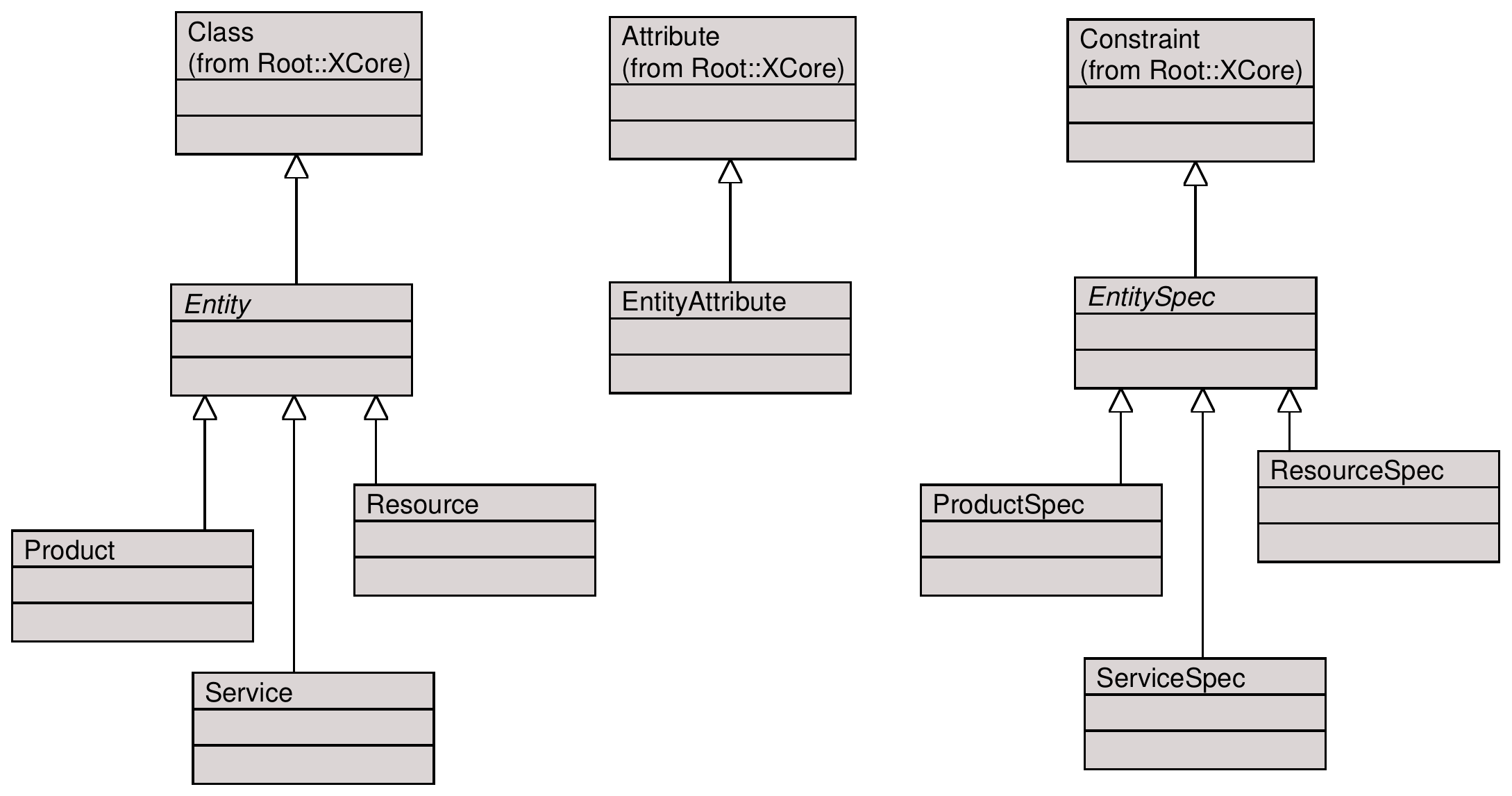}
\caption{The abstract syntax model for the inventory language}
\label{inventoryas}
\end{center}
\end{figure}

Figure \ref{inventoryas} shows the abstract syntax model for the
inventory language. As mentioned earlier, it includes concepts
from the OSS/J Core Business Entities, which are a subset of TMF's
NGOSS standard. The inventory language consists of the following
constructs:

\begin{itemize}
\item Entity, that represents any type of information included in
the inventory. According to the specification, three types of
inventory content are defined, namely, Product, Service and
Resource, which extend type Entity. \item   EntitySpecification,
that represents configurations of Entities, i.e. constraints, such
as range of values or preconfigured setting on features of the
Entity. Again, the API specification defines three subtypes of
EntitySpecification, namely, ProductSpecification,
ServiceSpecification and ResourceSpecification, each representing
specifications for Service, Product and Resource, respectively. ·
EntityAttribute, that represents relationships between Entity
types.
\end{itemize}

A number of concepts from the XCore package are specialised in
order to reuse their syntax and semantics:

\begin{itemize}
\item   Entity specialises the class XCore::Class, hence it can be
instantiated and contain attributes, operations and constraints.
\item EntitySpecification specialises XCore::Constraint. It can,
therefore, be owned by an Entity and contain an evaluate-able XOCL
expression. In the Inventory API specification document,
EntitySpecification is represented as a UML class, which has a
simple semantics, and thereby great modelling incapacity to
express in full potential the concept semantics as an Entity
configuration constraint. Therefore, by modelling
EntitySpecification as a pure constraint, rich expressive power is
conveyed to the concept enabling it to represent complex Entity
configurations. \item   EntityAttribute specialises the class
XCore::Attribute and is used to associate different Entity types.
\end{itemize}

\subsection{Well-formedness Rules}

A number of constraints (well-formedness rules) apply to the
inventory language. These are expressed in OCL. As an example, the
following OCL constraint states that if an Entity specialises
another Entity it must be of the same type as the parent entity.
That is, entity IPStream\_S of figure \ref{inventorymodel}, for
instance, can inherit from IPStream, as both are of type Service,
but cannot inherit from IPVPN that is of type Product. Here, of()
is an XOCL operation that returns the meta-class of the entity
(i.e. the class that the entity is an instance of).

\begin{lstlisting}
context Entity
  @Constraint SameParentType
    parents->select(p |
      p.isKindOf(Entities::Entity))->forAll(p |
        p.of() = self.of())
  end
\end{lstlisting}Another noteworthy constraint, formally delivering an important
semantic property of the OSS/J Inventory API specification,
involves the association of an Entity type with the correct type
of EntitySpecification. In other words, classes of type Service,
for instance, can only have specifications of type
ServiceSpecification and not of type ProductSpecification or
ResourceSpecification. The XOCL for the constraint follows:

\begin{lstlisting}
context Entity
  @Constraint CorrectSpecs
    self.constraints->forAll(c |
     let ctype = c.of()
     in @Case ctype of
        [ IML::Entities::ServiceSpec ] do
          self.isKindOf(IML::Entities::Service)
        end
        [ IML::Entities::ProductSpec ] do
          self.isKindOf(IML::Entities::Product)
        end
        [ IML::Entities::ResourceSpec ] do
          self.isKindOf(IML::Entities::Resource)
        end
      end
   end)
\end{lstlisting}\section{Concrete Syntax}

Because package extension and meta-packages (see section
\ref{metapackages}) will be used to introduce stereotyped diagram
elements for the language, there is no need to define a separate
concrete syntax model.

\section{Semantics}

Because all concepts in the inventory language specialise XCore
concepts that already have an executable semantics, and the
concepts add no further semantic properties, there is no need to
define a separate model of semantics.

\section{Instantiation}

\begin{figure}[htb]
\begin{center}
\includegraphics[width=9cm]{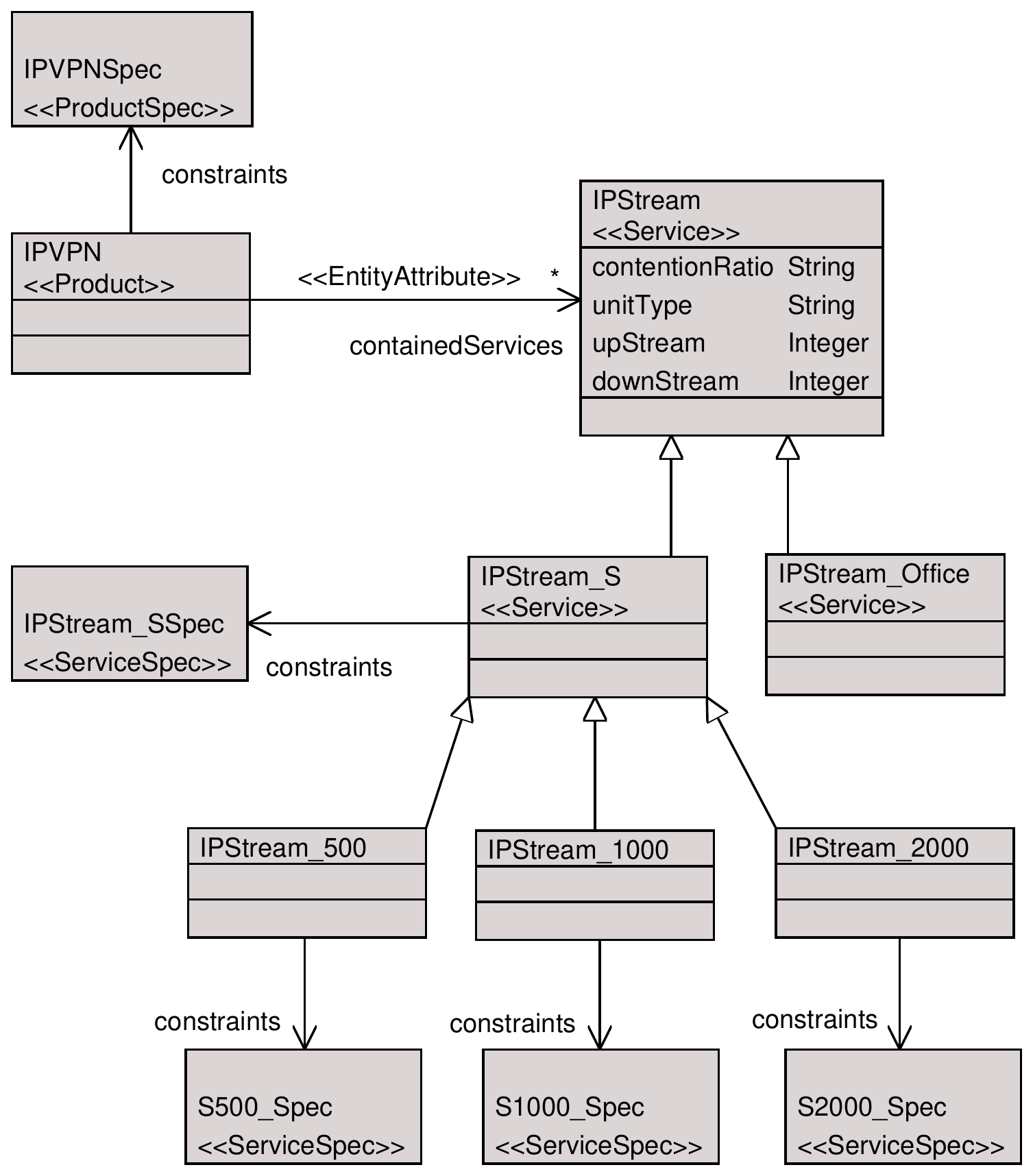}
\caption{An inventory model} \label{inventorymodel}
\end{center}
\end{figure}

In figure \ref{inventorymodel} an inventory model is presented,
which is an instance of the inventory specific metamodel (its
meta-package). It illustrates a subset of an IP Virtual Private
Network (IPVPN) product. The model shows an IPVPN containing
(containedServices attribute) many IPStream entities, an ADSL
service that comes in different offerings for home and for office
premises represented by IPStream\_S and IPStream\_Office,
respectively. IPStream\_S is further specialised by
IPStream\_S500, IPStream\_S1000 and IPStream\_S2000, entities
differentiating on the downstream bandwidth of the link that is,
respectively, 500, 1000 and 2000 kbps. Individual features of the
latter entities are defined in the accompanying ServiceSpec
constraints, namely, S500Spec, S1000Spec and S2000Spec. Similarly,
features of the IPVPN product and the IPStream\_S service are
specified in the IPVPNSpec and IPStream\_SSpec specification
constraints.

\begin{figure}[htb]
\begin{center}
\includegraphics[width=10cm]{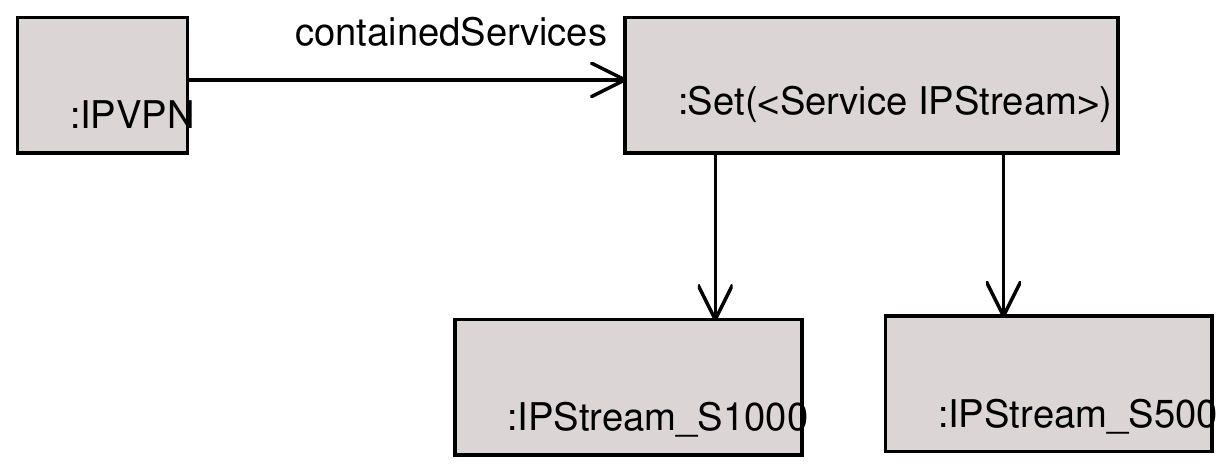}
\caption{A snapshot (instance) of the IPVPNModel}
\label{inventorysnapshot}
\end{center}
\end{figure}

Because all model entities of figure \ref{inventorymodel} are
instances of inventory meta-classes that specialise Entity, which,
in turn, extends class XCore::Class, they inherit the ability to
have constraints, attributes and operations (and their associated
specialisations, namely, Specifications and EntityAttribute). As
an example, the IPStream\_S2000 is associated with S2000Spec,
which has the following XOCL body:

\begin{lstlisting}
self.upStream = 250 and
self.downStream = 2000 and
self.unitType ="kbps"
\end{lstlisting}In addition, XOCL can be used to write operations on the inventory
model. XOCL extends OCL with a small number of action primitives,
thus turning it into a programming language at the modelling
level. As an example, the following operation creates an instance
of an IPStream and adds it as a containedServices attribute to an
IPVPN:

\begin{lstlisting}
context IPVPN
  @Operation addIPStream(up,dwn,unit,con)
    self.containedServices :=
      self.containedService->including(IPStream(up,dwn,unit,con))
  end
\end{lstlisting}Finally, because the entities in the model are themselves
instantiable, it is possible to create an instance of the
IPStreamModel and check that the instance satisfies the
constraints that are defined in the inventory model (see figure
\ref{inventorysnapshot}). This is a further level of instantiation
that is possible because of the metaPackage relationship between
the inventory model and the inventory language meta-model.
Furthermore, the operations on the model can be executed, allowing
all aspects of the model to be validated.

\section{Transformations}

Using XMap, two mappings have been defined from the inventory
language. The first generates EJBs, while the second focuses on
the generation of Java and a Java class tool. We concentrate on
the second one here.

\begin{figure}[htb]
\begin{center}
\includegraphics[width=15cm]{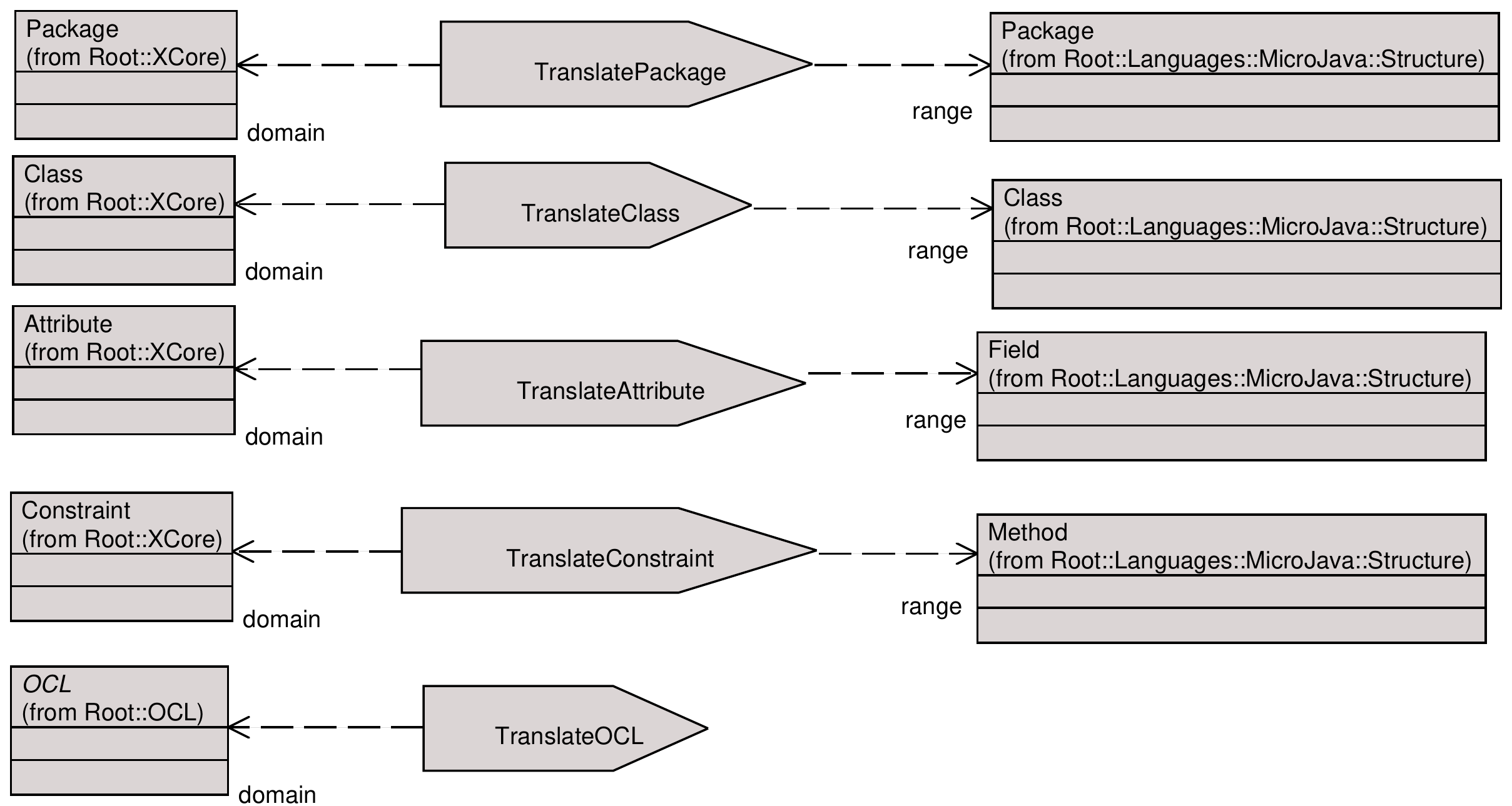}
\caption{Overview of the XCore to Java mapping}
\label{inventorymapping}
\end{center}
\end{figure}

The model of figure \ref{inventorymapping} shows the mappings used
to generate Java. Rather than mapping directly from the inventory
language meta-model, a more generic approach is taken in which the
mapping was defined from XCore classes. Because the inventory
language extends the XCore meta-model, they therefore also apply
to inventory models (and any other language specialisations
defined in the future).

Every element in the XCore package has a mapping to a
corresponding element in the Java meta-model. The following clause
describes a mapping from an XCore class to a Java class:

\begin{lstlisting}
context TranslateClass
    @Clause MapClass
      XCore::Class[name = name,
                   parents = P,
                   operations = O,
                   constraints = C,
                   attributes = A] do
      classToMicroJava(name,P,O,C,A)
    end
\end{lstlisting}Here, a Class is mapped to a Java Class, which is the result of
passing the class's name, parents, operations, constraints and
attributes into the operation classToMicroJava(). This operation
is shown below:

\begin{lstlisting}
context XCore::Class
  @Operation classToMicroJava(name,P,O,C,A)
    let K = constraintsToMicroJava(C);
        M = O->asSeq->collect(o | XCoretoMicroJava(o));
        F = A->asSeq->collect(a | XCoretoMicroJava(a))
    in if P = Set{Object}
       then [| @Java class <name> { <* F + K + M *> } end |]
       else
         let parent = P->sel.name.toString()
         in [| @Java class <name> extends <Seq{parent}> { <* F + K + M *> } end |]
         end
       end
    end
\end{lstlisting}Briefly, the operation makes use of quasi quotes (see chapter
\ref{concretechapter}) to 'drop' the name, parent and (once they
have been translated) the attributes, operations and constraints
into a syntactical definition of a Java class. This is possible
because an XBNF grammar has been defined for the MicroJava
language. The result will be an instance of the MicroJava language
metamodel, which can then outputted as a program in textual form.

An important point to make about the mapping is that it translates
all elements of an XCore model (and by specialisation and
inventory model) into Java. This includes the bodies of operations
and constraints, which are translated into Java operation. The
resulting Java program can be run and checked against the
behaviour of the original model running on the VM.

\section{Tool Mapping}

\begin{figure}[htb]
\begin{center}
\includegraphics[width=16cm]{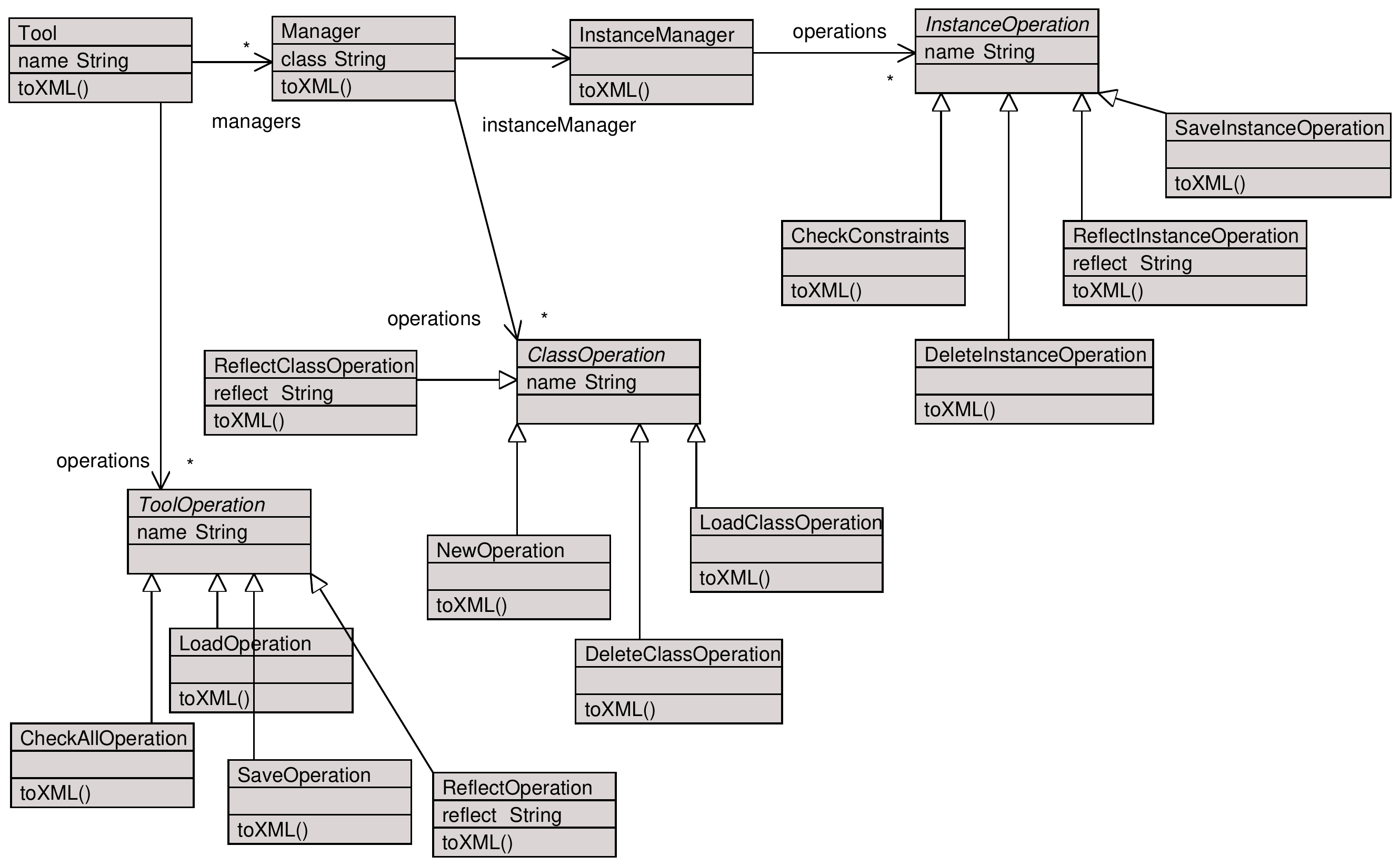}
\caption{The class tool metamodel} \label{toolmodel}
\end{center}
\end{figure}

While the above mapping generates a standalone Java program
corresponding to an inventory model, it would more useful to users
of the language if the model it represents could be interacted
with via a user interface. To achieve this, a mapping was
constructed from XCore to a meta-model of a tool interface for
managing object models. This represents a domain specific language
for tools. The meta-model of the class tool interface is shown in
figure \ref{toolmodel}. A class tool provides an interface that
supports a standard collection of operations on objects, such as
saving and loading objects and checking constraints on objects. In
addition, a class tool defines a number of managers on classes,
which enable instances of classes to be created and then checked
against their class's constraints or their operations run.

\begin{figure}[htb]
\begin{center}
\includegraphics[width=14cm]{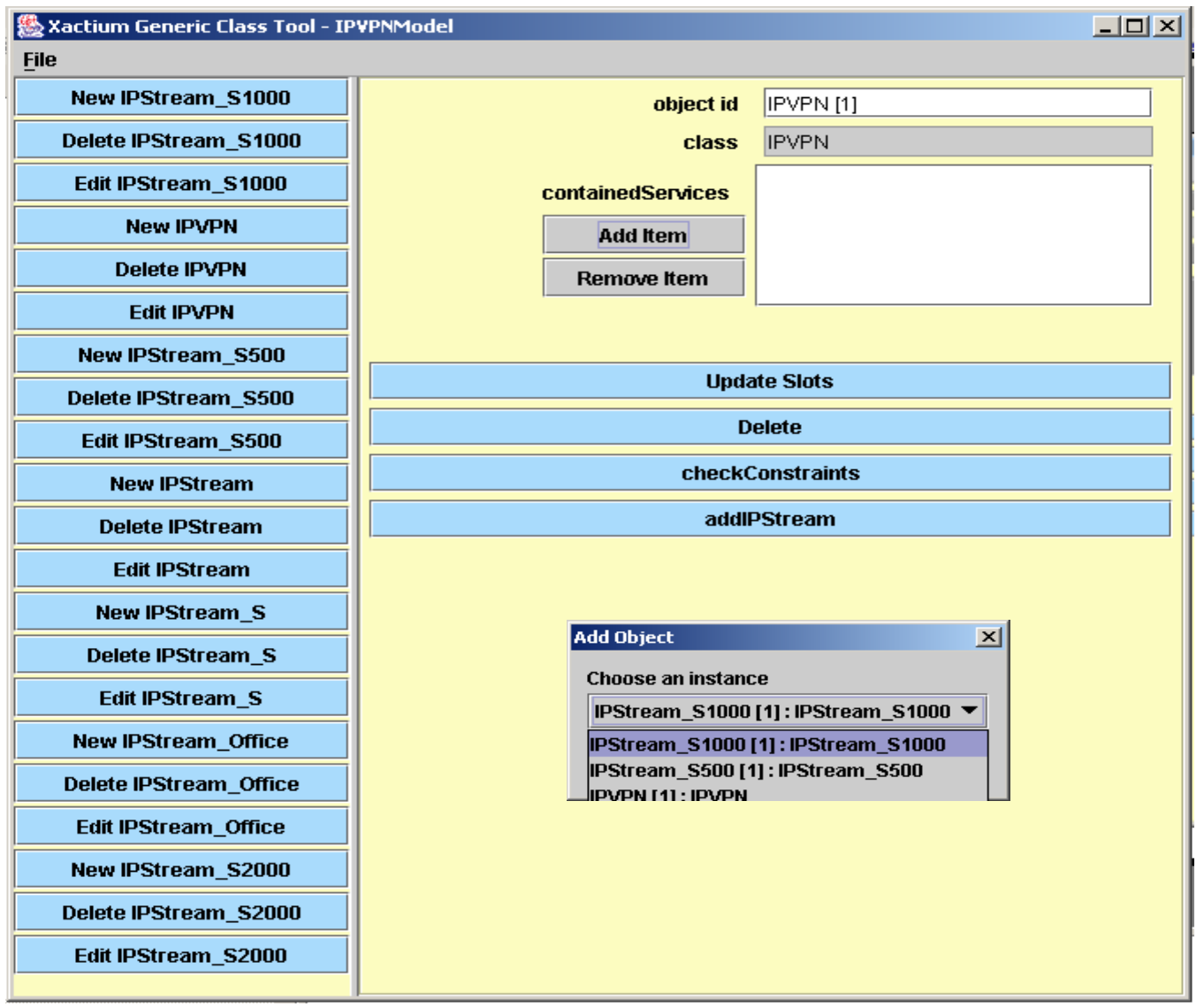}
\caption{The generated class tool} \label{classtoolexample}
\end{center}
\end{figure}

A mapping can be defined to the class tool meta-model (not shown
here), which generates a tailored user interface for creating and
manipulating instances of a meta-modelling language such as the
inventory language. Applying this mapping to the IPVPN model shown
in figure \ref{toolmodel} results in the generation of the class
tool in figure \ref{classtoolexample}. Here, buttons have been
generated for each of the entities in the model. These allow the
user to create new instances, edit their slot values and delete
instances. As the figure shows, a button for invoking the
addIPStream() method defined earlier has also been added in the
GUI executing functionality that implements in Java the method's
behaviour described in the model with XOCL.

\section{Conclusion}

This chapter has shown how a relatively light weight approach to
extending a metamodel can be used to define a domain specific
modelling language. Metapackages were used to ensure consistancy
of models against its metmodel. Because of the completeness of the
new language, it was then possible to generate a complete
deployment of domain specific language in Java.

\chapter{Case Study 3: A Simple Programming Language}

\section{Introduction}
\label{Intro} This chapter describes how XMF can be used to define
the syntax and semantics of a simple action based programming
language called XAction. It begins by presenting the XAction
language and then define a concrete grammar for it. Next a variety
of approaches, including operational and translational approaches
are used to define an executable semantics for the language. Note,
this chapter provides an in depth technical treatment of
semantics.

\section{XAction}

{\em XAction} is a simple action language with values that are
either records or are atomic. An atomic data value is a string,
integer or boolean. A record is a collection of named values.
XAction is block-structured where blocks contain type definitions
and value definitions. XAction has simple control structures:
conditional statements and loops. The following is a simple
example XAction program that builds a list of even numbers from 2
to 100:

\small
\begin{verbatim}
begin
  type Pair is head tail end
  type Nil is end
  value length is 100 end
  value list is new Nil end
  while length > 0 do
    begin
      if length % 2 = 0
      then
        begin
          value pair is new Pair end
          pair.head := length;
          pair.tail := list;
          list := pair;
        end
      end;
      length := length - 1;
    end
end
\end{verbatim}
\normalsize

The definition of XAction is structured as a collection of XMF
packages. The {\tt Values} package defines the semantic domain for
XAction; it contains classes for each type of program value.
Executable program phrases in XAction are divided into two
categories: {\tt Expressions} and {\tt Statements}. Expressions
evaluate to produce XAction values. Statements are used to control
the flow of execution and to update values.
\begin{lstlisting}
@Package XAction
  @Package Values end
  @Package Expressions end
  @Package Statements end
end
\end{lstlisting}
The rest of this section defines the syntax of XAction by giving
the basic class definitions and the XBNF grammar rules for the
language constructs.

\subsection{XAction Values}

XAction expressions evaluate to produce XAction values. Values are
defined in the {\tt Values} package and which is the {\em semantic
domain} for XAction. Values are either atomic: integers and
booleans, or are records. We use a simple representation for
records: a sequence of values indexed by names.

XAction records are created by instantiating XAction record types.
A record type is a sequence of names. Types raise an interesting
design issue: should the types be included as part of the semantic
domain since evaluation of certain XAction program phrases give
rise to types that are used later in the execution to produce
records. The answer to the question involves the phase distinction
that occurs between {\em static} analysis (or execution) and {\em
dynamic} execution. Types are often viewed as occurring only
during static analysis; although this is not always the case. We
will show how the semantics of XAction can be defined with and
without dynamic types.

All values are instances of sub-classes of the class {\tt Value}:
\begin{lstlisting}
context Values
 @Class Value
   @Attribute value : Element end
   @Constructor(value) ! end
 end
\end{lstlisting}
Atomic values are either booleans or integers. Each class defines
operations that the semantic domain provides for manipulating
XAction values. The classes below show the structure and a
representative sample of operations:
\begin{lstlisting}
context Values
  @Class Bool extends Value
    @Operation binAnd(Bool(b))
      Bool(value and b)
    end
    @Operation binOr(Bool(b))
      Bool(value or b)
    end
  end
\end{lstlisting}\begin{lstlisting}
context Values
  @Class Int extends Value
    @Operation binAdd(Int(n))
      Int(value + n)
    end
  end
\end{lstlisting}
Record types are sequences of names. A type provides a {\tt new}
operation that instantiates the type to produce a new record. This
operation is only meaningful if we have dynamic types:
\begin{lstlisting}
context Values
  @Class Type extends Value
    @Attribute names : Seq(String) end
    @Constructor(names) ! end
    @Operation new()
      Record(self,names->collect(n | Seq{n | null}))
    end
  end
\end{lstlisting}
Records are sequences of values indexed by names; the names are
found by navigating to the type of the record:
\begin{lstlisting}
context Values
  @Class Record extends Value
    @Attribute type : Type end
    @Attribute fields : Seq(Element) end
    @Constructor(type,fields) ! end
    @Operation lookup(name:String)
      fields->at(type.names->indexOf(name))
    end
    @Operation update(name:String,value:Element)
      fields->setAt(type.names->indexOf(name),value)
    end
  end
\end{lstlisting}\subsection{XAction Expressions}

XAction expressions are program phrases that evaluate to produce
XAction values. The following classes define the expression types:
\begin{lstlisting}
context Expressions
  @Class Exp
  end
end
\end{lstlisting}
A binary expression has a left and right sub-expression and an
operation. The name of the operation is represented as a string:
\begin{lstlisting}
context Expressions
  @Class BinExp extends Exp
    @Attribute op : String end
    @Attribute left : Exp end
    @Attribute right : Exp end
    @Constructor(op,left,right) ! end
  end
\end{lstlisting}
An atomic constant expression is either an integer or a boolean:
\begin{lstlisting}
context Expressions
  @Class Const extends Exp
    @Attribute value : Element end
    @Constructor(value) ! end
  end
end
\end{lstlisting}
A new record is produced by performing a {\tt new} expression. The
type to instantiate is given as a string. An alternative
representation for types in {\tt new} expressions would be to
permit an arbitrary expression that {\em evaluates} to produce a
type. This design choice would rule out static typing and force
the language to have dynamic types. We wish to use XAction to
illustrate the difference between dynamic and static types in
semantic definitions so we use strings to name types in {\tt new}
expressions:
\begin{lstlisting}
context Expressions
  @Class New extends Exp
    @Attribute type : String end
    @Constructor(type) ! end
  end
end
\end{lstlisting}
A variable is just a name:
\begin{lstlisting}
context Expressions
  @Class Var extends Exp
    @Attribute name : String end
    @Constructor(name) ! end
  end
\end{lstlisting}
A record field ref is:
\begin{lstlisting}
context Expressions
  @Class FieldRef extends Exp
    @Attribute value : Exp end
    @Attribute name : String end
    @Constructor(value,name) ! end
  end
\end{lstlisting}
The concrete syntax of expressions is defined by the XBNF grammar
for the class {\tt Exp}. The grammar parses the expression syntax
and synthesizes instances of the expression classes:
\begin{lstlisting}
context Exp
 @Grammar
   // Start at Exp. Logical operators bind weakest.
   Exp ::= e = ArithExp [ op = LogicalOp l = Exp { BinExp(op,e,l) } ].
   LogicalOp ::= 'and' { "and" } | 'or' { "or" }.
   // The '.' for field ref binds tighter than '+' etc.
   ArithExp ::= e = FieldRef [ op = ArithOp a = FieldRef { BinExp(op,e,a) } ].
   ArithOp ::= '+' { "+" }.
   // A field reference '.' optionally follows an atomic expression.
   FieldRef ::= e = Atom ('.' n = Name { FieldRef(e,n) } | { e }).
   // Atomic expressions can be arbitrary exps if in ( and ).
   Atom ::= Const | Var | New | '(' Exp ')'.
   Const ::= IntConst | BoolConst.
   IntConst ::= i = Int { Const(i) }.
   BoolConst ::= 'true' { Const(true) } | 'false' { Const(false) }.
   Var ::= n = Name { Var(n) }.
   New ::= 'new' n = Name { New(n) }.
 end
\end{lstlisting}\subsection{XAction Statements}

XAction statements are used to:
\begin{itemize}
\item Introduce new names associated with either types or values.
\item Control the flow of execution. \item Perform side effects on
records.
\end{itemize}
The following classes define the statement types for XAction:
\begin{lstlisting}
context Statements
  @Class Statement
  end
end
\end{lstlisting}
A block (as in Pascal or C) contains local definitions. Names
introduced in a block are available for the rest of the statements
in the block (including sub-blocks) but are not available when
control exits from the block:
\begin{lstlisting}
context Statements
  @Class Block extends Statement
    @Attribute statements : Seq(Statement) end
    @Constructor(statements) ! end
  end
end
\end{lstlisting}
A declaration introduces either a type or a value binding:
\begin{lstlisting}
context Statements
  @Class Declaration isabstract extends Statement
    @Attribute name : String end
  end
end
\end{lstlisting}
A type declaration associates a type name with a sequence of field
names. To keep things simple we don't associate fields with types:
\begin{lstlisting}
context Statements
  @Class TypeDeclaration extends Declaration
    @Attribute names : Seq(String) end
    @Constructor(name,names) ! end
  end
end
\end{lstlisting}
A value declaration associates a name with a new value. The value
is produced by performing an expression at run-time:
\begin{lstlisting}
context Statements
  @Class ValueDeclaration extends Declaration
    @Attribute value : Exp end
    @Constructor(name,value) ! end
  end
end
\end{lstlisting}
A while statement involves a test and a body:
\begin{lstlisting}
context Statements
  @Class While extends Declaration
    @Attribute test : Exp end
    @Attribute body : Statement end
    @Constructor(test,body) ! end
  end
end
\end{lstlisting}
An if statement involves a test, a then-part and an else-part:
\begin{lstlisting}
context Statements
  @Class If extends Declaration
    @Attribute test : Exp end
    @Attribute thenPart : Statement end
    @Attribute elsePart : Statement end
    @Constructor(test,elsePart) ! end
  end
end
\end{lstlisting}\begin{lstlisting}
context Statements
  @Class FieldUpdate extends Declaration
    @Attribute record : Exp end
    @Attribute name : Exp end
    @Attribute value : Exp end
    @Constructor(record,name,value) ! end
  end
end
\end{lstlisting}\begin{lstlisting}
context Statements
  @Class Update extends Declaration
    @Attribute name : String end
    @Attribute value : Exp end
    @Constructor(name,value) ! end
  end
end
\end{lstlisting}\begin{lstlisting}
context Statement
  @Grammar extends Exp.grammar
    Statement ::= Block | Declaration | While | If | Update | FieldUpdate.
    Block ::= 'begin' s = Statement* 'end' { Block(s) }.
    Declaration ::= TypeDeclaration | ValueDeclaration.
    TypeDeclaration ::= 'type' n = Name 'is' ns = Name* 'end' {
      TypeDeclaration(n,ns) }.
    ValueDeclaration ::= 'value' n = Name 'is' e = Exp 'end' {
      ValueDeclaration(n,e) }.
    FieldUpdate ::= e = Exp '.' n = Name ':=' v = Exp ';' {
      FieldUpdate(e,n,v) }.
    While ::= 'while' e = Exp 'do' s = Statement 'end' {
      While(e,s) }.
    If ::= 'if' e = Exp 'then' s1 = Statement 'else' s2 = Statement 'end' {
      If(e,s1,s2) }.
    Update ::= n = Name ':=' e = Exp ';' {
      Update(n,e) }.
  end
\end{lstlisting}\section{An Evaluator for XAction}

As described in the introducion we are interested in defining
XAction operational semantics. We will do this in a number of
different ways in the rest of this note. The first, and possibly
most straightforward, approach is to define an {\em interpreter}
for XAction in the XOCL language. This involves writing an {\tt
eval} operation for each of the XAction syntax classes. The {\tt
eval} operation must be parameterized with respect to any context
information that is required to perform the evaluation. An XAction
program {\tt p} is then evaluated in a context {\tt e} by: {\tt
p.eval(e)}.

\subsection{Evaluating Expressions}
Expression evaluation is defined by adding {\tt eval} operations
to each class in {\tt Expressions} as follows:
\begin{lstlisting}
context Exp
  @AbstractOp eval(env:Env):Value
  end
\end{lstlisting}
Evaluation of a constant produces the appropriate semantic domain
value:
\begin{lstlisting}
context Const
  @Operation eval(env)
    @TypeCase(value)
      Boolean do Bool(value) end
      Integer do Int(value) end
    end
  end
\end{lstlisting}
Evaluation of a variable involves looking up the current value.
The value is found in the current context of evaluation: this must
contain associations between variable names and their values. This
is the only thing required of the XAction evaluation context and
therefore we represent the context as an {\em environment} of
variable bindings:
\begin{lstlisting}
context Var
  @Operation eval(env)
    env.lookup(name)
  end
\end{lstlisting}
Evaluation of a binary expression involves evaluation of the
sub-expressions and then selecting an operation based on the
operation name. The following shows how XAction semantics is
completely based on XOCl semantics since {\tt +} in XAction is
performed by {\tt +} in XOCL.
\begin{lstlisting}
context BinExp
  @Operation eval(env)
    @Case op of
      "and" do left.eval(env).binAnd(right.eval(env)) end
      "or"  do left.eval(env).binOr(right.eval(env)) end
      "+"   do left.eval(env).binAdd(right.eval(env)) end
    end
  end
\end{lstlisting}
Creation of new records is performed by evaluaing a {\tt new}
expression. The interpreter has dynamic types so the type to
instantiate is found by looking up the type name in the current
environment:
\begin{lstlisting}
context New
  @Operation eval(env)
    env.lookup(type).new()
  end
\end{lstlisting}
Field reference is defined as follows:
\begin{lstlisting}
context FieldRef
  @Operation eval(env)
    value.eval(env).lookup(name)
  end
\end{lstlisting}\subsection{Evaluating Statements}

XAction statements are performed in order to introduce new names,
control flow or to update a record field. Statements are defined
to evaluate in a context and must observe the rules of scope that
require variables are local to the block that introduces them. The
context of execution is an environment; evaluation of a statement
may update the supplied environment, so statement evaluation
returns an environment:
\begin{lstlisting}
context Statement
  @AbstractOp eval(env):Env
  end
\end{lstlisting}
A value declaration evaluates the expression part and then extends
the supplied environment with a new binding:
\begin{lstlisting}
context ValueDeclaration
  @Operation eval(env)
    env.bind(name,value.eval(env))
  end
\end{lstlisting}
A type declaration extends the supplied environment with a new
type:
\begin{lstlisting}
context TypeDeclaration
  @Operation eval(env)
    env.bind(name,Type(names))
  end
\end{lstlisting}
A block must preserve the supplied environment when its evaluation
is complete. Each statement in the block is performed in turn and
may update the current environment:
\begin{lstlisting}
context Block
  @Operation eval(originalEnv)
    let env = originalEnv
    in @For statement in statements do
         env := statement.eval(env)
       end
    end;
    originalEnv
    end
\end{lstlisting}
A {\tt while} statement continually performs the body while the
test expression returns {\tt true}. A while body is equivalent to
a block; so any updates to the supplied environment that are
performed by the while body are discarded on exit:
\begin{lstlisting}
context While
  @Operation eval(originalEnv)
    let env = orginalEnv
    in @While test.eval(env).value do
          env := body.eval(env)
       end;
       originalEnv
    end
  end
\end{lstlisting}
An {\tt if} statement conditionally performs one of its
sub-statements:
\begin{lstlisting}
context If
  @Operation eval(env)
    if test.eval(env).value
    then thenPart.eval(env)
    else elsePart.eval(env)
    end
  end
\end{lstlisting}\begin{lstlisting}
context FieldUpdate
  @Operation eval(env)
    record.eval(env).update(name,value.eval(env))
  end
\end{lstlisting}\begin{lstlisting}
context Update
  @Operation eval(env)
    env.update(name,value.eval(env))
  end
\end{lstlisting}\section{A Translator for XAction with Run-Time Types}

The previous section defines an interpreter for XAction. This is
an appealing way to define the operational semantics of a language
because the rules of evaluation work directly on the abstract
syntax structures. However the resulting interpreter can often be
very inefficient. Furthermore, an interpreter can lead to an {\em
evaluation phase distinction}. Suppose that XAction is to be
embedded in XOCL. XOCL has its own interpretive mechanism (the XMF
VM); at the boundary between XOCL and XAction the XOCL
interpretive mechanism must hand over to the XAction interpreter
-- the XAction code that is performed is a data structure, a
completely alien format to the VM. This phase distinction can lead
to problems when using standard tools, such as save and load
mechanisms, with respect to the new language. For example a
mechanism that can save XOCL code to disk cannot be used to save
XAction code to disk (it can, however, be used to save the XAction
interpreter to disk).

An alternative strategy is to translate the source code of XAction
to a language for which we have an efficient implementation.  No
new interpretive mechanism is required and no phase distinction
arises. Translation provides the opportunity for static analysis
(since translation is performed prior to executing the program).
As we mentioned earlier, static analysis can translate out any
type information from XAction programs; the resulting program does
not require run-time types. Since static analysis requires a
little more work, this section describes a simple translation from
XAction to XOCL that results in run-time types; the subsequent
section shows how this can be extended to analyse types statically
and remove them from the semantic domain.

\subsection{Translating Expressions}

Translation is defined by adding a new operation {\tt desugar1} to
each sbatract syntax class. There is no static analysis, so the
operation does not require any arguments. The result of the
operation is a value of type {\tt Performable} which is the type
of elements that can be executed by the XMF execution engine.
\begin{lstlisting}
context Exp
  @AbstractOp desugar1():Performable
  end
\end{lstlisting}
An XAction constant is translated to an XOCL constant:
\begin{lstlisting}
context Const
  @Operation desugar1():Performable
    @TypeCase(value)
      Boolean do BoolExp(value) end
      Integer do IntExp(value) end
    end
  end
\end{lstlisting}
An XAction binary expression is translated to an XOCL binary
expression. Note that the sub-expressions are also translated:
\begin{lstlisting}
context BinExp
  @Operation desugar1():Performable
    @Case op of
      "and" do [| <left.desugar1()> and <right.desugar1()> |] end
      "or"  do [| <left.desugar1()> and <right.desugar1()> |] end
      "+"   do [| <left.desugar1()> + <right.desugar1()> |] end
    end
  end
\end{lstlisting}
An XAction {\tt new} expression involves a type name. Types will
be bound to the appropriate variable name in the resulting XOCL
program; so the result of translation is just a message {\tt new}
sent to the value of the variable whose name is the type name:
\begin{lstlisting}
context New
  @Operation desugar1():Performable
    [| <OCL::Var(type)>.new() |]
  end
\end{lstlisting}
XAction variables are translated to XOCL variables:
\begin{lstlisting}
context Var
  @Operation desugar1():Performable
    OCL::Var(name)
  end
\end{lstlisting}
XAction field references are translated to the appropriate call on
a record:
\begin{lstlisting}
context FieldRef
  @Operation desugar1():Performable
    [| <value.desugar1()>.ref(<StrExp(name)>) |]
  end
\end{lstlisting}\subsection{Translating Statements}

An XAction statement can involve local blocks. The equivalent XOCL
expression that provides local definitions is {\tt let}. A {\tt
let} expression consists of a name, a value expression and a body
expression. Thus, in order to translate an XAction declaration to
an XOCL {\tt let} we need to be passed the body of the {\tt let}.
This leads to a translational style for XAction actions called
{\em continuation passing} where each {\tt desugar1} operation is
supplied with the XOCL action that will be performed next:
\begin{lstlisting}
context Statement
  @AbstractOp desugar1(next:Performable):Performable
  end
\end{lstlisting}
A type declaration is translated to a local definition for the
type name. Note that the expression {\tt names.lift()} translates
the sequence of names to an expression that, when performed,
produces the same sequence of names: {\tt list} is a means of
performing evaluation in reverse:
\begin{lstlisting}
context TypeDeclaration
  @Operation desugar1(next:Performable):Performable
    [| let <name> = Type(<names.lift()>)
       in <next>
       end
    |]
  end
\end{lstlisting}
A value declaration is translated to a local decinition:
\begin{lstlisting}
context ValueDeclaration
  @Operation desugar1(next:Performable):Performable
    [| let <name> = <value.desugar1()>
       in <next>
       end
    |]
  end
\end{lstlisting}
A block requires each sub-statement to be translated in turn.
Continuation passing allows us to chain together the sequence of
statements and nest the local definitions appropriately. The
following auxiliary operation is used to implement
block-translation:
\begin{lstlisting}
context Statements
  @Operation desugar1(statements,next:Performable):Performable
    @Case statements of
      Seq{} do
        next
      end
      Seq{statement | statements} do
        statement.desugar1(Statements::desugar1(statements,next))
      end
    end
  end
\end{lstlisting}
Translation of a block requires that the XOCL local definitions
are kept local. Therefore, the sub-statements are translated by
chaining them together and with a final continuation of {\tt
null}. Placing the result in sequence with {\tt next} ensures that
any definitions are local to the block.
\begin{lstlisting}
context Block
  @Operation desugar1(next:Performable):Performable
    [| <Statements::desugar1(statements,[| null |])> ;
       <next>
    |]
  end
\end{lstlisting}
A {\tt while} statement is translated to the equivalent expression
in XOCL:
\begin{lstlisting}
context While
  @Operation desugar1(next:Performable):Performable
    [| @While <test.desugar1()>.value do
         <body.desugar1([|null|])>
       end;
       <next>
    |]
  end
\end{lstlisting}
An {\tt if} statement is translated to an equivalent expression in
XOCL:
\begin{lstlisting}
context If
  @Operation desugar1(next:Performable):Performable
    [| if <test.desugar1()>.value
       then <thenPart.desugar1(next)>
       else <elsePart.desugar1(next)>
       end
    |]
  end
\end{lstlisting}\begin{lstlisting}
context FieldUpdate
  @Operation desugar1(next:Performable):Performable
    [| <record.desugar1()>.update(<StrExp(name)>,<value.desugar1()>);
       <next>
    |]
  end
\end{lstlisting}\begin{lstlisting}
context Update
  @Operation desugar1(next:Performable):Performable
    [| <name> := <value.desugar1()>;
       <next>
    |]
  end
\end{lstlisting}\section{A Translator for XAction without Run-Time Types}

It is usual for languages to have a static (or {\em compile time})
phase and a dynamic (or {\em run time}) phase. Many operational
features of the language can be performed statically. This
includes type analysis: checking that types are defined before
they are used and allocating appropriate structures when instances
of types are created. This section shows how the translator for
XAction to XOCL from the previous section can be modified so that
type analysis is performed and so that types do not occur at
run-time.

\subsection{Translating Expressions}

Since types will no longer occur at run-time we will simplify the
semantic domain slightly and represent records as {\em a-lists}.
An a-list is a sequence of pairs, the first element of each pair
is a ket and the second element is a value. In this case a record
is an a-list where the keys are field name strings. XOCL provides
operations defined on sequences that are to be used as a-lists:
{\tt l->lookup(key)} and {\tt l->set(key,value)}.

The context for static analysis is a type environment. Types now
occur at translation time instead of run-time therefore that
portion of the run-time context that would contain associations
between type names and types occurs during translation:
\begin{lstlisting}
context Exp
  @AbstractOp desugar2(typeEnv:Env):Performable
  end
\end{lstlisting}
Translation of a constant is as for {\tt desugar1}:
\begin{lstlisting}
context Const
  @Operation desugar2(typeEnv:Env):Performable
    self.desugar1()
  end
\end{lstlisting}
Translation of binary expressions is as for {\tt desugar1} except
that all translation is performed by {\tt desugar2}:
\begin{lstlisting}
context BinExp
  @Operation desugar2(typeEnv:Env):Performable
    @Case op of
      "and" do [| <left.desugar2(typeEnv)> and
                  <right.desugar2(typeEnv)> |] end
      "or"  do [| <left.desugar2(typeEnv)> and
                  <right.desugar2(typeEnv)> |] end
      "+"   do [| <left.desugar2(typeEnv)> +
                  <right.desugar2(typeEnv)> |] end
    end
  end
\end{lstlisting}
Translation of a variable is as before:
\begin{lstlisting}
context Var
  @Operation desugar2(typeEnv:Env):Performable
    self.desugar1()
  end
\end{lstlisting}
A {\tt new} expression involves a reference to a type name. The
types occur at translation time and therefore part of the
evaluation of {\tt new} can occur during translation. The type
should occur in the supplied type environment; the type contains
the sequence of field names. The result of translation is an XOCL
expression that constructs an a-list based on the names of the
fields in the type. The initial value for each field is {\tt
null}:
\begin{lstlisting}
context New
  @Operation desugar2(typeEnv:Env):Performable
    if typeEnv.binds(type)
    then
      let type = typeEnv.lookup(type)
      in type.names->iterate(name exp = [| Seq{} |] |
           [| <exp>->bind(<StrExp(name)>,null) |])
      end
    else self.error("Unknown type " + type)
    end
  end
\end{lstlisting}
A field reference expression is translated to an a-list lookup
expression:
\begin{lstlisting}
context FieldRef
  @Operation desugar2(typeEnv:Env):Performable
    [| <value.desugar2(typeEnv)>->lookup(<StrExp(name)>) |]
  end
\end{lstlisting}\subsection{Translating Statements}

A statement may contain a local type definition. We have already
discussed continuation passing with respect to {\tt desugar1}
where the context for translation includes the next XOCL
expression to perform. The {\tt desugar2} operation cannot be
supplied with the next XOCL expression because this will depend on
whether or not the current statement extends the type environment.
Therefore, in {\tt desugar2} the continuation is an operation that
is awaiting a type environment and produces the next XOCL
expression:
\begin{lstlisting}
context Statement
  @AbstractOp desugar2(typeExp:Env,next:Operation):Performable
  end
\end{lstlisting}
A type declaration binds the type at translation time and supplies
the extended type environment to the continuation:
\begin{lstlisting}
context TypeDeclaration
  @Operation desugar2(typeEnv:Env,next:Operation):Performable
    next(typeEnv.bind(name,Type(names)))
  end
\end{lstlisting}
A value declaration introduces a new local definition; the body is
created by supplying the unchanged type environment to the
continuation:
\begin{lstlisting}
context ValueDeclaration
  @Operation desugar2(typeEnv:Env,next:Operation):Performable
    [| let <name> = <value.desugar2(typeEnv)>
       in <next(typeEnv)>
       end
    |]
  end
\end{lstlisting}
Translation of a block involves translation of a sequence of
sub-statements. The following auxiliary operation ensures that the
continuations are chained together correctly:
\begin{lstlisting}
context Statements
  @Operation desugar2(statements,typeEnv,next):Performable
      @Case statements of
        Seq{} do
          next(typeEnv)
        end
        Seq{statement | statements} do
          statement.desugar2(
            typeEnv,
            @Operation(typeEnv)
              Statements::desugar2(statements,typeEnv,next)
            end)
        end
      end
    end
\end{lstlisting}
A block is translated to a sequence of statements where local
definitions are implemented using nested {\tt let} expressions in
XOCL. The locality of the definitions is maintained by sequencing
the block statements and the continuation expression:
\begin{lstlisting}
context Block
  @Operation desugar2(typeEnv:Env,next:Operation):Performable
    [| <Statements::desugar2(
         statements,
         typeEnv,
         @Operation(ignore)
           [| null |]
         end)>;
       <next(typeEnv)>
    |]
  end
\end{lstlisting}
A {\tt while} statement is translated so that the XOCL expression
is in sequence with the expression produced by the contintuation:
\begin{lstlisting}
context While
  @Operation desugar2(typeEnv:Env,next:Operation):Performable
    [| @While <test.desugar2(typeEnv)>.value do
         <body.desugar2(typeEnv,@Operation(typeEnv) [| null |] end)>
         end;
         <next(typeEnv)>
       end
    |]
  end
\end{lstlisting}
The {\tt if} statement is translated to an equivalent XOCL
expression:
\begin{lstlisting}
context If
  @Operation desugar2(typeEnv:Env,next:Operation):Performable
    [| if <test.desugar2(typeEnv)>.value
       then <thenPart.desugar2(typeEnv,next)>
       else <elsePart.desugar2(typeEnv,next)>
       end
    |]
  end
\end{lstlisting}\begin{lstlisting}
context FieldUpdate
  @Operation desugar2(typeEnv:Env,next:Operation):Performable
    [| <record.desugar2(typeEnv)>.update(
          <StrExp(name)>,
          <value.desugar2(typeEnv)>);
       <next(typeEnv)>
    |]
  end
\end{lstlisting}\begin{lstlisting}
context Update
  @Operation desugar2(typeEnv:Env,next:Operation):Performable
    [| <name> := <value.desugar2(typeEnv)>;
       <next(typeEnv)>
    |]
  end
\end{lstlisting}\section{Compiling XAction}

The previous section shows how to perform static type anslysis
while translating XAction to XOCL. XOCL is then translated to XMF
VM instructions by the XOCL compiler (another translation
process). The result is that XAction cannot to anything that XOCL
cannot do. Whilst this is not a serious restriction, there may be
times where a new language wishes to translate directly to the XMF
VM without going through an existing XMF language. This may be in
order to produce highly efficient code, or because the language
has some unusual control constructs that XOCL does not support.
This section shows how XAction can be translated directly to XMF
VM instructions.

\subsection{Compiling Expressions}

\begin{lstlisting}
context Exp
  @AbstractOp compile(typeEnv:Env,valueEnv:Seq(String)):Seq(Instr)
  end
\end{lstlisting}\begin{lstlisting}
context Const
  @Operation compile(typeEnv,valueEnv)
    @TypeCase(value)
      Boolean do
        if value
        then Seq{PushTrue()}
        else Seq{PushFalse()}
        end
      end
      Integer do
        Seq{PushInteger(value)}
      end
    end
  end
\end{lstlisting}\begin{lstlisting}
context Var
  @Operation compile(typeEnv,valueEnv)
    let index = valueEnv->indexOf(name)
    in if index < 0
       then self.error("Unbound variable " + name)
       else Seq{LocalRef(index)}
       end
    end
  end
\end{lstlisting}\begin{lstlisting}
context BinExp
  @Operation compile(typeEnv,valueEnv):Seq(Instr)
    left.compile(typeEnv,valueEnv) +
    right.compile(typeEnv,valueEnv) +
    @Case op of
      "and" do Seq{And()} end
      "or"  do Seq{Or()} end
      "+"   do Seq{Add()} end
    end
  end
\end{lstlisting}\begin{lstlisting}
context New
  @Operation compile(typeEnv,valueEnv):Seq(Instr)
    self.desugar2(typeEnv).compile()
  end
\end{lstlisting}\begin{lstlisting}
context FieldRef
  @Operation compile(typeEnv,valueEnv):Seq(Instr)
    Seq{StartCall(),
        PushStr(name)}
    value.compile(typeExp,valueExp) +
    Seq{Send("lookup",1)}
  end
\end{lstlisting}\subsection{Compiling Statements}

\begin{lstlisting}
context Statement
  @AbstractOp compile(typeEnv:Env,varEnv:Seq(String),next:Operation):Seq(Instr)
  end
\end{lstlisting}\begin{lstlisting}
context TypeDeclaration
  @Operation compile(typeEnv,varEnv,next)
    next(typeEnv.bind(name,Type(names)),varEnv)
  end
\end{lstlisting}\begin{lstlisting}
context ValueDeclaration
  @Operation compile(typeEnv,varEnv,next)
    value.compile(typeEnv,varEnv) +
    Seq{SetLocal(name,varEnv->size),
        Pop()} +
    next(typeEnv,varEnv + Seq{name})
  end
\end{lstlisting}\begin{lstlisting}
context Statements
  @Operation compile(statements,typeEnv,varEnv,next)
    @Case statements of
      Seq{} do
        next(typeEnv,varEnv)
      end
      Seq{statement | statements} do
        statement.compile(
          typeEnv,
          varEnv,
          @Operation(typeEnv,varEnv)
            Statements::compile(statements,typeEnv,varEnv,next)
          end)
      end
    end
  end
\end{lstlisting}\begin{lstlisting}
context Block
  @Operation compile(typeEnv,varEnv,next)
    Statements::compile(
      statements,
      typeEnv,
      varEnv,
      @Operation(localTypeEnv,localVarEnv)
        next(typeEnv,varEnv)
        end)
  end
\end{lstlisting}\begin{lstlisting}
context While
  @Operation compile(typeEnv,varEnv,next)
     Seq{Noop("START")} +
     test.compile(typeEnv,varEnv) +
     Seq{SkipFalse("END")} +
     body.compile(typeEnv,varEnv,
       @Operation(typeEnv,varEnv)
         Seq{}
       end) +
     Seq{Skip("START")} +
     Seq{Noop("END")} +
     next(typeEnv,varEnv)
  end
\end{lstlisting}\begin{lstlisting}
context If
  @Operation compile(typeEnv,varEnv,next)
     test.compile(typeEnv,varEnv) +
     Seq{SkipFalse("ELSE")} +
     thenPart.compile(typeEnv,varEnv,
       @Operation(typeEnv,varEnv)
         Seq{Skip("END")}
       end) +
     Seq{Noop("ELSE")} +
     elsePart.compile(typeEnv,varEnv,
       @Operation(typeEnv,varEnv)
         Seq{Skip("END")}
       end) +
     Seq{Noop("END")} +
     next(typeEnv,varEnv)
  end
\end{lstlisting}\section{Abstract Syntax to Concrete Syntax}

We have shown how XAction is translated from concrete syntax to
abstract syntax by defining an XBNF grammar. It is often useful to
be able to translate in the opposite direction and produce
concrete syntax from abstract syntax. This can be done with or
without formatting. The latter is useful only when the concrete
syntax is to be consumed by a machine or when it can be supplied
to a pretty-printing tool.

Formatting of code can be performed in fairly sophisticated ways,
for example allowing the width of the page to be supplied as a
parameter to the formatter. This section shows how a simple code
formatter for XAction can be defined by attaching {\tt pprint}
operations to the abstract syntax classes.

An expression is formatted by supplying it with an output channel,
it is assumed that the channel is in the correct output column:
\begin{lstlisting}
@AbstractOp pprint(out:OutputChannel) end
\end{lstlisting}
A variable is pretty-printed by printing its name:
\begin{lstlisting}
context Var
  @Operation pprint(out)
    format(out,"~S",Seq{name})
  end
\end{lstlisting}
A constant is pretty-printed by printing its value:
\begin{lstlisting}
context Const
  @Operation pprint(out)
    format(out,"~S",Seq{value})
  end
\end{lstlisting}
A {\tt new} expression prepends the type with the keyword:
\begin{lstlisting}
context New
  @Operation pprint(out)
    format(out,"new ~S",Seq{type})
  end
\end{lstlisting}
A binary expression pretty-prints the left sub-expression, the
operator name and then the right sub-expression:
\begin{lstlisting}
context BinExp
  @Operation pprint(out)
    left.pprint(out);
    format(out," ~S ",Seq{op});
    right.pprint(out)
  end
\end{lstlisting}
A statement is pretty-printed by supplying it with the output
channel and the current level of indentation. The indentation
controls how many tab-stops must be output after each newline.
This is necessary because statements can be nested and indentation
is used to visualise the level of nesting.
\begin{lstlisting}
context Statement
  @AbstractOp pprint(out:OutputChannel,indent:Integer) end
\end{lstlisting}
A block is pretty-printed by incrementing the indentation for each
nested statement:
\begin{lstlisting}
context Block
  @Operation pprint(out,indent)
    format(out,"begin");
    @For s in statements do
      format(out,"~%~V",Seq{indent + 2});
      s.pprint(out,indent + 2)
    end;
    format(out,"~%~Vend",Seq{indent})
  end
\end{lstlisting}
An {\tt if} statement is pretty-printed as follows:
\begin{lstlisting}
context If
  @Operation pprint(out,indent)
    format(out,"if ");
    test.pprint(out);
    format(out,"~%~Vthen~%~V",Seq{indent,indent + 2});
    thenPart.pprint(out,indent+2);
    format(out,"~%~Velse~%~V",Seq{indent,indent + 2});
    elsePart.pprint(out,indent+2);
    format(out,"~%~Vend",Seq{indent})
  end
\end{lstlisting}
A type declaration is pretty-printed as follows, note the use of
{\tt ~\{} to iterate through the sequence of field names in the
{\tt format} control string:
\begin{lstlisting}
context TypeDeclaration
  @Operation pprint(out,indent)
    format(out,"type ~S is ~{,~;~S~} end",Seq{name,names})
  end
\end{lstlisting}
A value declaration:
\begin{lstlisting}
context ValueDeclaration
  @Operation pprint(out,indent)
    format(out,"value ~S is ",Seq{name});
    value.pprint(out);
    format(out," end")
  end
\end{lstlisting}
A {\tt while} statement:
\begin{lstlisting}
context While
  @Operation pprint(out,indent)
    format(out,"while ");
    test.pprint(out);
    format(out," do~%~V",Seq{indent+2});
    body.pprint(out,indent+2);
    format(out,"~%~Vend",Seq{indent})
  end
\end{lstlisting}\section{Conclusion}

This chapter has shown how XMF can be used to define the
operational semantics of languages. We have shown how to implement
an interpreter for a simple language and how to translate the
language to existing XMF languages. We have discussed a number of
different issues relating to language translation, in particular
how much work is performed statically and how much is left to
run-time.

\chapter{Case Study 4: Interactive TV}

\section{Introduction}

An increasing number of interactive applications can be downloaded
onto devices such as mobile phones, PDAs, web-browsers and TV set-top
boxes. The applications involve presenting the user with information,
options, menus and buttons. The user typically enters information
by typing text and choosing amongst alternatives. An event is generated
by the user clicking a button or selecting from a menu. Once an event
is generated an engine that services the interactive application processes
the event, updates its internal state and then produces a new dialog
to present to the user.

The dialogs required by the interactive applications tend to be fairly
simple and are often used in conjunction with other applications such
as being broadcast together with TV video and audio content. The technology
used to construct the interactive application should be accessible
to as broad a spectrum of users as possible, including users whose
primary skill is not in developing interactive software applications.

Technologies used for interactive displays include Java-based technologies
such as the Multimedia Home Platform (MHP)\cite{MHP} , HTML and JavaScript.
These technologies are very platform specific. They include a great
deal of technical detail and are certainly not approachable by a non-specialist.
Furthermore, the general low-level nature of the technologies does
not enforce any standard look-and-feel to interactive applications.
The applications developed for a given client (for example a single
TV company) should have a common look and feel that is difficult to
enforce at such a low-level.

A common way to abstract from technical detail and to enforce a common
look-and-feel for a suite of applications it to develop a \emph{domain-specific
language} (DSL) whose concepts match the expectations and skill-levels
of the intended users. A DSL for interactive applications will include
constructs for expressing textual content, buttons and choices. The
DSL leaves the rendering of the display features and the event generation
to a display engine that enforces a given look-and-feel. The display
engine can be replaced, changing the look-and-feel without changing
the DSL.

In addition to the DSL supporting high-level features for expressing
display content, it must provide some means for describing what the
application \emph{does}. Normally, application processing algorithms
are expressed at a low-level in program-code. If the DSL is designed
with an \emph{execution engine} then the same approach to abstraction
from rendering detail can be applied to abstraction from the operational
detail.

An executable DSL is a language for expressing complete applications
without detailed knowledge of implementation technologies. The xDSL
is a modelling language, instances of which are expressed as data.
An execution engine processes the data and runs the application. The
xDSL engine can be embedded within devices and other software applications
in order to run the models.

This chapter describes the design of an xDSL for interactive applications.
The xDSL has a textual syntax and is implemented in the tool XMF.
XMF is a platform for developing DSL-based applications. It provides
an extensive high-level language called XOCL for developing applications
and XOCL can be extended with new language constructs. The design
of XMF has been based on Common Lisp \cite{Lisp}, Smalltalk, Scheme
\cite{Scheme} and ObjVLisp \cite{objVlisp}. The DSL for interactive
applications described in this chapter can be developed on any platform,
but XMF is ideally suited to the task.

The rest of this chapter is structured as follows: section \ref{sec:Interactive-Application-Architecture}
describes an architecture for interactive applications based on an
xDSL; section \ref{sec:A-DSL-for-Interactive-Applications} describes
the xDSL from the point of view of the application developer, it outlines
the language features in terms of a simple application involving a
quiz; section \ref{sec:Implementation} describes how the xDSL is
implemented on XMF; section \ref{sec:Simulation} shows how the rendering
engine can be simulated and connected to the xDSL engine to simulate
the execution of an interactive application; section \ref{sec:XML-Representation-for}
shows how the application models can be serialized as XML; section
\ref{sec:Conclusion} concludes by reviewing the key features of the
approach and the technology used to develop the xDSL engine.

\section{Interactive Application Architecture\label{sec:Interactive-Application-Architecture}}

\begin{figure}
\begin{center}
\includegraphics[scale=0.5]{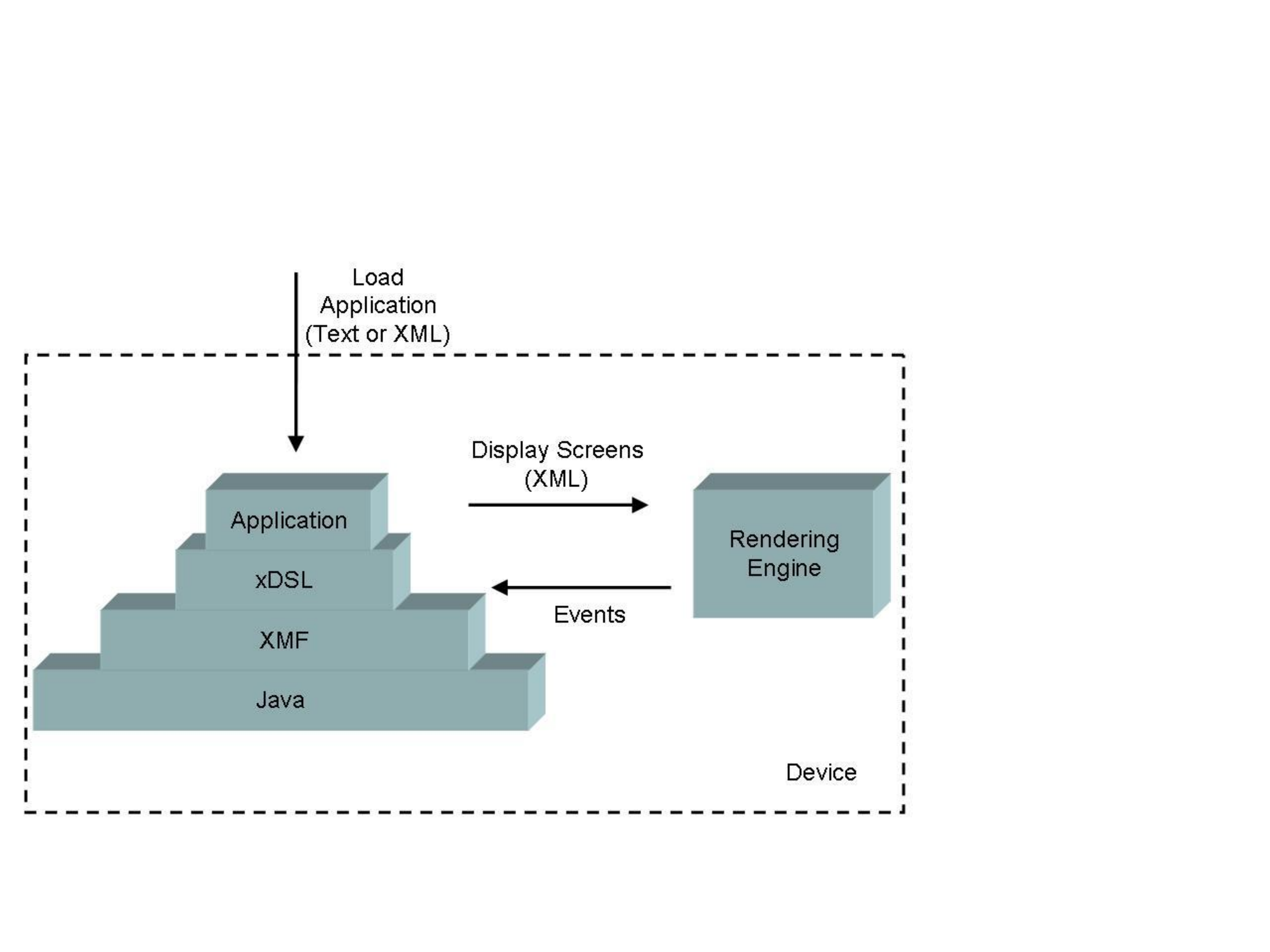}

\caption{Application Architecture\label{fig:Application-Architecture}}
\end{center}
\end{figure}

Figure \ref{fig:Application-Architecture} shows an overview of the
architecture for an interactive application xDSL. XMF is used as the
DSL implementation engine. XMF provides facilities for developing
text-based modelling languages and their associated execution engines.
The xDSL is written as a model in XMF, the applications are then loaded
onto the xDSL engine and executed.

The application generates display information in a general-purpose
format; in this case XML. The XML display information is sent to a
rendering engine. The engine understands the display features of the
DSL and interprets them in terms of the rendering technology. This
achieves a separation of concerns whereby the DSL can focus on the
information content and execution logic whereas the rendering engine
can focus on a standard way of displaying the information without
necessarily having to understand anything about the application that
is being rendered.

The rendering engine controls the hardware that interacts with the
user. The user generates events that are sent back to the xDSL engine.
In principle the events can be very detailed and can be encoded in
a suitable DSL. The example presented in this chapter uses a very simple
encoding of events.

When the xDSL receives an event, it must process the data in an appropriate
way to produce more display information for the rendering engine.
The processing information is expressed in the application model running
on the xDSL engine. The display/event loop continues until the application
is terminated.

The architecture shown in figure \ref{fig:Application-Architecture}
has been used a number of times based on the XMF processing engine.
The development environment XMF-Mosaic is completely based upon a
number of rendering engines based on various features of Eclipse:
browsers, property editors and diagram editors. The architecture has
also been successfully used where XMF is deployed as a web-application
and the rendering engine is a standard web-browser (using various
combinations of HTML and the Google Web Toolkit).

\section{A DSL for Interactive Applications\label{sec:A-DSL-for-Interactive-Applications}}

This section presents a simple interactive application expressed using
the xDSL and then shows it running. The following section shows how
the xDSL is implemented in XMF.

Textual languages are developed in XMF by extending the basic language
with new language features. XMF has a small basic language; the rest
is developed by extension. Each new language feature starts with an
'@' character; the feature may be used wherever any normal language
construct is expected. In this way, the XMF engine can be developed
into any special purpose DSL engine.

\newpage{}

The following fragment shows the start of an interactive application:

\begin{lstlisting}
@Model Quiz
  // The Quiz model describes an interactive application
  // for a TV quiz. Viewers are presented with a sequence
  // of questions and get a final score...
  score : Integer;
  // Screen definitions...
end
\end{lstlisting}Each model consists of a collection of screens. Each screen describes
how it is to be rendered and how it responds to events. For example,
the following screen starts the application. It places some text above
a button named Start. When the engine receives a Start event then
the application makes a transition to the screen named Question1:

\begin{lstlisting}
screen START()
  vertical
    text Welcome to the Quiz. Click the button to Start end
    button Start
      go Question1()
    end
  end
end
\end{lstlisting}Options are offered using a construct that lists the options (how
they are displayed is up to the rendering engine). The option group
is named; the name is used to refer to the particular option value
that the user selects when the event returns to the xDSL engine. This
is a typical way of encoding variable information during dialogs:
http does this and can be used to determine the values of fields and
choices on HTML screens. The first part of the Question1 screen uses
options as shown below:

\begin{lstlisting}
screen Question1()
  vertical
    text What is the capital of England? end
    options Choice
      option London; 
      option Paris;
      option Madrid;
    end
    // Question1 continues...
\end{lstlisting}Layout can be handled using many sophisticated schemes. A useful,
simple way to specify layout is to use horizontal and vertical flow,
where these can be nested. The options are displayed below the text
in the fragment given above. In the fragment below, the buttons for
Next and Quit are displayed next to each other (but below the options):

\begin{lstlisting}
    // ... Question1 continues...
    horizontal
      button Next
        // Next action continues...
      end
      button Quit
       go Quit()
      end
    end
  end
end
\end{lstlisting}The Next event is received after the user has chosen an option. If
the option is correct then the user is congratulated and the score
is incremented, otherwise the user is told the correct answer. In
both cases the dialog continues with the next question. 

Actions may be conditional, the conditional expression may refer to
choice variables, values of variables passed to screens and the current
state of the model instance. Actions may also produce displays (without
having to go to a new screen) which allows variables to be scoped
locally within an action %
\footnote{We really should have a let-construct and some local variables here
to show that the nested display has access to locally-scoped variables
over a user-transaction.%
}. In the following, the Next action has two local displays that are
used to respond to the choice:

\begin{lstlisting}
        // ...Next action continues...
        if Choice = "London" 
        then 
          display
            text Well Done end
            button Next
              score := score + 1;
              go Question2()
            end
          end
        else
          display
            text Wrong! Answer is London. end
            button Next
              go Question2()
          end
        end
      end
\end{lstlisting}The DSL is simple and closely matches the concepts required to define
an interactive application. It could be extended in a variety of ways,
for example pattern matching event data and further display concepts.
It includes execution by encoding event handlers. It deals with complexity
by being simple and supporting nesting with local scope and modularity.
A non-expert in interactive software applications should have no problems
writing an application in this DSL.

The following shows a partial execution of this application. Since
there is no rendering engine attached, the XML is printed and the
responses encoded by hand:

\begin{lstlisting}
<Screen>
  <Vertical>
    <Text text=' Welcome to the Quiz. Click the button to Start '/>
    <Button name='Start'/>
  </Vertical>
</Screen>
Start                <-- Event from rendering engine
<Screen>
  <Vertical>
    <Text text=' What is the capital of England? '/>
    <Options name='Choice'>
      <Option name='London'/>
      <Option name='Paris'/>
      <Option name='Madrid'/>
    </Options>
    <Horizontal>
      <Button name='Next'/>
      <Button name='Quit'/>
    </Horizontal>
  </Vertical>
</Screen>
Next Choice=London   <-- Event from rendering engine
<Screen>
  <Text text=' Well Done '/>
  <Button name='Next'/>
</Screen>
Next                 <-- Event from rendering engine
\end{lstlisting}
\section{Implementation\label{sec:Implementation}}

The implementation of the xDSL has two main components: a syntax model
and a grammar that transforms text to instances of the syntax model,
and a semantics model that defines an execution engine. The syntax
model defines a modeling language that defines an application \emph{type};
an instance of the type is defined by the semantics model. This is
a typical way to define a language: models represent things that can
be \emph{performed} in some sense. Performing the models produces
instances whose behaviour is expressed by the model. Another way to
think about this is that we aim to produce libraries of reusable interactive
applications (instances of the syntax model). A run-time occurrence
of an application is described by the semantic model.

XMF provides facilities for working with text including grammars,
XML parsers and XML formatters. The models have been developed using
the XMF development engine XMF-Mosaic and then run on the basic engine.

\subsection{Syntax}

The syntax for the DSL has two parts: the abstract syntax and the
concrete syntax. The abstract syntax consists of a collection of models
that are described below. The concrete syntax is defined by a collection
of grammars that are defined at the end of this section.

\begin{figure}
\begin{center}
\includegraphics[scale=0.6]{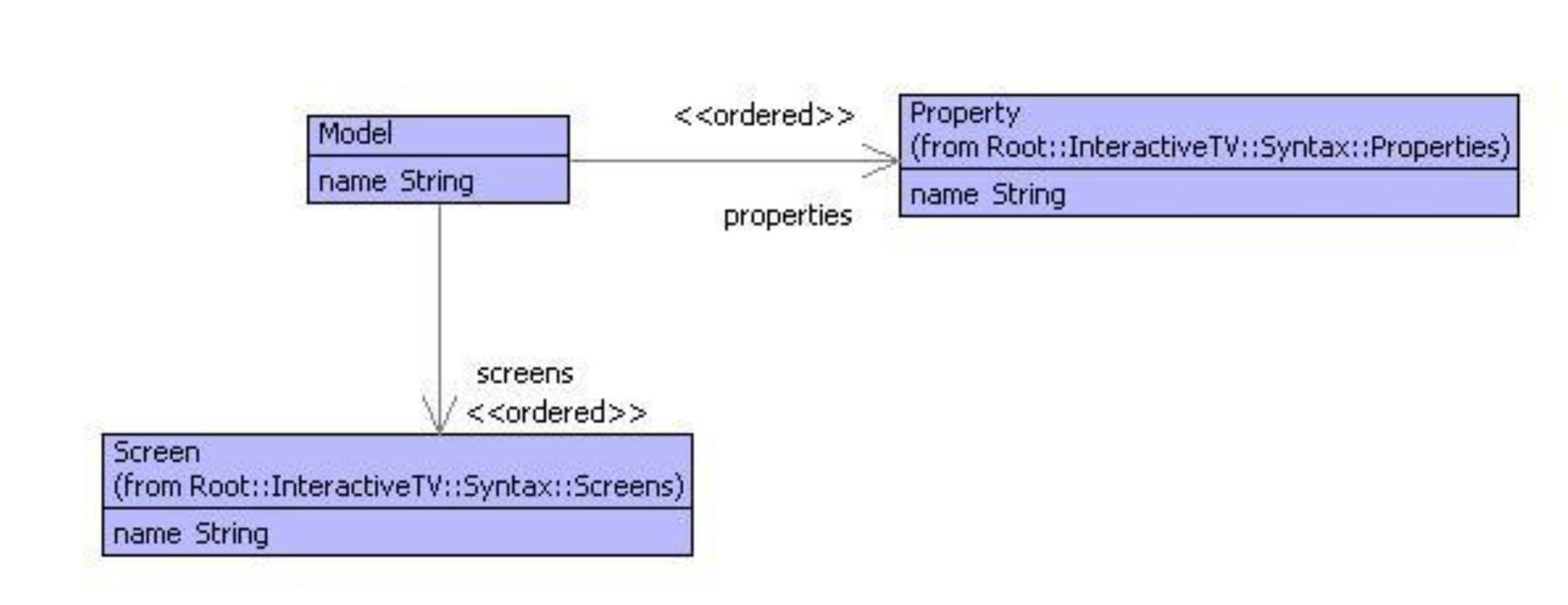}

\caption{Interactive Models\label{fig:Interactive-Models}}

\end{center}
\end{figure}

\begin{figure}
\begin{center}
\includegraphics[scale=0.75]{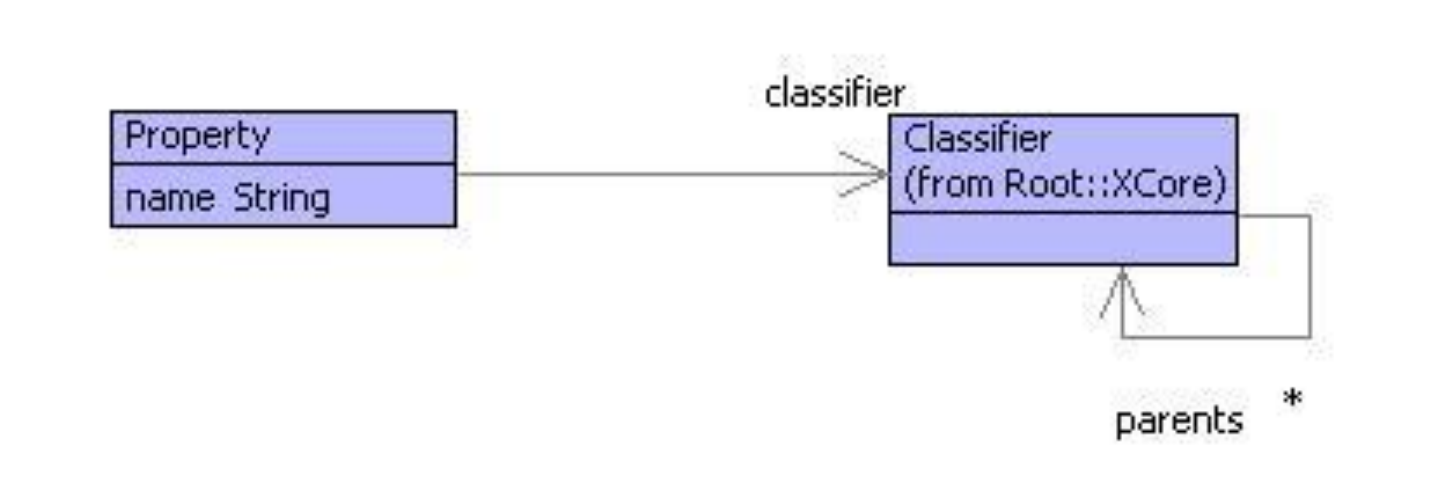}

\caption{Properties\label{fig:Properties}}

\end{center}
\end{figure}

Figure \ref{fig:Interactive-Models} shows the top-level model for
the interactive application language. A model consists of a collection
of properties and a collection of screens. Each property is defined
by the model in figure \ref{fig:Properties}; it has a name and a
classifier. XMF has a meta-model that provides features such as Class
and Object. A type is called a Classifier in XMF; Integer, String,
Set(Object) are all XMF classifiers.

\begin{figure}
\begin{center}
\includegraphics[scale=0.75]{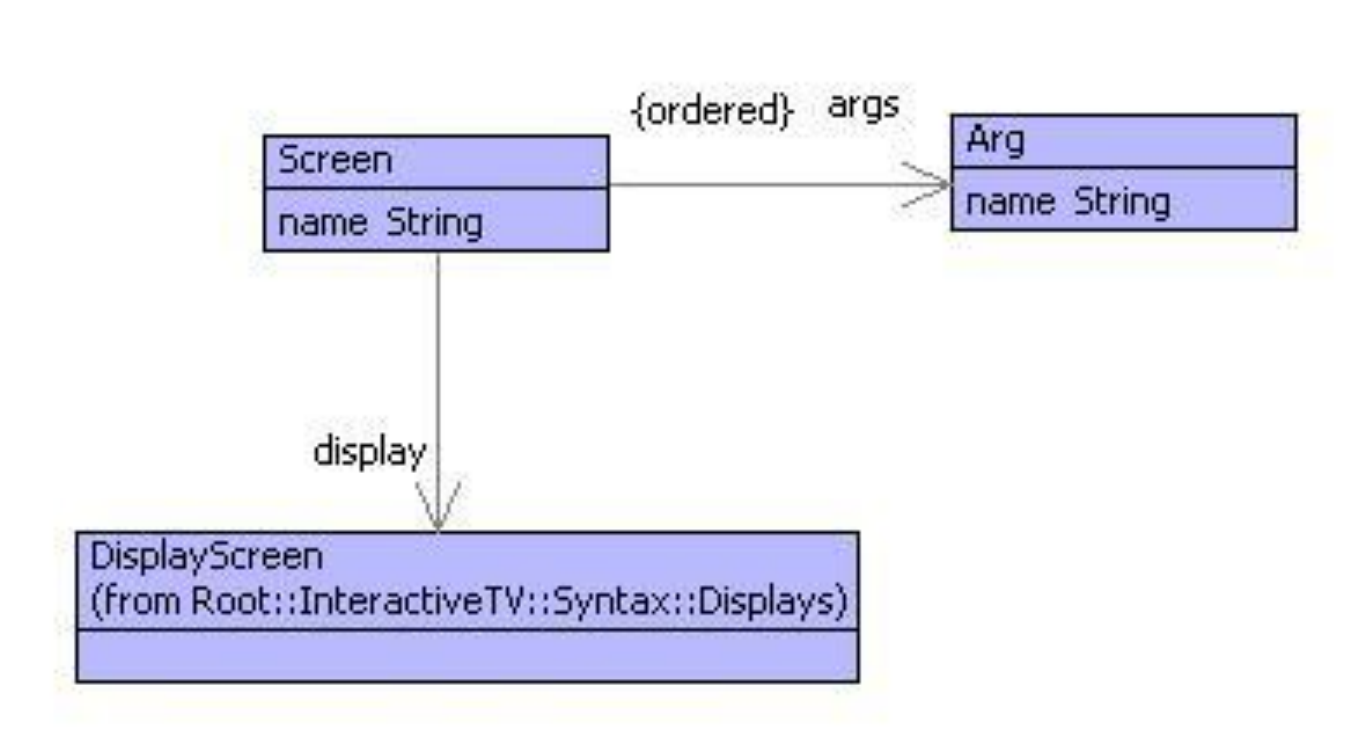}

\caption{Screens\label{fig:Screens}}
\end{center}
\end{figure}

\begin{figure}
\begin{center}
\includegraphics[scale=0.4]{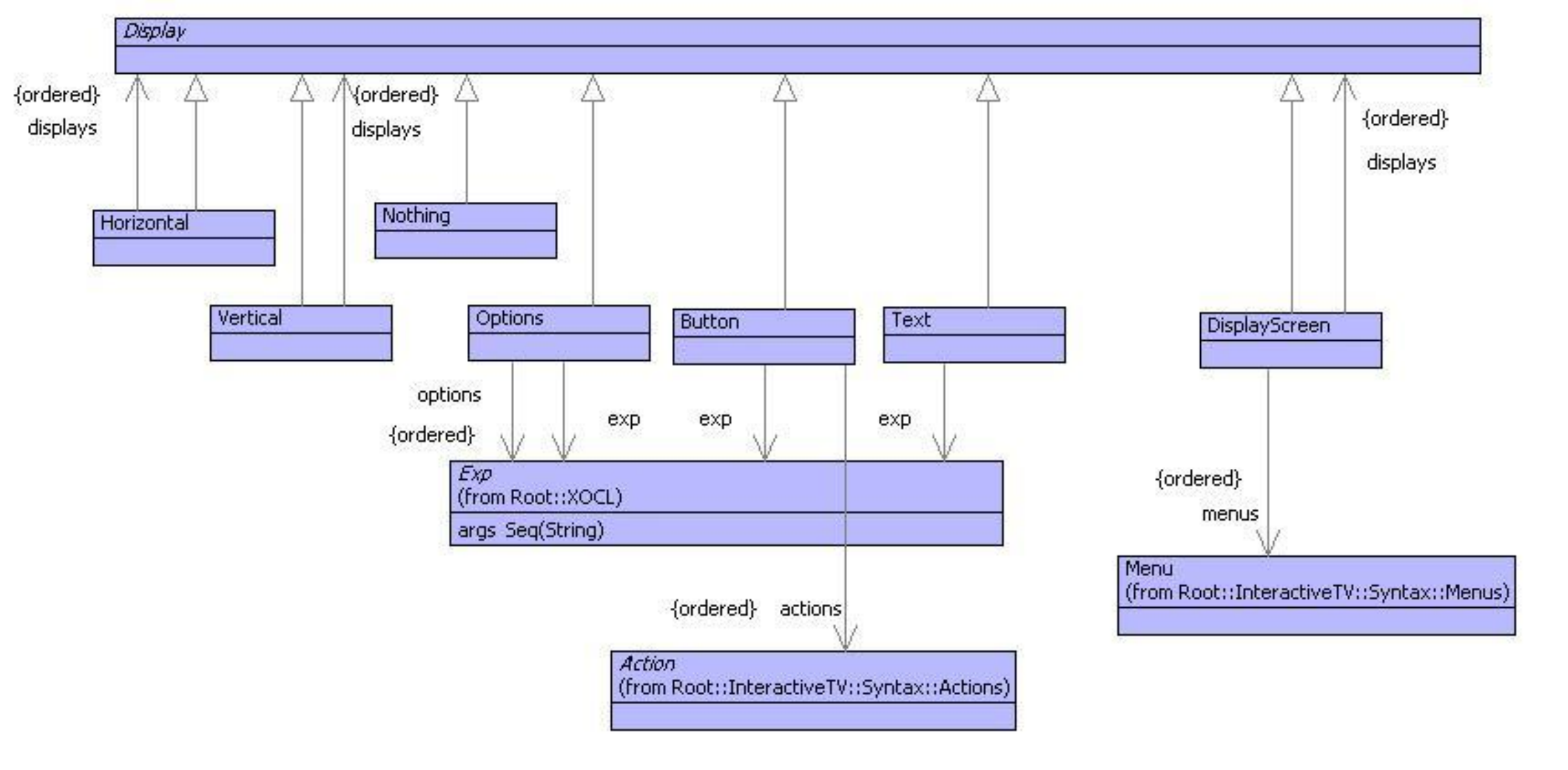}

\caption{Displays\label{fig:Displays}}
\end{center}
\end{figure}

Screens are shown in figure \ref{fig:Screens}. A screen has a collection
of arguments. An action may cause the application to make a transition
to a screen in which case the transition can supply argument values
to be used when calculating the display for the screen. Figure \ref{fig:Displays}
shows the model for displays. Each element of the display model can
produce XML output that is understood by the rendering engine. In
addition, some of the display elements are associated with actions
that will be performed when the appropriate event is received by the
xDSL engine.

The display elements of an application model refer to an XOCL class
called Exp. This is used wherever an executable fragment of code is
required. It allows features of the displays to be computed dynamically.
For example when a transition to a screen is made, the names of the
buttons may depend on the values of the arguments that are passed
to the screen. This is the reason why a button has an exp: it is used
to calculate the label on the button in terms of the variables that
are in scope at the time%
\footnote{Unfortunately no examples of this feature are given in the document.
However, imagine a list of voting options that will depend on the
current state of the system.%
}. The Exp class is a way of importing the models for XMF expressions
into the displays model.

\begin{figure}
\begin{center}
\includegraphics[scale=0.75]{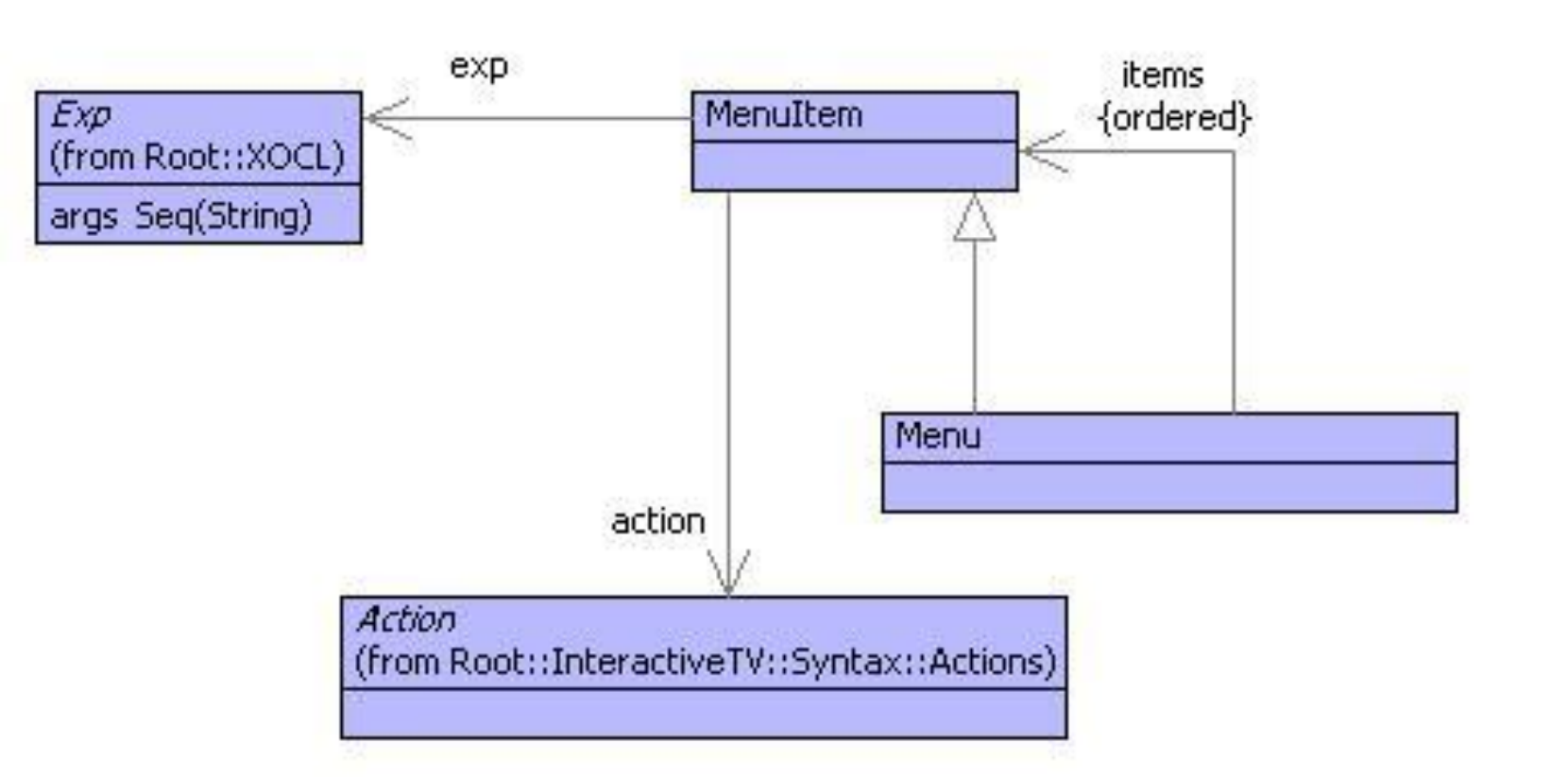}

\caption{Menus\label{fig:Menus}}
\end{center}
\end{figure}

\begin{figure}[htb]
\begin{center}
\includegraphics[scale=0.75]{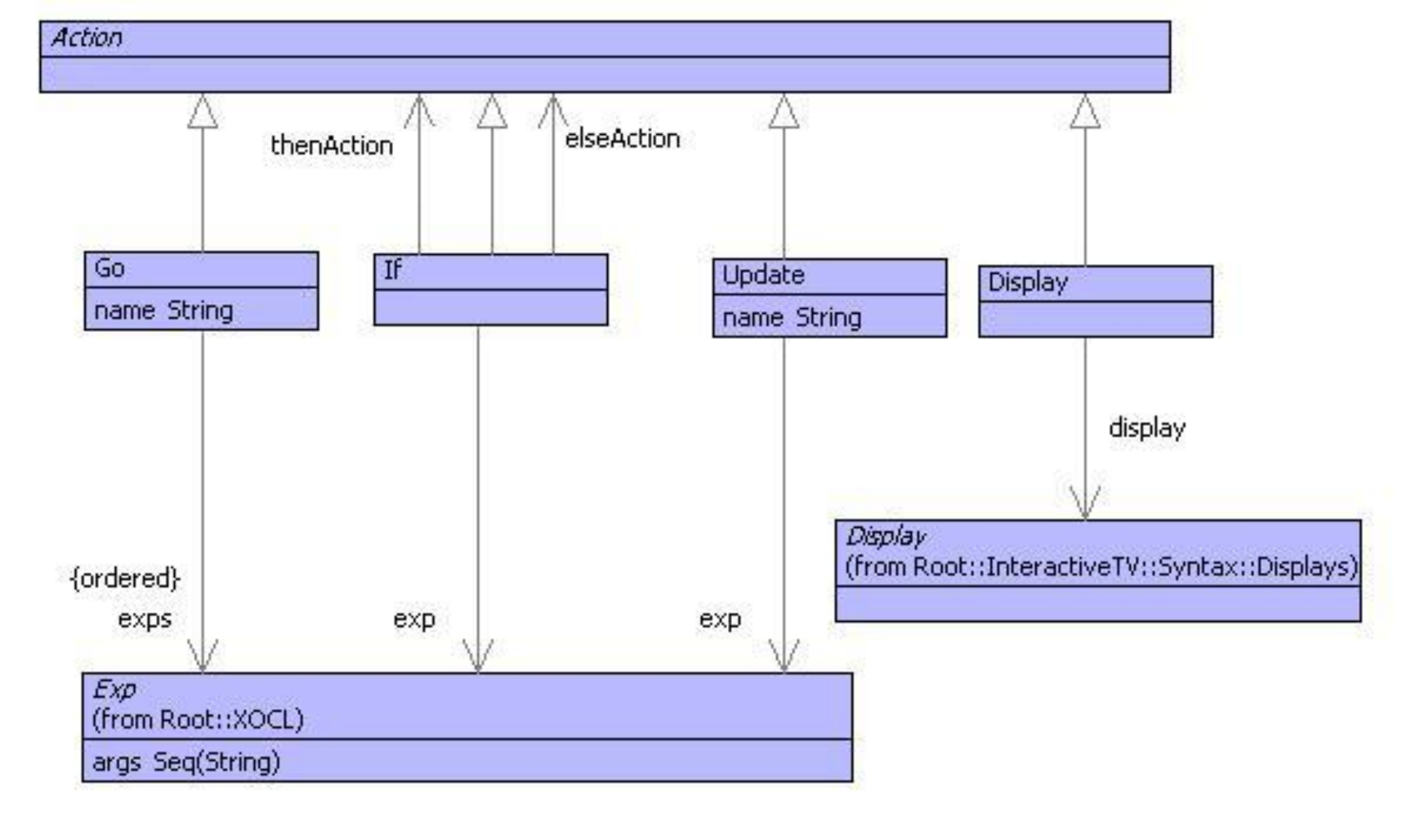}

\caption{Actions\label{fig:Actions}}
\end{center}
\end{figure}

Menus are shown in figure \ref{fig:Menus}. Each menu has a label
that is computed (again depending on the context) and an action. Actions
are defined in figure \ref{fig:Actions}. An action is either a transition
to a new screen (Go), a conditional action (If), an update to a variable
currently in scope (Update) or a local display.

The developer of an interactive application does not directly create
instances of the abstract syntax model. The idea is that they write
in a text language that is parsed to synthesize instances of the syntax
model%
\footnote{Another way of doing this is to use some form of graphical notation.
XMF is designed to interface to EMF \cite{emf}and GMF \cite{gmf}and
therefore provide execution engines for EMF models. %
}.

XMF allows any class to be extended with a grammar. The class then
defines a new syntax construct that is completely integrated into
the base language of XMF. Any number of classes can be added in this
way and new syntax classes can build on existing syntax classes. The
following is a simple example of a class definition that implements
a simple guarded expression:

\begin{lstlisting}
@NotNull [e1].m(e2,e3,e4)
  else e5
end
\end{lstlisting}where e1 evaluates to produce an object to which we want to send the
message m with args e2,e3 and e4. However, e1 might produce null in
which case we don't want to send the message, we want to do e5 instead.
This language construct is implemented as follows in XMF:

\newpage{}

\begin{lstlisting}
@Class NotNull extends Sugar

  @Attribute exp       : String            end
  @Attribute name      : String            end
  @Attribute isMessage : Boolean           end
  @Attribute args      : Seq(Performable)  end
  @Attribute error     : Performable       end
    
  @Constructor(exp,name,error) end
    
  @Constructor(exp,name,args,error) 
    self.isMessage := true
  end
    
  @Grammar extends OCL.grammar
    NotNull ::= 
      '[' e = Exp ']' '.' n = Name NotNullTail^(e,n) 'end'.
      
    NotNullTail(e,n) ::= 
      '(' as = NotNullArgs ')' x = NotNullElse { NotNull(e,n,as,x) }
    | x = NotNullElse { NotNull(e,n,x) }.
      
    NotNullArgs ::=
      e = Exp es = (',' Exp)* { Seq{e|es} }
    | { Seq{} }.
      
    NotNullElse ::=
      'else' Exp 
    | { [| null |] }.
      
  end
    
  @Operation desugar():Performable
    [| let notNullValue = <exp>
       in if notNullValue = null
          then <error>
          else <if isMessage
                then Send([| notNullValue |],name,args)
                else [| notNullValue.<name> |]
                end>
          end
       end
    |]
  end
    
end
\end{lstlisting}The key features of the NotNull class are as follows:

\begin{itemize}
\item The class extends Sugar which means that it is defining a new syntax
construct by providing an operation called 'desugar' whose responsibility
is to turn an instance of NotNull into program code.
\item The grammar definition extends the OCL grammar and thereby imports
all of the basic grammar-rule definitions. This provides the rule
for Exp which is the top-level grammar-rule for all language constructs.
\item Each grammar-rule consists of a name and a body. The rule may optionally
have arguments. The body consists of terminals (in ' and'), builtins
such as Name, rule-calls (possibly with arguments) and actions (inside
\{ and \}). The rule actions are any program expression, in most cases
they use class-constructors to create an instance of a named class.
\item The desugar operation uses lifting-quotes ({[}| and |]) to create
an instance of syntax-classes. The opposite of \emph{lifting} is \emph{dropping}
(< and >) used to calculate syntax by evaluating a program expression.
\end{itemize}
The rest of this section describes how the grammar feature of XMF
can be used to define the interaction language. A model consists of
a name followed by properties and screen definitions:

\begin{lstlisting}
context Model
  @Grammar extends Property.grammar, Screen.grammar
     Model ::= n = Name ps = Property* ss = Screen* 'end' {
        Model(n,ps,ss)
      }.
  end
\end{lstlisting}A property is a name and a simple expression (that cannot include
the ';' operator). The property-rule action uses an interesting feature
of syntax classes that allows the expression to be dropped into the
syntax element and is thereby evaluated to produce a classifier for
the property type:

\begin{lstlisting}
context Property extends OCL::OCL.grammar
  @Grammar extends OCL.grammar
    Property ::= n = Name ':' e = SimpleExp ';' {
      Property(n,Drop(e))
    }.
  end
\end{lstlisting}A screen has a name, arguments, menus and display elements. The rule
for screen arguments shows how optional elements are processed: it
returns either a non-empty sequence of names Seq\{a|as\} (head followed
by tail) or the empty sequence Seq\{\}:

\begin{lstlisting}
context Screen
  @Grammar extends Menu.grammar, Display.grammar
    Screen ::= 
      'screen' n = Name '(' as = ScreenArgs ')' 
         ms = Menu* 
         ds = Display* 
      'end' { Screen(n,as,DisplayScreen(ms,ds)) }.
    ScreenArgs ::=
      a = Name as = (',' Name)* { Seq{a|as} }
    | { Seq{} }.
  end
\end{lstlisting}A menu is shown below. This shows how expressions are captured in
data. The rule for a menu item name returns an instance of the class
Exp that is used in data to wrap an evaluable expression. There are
two forms of construction for Exp: Exp(e) and Exp(e,V,null). In both
cases e is an instance of a syntax class. In the second case V is
a collection of variable names that occur freely in e. The values
of variables in V can be supplied when the expression is evaluated
(via keyApply as shown below).

Another interesting feature of the menu item name rule is the use
of 'lift' to transform a data element (in this case a string n) into
an expression whose evaluation produces the original data element:

\begin{lstlisting}
context Menu
  @Grammar extends OCL.grammar
    Menu ::= 'menu' n = MenuItemName is = MenuItem* 'end' {
      Menu(n,is)
    }.
    MenuItemName ::= 
      n = Name { Exp(n.lift()) }
    | e = SimpleExp { Exp(e,e.FV(),null) }.
    MenuItem ::=
      Menu
    | 'item' n = MenuItemName a = Action 'end' { MenuItem(n,a) }.
  end
\end{lstlisting}\newpage{}

Since Display is an abstract class, the grammar-rule for Display is
a list of concrete alternatives:

\begin{lstlisting}
context Display
  @Grammar extends Action.grammar, OCL.grammar
    Display ::=
        Text
      | Button
      | Options
      | Horizontal
      | Vertical.
    Text ::= 'text' t = Char* 'end' { 
      Text(Exp(t.asString().lift())) 
    }.
    Button ::= 
      'button' n = ComputedName 
        as = Action* 
      'end' { Button(n,as) }.
    ComputedName ::=
      n = Name { Exp(n.lift()) }
    | e = SimpleExp { Exp(e,e.FV(),null) }.
    Options ::= 
      'options' n = ComputedName 
        os = Option* 
      'end' { Options(n,os) }.
    Option ::=
      'option' n = Name ';' { n }.
    Horizontal ::=
      'horizontal'
        ds = Display*
      'end' { Horizontal(ds) }.
    Vertical ::=
      'vertical'
        ds = Display*
      'end' { Vertical(ds) }.
  end
\end{lstlisting}\newpage{}

Action is another example of an abstract class:

\begin{lstlisting}
contxt Action
  @Grammar extends OCL.grammar
    Action ::=
        UpdateProperty
      | IfAction
      | Go
      | DisplayScreen.
      DisplayScreen ::= 
        'display' 
          ms = Menu* 
          ds = Display* 
        'end' { DisplayScreen(ms,ds) }.
      UpdateProperty ::=
        n = Name ':=' e = SimpleExp ';' {
          Update(n,Exp(e,e.FV(),null))  
      }.
      Go ::= 'go' n = Name '(' as = GoArgs ')' { Go(n,as) }.
      GoArgs ::= 
        e = GoArg es = (',' GoArg)* { Seq{e|es} }
      | { Seq{} }.
      GoArg ::= e = SimpleExp { Exp(e) }.
      IfAction ::=
        'if' e = SimpleExp
        'then' d = Action
        IfActionTail^(e,d).
      IfActionTail(e,d1) ::=
        'else' d2 = Action 'end' { If(Exp(e,e.FV(),null),d1,d2) }
      | 'end' { If(Exp(e,e.FV(),null),d1,null) }. 
  end
\end{lstlisting}
\subsection{Semantics}

\begin{figure}
\begin{center}
\includegraphics[scale=0.6]{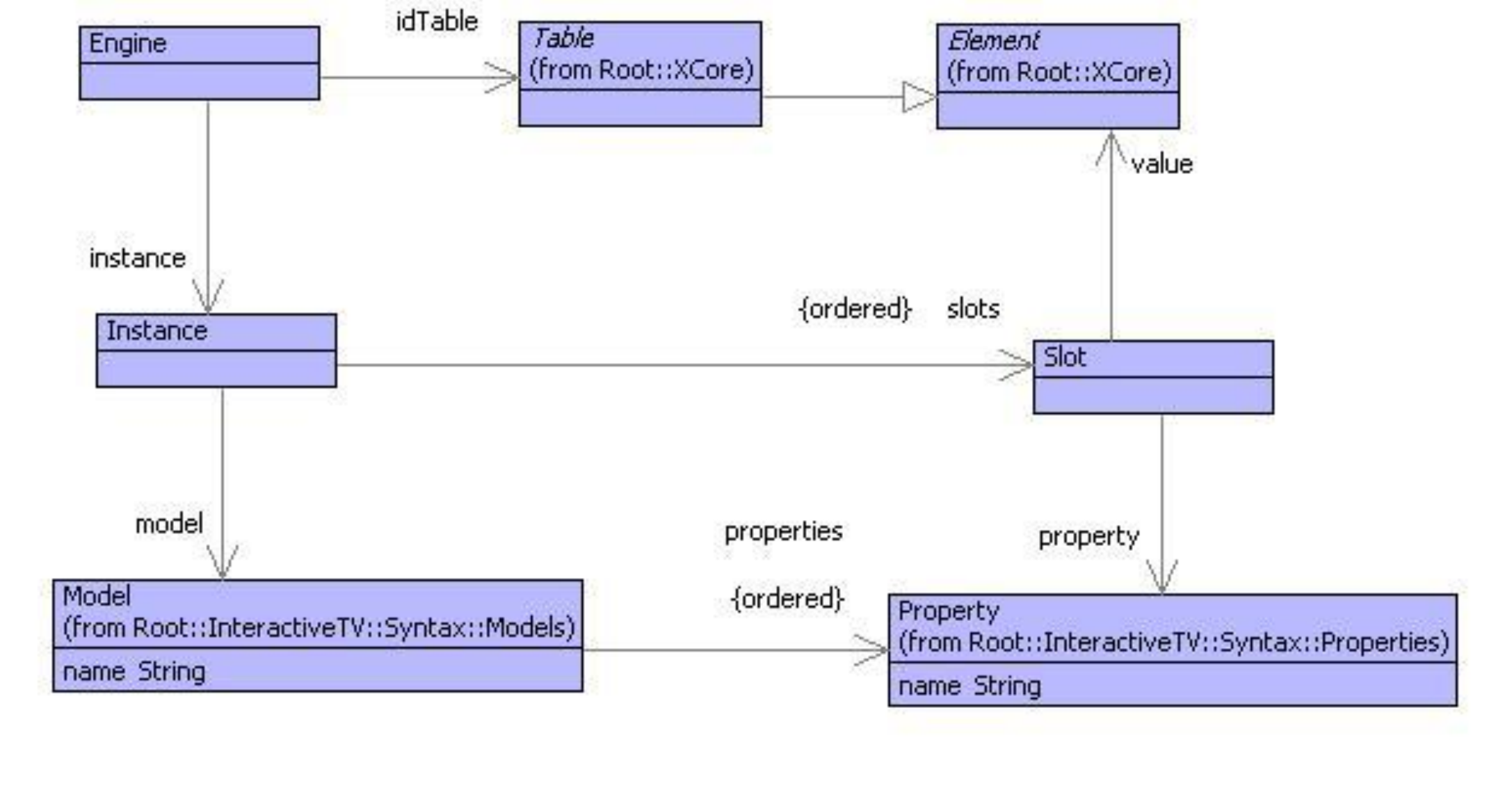}

\caption{Semantics Model\label{fig:Semantics-Model}}
\end{center}
\end{figure}

The semantics of the interactive application modelling language defines
an instance model for the syntax and also defines an execution engine
for the instances. The instance model is shown in figure \ref{fig:Semantics-Model}.
An instance of a Model is defined by the class Instance. It should
have a slot for each property of the model. The value of each slot
should have a type that is defined by the classifier of the corresponding
property.

The class Engine defines the execution engine. An engine controls
an instance and maintains an id-table that maps event ids to handlers
as shown below. The idea is that each time an event occurs, a handler
from the id-table is used to produce a new XML display that is sent
to the rendering engine. The XML data is calculated from the information
in the model and the current state of the instance slots.

The rest of this section defines the engine execution semantics. The
following section shows how the engine is connected to a rendering
engine. The engine processes a screen-transition using the 'go' operation
defined below. XMF supports many types of input and output channel.
The 'go' operation shows an example of a string output channel used
to capture and then return the XML output data as a string:

\begin{lstlisting}
context Engine
  @Operation go(screen:String,args:Seq(Element))
    let sout = StringOutputChannel()
    in instance.go(screen,args,self,sout);
       sout.getString()
    end
  end
\end{lstlisting}An instance handles a screen-transition by looking up the screen in
its model. If the screen exists then it is requested to display itself
with the supplied argument values:

\begin{lstlisting}
context Instance
  @Operation go(screen:String,args:Seq(Element),engine:Engine,out:OutputChannel)
    @NotNull [model.indexScreensByName(screen,null)].display(self,args,engine,out)
      else self.error("No screen called " + screen)
    end
  end
\end{lstlisting}A screen delegates the 'display' message to its underlying display
component. The screen argument names are bound to the argument values
to produce an \emph{environment of bindings} using the 'env' operation:

\begin{lstlisting}
context Screen
  @Operation display(instance,args,engine,out)
    display.display(instance,self.env(instance,args),engine,out)
  end
context Screen    
  @Operation env(instance,values)
    let env = args.zip(values)
    in instance.slots()->iterate(slot env = env | 
         env.bind(slot.property().name(),slot.value()))
    end
  end
\end{lstlisting}Each display element type : Button; Text; Horizontal; Vertical; and,
Options implements a 'display' operation that writes XML data to the
supplied output channel. As a side effect, if any of the elements
have event handling actions then the engine is updated with an appropriate
handler for the event when it is received from the rendering engine.

Each 'display' operation shows the use of an XMF language feature
@XML ... end that is used to write XML output to a channel. The construct
has the form:

\begin{lstlisting}
@XML(out)
  <TAG ATTS>
    .. program code ...
  </TAG>
end
\end{lstlisting}where the literal XML data is written to the supplied output channel.
In-between writing the starting tag and ending tag, an arbitrary program
is processed.

\begin{lstlisting}
context DisplayScreen
  @Operation display(instance,env,engine,out)
    @XML(out)
      <Screen>
        @For menu in menus do
          menu.display(instance,env,engine,out)
        end
        @For display in displays do
          display.display(instance,env,engine,out)
        end
      </Screen>
    end
  end
\end{lstlisting}The 'display' operation for text shows an example of the shortened
form of the XML construct with no body, and also the use of the 'keyApply'
operation of the Exp class. The 'env' argument supplied to 'display'
contains bindings for all variables in scope. The 'keyApply' operation
performs the expression in the context of these variables:

\begin{lstlisting}
context Text
  @Operation display(instance,env,engine,out)
    @XML(out)
      <"Text" text=exp.keyApply(env)/>
    end
  end
\end{lstlisting}A button contains an action that is used to handle the event arising
from the user pressing the button in the rendering engine. The 'display'
operation for Button shows how an event handler is registered in the
engine. The arguments passed to 'registerActions' are the context
required to perform the actions when the event associated with 'id'
is received:

\begin{lstlisting}
context Button
  @Operation display(instance,env,engine,out)
    let id = exp.keyApply(env)
    in engine.registerActions(id,instance,env,actions);
       @XML(out)
         <Button name=id/>
       end
    end
  end
\end{lstlisting}Horizontal and Vertical are similar:

\begin{lstlisting}
context Horizontal
  @Operation display(instance,env,engine,out)
    @XML(out)
      <Horizontal>
        @For display in displays do
          display.display(instance,env,engine,out)
        end
      </Horizontal>
    end
  end
\end{lstlisting}\newpage{}

The 'display' operation for Options shows an example of interleaving
of XML and program code:

\begin{lstlisting}
context Options
  @Operation display(instance,env,engine,out)
    @XML(out)
      <Options name=exp.keyApply(env)>
        @For option in options do
          @XML(out)
            <Option name=option/>
          end
        end
      </Options>
    end
  end
\end{lstlisting}The 'registerActions' operation of Engine must define a handler for
an event. The definition associates the event identifier 'id' with
an operation in the id-table of the engine. Actions are performed
using their 'perform' operation which expects to receive arguments
that include the current environment of variable bindings. The variables
available to an action include all those bound by selecting options
on the display. These display-bound variables are supplied to the
handler (in the same way that http works) as an environment 'env2':

\begin{lstlisting}
contxt Engine
  @Operation registerActions(id,instance,env1,actions)
    idTable.put(id,
      @Operation(env2)
        let value = null
        in @For action in actions do
             value := action.perform(instance,env2 + env1,self)
           end;
           value
        end
      end)
  end
\end{lstlisting}There are four types of action: If; Update; Go; and, Display. Each
action produces a result and the last action performed should return
an XML string to be sent to the rendering engine. If performs one
of two actions (or nothing) depending on the outcome of a test:

\begin{lstlisting}
context If
  @Operation perform(instance,env,engine)
    if exp.keyApply(env + instance.env())
    then thenAction.perform(instance,env,engine)
    else @NotNull [elseAction].perform(instance,env,engine) end
    end
  end
\end{lstlisting}An update changes the value of a variable currently in scope. The
variables in scope are: the slots of the instance; the current screen
arguments. The following operation checks whether the named variable
is a slot and updates the instance appropriately, or updates the current
environment:

\begin{lstlisting}
context Update
  @Operation perform(instance,env,engine)
    @NotNull [instance.getSlot(name)].setValue(exp.keyApply(env + instance.env()))
      else env.set(name,exp.keyApply(env + instance.env()))
    end
  end
\end{lstlisting}Go makes a transition to a new screen. The screen will produce the
XML output. Notice that the current 'env' is not supplied to the 'go'
operation; therefore any variables currently in scope are not available
to the target screen unless their values are passed as arguments:

\begin{lstlisting}
context Go
  @Operation perform(instance,env,engine)
    engine.go(name,exps->collect(exp | exp.keyApply(env)))
  end
\end{lstlisting}Display is a way of locally displaying a screen without losing the
variables that are currently in scope:

\begin{lstlisting}
context Display
  @Operation perform(instance,env,engine)
    let sout = StringOutputChannel()
    in display.perform(instance,env,engine,sout);
       sout.getString()
    end
  end
\end{lstlisting}
\subsection{Handling Events}

Events occur when the user interacts with the rendering engine, for
example by pressing a button. When the event occurs, the current screen
may contain any number of option groups. Each option group is named
and offers a number of alternative values. The selected option may
affect the behaviour of the engine in terms of variable updates and
screen transitions. Therefore, the event sent from the rendering engine
to the xDSL engine must encode the value of any option variables currently
displayed.

In addition there may be any number of ways an event can be raised:
menu selection or button press. Each must be uniquely identified and
the event must supply the identifier of the event that occurred. 

An event is defined to have a format that starts with the event id
and is followed by any number of option variable/value pairs:

\begin{lstlisting}
<ID> <VAR>=<VALUE> ... <VAR>=<VALUE>
\end{lstlisting}The event is encoded as a string and must be decoded by the engine.
This is easily done by defining an event as a grammar-rule:

\begin{lstlisting}
context Engine
  @Grammar
    Event ::= n = Name e = Binding* { Seq{n|e} }.
    Binding ::= n = Name '=' v = Name { Seq{n|v} }.
  end
\end{lstlisting}When an event is received by the engine it is supplied to 'getDisplay'
which calculates a new XML display string for the rendering engine.
The operation uses the grammar defined above to synthesize a pair
Seq\{id|env\} containing the event id and an environment of option-group
variable bindings. If the id is bound in the id-table then the handler
is supplied with the environment:

\begin{lstlisting}
context Engine
  @Operation getDisplay(event:String)
    let pair = Engine.grammar.parseString(event,"Event",Seq{}) then
        id = pair->head;
        env = pair->tail
    in @TableGet handler = idTable[id] do
         idTable.clear();
         handler(env)
       else self.error("No handler for " + name)
       end
    end
  end
\end{lstlisting}
\section{Simulation\label{sec:Simulation}}

Figure \ref{fig:Application-Architecture} shows the architecture
of an interactive application. The rendering engine is external to
the design of an xDSL; the relationship between the two is defined
by the XML schema for the display language and the format of event
strings. However, it is useful to be able to simulate the rendering
engine in order to test the xDSL engine. This can be done by setting
up a simple test harness for a pair of data consumers and linking
the xDSL engine with a rendering engine simulation that reads events
strings in from the terminal.

\newpage{}

The following class implements a data producer-consumer pair:

\begin{lstlisting}
@Class Consumer

  @Attribute filter : Operation end
  @Attribute other  : Consumer (!) end
    
  @Constructor(filter) ! end
    
  @Operation consume(data)
    other.consume(filter(data))
  end
end
\end{lstlisting}The filter operation is used to generate data that is supplied to
the other consumer. If a pair of Consumer instances are linked together
then the data will bounce back and forth as required. The following
operation creates a filter for the xDSL engine:

\begin{lstlisting}
@Operation mk_xDSL_filter(model:Model)
  let engine = Engine(model.new())
  in @Operation(event)
       engine.getDisplay(event)
     end
  end
end
\end{lstlisting}The following filter operation simulates the rendering engine. It
does so by pretty-printing the XML to the standard-output. An XML
string can be transformed into an XML tree using the 'asXML' operation
defined for String. The standard-input is flushed and a line containing
the event is read and returned:

\begin{lstlisting}
@Operation renderFilter(xml:String)
  xml.asXML().pprint(stdout);
  "".println();
  stdin.skipWhiteSpace();
  stdin.readLine().stripTrailingWhiteSpace()
end 
\end{lstlisting}Given a model 'model', the following code produces, and starts, a
simulation:

\begin{lstlisting}
@Operation simulate(model:Model)
  let eConsumer = Consumer(mk_xDSL_filter(model));
      dConsumer = Consumer(renderFilter)
  in eConsumer.setOther(dConsumer);
     dConsumer.setOther(eConsumer);
     eConsumer.consume("START")
  end
end
\end{lstlisting}
\section{XML Representation for Applications\label{sec:XML-Representation-for}}

A requirement for interactive applications is to be able to dynamically
update the content and to be able to transfer the content from remote
locations in a standard format. The application describes in this
chapter is completely represented in data. This means that, although
the application is executable, it can easily be serialized, sent over
a communications link, and then uploaded onto the device that is running
the xDSL engine.

XMF provides support for encoding any data elements as XML. There
is a basic markup provided for all XMF data; the mechanisms for which
can easily be extended to provide bespoke XML encoding. Using the
basic mechanisms, a model can be encoded as follows:

\begin{lstlisting}
@WithOpenFile(fout -> "c:/model.xml")
  let xout = XMLOutputChannel(fout,NameSpaceXMLFormatter())
  in xout.writeValue(model)
  end
end
\end{lstlisting}%
\begin{figure}
\begin{center}
\includegraphics[scale=0.6]{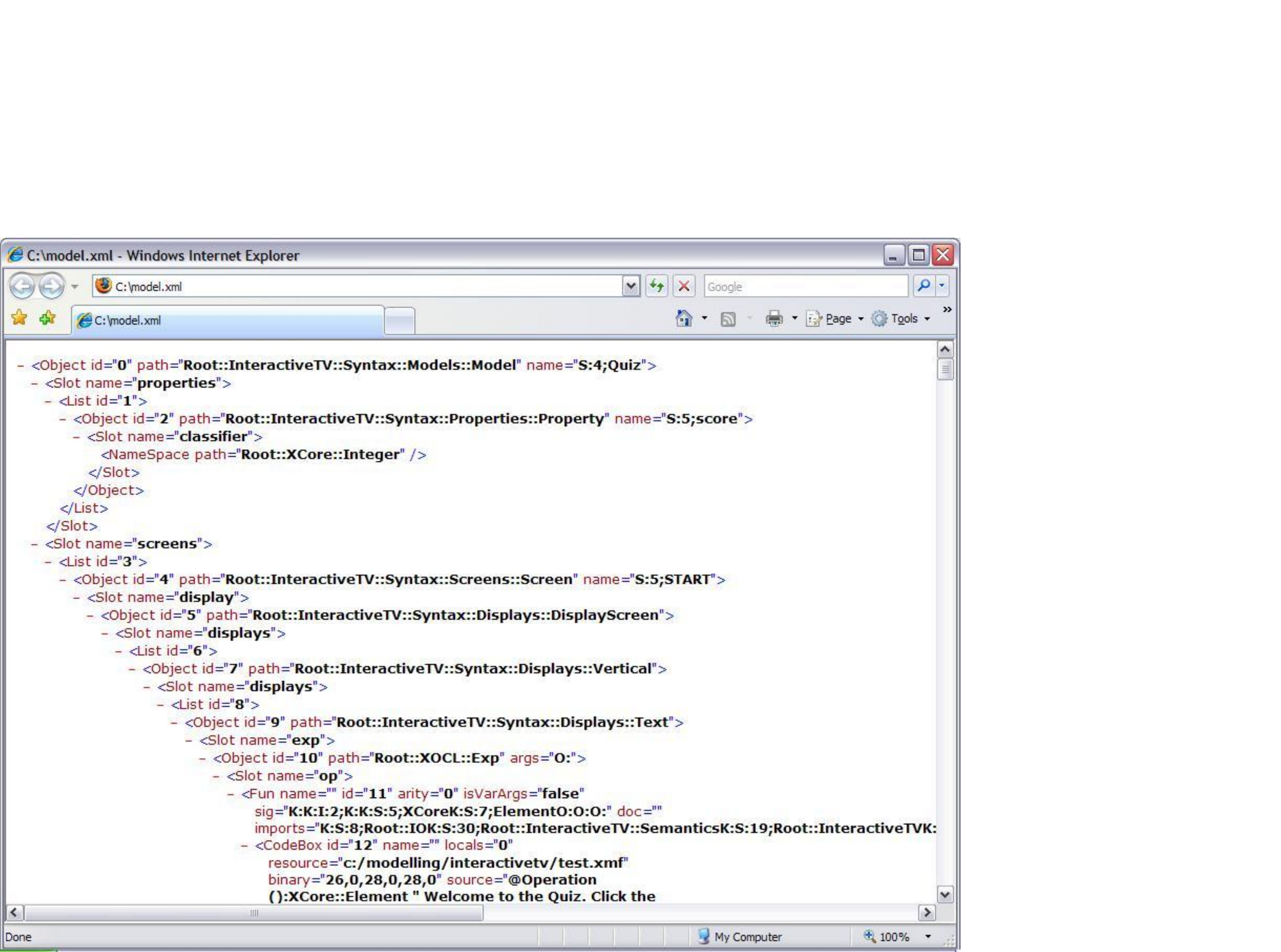}

\caption{A Serialized Model\label{fig:A-Serialized-Model}}
\end{center}
\end{figure}

The resulting output is produced in an XML file that is shown in figure
\ref{fig:A-Serialized-Model}. The XML markup shown in the figure
is the default vanilla-flavour markup provided by XMF. It is possible
to add adapters to an XML output channel that filter the data as it
is processed and thereby use domain-specific markup. So instead of
seeing Object and Slot as element tags, we might use Screen and Button.

The XML can be read back in using the following code:

\begin{lstlisting}
@WithOpenFile(fin <- ModelFile)
  let xin = XMLInputChannel(fin,NameSpaceXMLInflater())
  in xin.parse()
  end
end
\end{lstlisting}
\section{Conclusion\label{sec:Conclusion}}

This chapter has described an approach to modelling systems whereby
an engine for an executable domain-specific language (xDSL) is used
to develop and run the applications. The xDSL is designed to use concepts
that are suited to the application domain which allows the language
to abstract away from implementation details; the language is then
executable, can be used with different implementation platform technologies,
and is suitable for use by people whose primary skill lies in the
application domain rather than the implementation technology.

A method for developing xDSLs has been shown that involves a separation
of concerns between syntax elements that describe type-level features
of a model, and semantics elements that define run-time features of
an application. Experience has shown that this separation is natural
and allows the xDSL developer to effectively manage the process.

We have shown how a textual syntax can be added to an xDSL. In practice,
most xDSLs will require a concrete syntax. The precise form of this
will depend on the nature of the application and who the intended
users are. Sometimes, a purely graphical syntax is appropriate (for
example UML class-diagrams). Other times a purely textual syntax works
best, especially where executable features are involved and when complexity
can be controlled effectively using textual nesting. Often there is
scope for a mixture of the two where different aspects of the models
are constructed using graphical or textual syntax.

Modelling all features of a language has a number of benefits that
arise because everything is \emph{reified}. Reification involves representing
everything as data as opposed to program code or transient non-tangible
artifacts such as system events. Once everything is reified, many
types of analysis are possible including well-formedness checking,
type checking, application of style rules. It becomes easy to capture
and apply patterns and to perform refactoring. All features of an
application can be transformed to a target technology. 

Modelling actions is particularly important here; often actions are
left as unprocessed strings of program code which makes it very difficult
to analyze and run as part of an xDSL engine. The application given
in this paper has shown that it is straightforward to model actions
and to integrate them into the execution engine for an xDSL. By following
a few basic guidelines in terms of variable scope and control flow,
actions are easy to implement and are completely integrated into the
xDSL, its analysis and transformation.

The approach models the xDSL and executes it directly using an engine
(in this case XMF). This is attractive because it provides a high
degree of control over the language. It should be contrasted with
a translational approach to implementing a DSL whereby the model is
translated to the source code of a target language (such as Java or
C++) for which there is an implementation platform. This is an approach
taken by Swul \cite{SWUL} and GMF \cite{gmf} for GUI applications.
Translational approaches have some advantages: notably open architectures;
efficiency; arbitrary extensibility. However, there are some significant
disadvantages relating to the complexity of the generated code including
maintainability and understandability. It should be noted that an
xDSL engine-based approach does not preclude a translational approach.

We have given a complete implementation of an interactive application
xDSL using the features of XMF. XMF is an engine that is specifically
designed to support this kind of application development. It has very
high-level language features that support modelling concepts, it is
executable, and is designed to support textual language extension
through the use of extensible grammars. XMF directly supports textual
xDSLs and provides native interfaces to Java and EMF/GMF for use with
other concrete syntaxes.

XMF may be used to develop an xDSL and then deploy the language as
a stand-alone engine as shown in figure \ref{fig:Application-Architecture}.
XMF runs its own virtual machine and has a number of interface features
that allow it to connect to external applications. XMF may also be
used to develop an executable design of an application which is then
exported on to another implementation platform.
\chapter{Case Study 5: Graphs}

\section{Introduction}

Unfortunately, information is not always convniently organised in
a tree structure. A tree structure does not make allowances for relationships
that span the tree or where cycles occur in the data. For example,
what happens when a company employee fills two roles within the company
in different departments? It would be approprate for the employee
to occur underneath both departments in the tree; the employee is
\textit{shared} between the departments or equivalently there are
two different \textit{paths} from the root of the tree to the employee.

Trees do not represent sharing and multiple paths very well. There
are strategies; for example, XML is a tree structured data format
where labels are used to mark up the elements in order to represent
sharing. When data representation requires sharing, it is usually
because the data naturally forms a \textit{graph}. Graphs can be encoded
in trees (and other forms of data representation), but if the data
requires a graph then it is probably best to use one and be done with
it.

\begin{figure}
\begin{center}
\includegraphics[scale=0.75]{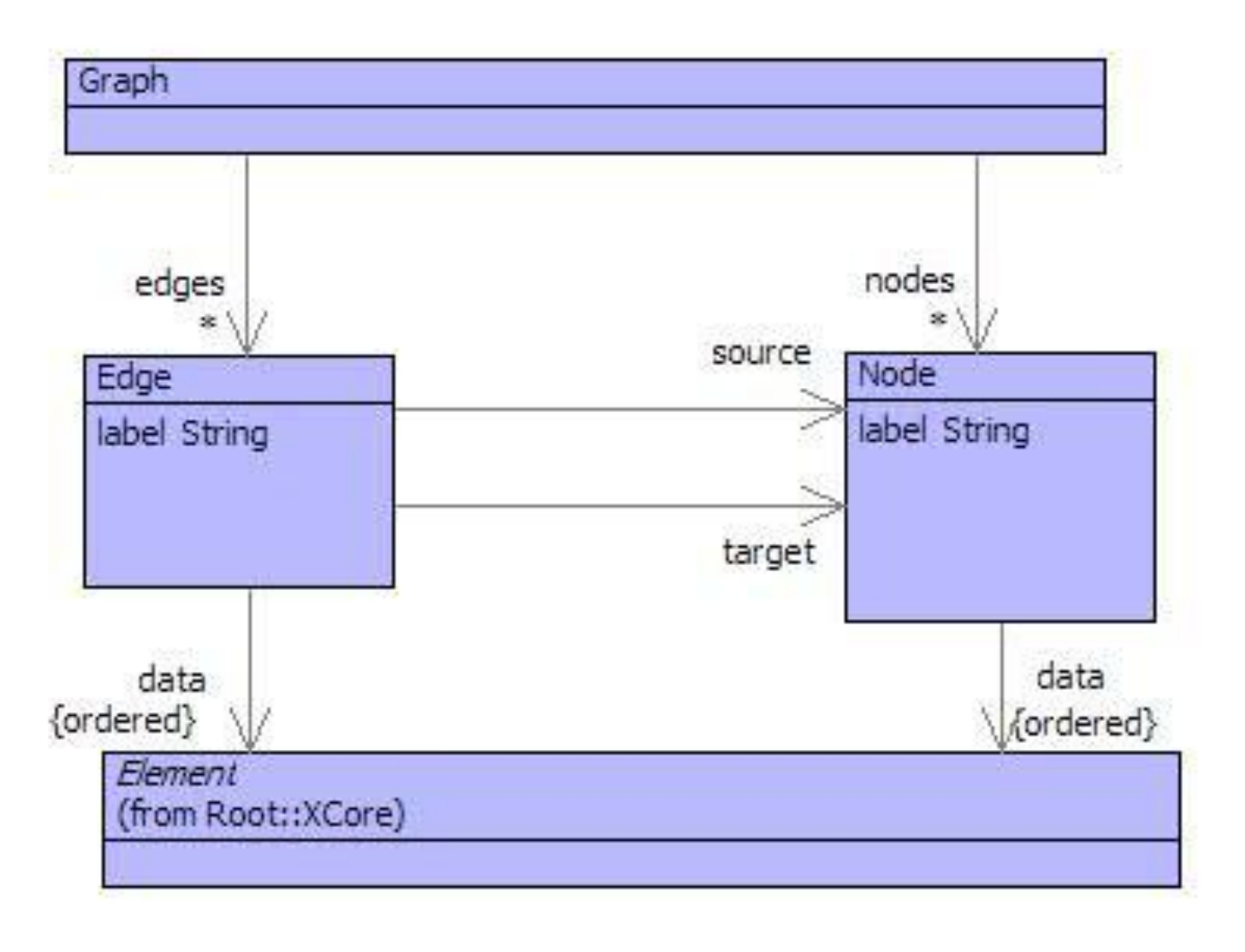}

\caption{Graphs\label{fig:Graphs}}
\end{center}
\end{figure}

\section{A Model of Graphs}

Figure \ref{fig:Graphs} shows a model of graphs. A graph consists
of a set of nodes and a set of edges. Each node has a label and a
sequence of data. The label is used to identify the node and the data
is used by applications for whatever purpose they require.

Each edge has a label and data, and also has a source and target node.
Edges go between nodes and are directed. The diagram shown in figure
\ref{fig:Graphs} is itself an example of a graph where the nodes
are displayed as class boxes and the edges are shown as attribute
edges. Notice that the node labelled Element is shared (parent-wise)
between Edge and Node; equivalently there are two paths from the rot
node (labelled Graph) to the node labelled Element: Graph, Edge, Element
and Graph, Node, Element. Such sharing is difficult to represent using
trees.

Graphs are a very rich form of data representation. There is a wealth
of material written about graphs and how to process them. Here are
some useful operations on graphs defined in XOCL:

\begin{lstlisting}
context Graph
  @Operation nodeFor(label:String):Node
    @Find(node,nodes)
      when node.label() = label
      else null
    end
  end
\end{lstlisting}The operation nodeFor returns the node with the supplied label, or
null if no such node exists. The operation edgesFrom returns the set
of edges from a given node:

\begin{lstlisting}
context Graph
  @Operation edgesFrom(n:Node):Set(Edge)
    edges->select(e | e.source() = n)
  end
\end{lstlisting}Graphs are generally useful and therefore it is appropriate to have
a general purpose language construct to define a graph. As mentioned
above, each use of a graph structure will attach different amounts
of data to the nodes and labels. The data is used to support the application
specific processing of the graph. Therefore, a general purpose language
construct for graph construction should support:

\begin{enumerate}
\item Arbitrary node and edge data.
\item Plug-points for the sub-classes of Graph, Node and Edge that are used
to represent the graph.
\end{enumerate}

\section{Graph Applications}

Here are two examples of different graph applications:

\begin{lstlisting}
@Graph(Routes,Location,Road)
  London()
    M1(200) -> Leeds
    A1(50)  -> Cambridge
end

@Graph(Plan,Task,Dependency)
  Start("January")
    -> Contractors
    -> Plans
  Contractors("March")
  Plans("April")
end
\end{lstlisting}The first graph is represented as an instance of the class Routes
where the nodes and edges are instances of the classes Location and
Road. These classes are sub-classes of Graph, Node and Edge repsectively.
Locations have no data; the three locations have labels London, Leeds
and Cambridge. 

An edge is listed below the source node. In the first example graph,
there are two edges with labels M1 and A1. The edges have data 100
and 50 (being the distance in miles) and the label of the edge target
is givn after ->.

The second example is a plan graph. The nodes have data that represents
the month at which the task is completed. Edges have no labels or
data (they just represent dependencies).

The proposed structure for a graph definition has plug-points for
the graph, node and edge classes and a body consisting of a sequence
of node definitions. A node definition n is a node label, followed
by node data in parentheses followed by a series of edge definitions
for which the source of the node is n. An edge definition is an optional
edge label, optional edge data in parentheses, an arrow and then the
label of the target node. Here is an example:

\begin{lstlisting}
@Graph(G,N,E)
  n1(a,b,c)
    e1(d) -> n2
    e2() -> n3
  n2()
    e3() -> n1
  n3()
\end{lstlisting}When the parser encounters a graph definition it will synthesize program
code that, when evaluated, produces the required graph. Are there
any rules that need to be observed when this synthesis takes place?
Given the model of graphs in figure \ref{fig:Graphs}, a graph contains
nodes and edges, and edges link nodes. Here is a possible program
that produces the graph above\begin{lstlisting}
(1) let g = G()
(2) in g.addToNodes(N("n1",Seq{a,b,c}));
       g.addToNodes(N("n2"));
(3)    g.addToNodes(N("n3"));
(4)    g.addToEdges(E("e1",Seq{d},g.nodeFor("n1"),g.nodeFor("n2")));
       g.addToEdges(E("e2",g.nodeFor("n1"),g.nodeFor("n3")));
(5)    g.addToEdges(E("e3",g.nodeFor("n2"),g.nodeFor("n1")));
(6)    g
    end
\end{lstlisting}Line (1) creates the graph using the supplied class G (a sub-class
of Graph). Each node must be added first in lines (2-3) so that edges
can then be created \textit{between} the nodes in lines (4-5). Note
that the supplied classes N and E are used to create the nodes and
edges. Finally the graph is returned in line (6).

The rules for graph construction are: create the graph, add the nodes
and then add the edges. Unfortunately, the graph definition construct
does not follow this pattern; it interleaves node and edge definitions.
A strategy is required to untangle this interleaving.

One way to address the interleaving is to have the parser synthesize
an intermediate graph definition that is processes using two or more
passes. This is perfectly respectable, and often a sensible way forward
when the required processing is fairly complex. 

In this case, the processing is not that complex, so another strategy
is used. To see how this works, a few definitions are required. An
\textit{edge constructor} expects a graph and an edge class; it adds
some edges to the supplied graph. A \textit{node constructor} expects
a graph, a node class, an edge class and a collection of edge constructors;
it adds some nodes to the supplied graph and then uses the edge constructor
to add some edges.

Node definitions are synthesized into node constructors and edge definitions
into edge constructors. The trick is to build up the edge constuctors
so that they are performed after all the node constructors. Since
the edge constructors are supplied to the node constructors, this
should be easy. Using the running example from above:

\begin{lstlisting}
n3 =
  @Operation(nodeConstructor)
    @Operation(g,N,E,edgeConstructor)
      g.addToNodes(N("n3"));
      nodeConstructor(g,N,E,edgeConstructor)
    end
  end
\end{lstlisting}The node definition for n3 is transformed into an operation that is
supplied with a node constructor and returns a node constructor. This
construction allows n3 to be linked with other noe constuctors without
knowing any details -- i.e. n3 can be defined \textit{in isolation}. 

The node constructor for n2 is similar, but involves the addition
of an edge constructor:

\begin{lstlisting}
n2 =
  @Operation(nodeConstructor)
    @Operation(g,N,E,edgeConstructor)
      let e3 = 
        @Operation(g,E)
          g.addToEdges(E("e3",g.nodeFor("n2"),g.nodeFor("n1")))
        end
      in g.addToNodes(N("n2"));
         nodeConstructor(g,N,E,addEdges(edgeConstructor,e3))
      end
    end
  end
\end{lstlisting}Note how the edge constructor for e3 is added to the supplied edge
contructor (using the yet-to-be-defined addEdges) when the supplid
node constructor is activated. This is the key to deferring the construction
of edges until all the nodes have been defined.

What should addEdges do? It is used to link all the edge constructors
together so that they all get activated. It takes two edge constructors
and returns an edge constructor:

\begin{lstlisting}
@Operation addEdges(ec1,ec2)
  @Operation(g,E)
    ec1(g,E);
    ec2(g,E)
  end
end
\end{lstlisting}The noe constructor for n1 is similar, but two edge constructors are
required:

\begin{lstlisting}
n1 =
  @Operation(nodeConstructor)
    @Operation(g,N,E,edgeConstructor)
      let e1 = 
        @Operation(g,E)
          g.addToEdges(E("e1",g.nodeFor("n1")Seq{d},g.nodeFor("n2")))
        end;
          e2 = 
        @Operation(g,E)
          g.addToEdges(E("e1",g.nodeFor("n1"),g.nodeFor("n3")))
        end then
          edges = addEdges(e1,e2)
      in g.addToNodes(N("n2"));
         nodeConstructor(g,N,E,addEdges(edgeConstructor,edges))
      end
    end
  end
\end{lstlisting}The complete graph can now be defined by linking the node constructors
together and supplying a graph:

\begin{lstlisting}
let nc = addNodes(n1,addNodes(n2,addNodes(n3,noNodes)))
in nc(G(),N,E,@Operation(g,E) g end)
end
\end{lstlisting}Each of the node constructors are linked via an operation addNodes.
The left-hand argument of addNodes is an operation that maps a node
constructor to a node constructor. The right-hand argument is a node
constructor. It is easier to see how this works from the definition:

\begin{lstlisting}
@Operation addNodes(nodeCnstrCnstr,nodeCnstr2)
  @Operation(g,N,E,edgeConstructor)
    let nodeCnstr1 = nodeCnstrCnstr(nodeCnstr2)
    in nodeCnstr1(g,N,E,edgeConstructor)
    end
  end
end
\end{lstlisting}The mechanism used by addNodes is an example of a typical pattern
that threads sequences of operations together. It allows the node
constuctor encoded in nodeCnstrCnstr to occur before that encoded
in nodeCnstr2 while also allowing the edge constructors produced by
the first to be handed on to the second (because they are to be deferrred
until all the nodes are added to the graph).

There are two types of constructor, each of which can occur repeatedly
in a sequence: nodes and edges. When this occurs, it is usual to have
some way to encode an empty sequence; in this case there are noNodes
and noEdges. Both of these are constructors:

\begin{lstlisting}
@Operation noNodes(g,N,E,edgeConstuctor)
  edgeConstructor(g,E)
end

@Operation noEdges(g,E)
  null
end
\end{lstlisting}
noNodes is a node constructor that starts edge construction. Therefore,
noNodes should be the right-most node constuctor in a sequence that
is combined using addNodes. noEdges does nothing, and can occur anywhere
in a sequence.

The grammar for graph definition synthesizes node and edge constructors
combined usin addNodes and addEdge. When an empty sequence is encountered,
the gramar synthesizes noNodes and noEdges respectively. The grammar
is defined below:

\begin{lstlisting}
  @Grammar extends OCL::OCL.grammar
    Data ::= '(' es = CommaSepExps ')' { SetExp("Seq",es) }.
    Edges(s) ::= e = Edge^(s) es = Edges^(s) 
      { [| addEdges(<e>,<es>) |] } 
    | { [| noEdges |] }.
    Edge(s) ::= l = Label d = OptData '->' t = Label { [| 
      @Operation(g,E)
        g.addToEdges(E(<l>,<d>,g.nodeFor(<s>),g.nodeFor(<t>)))
      end 
    |] }.
    Graph ::= '(' mkGraph = Exp ',' mkNode = Exp ',' mkEdge = Exp')' 
      GraphBody^(mkGraph,mkNode,mkEdge).
    GraphBody(mkGraph,mkNode,mkEdge) ::= ns = Nodes 'end' { [| 
      <ns>(<mkGraph>(),<mkNode>,<mkEdge>,@Operation(g,E) g end) 
    |] }.
    Label ::= NameExp | { "".lift() }.
    NameExp ::= n = Name { n.lift() }.
    Nodes ::= n = Node ns = Nodes 
      { [| addNodes(<n>,<ns>) |] } 
    | { [| noNodes |] }.
    Node ::= l = Label d = Data e = Edges^(l) { [|
      @Operation(Cn)
        @Operation(g,N,E,Ce)
          g.addToNodes(N(<l>,<d>));
          Cn(g,N,E,addEdges(<e>,Ce))
        end
      end
    |] }.
    OptData ::= Data | { [| Seq{} |] }.
  end
\end{lstlisting}

\backmatter
\bibliographystyle{alpha}
\bibliography{bib_willans,bib_sammut,bib_tony}
\end{document}